\documentclass[11pt]{article}

\usepackage{amsmath, amsthm}
\usepackage{amsmath, amsfonts}
\usepackage{amsmath, amssymb}
\usepackage{amsmath}
\usepackage{graphics}
\usepackage{graphicx}
\usepackage{color}
\usepackage{hyperref}
\usepackage[all]{xy}

\newtheorem{theorem}{Theorem}[subsection]
\newtheorem{definition}[theorem]{Definition}
\newtheorem{definition-lemma}[theorem]{Definition/Lemma}
\newtheorem{definition-explanation}[theorem]{Definition/Explanation}
\newtheorem{explanation-definition}[theorem]{Explanation/Definition}
\newtheorem{definition-fact}[theorem]{Definition/Fact}
\newtheorem{definition-notation}[theorem]{Definition/Notation}
\newtheorem{definition-conjecture}[theorem]{Definition/Conjecture}
\newtheorem{lemma}[theorem]{Lemma}
\newtheorem{lemma-definition}[theorem]{Lemma/Definition}
\newtheorem{proposition}[theorem]{Proposition}
\newtheorem{corollary}[theorem]{Corollary}
\newtheorem{remark}[theorem]{\it Remark}
\newtheorem{remark-notation}[theorem]{\it Remark/Notation}

\newtheorem{application-lemma}[theorem]{Application/Lemma}

\newtheorem{example}[theorem]{Example}
\newtheorem{example-definition}[theorem]{Example/Definition}

\newtheorem{definition-prototype}[theorem]{Definition-Prototype}

\numberwithin{equation}{subsection}

\newtheorem{stheorem}{Theorem}[section]
\newtheorem{sdefinition}[stheorem]{Definition}
\newtheorem{sdefinition-lemma}[stheorem]{Definition/Lemma}
\newtheorem{sdefinition-explanation}[stheorem]{Definition/Explanation}
\newtheorem{sexplanation-definition}[stheorem]{Explanation/Definition}
\newtheorem{sdefinition-fact}[stheorem]{Definition/Fact}
\newtheorem{sdefinition-notation}[stheorem]{Definition/Notation}
\newtheorem{sdefinition-conjecture}[stheorem]{Definition/Conjecture}
\newtheorem{slemma}[stheorem]{Lemma}
\newtheorem{slemma-definition}[stheorem]{Lemma/Definition}

\newtheorem{sremark}[stheorem]{\it Remark}
\newtheorem{sremark-notation}[stheorem]{\it Remark/Notation}

\newtheorem{sapplication-lemma}[stheorem]{Application/Lemma}

\newtheorem{sexample}[stheorem]{Example}
\newtheorem{sexample-definition}[stheorem]{Example/Definition}

\newtheorem{sdefinition-prototype}[stheorem]{Definition-Prototype}

\newtheorem{sstheorem}{Theorem}[subsubsection]
\newtheorem{ssdefinition}[sstheorem]{Definition}
\newtheorem{ssdefinition-lemma}[sstheorem]{Definition/Lemma}
\newtheorem{ssdefinition-explanation}[sstheorem]{Definition/Explanation}
\newtheorem{ssexplanation-definition}[sstheorem]{Explanation/Definition}
\newtheorem{ssdefinition-fact}[sstheorem]{Definition/Fact}
\newtheorem{ssdefinition-notation}[sstheorem]{Definition/Notation}
\newtheorem{ssdefinition-conjecture}[sstheorem]{Definition/Conjecture}
\newtheorem{sslemma}[sstheorem]{Lemma}
\newtheorem{sslemma-definition}[sstheorem]{Lemma/Definition}
\newtheorem{ssproposition}[sstheorem]{Proposition}

\newtheorem{ssremark-notation}[sstheorem]{\it Remark/Notation}

\newtheorem{ssapplication-lemma}[sstheorem]{Application/Lemma}

\newtheorem{ssexample-definition}[sstheorem]{Example/Definition}

\newtheorem{ssdefinition-prototype}[sstheorem]{Definition-Prototype}

\newcommand{\Aut}{\mbox{\it Aut}\,}

\newcommand{\Comm}{\mbox{\it Comm}\,}

\newcommand{\CSWZ}{\mbox{\scriptsize\it CS/WZ}\,}
 \newcommand{\tinyCSWZ}{\mbox{\tiny\it CS/WZ}\,}

 \newcommand{\tinyDBI}{\mbox{\it\tiny DBI}\,}

\newcommand{\Der}{\mbox{\it Der}\,}

\newcommand{\End}{\mbox{\it End}\,}

\newcommand{\Endsheaf}{\mbox{\it ${\cal E}\!$nd}\,}

\newcommand{\GL}{\mbox{\it GL}}

\newcommand{\Id}{\mbox{\it Id}\,}

\newcommand{\Image}{\mbox{\it Im}\,}

\newcommand{\Inn}{\mbox{\it Inn}\,}

\newcommand{\Kahlerscriptsize}{\mbox{\scriptsize\it K\"{a}hler}}

\newcommand{\ModCategory}{\mbox{\it ${\cal M}$\!od}\,}

\newcommand{\nAGSM}{\mbox{\scriptsize\it nAGSM}}

\newcommand{\tinyNG}{\mbox{\it\tiny NG}\,}

\newcommand{\tinyPolyakov}{\mbox{\it\tiny Polyakov}\,}

\newcommand{\Real}{\mbox{\it Re}\,}

\newcommand{\Space}{\mbox{\it Space}\,}

\newcommand{\Span}{\mbox{\it Span}\,}
\newcommand{\Spec}{\mbox{\it Spec}\,}
 \newcommand{\boldSpec}{\mbox{\it\bf Spec}\,}

\newcommand{\Supp}{\mbox{\it Supp}\,}

 \newcommand{\scriptsizeSym}{\mbox{\scriptsize\it Sym}}
 
\newcommand{\SymDet}{\mbox{\it SymDet}\,}

\newcommand{\Tor}{\mbox{\it Tor}\,}
\newcommand{\Tr}{\mbox{\it Tr}\,}

\newcommand{\YMscriptsize}{\mbox{\scriptsize\it YM}}
 \newcommand{\YMtiny}{\mbox{\tiny\it YM}}

\newcommand{\ad}{\mbox{\it ad$_{\,}$}}

\newcommand{\determinant}{\mbox{\it det}\,}
\newcommand{\dilatonscriptsize}{\mbox{\scriptsize\it dilaton}}
 \newcommand{\dilatontiny}{\mbox{\tiny\it dilaton}}
\newcommand{\dimm}{\mbox{\it dim}\,}

\newcommand{\divv}{\mbox{\it div}\,}

\newcommand{\gaugescriptsize}{\mbox{\scriptsize\it gauge}\,}
 \newcommand{\gaugetiny}{\mbox{\tiny\it gauge}\,}

\newcommand{\kineticscriptsize}{\mbox{\scriptsize\it kinetic}}

\newcommand{\mapscriptsize}{\mbox{\scriptsize\it map}}

\newcommand{\nc}{\mbox{\scriptsize\it nc}}

\newcommand{\pr}{\mbox{\it pr}}

\newcommand{\redscriptsize}{\mbox{\scriptsize\rm red}\,}

\newcommand{\standardscriptsize}{\mbox{\scriptsize\it standard}\,}

\newcommand{\vol}{\mbox{\it vol}\,}

\newcommand{\boldd}{\mbox{\boldmath $d$}}
  \newcommand{\scriptsizeboldd}{\mbox{\scriptsize\boldmath $d$}}
  \newcommand{\tinyboldd}{\mbox{\tiny\boldmath $d$}}

\newcommand{\boldx}{\mbox{\boldmath $x$}}
  
\newcommand{\boldy}{\mbox{\boldmath $y$}}
  \newcommand{\scriptsizeboldy}{\mbox{\scriptsize\boldmath $y$}}
\newcommand{\boldz}{\mbox{\boldmath $z$}}
  \newcommand{\scriptsizeboldz}{\mbox{\scriptsize\boldmath $z$}}

\newcommand{\longrightaarrow}{\longrightarrow\hspace{-3ex}\longrightarrow}

\newcommand{\odotwedge}{\stackrel{\mbox{\tiny $\odot$}}{\wedge}}

\begin{document}

\enlargethispage{24cm}

\begin{titlepage}

$ $

\vspace{-1.5cm} 

\noindent\hspace{-1cm}
\parbox{6cm}{\small February 2017}\
   \hspace{7cm}\
   \parbox[t]{6cm}{arXiv:yymm.nnnnn [hep-th] \\
                D(13.3): standard  
				}

\vspace{2cm}

\centerline{\large\bf
 Dynamics of D-branes$\;$ II. The standard action}
\vspace{1ex}   
\centerline{\bf 
 --- an analogue of the Polyakov action for (fundamental, stacked) D-branes
}

\bigskip

\vspace{3em}

\centerline{\large
  Chien-Hao Liu   
            \hspace{1ex} and \hspace{1ex}
  Shing-Tung Yau
}

\vspace{4em}

\begin{quotation}
\centerline{\bf Abstract}

\vspace{0.3cm}

\baselineskip 12pt  
{\small
 We introduce a new action $S_{standard}^{(\rho,h; \Phi,g,B,C)}$ for D-branes that is to D-branes 
  as the Polyakov action is to fundamental strings. 
 This `standard action' is abstractly a non-Abelian gauged sigma model 
  --- based on maps $\varphi: (X^{\!A\!z},E;\nabla)\rightarrow Y$ 
  from an Azumaya/matrix manifold $X^{\!A\!z}$ with a fundamental module $E$ with a connection $\nabla$ to $Y$ --- 
  enhanced by the dilaton term, the gauge-theory term, 
  and the Chern-Simons/Wess-Zumino term that couples $(\varphi,\nabla)$ to Ramond-Ramond field. 
  In a special situation, this new theory merges the theory of harmonic maps and a gauge theory, 
   with a nilpotent type fuzzy extension. 
 With the analysis developed in D(13.1) (arXiv:1606.08529 [hep-th]) for such maps and
   an improved understanding of the hierarchy of various admissible conditions on the pairs $(\varphi,\nabla)$ 
   beyond D(13.2.1) (arXiv:1611.09439 [hep-th]) and 
   how they resolve the built-in obstruction to pull-push of covariant tensors 
   under a map from a noncommutative manifold to a commutative manifold, 
 we develop further in this note some covariant differential calculus needed and 
  apply them to work out 
   the first variation --- and hence the corresponding equations of motion for D-branes --- of the standard action   and 
   the second variation of the kinetic term for maps and the dilaton term in this action. 
 Compared with the non-Abelian Dirac-Born-Infeld action constructed in D(13.1) along the same line, 
  the current note brings the Nambu-Goto-string-to-Polyakov-string analogue to D-branes. 
 The current bosonic setting is the first step toward the dynamics of fermionic D-branes
   (cf. D(11.2): arXiv:1412.0771 [hep-th]) and their quantization 
  as fundamental dynamical objects,
  in parallel to what happened to the theory of fundamental strings during years 1976--1981.
 } 
\end{quotation}

\vspace{8em}

\baselineskip 12pt
{\footnotesize
\noindent
{\bf Key words:} \parbox[t]{14cm}{D-brane; admissible condition;
     standard action, enhanced non-Abelian gauged sigma model;\\
     Azumaya manifold, $C^{\infty}$-scheme, harmonic map; first and second variation, equations of motion
 }} 

 \bigskip

\noindent {\small MSC number 2010:  81T30, 35J20; 16S50, 14A22, 35R01
} 

\bigskip

\baselineskip 10pt
{\scriptsize
\noindent{\bf Acknowledgements.}
We thank
 Andrew Strominger, Cumrun Vafa
   for influence to our understanding of strings, branes, and gravity.
C.-H.L.\ thanks in addition
 Pei-Ming Ho for a discussion on Ramond-Ramond fields and literature guide and Chenglong Yu for a discussion
   on admissible conditions;
 Artan Sheshmani, Brooke Ullery,  Ashvin Vishwanath for special/topic/basic courses, spring 2017;
 Ling-Miao Chou
   for comments that improve the illustrations and moral support.
The project is supported by NSF grants DMS-9803347 and DMS-0074329.
} 

\end{titlepage}

\newpage

\enlargethispage{24cm}
\begin{titlepage}

$ $

\vspace{2em}

\centerline{\small\it
 Chien-Hao Liu dedicates this work to}
\centerline{\small\it Noel Brady, Hung-Wen Chang, Chongsun Chu, William Grosso, Pei-Ming Ho, Inkang Kim,}
\centerline{\small\it who enriched his years at U.C.\ Berkeley tremendously.$^{\ast}$}

\vspace{5em}

\baselineskip 11pt

 \noindent
 $^{\ast}${\scriptsize 
 (From C.H.L.)\hspace{1em}
It was an amazing time when I landed at Berkeley in the 1990s,
 following Thurston's transition to Mathematical Sciences Research Institute (M.S.R.I.).
On the mathematical side,
  representing figures on geometry and topology ---
   $2$- and $3 $-dimensional geometry and topology
    and related dynamical system (Andrew Casson, Curtis McMullen, William Thurston),	
   $4$-dimensional geometry and topology (Robion Kirby),
   $5$-and-above dimensional topology (Morris Hirsch, Stephen Smale),
   algebraic geometry (Robin Hartshorne),
   differential and complex geometry (Wu-Yi Hsiang, Shoshichi Kobayashi, Hong-Shi Wu),
   symplectic geometry and geometric quantization (Alan Weinstein),
   combinatorial and geometric group theory
   (relevant to $3$-manifold study via the fundamental groups, John Stallings)    ---
 seem to converge at Berkeley.
 On the physics side,
  several first-or-second generation string-theorists
   (Orlando Alvarez, Korkut Bardakci, Martin Halpern)
  and one of the creators of supersymmetry (Bruno Zumino)
  were there.
 It was also a time when enumerative geometry and topology motivated by quantum field and string theory
  started to emerge
  and for that there were quantum invariants of $3$-manifolds (Nicolai Reshetikhin)
  and mirror symmetry (Alexandre Givental) at Berkeley.
 Through the topic courses almost all of them gave during these years on their related subject
    and the timed homeworks one at least attempted,
   one may acquire a very broad foundation toward a cross field between mathematics and physics
   if one is ambitious and diligent enough.

 However,
  despite such an amazing time and an intellectually enriching environment,
  it would be extremely difficult, if not impossible, to learn at least some good part of them without a friend
  --- well, of course, unless one is a genius, which I am painfully not.
 For that, on the mathematics side,
    I thank
	  Hung-Wen (Weinstein and Givental's student)
	   who said hello to me in our first encounter at the elevator at the 9th floor in Evans Hall
	    and influenced my understanding of many topics ---
		 particularly symplectic geometry, quantum mechanics and gauge theory ---
		due to his unconventional background from physics, electrical engineering, to mathematics; 
      Bill (Stallings' student) and Inkang (Casson's student)
	   for suggesting me a weekly group meeting on Thurston's lecture notes at Princeton on $3$-manifolds
	   \begin{itemize}
	    \item[$\mbox{[Th]}$] \parbox[t]{49em}{W.P.\ Thurston,
		   {\sl The geometry and topology of three-manifolds},
		   typed manuscript, Department of Mathematics, Princeton University, 1979.}
	   \end{itemize}
        which lasted for one and a half year	and tremendously influenced the depth of my understanding
		of Thurston's work;
     Noel (Stallings' student)
	   for a reading seminar on the very thought-provoking French book
	   \begin{itemize}
	    \item[$\mbox{[Gr]}$] \parbox[t]{49em}{M.\ Gromov, 		
	   {\sl Structures m\'{e}trique pour les vari\'{e}t\'{e}s riemanniennes},
		r\'{e}dig\'{e} par J.\ Lafontaine et P.\ Pansu,
        Text Math.\ 1, Cedic/Fernand-Nathan, Paris, 1980.}		
	   \end{itemize}
	   for a summer.
 The numerous other after-class discussions for classes some of them and I happened to sit in
   shaped a large part of my knowledge pool, even for today.
 Incidentally,
  thanks to Prof.\ Kirby, who served as the Chair for the Graduate Students in the Department
  at that time.
  I remember his remark to his staff when I arrived at Berkeley and reported to him after meeting Prof.\ Thurston:
  ``We have to treat them [referring to all Thurston's then students] as nice as our own"

 On the physics side, thanks to 	
  Chongsun and Pei-Ming (both Zumino's students) 	
   for helping me understand the very challenging topic: quantum field theory,
    first when we all attended Prof.\ Bardakci's course Phys 230A, Quantum Field Theory,
	based closely on the book
	 \begin{itemize}
	  \item[$\mbox{[Ry]}$]	L.H.\ Ryder,
        {\sl Quantum field theory}, Cambridge Univ.\ Press, 1985.	
	 \end{itemize}
  and a second time a year later
      when we repeated it through the same course given by Prof.\ Alvarez,
      followed by a reading group meeting on Quantum Field Theory guided by Prof.\ Alvarez
	  --- another person who I forever have to thank and another event which changed the course of my life permanently.
 The former with Prof.\ Bardakci was
  a semester I had to spend at least three-to-four days a week just on this course:
    attending lectures, understanding the notes, reading the corresponding chapters or sections of the book,
	doing the homeworks, occasionally looking into literatures to figure out some of the homeworks, and
	correcting the mistakes I had made on the returned homework.
  I even turned in the take-home final.
  Amusingly, due to the free style of the Department of Mathematics at Princeton University
   and the visiting student status at U.C.\ Berkeley,
    that is the first of the only three courses and only two semesters throughout my graduate student years
	 for which I ever honestly did like a student:
	  do the homework, turn in to get graded, do the final, and in the end get a semester grade back.
 Special thanks to Prof.\ Bardakci and the TA for this course, Bogdan Morariu,
   for grading whatever I turned in, though I was not an officially registered student in that course.
 
 That was a time when I could study something purely for the beauty, mystery, and/or joy of it.
 That was a time when the future before me seemed unbounded.
 That was a time when I did not  think too much about the less pleasant side of doing research:
    competitions, publications, credits, $\cdots$.
 That was a time I was surrounded by friends, though only limitedly many,
    in all the best senses the word `friend' can carry.
 Alas, that wonderful time, with such a luxurious leisure, is gone forever!
} 

\end{titlepage}


\newpage
$ $

\vspace{-3em}

\centerline{\sc
 Dynamics of D-branes, II: The Standard Action
 } %

\vspace{2em}


\begin{flushleft}
{\Large\bf 0. Introduction and outline}
\end{flushleft}
In this sequel to
   D(11.1)    (arXiv:1406.0929 [math.DG]),
   D(11.3.1) (arXiv:1508.02347 [math.DG]),
   D(13.1)    (arXiv:1606.08529 [hep-th]) and
   D(13.2.1) (arXiv:1611.09439 [hep-th])    and
 along the line of our understanding of the basic structures on D-branes in Polchinski's TASI 1996 Lecture Notes
  from the aspect of Grothendieck's modern Algebraic Geometry
  initiated in D(1) (arXiv:0709.1515 [math.AG]),
 we introduce a new action $S_{standard}^{(\rho,h; \Phi,g,B,C)}$ for D-branes
  that is to D-branes as the (Brink-Di Vecchia-Howe/Deser-Zumino/)Polyakov action is to fundamental strings.
 This action depends
   both on the (dilaton field $\rho$, metric $h$) on the underlying topology $X$ of the D-brane world-volume
   and on the background (dilaton field $\Phi$, metric $g$, $B$-field $B$, Ramond-Ramond field $C$) 
     on the target space-time $Y$;
   and is naturally a non-Abelian gauged sigma model ---
     based on maps $\varphi: (X^{\!A\!z},E;\nabla)\rightarrow Y$ from an Azumaya/matrix manifold
	 $X^{\!A\!z}$ with a fundamental module $E$ with a connection $\nabla$ to $Y$ ---
    enhanced by
	  the dilaton term that couples $(\varphi,\nabla)$ to $(\rho, \Phi)$,
      the $B$-coupled gauge-theory term that couples $\nabla$ to $B$,   and
	  the Chern-Simons/Wess-Zumino term that couples $(\varphi,\nabla)$ to $(B,C)$
	 in our standard action $S_{standard}^{(\rho,h; \Phi,g,B,C)}$.
	
Before one can do so,
   one needs to resolve the built-in obstruction of pull-push of covariant tensors
    under a map from a noncommutative manifold to a commutative manifold.
 Such issue already appeared in the construction of the non-Abelian Dirac-Born-Infeld action  (D(13.1) ).
 In this note, we give a hierarchy of various admissible conditions on the pairs $(\varphi,\nabla)$
   that are enough to resolve the issue while being open-string compatible (Sec.$\,$2).
 This improves our understanding of admissible conditions beyond D(13.2.1).
With the noncommutative analysis developed in D(13.1),
  we develop further in this note some covariant differential calculus for such maps (Sec.$\,$3) and
  use it to define the standard action for D-branes (Sec.$\,$4).
After promoting the setting to a family version (Sec.$\,$5),
  we work out the first variation --- and hence the corresponding equations of motion for D-branes ---
   of the standard action (Sec.$\,$6)
  and the second variation of the kinetic term for maps and the dilaton term in this action (Sec.$\,$7).
  
Compared with the non-Abelian Dirac-Born-Infeld action constructed in D(13.1) along the same line, 
   the current standard action is clearly much more manageable.
Classically and mathematically and
   in the special case where
     the background $(\Phi, B, C)$ on $Y$ is set to vanish,
   this new theory is a merging of the theory of harmonic maps and a gauge theory
   (free to choose either a Yang-Mills theory or other kinds of applicable gauge theory)
   with a nilpotent type fuzzy extension.
The current bosonic setting is the first step toward fermionic D-branes
  (cf.\ D(11.2): arXiv:1412.0771 [hep-th]) and their quantization
   as fundamental dynamical objects,
   in  parallel to what happened for fundamental superstrings during 1976--1981;
(the road-map at the end: `{\sl Where we are}').

\bigskip
\bigskip

\noindent
{\bf Convention.}
 References for standard notations, terminology, operations and facts are\\
  (1) Azumaya/matrix  algebra: [Ar], [Az], [A-N-T]; \hspace{.6em}
  (2) sheaves and bundles: [H-L];$\;$ with connection: [Bl], [B-B], [D-K], [Ko];  \hspace{.6em}
  (3) algebraic geometry: [Ha];$\;$  $C^{\infty}$ algebraic geometry: [Jo];  \hspace{.6em}
  (4) differential geometry: [Eis], [G-H-L], [Hi], [H-E], [K-N];  \hspace{.6em}
  (5) noncommutative differential geometry: [GB-V-F];  \hspace{.6em}
  (6) string theory and D-branes: [G-S-W], [Po2], [Po3].
 \begin{itemize}
  \item[$\cdot$]
   For clarity, the {\it real line} as a real $1$-dimensional manifold is denoted by ${\Bbb R}^1$,
    while the {\it field of real numbers} is denoted by ${\Bbb R}$.
   Similarly, the {\it complex line} as a complex $1$-dimensional manifold is denoted by ${\Bbb C}^1$,
    while the {\it field of complex numbers} is denoted by ${\Bbb C}$.
	
  \item[$\cdot$]	
  The inclusion `${\Bbb R}\subset{\Bbb C}$' is referred to the {\it field extension
   of ${\Bbb R}$ to ${\Bbb C}$} by adding $\sqrt{-1}$, unless otherwise noted.

   
 \item[$\cdot$]
  All manifolds are paracompact, Hausdorff, and admitting a (locally finite) partition of unity.
  We adopt the {\it index convention for tensors} from differential geometry.
   In particular, the tuple coordinate functions on an $n$-manifold is denoted by, for example,
   $(y^1,\,\cdots\,y^n)$.
  However, no up-low index summation convention is used.
   
  
  \item[$\cdot$]
  For this note, `{\it differentiable}', `{\it smooth}', and $C^{\infty}$ are taken as synonyms.
  
  \item[$\cdot$]
   {\it matrix} $m$ vs.\ manifold of dimension $m$
   
  \item[$\cdot$]
   the Regge slope $\alpha^{\prime}$
   vs.\ dummy labelling index $\alpha$
   vs.\ covariant tensor $\alpha$

  \item[$\cdot$]
   {\it section} $s$ of a sheaf or vector bundle vs.\ dummy labelling index $s$

   
  \item[$\cdot$]
   {\it algebra} ${\cal A}_{\varphi}$ vs.\ {\it connection $1$-form} $A_{\mu}$

  \item[$\cdot$]
   {\it ring} $R$ vs.\ $k$-th {\it remainder} $R[k]$ vs.\ {\it Riemann curvature tensor} $R_{ijkl}$
   
  \item[$\cdot$]
   {\it boundary} $\partial U$ of an open set $U$ vs.\
     {\it partial differentiations} $\partial_t$, $\partial /\partial y^i$
   
  \item[$\cdot$]
   $\Spec R $ ($:=\{\mbox{prime ideals of $R$}\}$)
         of a commutative Noetherian ring $R$  in algebraic geometry\\
   vs.\ $\Spec R$ of a $C^k$-ring $R$
  ($:=\Spec^{\Bbb R}R :=\{\mbox{$C^k$-ring homomorphisms $R\rightarrow {\Bbb R}$}\}$)

  \item[$\cdot$]
  {\it morphism} between schemes in algebraic geometry
    vs.\ {\it $C^{\infty}$-map} between $C^{\infty}$-manifolds or $C^{\infty}$-schemes
         	in differential topology and geometry or $C^{\infty}$-algebraic geometry
			
  \item[$\cdot$]
   group {\it action} vs.\  {\it action} functional for D-branes

  \item[$\cdot$]
   {\it metric tensor} $g$ vs.\ element $g^{\prime}$   in a {\it group} $G$ vs.\
      gauge coupling constant $g_{\gaugescriptsize}$

  \item[$\cdot$]
   {\it sheaves} ${\cal F}$, ${\cal G}$ vs.\
    {\it curvature tensor} $F_{\nabla}$,
	{\it gauge-symmetry group} ${\cal G}_{\gaugescriptsize}$
  
  \item[$\cdot$]
   {\it dilaton field $\rho$} vs.\ {\it representation} $\rho_{\gaugescriptsize}$
     of a gauge-symmetry group ${\cal G}_{\gaugescriptsize}$
   			
  \item[$\cdot$]
   The `{\it support}' $\Supp({\cal F})$
    of a quasi-coherent sheaf ${\cal F}$ on a scheme $Y$ in algebraic geometry
     	or on a $C^k$-scheme in $C^k$-algebraic geometry
    means the {\it scheme-theoretical support} of ${\cal F}$
   unless otherwise noted;
    ${\cal I}_Z$ denotes the {\it ideal sheaf} of
    a (resp.\ $C^k$-)subscheme of $Z$ of a (resp.\ $C^k$-)scheme $Y$;
    $l({\cal F})$ denotes the {\it length} of a coherent sheaf ${\cal F}$ of dimension $0$.

  \item[$\cdot$]
   For a sheaf ${\cal F}$ on a topological space $X$,
   the notation `$s\in{\cal F}$' means a local section $s\in {\cal F}(U)$
      for some open set $U\subset X$. 

  \item[$\cdot$]	  
   For an ${\cal O}_X$-module ${\cal F}$, 
    the {\it fiber} of ${\cal F}$ at $x\in X$	 is denoted ${\cal F}|_x$ 
	while the {\it stalk} of ${\cal F}$ at $x\in X$ is denoted ${\cal F}_x$.
  
  \item[$\cdot$]
   {\it coordinate-function index}, e.g.\ $(y^1,\,\cdots\,,\, y^n)$ for a real manifold
      vs.\  the {\it exponent of a power},
	  e.g.\  $a_0y^r+a_1y^{r-1}+\,\cdots\,+a_{r-1}y+a_r\in {\Bbb R}[y]$.
	
 


 
  \item[$\cdot$]
   The current Note D(13.3) continues the study in
	  \begin{itemize}
	   \item[]  \hspace{-2em} [L-Y8]\hspace{1em}\parbox[t]{34em}{{\it
	    Dynamics of D-branes I.
          The non-Abelian Dirac-Born-Infeld action, its first variation, and the equations of motion for D-branes
		  --- with remarks on the non-Abelian Chern-Simons/Wess-Zumino term},
        arXiv:1606.08529 [hep-th]. (D(13.1))
		 }
	  \end{itemize}  	
   Notations and conventions follow ibidem when applicable.
 \end{itemize}

\newpage
   
\begin{flushleft}
{\bf Outline}
\end{flushleft}
\nopagebreak
{\small
\baselineskip 12pt  
\begin{itemize}
 \item[0.]
  Introduction.
   
 \item[1.]
 Azumaya/matrix manifolds with a fundamental module and differentiable maps therefrom
  \vspace{-.6ex}
  \begin{itemize}
   \item[\Large $\cdot$]
    Azumaya/matrix manifolds with a fundamental module $(X^{\!A\!z},\,{\cal E})$
   
   \item[\Large $\cdot$]
    When ${\cal E}$ is equipped with a connection $\nabla$
    
   \item[\Large $\cdot$]
    Differentiable maps from $(X^{\!A\!z},\,{\cal E})$

   \item[\Large $\cdot$]
    Compatibility between the map $\varphi$ and the connection $\nabla$
  \end{itemize}

 \item[2.]
 Pull-push of tensors and admissible conditions on $(\varphi,\nabla)$
   \vspace{-.6ex}
   \begin{itemize}
    \item[2.1]
	Admissible conditions on $(\varphi,\nabla)$ and the resolution of the pull-push issue

    \item[2.2]
	Admissible conditions from the aspect of open strings
   \end{itemize}

 \item[3.]
 The differential $d\varphi$ of $\varphi$ and its decomposition,
 the three basic ${\cal O}_X^{\,\Bbb C}$-modules, induced structures, and some covariant calculus
  \vspace{-.6ex}
  \begin{itemize}
   \item[3.1]
	The differential $d\varphi$ of $\varphi$ and its decomposition induced by $\nabla$
  
   \item[3.2]
   The three basic ${\cal O}_X^{\,\Bbb C}$-modules relevant to $D\varphi$, with induced structures
	\begin{itemize}
	 \item[3.2.1]
     The ${\cal O}_X^{A\!z}$-valued cotangent sheaf
      ${\cal T}^{\ast}X\otimes_{{\cal O}_X}{\cal O}_X^{A\!z}$ of $X$, and beyond
  
	 \item[3.2.2]
	 The pull-back tangent sheaf $\varphi^{\ast}{\cal T}_{\ast}Y$
	
    \item[3.2.3]	
     The ${\cal O}_X^{\,\Bbb C}$-module
	  ${\cal T}^{\ast}X\otimes_{{\cal O}_X}\varphi^{\ast}{\cal T}_{\ast}Y$,
	  where $D\varphi$ lives
    \end{itemize}	
  \end{itemize}
    
 \item[4.]
 The standard action for D-branes
   \vspace{-.6ex}
   \begin{itemize}
    \item[\Large $\cdot$]
     The gauge-symmetry group $C^{\infty}(\Aut_{\Bbb C}(E))$
%
	
    \item[\Large $\cdot$]
	 The standard action for D-branes

    \item[\Large $\cdot$]
     The standard action as an enhanced non-Abelian gauged sigma model
   \end{itemize}
  
 \item[5.]
  Admissible family of admissible pairs $(\varphi_T,\nabla^T)$
   \vspace{-.6ex}
   \begin{itemize}
    \item[\Large $\cdot$]
     Basic setup and the notion of admissible families of admissible pairs $(\varphi_T, \nabla^T)$
	
    \item[\Large $\cdot$]
	 Three basic ${\cal O}_{X_T}$-modules with induced structures

    \item[\Large $\cdot$]
     Curvature tensors with $\partial_t$ and other order-switching formulae
	
	\item[\Large $\cdot$]
	 Two-parameter admissible families of admissible pairs
   \end{itemize}
   
 \item[6.]
 The first variation of the enhanced kinetic term for maps and ......
  %
  \vspace{-.6ex}
  \begin{itemize}	 	
    \item[6.1]	
     The first variation of the kinetic term for maps
	
	\item[6.2]
	 The first variation of the dilaton term
	
	\item[6.3]
	 The first variation of the gauge/Yang-Mills term and the Chern-Simons/Wess-Zumino term	
	 %
	  \begin{itemize}
	   \item[6.3.1]
	    The first variation of the gauge/Yang-Mills term
	
	   \item[6.3.2]
		The first variation of the Chern-Simons/Wess-Zumino term for lower-dimensional\\ D-branes
		\begin{itemize}
		  \item[6.3.2.1]
            D$(-1)$-brane world-point $(m=0)$

          \item[6.3.2.2]
            D-particle world-line $(m=1)$

          \item[6.3.2.3]
	        D-string world-sheet $(m=2)$

          \item[6.3.2.4]
	        D-membrane world-volume $(m=3)$			
		\end{itemize}
	  \end{itemize}
  \end{itemize}

 \item[7.]
 The second variation of the enhanced kinetic term for maps
  \vspace{-.6ex}
  \begin{itemize}
	\item[6.1]
	 The second variation of the kinetic term for maps
	
    \item[6.2]	
     The second variation of the dilaton term
  \end{itemize}

 \item[\LARGE $\cdot$]
  Where we are
 %
 %
 %
 %
 %
 %
 %
 
\end{itemize}
} 

\newpage

\section{Azumaya/matrix manifolds with a fundamental module and\\ differentiable maps therefrom}

Basics of maps from an Azumaya/matrix manifold with a fundamental module needed for the current note
 are collected in this section to fix terminology, notations, and conventions.
Readers are referred to
   [L-Y1] (D(1)),  [L-L-S-Y] (D(2)), [L-Y5] (D(11.1)) and [L-Y7] (D(11.3.1))
 for details;
 in particular, why this is a most natural description of D-branes when Polchinski's TASI 1996 Lecture Note
  is read from the aspect of Grothendieck's modern Algebraic Geometry.
See also [H-W] and [Wi2].
  
\bigskip

\begin{flushleft}
{\bf Azumaya/matrix manifolds with a fundamental module $(X^{\!A\!z},\,{\cal E})$}
\end{flushleft}
From the viewpoint of Algebraic Geometry, a D-brane world-volume
 is a manifold equipped with a noncommutative structure sheaf of a special type dictated by (oriented) open strings.

\bigskip

\begin{sdefinition} {\bf [Azumaya/matrix manifold with fundamental module]}$\;$ {\rm
 Let
   $X$ be a (real, smooth) manifold and
   $E$ be a (smooth) complex vector bundle over $X$.
 Let
    \begin{itemize}	
	 \item[\LARGE $\cdot$]
      ${\cal O}_X$ be the {\it structure sheaf} of (smooth functions on) $X$,
	
     \item[\LARGE $\cdot$]	
	  ${\cal O}_X^{\,\Bbb C}:= {\cal O}_X\otimes_{\Bbb R}{\Bbb C}$ be its complexification, 	
	
	 \item[\LARGE $\cdot$]
      ${\cal E}$ be the sheaf of (smooth) sections of $E$, (it's an ${\cal O}_X^{\,\Bbb C}$-module),
	   and
	
	 \item[\LARGE $\cdot$]
      $\Endsheaf_{{\cal O}_X^{\,\Bbb C}}({\cal E}) $
        be the {\it endomorphism sheaf} of ${\cal E}$ as an ${\cal O}_X^{\,\Bbb C}$-module\\
      (i.e.\ the sheaf of sections of the endomorphism bundle $\End_{\Bbb C}(E)$ of $E$).		
    \end{itemize}  	
 Then, the (noncommutative-)ringed topological space
   $$
     X^{\!A\!z}\;
	    :=\;   (X, {\cal O}_X^{A\!z}:= \Endsheaf_{{\cal O}_X^{\,\Bbb C}}({\cal E}))
   $$
   is called an {\it Azumaya manifold}\footnote{Unfamiliar
                                                                  physicists may consult [Ar] for basics of Azumaya algebras;
																    see also [Az] and [A-N-T].
																  Simply put, an {\it Azumaya manifold} is topologically a smooth manifold
																    but with a structure sheaf that has fibers Azumaya algebras over ${\Bbb C}$.
																  These fibers are all isomorphic to a matrix ring $M_{r\times r}({\Bbb C})$
																   (and hence the synonym {\it matrix manifold}) for some fixed $r$
																   but the isomorphisms involved are not canonical (and hence
                                                                      why the term `Azumaya manifold' is more appropriate mathematically).
																  }
   (or synonymously, a {\it matrix manifold} to be more concrete to string-theorists.)
 It is important to note that non-isomorphic complex vector bundles may give rise to isomorphic endomorphism bundles
   and from the string-theory origin of the setting,
     in which $E$ plays the role of a Chan-Paton bundle on a D-brane world-volume,
  we always want to record $E$	as a part of the data in defining $X^{\!A\!z}$.
 Thus, we call the pair
   $(X^{\!A\!z},{\cal E})$  (or $(X^{\!A\!z},E)$ in bundle notation)
   an {\it Azumaya/matrix manifold with a fundamental module}.
}\end{sdefinition}
  
\bigskip

While it may be hard to visualize $X^{\!A\!z}$ geometrically,
 there in general is an abundant family of
 commutative ${\cal O}_{X}$-subalgebras
  $$
   {\cal O}_X\; \subset {\cal A}\; \subset\; {\cal O}_X^{A\!z}
  $$
 that define an abundant family of $C^{\infty}$-schemes
  $$
     X_{\cal A}\;:=\; \boldSpec^{\Bbb R}({\cal A})\,,
  $$
 each finite and germwise algebraic over $X$.
They may help visualize $X^{\!A\!z}$ geometrically.
 
\bigskip

\begin{sdefinition} {\bf [(commutative) surrogate of $X^{\!A\!z}$]}$\;$ {\rm
 Such $X_{\cal A}$ is called a (commutative) {\it surrogate} of
  (the noncommutative manifold) $X^{\!A\!z}$.
 Cf.$\,${\sc Figure} 1-1.
}\end{sdefinition}
%
%
 \begin{figure} [htbp]
  \bigskip
  \centering

  \includegraphics[width=0.80\textwidth]{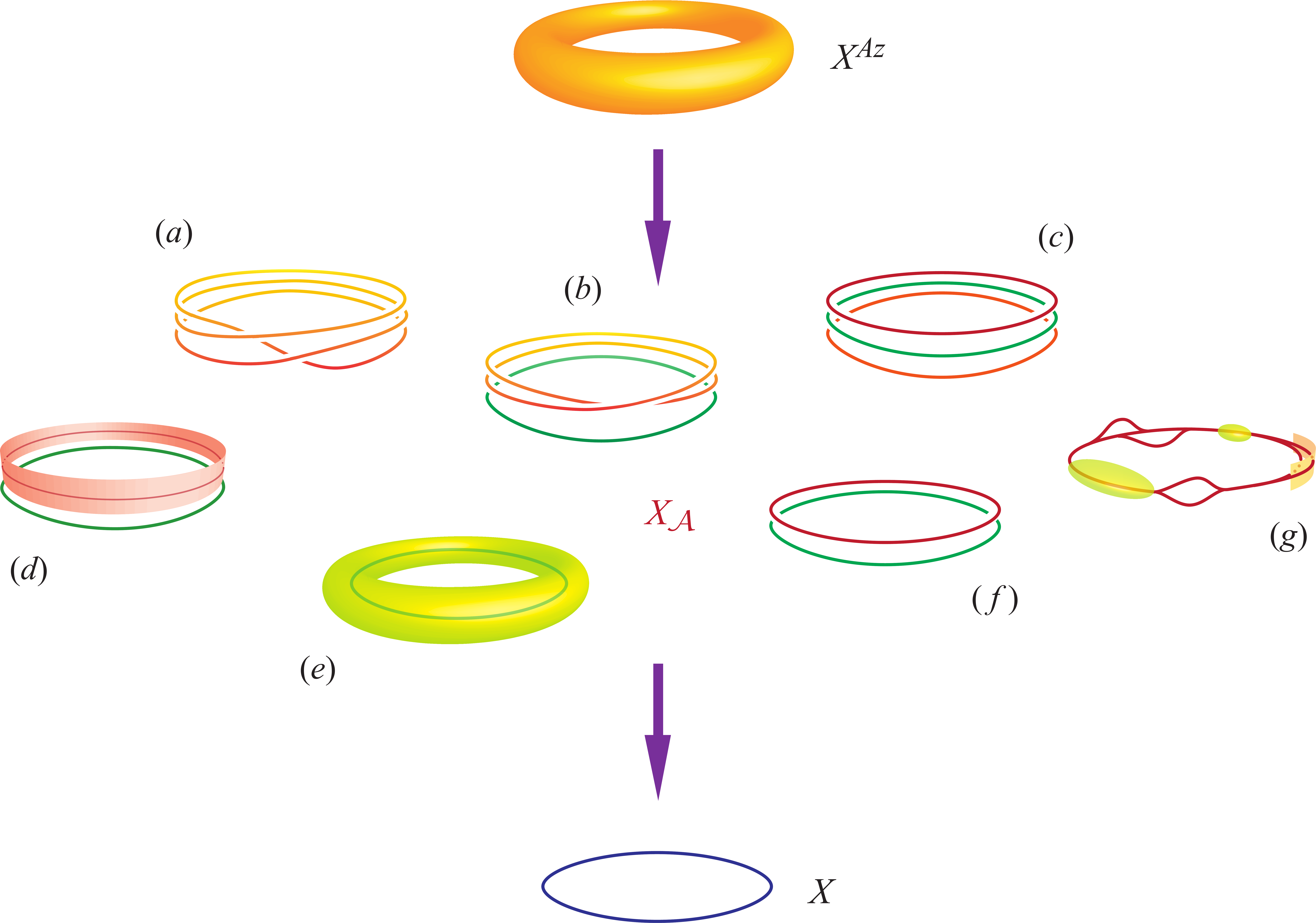}

  \bigskip
  \bigskip
  \centerline{\parbox{13cm}{\small\baselineskip 12pt
   {\sc Figure} 1-1.
    The noncommutative manifold $X^{\!A\!z}$ has an abundant collection of
	  $C^{\infty}$-schemes as its commutative surrogates.
	 See [L-Y8: {\sc Figure} 2-1-1: caption] (D(13.1)) for more details.
       }}
  \bigskip
 \end{figure}	

\bigskip

Without loss of generality, one may assume that $X$ is connected.
However even so, a surrogate $X_{\cal A}$ of $X^{\!A\!z}$ in general is disconnected locally over $X$
 (and can be disconnected globally as well; cf. {\sc Figure} 1-1).
To keep track of this algebraically, recall the following definition:
 
\bigskip

\begin{sdefinition} {\bf [complete set of orthogonal idempotents]}$\;$ {\rm (Cf.\ e.g.\ [Ei].)
 Let $R$ be an (associative, unital) ring, with the identity element  $1$.
 A set of elements $\{e_1, \,\cdots\,,\, e_s\}\subset R$ is called
  a {\it complete set of orthogonal idempotents} if the following three conditions are satisfied
  \begin{itemize}
   \item[(1)]  \parbox[t]{10em}{\it idempotent}\hspace{1em}
    $e_i^2=e_i$, $\;i=1,\,\cdots\,,\, s$.
	
   \item[(2)] \parbox[t]{10em}{\it orthogonal}\hspace{1em}
    $e_ie_j=0\;$ for $i\ne j$.

   \item[(3)] \parbox[t]{10em}{\it complete}\hspace{1em}
    $e_1 +\,\cdots\,+e_s=1$.
  \end{itemize}
 A complete set orthogonal idempotents $\{e_1,\,\cdots\,,\, e_s\}$ is called {\it maximal}
  if no $e_i$ in the set can be further decomposed into a summation
   $e_i= e^{\prime}+e^{\prime\prime}$ of two orthogonal idempotents.
}\end{sdefinition}

\bigskip

Let
 ${\cal O}_X\subset {\cal A}\subset {\cal O}_X^{A\!z}$
   be a commutative ${\cal O}_X$-subalgebra of ${\cal O}_X^{A\!z}$  and
 $X_{\!\cal A}$ the associate surrogate of $X^{\!A\!z}$.
Then,
 for $U\subset X$ an open set,
 there is a unique maximal complete set of orthogonal idempotents
 $\{e_1,\,\cdots\,,\, e_s\}$ of the $C^{\infty}$-ring ${\cal A}(U)$
 and it corresponds to the set of connected components of
   $X_{\!\cal A}|_U =\Spec^{\Bbb R}({\cal A}(U))$.
Up to a relabelling,
 $e_i$ corresponds the function on $X_{\!\cal A}(U)$
    that is constant $1$ on the $i$-th connected component and $0$ on all other connected components.

\bigskip

Finally, we recall also the tangent sheaf and the cotangent sheaf of $X^{\!A\!z}$.

\bigskip

\begin{sdefinition} {\bf [tangent sheaf, cotangent sheaf, inner derivations on $X^{\!A\!z}$]}$\;$ {\rm
 The sheaf of (left) derivations on ${\cal O}_X^{A\!z}$ is denoted by
   ${\cal T}_{\ast}X^{\!A\!z}$ and is called the {\it tangent sheaf} of $X^{\!A\!z}$.
 The sheaf of K\"{a}hler differentials of ${\cal O}_X^{A\!z}$ is dented by
   ${\cal T}^{\ast}X^{\!A\!z}$ and is called the {\it cotangent sheaf} of $X^{\!A\!z}$.
 ${\cal T}_{\ast}X^{\!A\!z}$ is naturally a (left) ${\cal O}_X^{\,\Bbb C}$-module
   while ${\cal T}^{\ast}X^{\!A\!z}$ is naturally a (left) ${\cal O}_X^{A\!z}$-module.
 For our purpose, we treat both as ${\cal O}_X^{\,\Bbb C}$-modules.
 There is a natural ${\cal O}_X^{\,\Bbb C}$-module homomorphism
   $$
    \begin{array}{ccc}
	   {\cal O}_X^{A\!z}  & \longrightarrow  & {\cal T}_{\ast}X^{\!A\!z}\\
	        m   & \longmapsto  & [m,\,\mbox{\Large $\cdot$}\,]\,,
	\end{array}	
   $$
   where $[m,\,\mbox{\Large $\cdot$}\,]$ acts on ${\cal O}_X^{A\!z}$ by
     $m^{\prime}\mapsto [m, m^{\prime}]:= mm^{\prime}-m^{\prime}m$.
 The image of this homomorphism is called the sheaf/${\cal O}_X^{\,\Bbb C}$-module of
  {\it inner derivations} on ${\cal O}_X^{A\!z}$
   and is denoted by $\Inn({\cal O}_X^{A\!z})$ or $\Inn(X^{\!A\!z})$.
 The kernel of the above map is exactly the center
  ${\cal O}_X^{\,\Bbb C}\cdot \Id_{\cal E}$, canonically identified with ${\cal O}_X^{\,\Bbb C}$,
  of ${\cal O}_X^{A\!z}$.
 When the choice of a representative of an element of $\Inn({\cal O}_X^{A\!z})$
   by an element in ${\cal O}_X^{A\!z}$ is irrelevant to an issue,
 we'll represent elements of $\Inn({\cal O}_X^{A\!z})$ simply by elements in ${\cal O}_X^{A\!z}$.
}\end{sdefinition}

\bigskip

\begin{flushleft}
{\bf When ${\cal E}$ is equipped with a connection $\nabla$}
\end{flushleft}
From the stringy origin of the setting with $E$ serving as the Chan-Paton bundle on the D-brane world-volume,
 $E$ is equipped with a gauge field (i.e.\ a connection) created by massless excitations of open strings.
Thus, let $\nabla$ be a connection on ${\cal E}$.
Then $\nabla$ induces a connection $D$ on
  ${\cal O}_X^{A\!z}:=\Endsheaf_{{\cal O}_X^{\,\Bbb C}}({\cal E})$.
With respect to a local trivialization of ${\cal E}$,
 $\nabla = d+ A$,
    where $A$  is an $\Endsheaf_{{\cal O}_X^{\,\Bbb C}}({\cal E})$-valued $1$-form on $X$.
Then $D= d+ [A\,,\,\,\mbox{\Large $\cdot$}\,]$ on ${\cal O}_X^{A\!z}$
  under the induced local trivialization.
As a consequence,
  $D$ leaves the center ${\cal O}_X^{\,\Bbb C}$ of ${\cal O}_X^{A\!z}$ invariant
 and restrict to the usual differential $d$ on ${\cal O}_X^{\,\Bbb C}$.
 
Once having the induced connection $D$ on ${\cal O}_X^{A\!z}$,
 one has then ${\cal O}_X^{\,\Bbb C}$-module homomorphism
 $$
   \begin{array}{cccl}
   {\cal T}_{\ast}X^{\Bbb C}  & \longrightarrow & {\cal T}_{\ast}X^{\!A\!z} \\
       \xi  & \longmapsto    & D_{\xi}   &.
   \end{array}
 $$

\bigskip

\begin{slemma} {\bf [$D$-induced decomposition of ${\cal T}_{\ast}X^{\!A\!z}$]}
 {\rm ([DV-M].)}
 One has the short exact sequence
   $$
      0\; \longrightarrow\;
	     \Inn({\cal O}_X^{A\!z})\; \longrightarrow\; {\cal T}_{\ast}X^{\!A\!z}\;
		  \longrightarrow\;  {\cal T}_{\ast}X^{\Bbb C}\; \longrightarrow\; 0
   $$
  split by the above map.
\end{slemma}

\bigskip

The following two lemmas address the issue of when an idempotent in ${\cal O}_X^{A\!z}$ can be
 constant under a derivation $\in {\cal T}_{\ast}X^{\!A\!z}$.

\bigskip

\begin{slemma} {\bf [(local) idempotent under $D_{\,}$]}$\;$
 With the above notations,
 let
  $U\subset X$ be an open set,
  $\xi$ a vector field on $U$,  and
  $\{e_1,\,\cdots\,,\, e_s\}$
        be a complete set  of orthogonal idempotents of ${\cal O}_X^{A\!z}(U)$.
 Assume that, say, $D_{\xi}e_1$ commutes with all $e_i$, $i=1,\,\cdots\,, \,s$.
 Then $D_{\xi}e_1=0$.
\end{slemma}

\smallskip

\begin{proof}
 Since $e_1^2=e_1$,
  $(D_{\xi}e_1)e_1 + e_1D_{\xi}e_1=D_{\xi}e_1$.
 If in addition $D_{\xi}e_1$ and $e_1$ commute, then one has $2(D_{\xi}e_1)e_1=D_{\xi}e_1$.
 The multiplication from the left by $e_1$ gives then
 $2(D_{\xi}e_1)e_1=(D_{\xi}e_1)e_1$; i.e.\ $(D_{\xi}e_1)e_1=0$.
 If, furthermore, $D_{\xi}e_1$ commutes also with all $e_2,\,\cdots\,,\, e_s$,
  then
    $0=D_{\xi}(e_je_1)=(D_{\xi}e_j)e_1 + e_j D_{\xi}e_1
	    = (D_{\xi}e_j)e_1+ (D_{\xi}e_1)e_j$, for $j=2,\,\cdots\,,\, s$.
 The multiplication from the left by $e_j$ gives then $(D_{\xi}e_1)e_j=0$, for $j=2,\,\cdots\,,\, s$.
 It follows that $D_{\xi}e_1 = (D_{\xi}e_1)(e_1+\,\cdots\,+e_s)=0$.
  
\end{proof}

\bigskip

\begin{slemma} {\bf [(local) idempotent under inner derivation]}$\;$
 With the above notations,
 let
  $U\subset X$ be an open set,
  $m\in {\cal O}_X^{A\!z}(U)$ represent an inner derivation of ${\cal O}_X^{A\!z}(U)$,  and
  $\{e_1,\,\cdots\,,\, e_s\}$
        be a complete set  of orthogonal idempotents of ${\cal O}_X^{A\!z}(U)$.
 Assume that, say, $[m, e_1]$ commutes with all $e_i$, $i=1,\,\cdots\,, \,s$.
 Then $[m, e_1]=0$.
\end{slemma}

\smallskip

\begin{proof}
 Note that the proof of Lemma~1.6
   uses only
  the Leibnitz rule property of $D_{\xi}$ on ${\cal O}_X^{A\!z}(U)$ and
  the commutativity property of $D_{\xi}e_1$ with $e_1,\,\cdots\,,\, e_s$.
 Since $[m,\,\mbox{\Large $\cdot$}\,]$ satisfies also
   the Leibniz rule property on ${\cal O}_X^{A\!z}(U)$ and
   by assumption $[m,e_1]$ commutes with $e_1,\,\cdots\,,\, e_s$,
  the same proof goes through.
   
\end{proof}

The contraction
  $\Endsheaf _{{\cal O}_X^{\,\Bbb C}}({\cal E})
     ={\cal E}\otimes_{{\cal O}_X^{\,\Bbb C}}{\cal E}^{\vee}
	 \rightarrow {\cal O}_X^{\,\Bbb C}$
 defines a {\it trace map}
 $$
   \Tr\;  :\;  {\cal O}_X^{A\!z}\;\longrightarrow \; {\cal O}_X^{\,\Bbb C}\,.
 $$
 One has
 $$
   d \Tr\;=\; \Tr D\,,
 $$
 where $d$ is the ordinary differential on ${\cal O}_X^{\,\Bbb C}$.

\bigskip

\begin{flushleft}
{\bf Differentiable maps from $(X^{\!A\!z},\,{\cal E})$}
\end{flushleft}
As a dynamical object in space-time,
 a D-brane moving in a space-time $Y$ is realized by a map from a D-brane world-volume to $Y$.
Back to our language, we need thus a notion of a `map from $(X^{\!A\!z},{\cal E};\nabla)$ to $Y$'
 that is compatible with the behavior of D-branes in string theory.

\bigskip

\begin{sdefinition} {\bf [map from Azumaya/matrix manifold]}$\;$ {\rm
 Let
  $X$ be a (real, smooth) manifold,
  $E$ be a complex vector bundle of rank $r$ over $X$,  and
  $(X^{\!A\!z}, E) := (X, C^{\infty}(\End_{\Bbb C}(E)), E)$
     be the associated Azumaya/matrix manifold with a fundamental module.
 A {\it map} (synonymously, {\it differentiable map}, {\it smooth map})
   $$
     \varphi\,:\, (X^{\!A\!z},E)\; \longrightarrow\; Y
   $$
   from $(X^{\!A\!z}, E)$ to a (real, smooth) manifold $Y$
   is defined contravariantly by a ring-homomorphism
   $$
     \varphi^{\sharp}\;:\; C^{\infty}(Y)\;\longrightarrow\; C^{\infty}(\End_{\Bbb C}(E))\,.
   $$
   
  Equivalently in terms of sheaf language,
   let ${\cal O}_Y$ be the structure sheaf of $Y$.
  Regard both ${\cal O}_Y$ and ${\cal O}_X^{A\!z}$ as equivalence classes of gluing system of rings
    over the topological space $Y$ and $X$ respectively.
  Then the above $\varphi^{\sharp}$ specifies
   an {\it equivalence class of gluing systems of ring-homomorphisms} over ${\Bbb R}\subset {\Bbb C}$
    $$
       {\cal O}_Y\; \longrightarrow \; {\cal O}_X^{A\!z}\,,
    $$
   which we will still denote by $\varphi^{\sharp}$.
}\end{sdefinition}

\bigskip

Through the Generalized Division Lemma \`{a} la Malgrange,
 one can show that $\varphi^{\sharp}$ extends to a commutative diagram
  $$
    \xymatrix{
	  \;{\cal O}_X^{A\!z}\;		
	     && \;{\cal O}_Y\;
		           \ar[ll]_-{\varphi^{\sharp}}
				   \ar@{_{(}->}[d]^-{pr^{\sharp}_{Y}}   \\
	   \;{\cal O}_X\rule{0ex}{1em}\;
	       \ar@{^{(}->}[rr]_-{pr^{\sharp}_X}      \ar@{^{(}->}[u]
	     && \;\;{\cal O}_{X\times Y}\;
		          \ar[llu]_-{\tilde{\varphi}^{\sharp}}\;,
    }
  $$
 of equivalence  classes of ring-homomorphisms
  (over ${\Bbb R}$ or ${\Bbb R}\subset{\Bbb C}$, whichever is applicable)
  between equivalence classes of gluing systems of rings,
  with
  $$
    \xymatrix{
	  \;{\cal A}_{\varphi}\;		
	     && \;{\cal O}_Y\;
		           \ar[ll]_-{f_{\varphi}^{\sharp}}
				   \ar@{_{(}->}[d]^-{pr^{\sharp}_{Y}}   \\
	  \;{\cal O}_X\rule{0ex}{1em}\;
	       \ar@{^{(}->}[rr]_-{pr^{\sharp}_X}      \ar@{^{(}->}[u]^-{\pi_{\varphi}^{\sharp}}
	     && \;\;{\cal O}_{X\times Y}\;
		          \ar[llu]_-{\tilde{f}_{\varphi}^{\sharp}}\;,
    }
  $$
 a commutative diagram of equivalence  classes of ring-homomorphisms
  between equivalence classes of gluing systems of $C^{\infty}$-rings.
Here,
 $\pr_X:X\times Y \rightarrow X$ and $\pr_Y:X\times Y\rightarrow Y$ are the projection maps,
 ${\cal O}_X\hookrightarrow {\cal O}_X^{A\!z}$
	 follows from the inclusion of the center ${\cal O}_X^{\,\Bbb C}$ of
	 ${\cal O}_X^{A\!z}$,  and
 $$
   {\cal A}_{\varphi}\; := {\cal O}_X\langle  \Image\varphi^{\sharp}\rangle\;
      =\;  \Image \tilde{\varphi}^{\sharp}\,.
 $$	   	
 {\rm (Cf.\ [L-Y7: Theorem 3.1.1] (D(11.3.1)).)}
  
In terms of spaces, one has the following equivalent diagram of maps
 $$
   \xymatrix{
    & {\cal E} \ar@{.>}[rd]     \ar@{.>}@/_1ex/[rdd]    \ar@{.>}@/_2ex/[rddd]   \\
    &  & X^{\!A\!z}\ar[rrrd]^-{\varphi}\ar@{->>}[d]^-{\sigma_{\varphi}}   \\
    &  & X_{\varphi}\ar[rrr]_-{f_{\varphi}}
	                                   \ar@{_{(}->} [rrrd]_{\tilde{f}_{\varphi}}
                                	   \ar@{->>}[d]^-{\pi_{\varphi}} &&& Y \\
	&  & X     &&& X\times Y \ar@{->>}[u]_-{pr_Y} \ar@{->>}[lll]^-{pr_X}  &,
    }
  $$
 where $X_{\varphi}$ is the $C^{\infty}$-scheme
  $$
    X_{\varphi}\;:=\; \boldSpec^{\Bbb R}{\cal A}_{\varphi}
  $$
  associated to ${\cal A}_{\varphi}$.
    
\bigskip

\begin{sdefinition} {\bf [graph of $\varphi$]}$\;$ {\rm
 The push-forward $\tilde{\varphi}_{\ast}{\cal E}=:\tilde{\cal E}_{\varphi}$
   of ${\cal E}$ under $\tilde{\varphi}$
   is called the {\it graph} of $\varphi$.
  It is an ${\cal O}_{X\times Y}^{\,\Bbb C}$-module.
 Its {\it $C^{\infty}$-scheme-theoretical support}
  is denoted by $\Supp(\tilde{\cal E}_{\varphi})$.
}\end{sdefinition}

\bigskip

\begin{sdefinition} {\bf [surrogate of $X^{\!A\!z}$ specified by $\varphi$]}$\;$ {\rm
 The $C^{\infty}$-scheme $X_{\varphi}$ is called
   the {\it surrogate} of $X^{\!A\!z}$ specified by $\varphi$.
}\end{sdefinition}

\bigskip

$X_{\varphi}$ is finite and germwise algebraic over $X$ and,
 by construction,
   it admits a canonical embedding
   $\tilde{f}_{\varphi}:X_{\varphi}\rightarrow X\times Y$
   into $X\times Y$ as a $C^{\infty}$-subscheme.
 The image is identical to $\Supp(\tilde{\cal E}_{\varphi})$.
  Cf.\ {\sc Figure} 1-2 and {\sc Figure} 1-3.
  %
  
\begin{figure}[htbp]
 \bigskip
  \centering
   \includegraphics[width=0.80\textwidth]{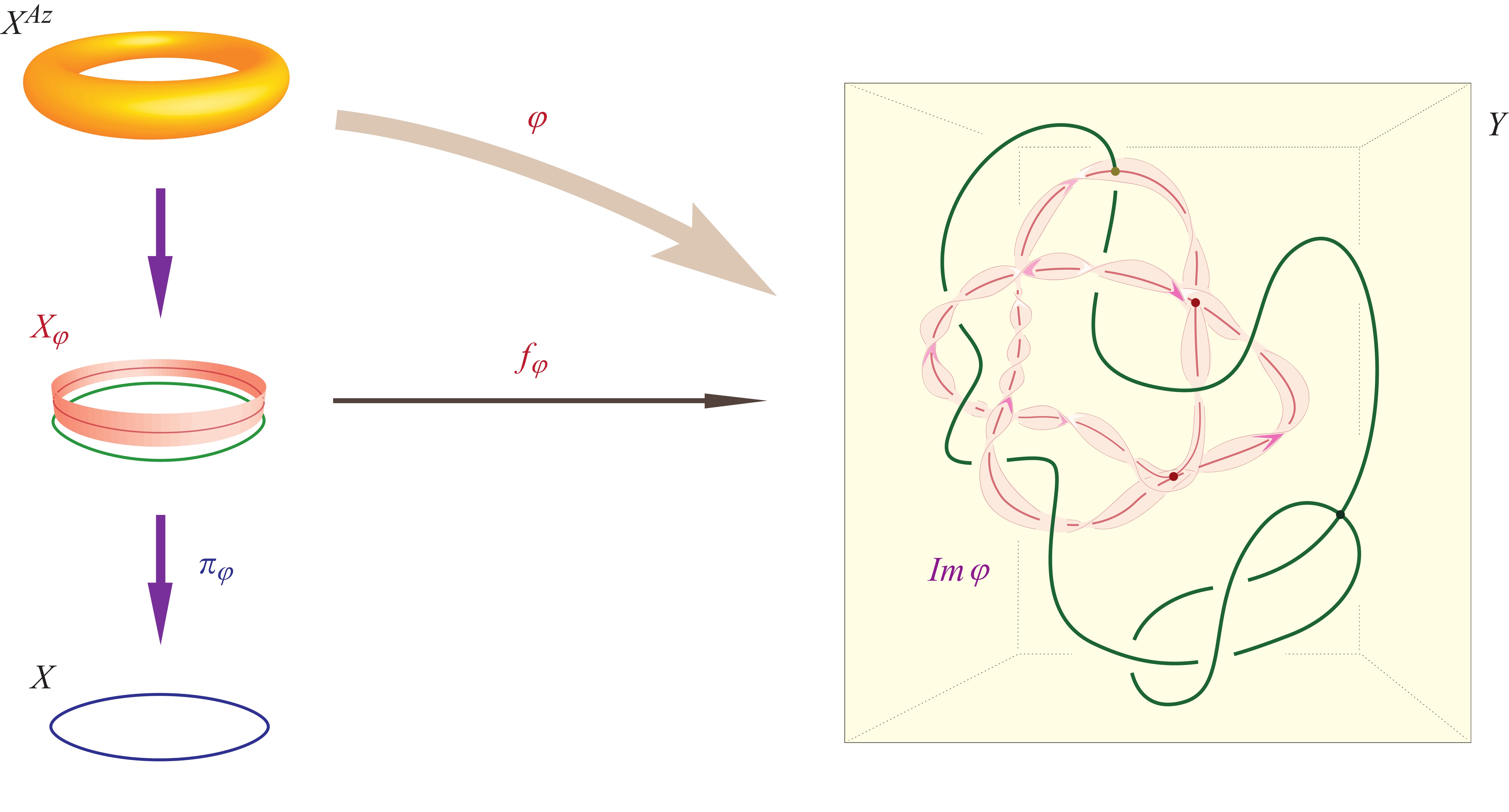}
 
  \bigskip
  
  \bigskip
  
 \centerline{\parbox{13cm}{\small\baselineskip 12pt
  {\sc Figure}~1-2.
  A map $\varphi:(X^{\!A\!z},E)\rightarrow Y$ specifies a surrogate $X_{\varphi}$
   of  $X^{\!A\!z}$ over $X$.
  $X_{\varphi}$ is a $C^{\infty}$-scheme that may not be reduced
   (i.e.\ it may have some nilpotent fuzzy structure thereon).
  It on one hand is dominated by $X^{\!A\!z}$ and on the other dominates and is finte
   and germwise algebraic over $X$.
  }}
\end{figure}		

\begin{figure} [htbp]
 \bigskip
 \centering
 \includegraphics[width=0.80\textwidth]{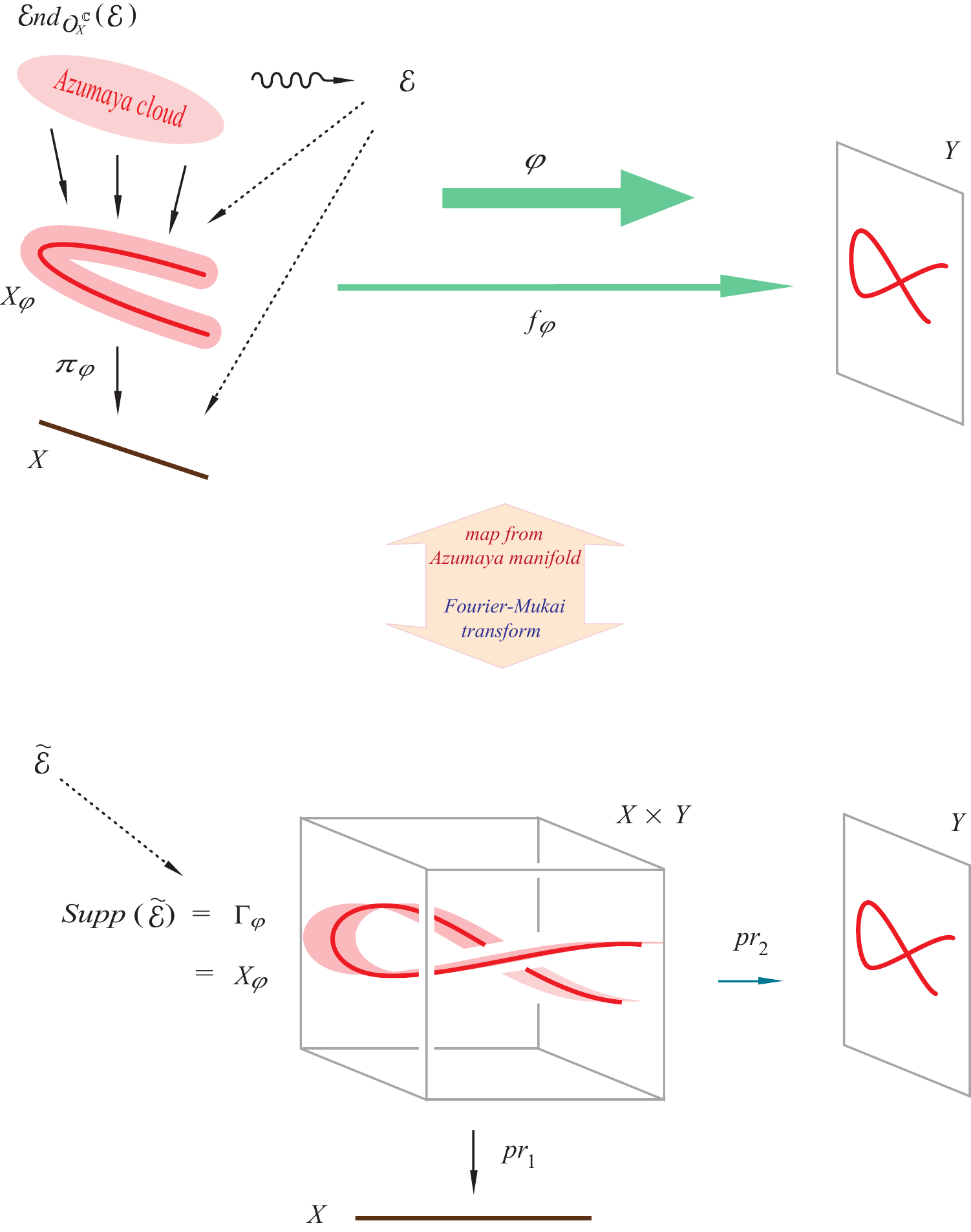}
 
 \vspace{4em}
 \centerline{\parbox{13cm}{\small\baselineskip 12pt
  {\sc Figure}~1-3.
    The equivalence between
      a map $\varphi$ from an Azumaya manifold with a fundamental module
	    $(X,{\cal O}_X^{Az}:=\Endsheaf_{{\cal O}_X^{\,\Bbb C}}({\cal E}),
		          {\cal E})$
	    to a manifold $Y$    and
	  a special kind of Fourier-Mukai transform
	    $\tilde{\cal E}\in \ModCategory^{\Bbb C}(X\times Y)$ from $X$ to $Y$. 		
   Here, $\ModCategory^{\Bbb C}(X\times Y)$
        is the category of ${\cal O}_{X\times Y}^{\,\Bbb C}$-modules. 		
      }}
 \bigskip
\end{figure}

\bigskip

\begin{flushleft}
{\bf Compatibility between the map $\varphi$ and the connection $\nabla$}
\end{flushleft}
Up to this point,
 the map $\varphi:(X^{\!A\!z},{\cal E})\rightarrow Y$ to $Y$
  and the connection $\nabla$ on ${\cal E}$
 are quite independent objects.
A priori, there doesn't seem to be any reason
 why they should constrain or influence each other at the current purely differential-topological level.
However, when one moves on to address the issue of constructing an action functional
   for $(\varphi,\nabla)$ as in [L-Y8] (D(13.1)),
one immediately realizes that,
  \begin{itemize}
   \item[\LARGE $\cdot$] \parbox[t]{38em}{\it
    Due to a built-in mathematical obstruction in the problem,
      one needs some compatibility condition between $\varphi$ and $\nabla$
     before one can even begin the attempt to construct an action functional for $(\varphi,\nabla)$.}
  \end{itemize}
Furthermore, as a hindsight, that there needs to be a compatibility condition on $(\varphi,\nabla)$  
 is also implied by string theory:
  \begin{itemize}
   \item[\LARGE $\cdot$] \parbox[t]{38em}{\it
    We need a condition on $(\varphi,\nabla)$ to encode the stringy fact  that
	the gauge field $\nabla$ on the D-brane world-volume as `seen' by open strings in $Y$ through $\varphi$
    should be massless.}
  \end{itemize}
 
We address such compatibility condition on $(\varphi,\nabla)$ systematically in the next section.

\bigskip

\section{Pull-push of tensors and admissible conditions on $(\varphi,\nabla)$}

When one attempts to construct an action functional for a theory that involves maps from a world-volume
 to a target space-time, one unavoidably has to come across the notion of
 `{\it pulling back a (covariant) tensor},  for example, the metric tensor or a differential form
 {\it on the target space-time to the world-volume}'.
In the case where only maps from a commutative world-volume to a (commutative)   space-time are involved,
 this is a well-established standard notion from differential topology.
However, in a case, like ours,
 where maps from a noncommutative world-volume to a (commutative) space-time is involved,
  $$
    \varphi\;: \Space(S)\; \longrightarrow\; Space(R)\,,
  $$
 with the accompanying contravariant ring-homomorphism
  $$
     \hspace{3em}S\;\longleftarrow\; R\;:\; \varphi^{\sharp}\hspace{3em},
  $$
 there is a built-in mathematical obstruction to such a notion.
Here,
 $S$ is an (associative, unital) noncommutative ring, $R$ is a (associative, unital) commutative ring,  and
 $\Space(S)$ and $\Space(R)$ are the topological spaces whose function rings are $S$ and $R$
 respectively.
 
For the noncommutative ring $S$, its (standard and functorial-in-the-category-rings)
 bi-$S$-module of K\"{a}hler differentials is naturally defined to be
 $$
   \Omega^{\Kahlerscriptsize}_S\;
    :=\;  \Span_{(S,S)}\{ ds\,|\, s\in S)\}/
	   (d(ss^{\prime})-(ds)s^{\prime}-s ds^{\prime}\,|\, s, s^{\prime}\in S)
 $$
 while for the commutative ring $R$ its (standard and functorial-in-the-category-of-{\it commutative}-rings)
 (left) $R$-module of K\"{a}hler differentials is naturally
 defined to be
 $$
   \Omega^{\Kahlerscriptsize}_R\; :=\;  \Span_R\{dr\, |\, r \in R\}/
      (d(rr^{\prime})-r^{\prime}dr-r dr^{\prime}\,|\, r, r^{\prime}\in R)\,,
 $$
 with the convention that $rdr^{\prime}=(dr^{\prime})r$ to turn it to a bi-$R$-module as well.
Treating $R$ as a ring (that happens to be commutative), it has also the
 (standard and functorial-in-the-category-rings) bi-$R$-module of K\"{a}hler differentials
 $$
   \Omega_R^{\nc, \Kahlerscriptsize}\;
    :=\;  \Span_{(R,R)}\{ dr\,|\, r\in R)\}/
	   (d(rr^{\prime})-(dr)r^{\prime}-r dr^{\prime}\,|\, r, r^{\prime}\in R)
 $$
 exactly like $\Omega^{\Kahlerscriptsize}_S$ for $S$.
There is a built-in tautological quotient homomorphism as bi-$R$-modules
 $$
  \begin{array}{rcccl}
   & \Omega_R^{\nc,\Kahlerscriptsize}   & \longrightaarrow   & \Omega^{\Kahlerscriptsize}_R \\
   & r_1 (dr) r_2            & \longmapsto           & r_1r_2\,dr   &,
  \end{array}
 $$
  whose kernel is generated by $\{rdr^{\prime}-(dr^{\prime})r\,|\,r,r^{\prime}\in R\}$.
Given the map $\varphi:\Space(S)\rightarrow \Space(R)$, one has the following built-in diagram
 $$
  \xymatrix{
   & \Omega^{\Kahlerscriptsize}_S
     && \Omega_R^{\nc,\Kahlerscriptsize}\ar[ll]_-{\varphi^{\ast}}  \ar@{->>}[d]\\
   &      && \Omega^{\Kahlerscriptsize}_R     &,
    }
 $$
 where $\varphi^{\ast}(r_1 (dr) r_2)
                  = \varphi^{\sharp}(r_1) d(\varphi^{\sharp}(r)) \varphi^{\sharp}(r_2)$.
The issue is now whether one can extend the above diagram to the following commutative diagram
  $$
  \xymatrix{
   & \Omega^{\Kahlerscriptsize}_S
     && \Omega_R^{\nc,\Kahlerscriptsize}\ar[ll]_-{\varphi^{\ast}}  \ar@{->>}[d]\\
   &&& \Omega_R^{\Kahlerscriptsize}
             \ar@{.>}[llu]^{\underline{\varphi}^{\ast}\raisebox{-.6ex}{?}}&.
    }
 $$
The answer is No, in general.
See, e.g., [L-Y5: Example 4.1.20] (D(11.1)) for an explicit counterexample.
When $R$ is a $C^{\infty}$-ring, e.g.\ the function-ring $C^{\infty}(Y)$ of a smooth manifold $Y$, 
 then the $R$-module $\Omega_R$ of differentials of $R$
   is a further quotient of the above module $\Omega_R^{\Kahlerscriptsize}$ of K\"{a}hler differentials
 by additional relations generated by applications of the chain rule on the transcendental smooth operations
  in the $C^{\infty}$-ring structure of $R$ ([Jo]; cf.\ [L-Y5: Sec.$\,$4.1] (D(11.1))).
The issue becomes even more involved.
In particular, as the counterexample ibidem shows
 \begin{itemize}
  \item[\Large$\cdot$] \parbox[t]{38em}{\it
   {\bf [built-in mathematical obstruction of pullback]}$\;$
    For a map $\varphi:(X^{\!A\!z},{\cal E})\rightarrow Y$,
	   there is no way to define functorially a pull-back map
       ${\cal T}^{\ast}Y \rightarrow {\cal T}^{\ast}X^{\!A\!z}$
	   that takes a (covariant) $1$-tensor on $Y$ to a $1$-tensor on $X^{\!A\!z}$.
	As a consequence, there is no functorial way to pull back a (covariant) tensor on $Y$
  	 to a tensor on $X^{\!A\!z}$.}
 \end{itemize}
 
Before the attempt to construct an action functional that involves such maps $\varphi$,
 one has to resolve the above obstruction first.
In [L-Y8] (D(13.1)), we learned how to use the connection $\nabla$
 to impose a natural admissible condition on $\varphi$ so that the above obstruction is bypassed
 through the surrogate $X_{\varphi}$ of $X^{\!A\!z}$ specified by $\varphi$.
With the lesson learned therefrom and further thought beyond [L-Y9] (D(13.2.1)),
 we propose (Sec.$\,$2.1) in this section a still-natural-but-much-weaker admissible condition
 on $(\varphi,\nabla)$ that bypasses even the surrogate $X_{\varphi}$
    but is still robust enough to construct naturally a pull-push map we need on tensors.
It turns out that this much weaker admissible condition remains to be compatible with open strings (Sec.$\,$2.2).

\bigskip

\subsection{Admissible conditions on $(\varphi,\nabla)$ and the resolution of the pull-push issue}

A hierarchy of admissible conditions on $(\varphi,\nabla)$ is introduced.
A theorem on
  how even the weakest admissible condition in the hierarchy can resolve the above obstruction in our case
 is proved.

\bigskip

\begin{flushleft}
{\bf Three hierarchical admissible conditions}
\end{flushleft}
\begin{definition} {\bf [admissible connection on ${\cal E}$]}$\;$ {\rm
 Let $\varphi:(X^{A\!z},{\cal E})\rightarrow Y$ be a map.
 For a connection $\nabla$ on ${\cal E}$,
   let $D$ be its induced connection on ${\cal O}_X^{A\!z}$.
 A connection $\nabla$ on ${\cal E}$ is called
   \begin{itemize}
    \item[]
    {\it $(\ast_1)$-admissible to $\varphi$}\hspace{1em}
     if $\;D_{\xi}{\cal A}_{\varphi}
	                 \subset \Comm({\cal A}_{\varphi})$;

    \vspace{-1.6ex}					
	\item[]	
    {\it $(\ast_2)$-admissible to $\varphi$}\hspace{1em}
     if $\;D_{\xi}\Comm({\cal A}_{\varphi})
	         \subset \Comm({\cal A}_{\varphi})$;

    \vspace{-1.6ex}			
	\item[]
    {\it $(\ast_3)$-admissible to $\varphi$}\hspace{1em}
      if $\;D_{\xi}{\cal A}_{\varphi}\subset  {\cal A}_{\varphi}$
  \end{itemize}	
  for all $\xi\in{\cal T}_{\ast}X$.
  Here,
   $\Comm({\cal A}_{\varphi})$ denotes the commutant of
      ${\cal A}_{\varphi}$ in ${\cal O}_X^{A\!z}$.
	
 When $\nabla$ is $(\ast_1)$-admissible to $\varphi$,
  we will take the following as synonyms:
   \begin{itemize}
    \item[\LARGE $\cdot$]
      $(\varphi,\nabla)$ is an {\it $(\ast_1)$-admissible pair},

	\vspace{-1.6ex}
    \item[\LARGE $\cdot$]	
    {\it $\varphi$ is $(\ast_1)$-admissible to $\nabla$},
	
	\vspace{-1.6ex}
	\item[\LARGE $\cdot$]
	 $\varphi:(X^{\!A\!z},{\cal E};\nabla)\rightarrow Y $ is $(\ast_1)$-admissible.
  \end{itemize}
 Similarly, for {\it $(\ast_2)$-admissible pair} $(\varphi,\nabla)$	
   and {\it $(\ast_3)$-admissible pair} $(\varphi,\nabla)$, ... , etc..
}\end{definition}

\medskip

\begin{lemma} {\bf [hierarchy of admissible conditions]}$\;$
 $$	
  \mbox{\it Admissible Condition $(\ast_3)$}\;
   \Longrightarrow\; \mbox{Admissible Condition $(\ast_2)$}\;
   \Longrightarrow\; \mbox{\it Admissible Condition $(\ast_1)$}\,.
 $$
\end{lemma}

\smallskip
 
\begin{proof}
 Admissible Condition $(\ast_3)$ says that
   the ${\cal O}_X$-subalgebra ${\cal A}_{\varphi}\subset {\cal O}_X^{A\!z}$
    is invariant under $D$-parallel transports along paths on $X$.
 Since $D$-parallel transports on ${\cal O}_X^{A\!z}$ are algebra-isomorphisms,
  if ${\cal A}_{\varphi}$ is $D$-invariant,
    the ${\cal O}_X$-subalgebra $\Comm({\cal A}_{\varphi})$ of ${\cal O}_X^{A\!z}$
	 must also be  $D$-invariant since it is determined by ${\cal A}_{\varphi}$ fiberwise algebraically.
 In other words, Admissible Condition $(\ast_3)$ $\Longrightarrow$ Admissible Condition $(\ast_2)$.
 
 Since ${\cal A}_{\varphi}$ is commutative,
  ${\cal A}_{\varphi}\subset \Comm({\cal A}_{\varphi.})$.
 Thus, the inclusion
   $D_{\mbox{\LARGE $\cdot$}}{\cal A}_{\varphi}
     \subset D_{\mbox{\LARGE $\cdot$}}\Comm({\cal A}_{\varphi})$
  always holds.
 This implies that
   Admissible Condition $(\ast_2)$ $\Longrightarrow$ Admissible Condition $(\ast_1)$.

\end{proof}
	
\bigskip
	
\begin{definition} {\bf [strict admissible connection on ${\cal E}$]}$\;$ {\rm
 Continuing Definition~2.1.1.
  Let $F_{\nabla}$ be the curvature tensor of $\nabla$.
  It is an ${\cal O}_X^{A\!z}$-valued $2$-form on $X$.
  Then, for {\Large $\cdot$}$=1,2,3$,
    $\nabla$ is called {\it strictly $(\ast_{\mbox{\Large $\cdot$}})$-admissible to $\varphi$}
	if
     \begin{itemize}	
	  \item[\LARGE $\cdot$]
	    $\nabla$ is $(\ast_{\mbox{\Large $\cdot$}})$-admissible to $\varphi$  and
	    $F_{\nabla}$ takes values in $\Comm({\cal A}_{\varphi})\subset {\cal O}_X^{A\!z}$.
	 \end{itemize}
 In this case,
   $(\varphi,\nabla)$	is said to be a {\it strictly $(\ast_{\mbox{\Large $\cdot$}})$-admissible pair}.
}\end{definition}

\bigskip

Clearly, the same hierarchy holds for strict admissible conditions:
 $$	
  \mbox{\it strict $(\ast_3)$}\;\;
   \Longrightarrow\;\; \mbox{\it strict $(\ast_2)$}\;\;
   \Longrightarrow\; \mbox{\it strict $(\ast_1)$}\,.
 $$
The Strict $(\ast_3)$-Admissible Condition on $(\varphi,\nabla)$ was introduced in
  [L-Y8: Definition~2.2.1] (D(13.1))
 to define the Dirac-Born-Infeld action for $(\varphi,\nabla)$.

\bigskip

\begin{lemma} {\bf [commutativity under admissible condition]}$\;$
 Let $\varphi:(X^{A\!z},{\cal E};\nabla)\rightarrow Y$ be a map.
 $\;(1)$
 If $(\varphi,\nabla)$ is $(\ast_1)$-admissible,
 then $[D_{\xi}\varphi^{\sharp}(f_1), \varphi^{\sharp}(f_2)]=0$
    for all $f_1,f_2\in C^{\infty}(Y)$ and $\xi\in {\cal T}_{\ast}X$.
 $\;\;(2)$
  If $(\varphi,\nabla)$ is $(\ast_2)$-admissible,
    then $[D_{\xi_1}\varphi^{\sharp}(f_1), D_{\xi_2}\varphi^{\sharp}(f_2)]=0$
      for all $f_1,f_2\in C^{\infty}(Y)$ and $\xi_1,\xi_2\in {\cal T}_{\ast}X$.	
\end{lemma}

\smallskip

\begin{proof}
 Statement (1) is the $(\ast_1)$-Admissible Condition itself.
 
 For Statement (2),
 let $f_1,f_2\in C^{\infty}(Y)$ and $\xi_1,\xi_2\in{\cal T}_{\ast}X$.
 Then $[D_{\xi}\varphi^{\sharp}(f_1),\varphi^{\sharp}(f_2)]=0$
   since ${\cal A}_{\varphi}\subset \Comm({\cal A}_{\varphi})$.
 Thus,  applying $D_{\xi_2}$ to both sides,
   $$
     [D_{\xi_2}D_{\xi_1}\varphi^{\sharp}(f_1),\varphi^{\sharp}(f_2)]\,
	  +\, [D_{\xi_1}\varphi^{\sharp}(f_1), D_{\xi_2}\varphi^{\sharp}(f_2)]\;=\; 0\,.
   $$
 The $(\ast_2)$-Admissible Condition implies that
   $$
      D_{\xi_2}D_{\xi_1}\varphi^{\sharp}(f_1)\in \Comm({\cal A}_{\varphi})\,.
   $$
 And, hence,
   $[D_{\xi_2}D_{\xi_1}\varphi^{\sharp}(f_1),\varphi^{\sharp}(f_2)]=0$.
  Statement (2) follows.

\end{proof}

\bigskip

\begin{flushleft}
{\bf Resolution of the pull-push issue under Admissible Condition $(\ast_1)$}
\end{flushleft}
The current theme is devoted to the proof of the following theorem:

\bigskip

\begin{theorem} {\bf [pull-push under $(\ast_1)$-admissible $(\varphi,\nabla)$]}$\;$
 Let $(\varphi,\nabla)$ be $(\ast_1)$-admissible.
 Then
 the assignment
 $$
   \begin{array}{ccccc}
   \varphi^{\diamond}& : & \Omega_{C^{\infty}(Y)}  & \longrightarrow
      & \Omega_{C^{\infty}(X)}
	       \otimes_{C^{\infty}(X)}C^{\infty}(\End_{\Bbb C}(E))\\[.8ex]
	&& f_1df_2  & \longmapsto   &    \varphi^{\sharp}(f_1)\, D\varphi^{\sharp}(f_2)
   \end{array}	
 $$
 is well-defined.
\end{theorem}

\bigskip

The study in [L-Y8: Sec.\ 4] (D(13.1)) allows one to express $\varphi^{\diamond}(df)$
 locally explicit enough so that one can check that $\varphi^{\diamond}$ is well-defined
 when $(\varphi,\nabla)$ is $(\ast_1)$-admissible.
Note that, with Lemma~2.1.2,
 this implies that
  if $(\varphi,\nabla)$ is either $(\ast_2)$- or $(\ast_3)$-admissible,
  then $\varphi^{\diamond}$ is also well-defined.
We now proceed to prove the theorem.

\bigskip

\begin{lemma}
{\bf [local expression of $\varphi^{\diamond}(df)$
           for $(\ast_1)$-admissible $(\varphi, \nabla)$, I]}$\;$
 Let $(\varphi, \nabla)$ be $(\ast_1)$-admissible;
   i.e.\ $D_{\xi}{\cal A}_{\varphi}
              \subset \Comm({\cal A}_{\varphi})$ for all $\xi\in{\cal T}_{\ast}X$.
 Let  $U\subset X$ be a small enough open set
   so that $\varphi(U^{A\!z})$ is contained in a coordinate chart of $Y$,
   with coordinate $\boldy =(y^1, \,\cdots\,,\, y^n)$.
 For $f\in C^{\infty}(Y)$,
  recall the germwise-over-$U$ polynomial $R^f[1]$ in $(y^1,\,\cdots\,,\, y^n)$
   with coefficients in ${\cal O}_U^{A\!z}$ from
   {\rm $\,$[L-Y8: Sec.\ 4 \& Remark/Notation 4.2.3.5] (D.13.1)}.
 Then, for $\xi$ a vector field on $U$ and $f\in C^{\infty}(Y)$, and at the level of germs over $U$,
  $$
     (\varphi^{\diamond}(df))(\xi)\;
	   =\; R^f[1]\!\!\left.\rule{0ex}{0.8em}\right|
			    _{\scriptsizeboldy^{\tinyboldd}
 			                    \rightsquigarrow\,
								D_{\xi}
								   (\varphi^{\sharp}(\scriptsizeboldy^{\tinyboldd})),\;
								 \mbox{\scriptsize for all multi-degree $\scriptsizeboldd$ in $R^f[1]$}}\,.
  $$
 Here,
  for a multiple degree $\boldd =(d_1,,\cdots\,,\, d_n)$, $d_i\in {\Bbb Z}_{\ge 0}$,
  $\boldy^{\scriptsizeboldd} := (y^1)^{d_1}\,\cdots\, (y^n)^{d_n}$ and\\
  $\boldy^{\scriptsizeboldd}
 			                    \rightsquigarrow\,
								D_{\xi}
								   (\varphi^{\sharp}(\boldy^{\scriptsizeboldd}))$
    means
     `replacing $\boldy^{\scriptsizeboldd}$
      by	$D_{\xi}(\varphi^{\sharp}(\boldy^{\scriptsizeboldd}))$'.	
\end{lemma}

\smallskip

\begin{proof}
 Denote the coordinate chart of $Y$ in the Statement by $V$.
 Let $\pr_X:X\times Y\rightarrow X$, $\pr_Y:X\times Y\rightarrow Y$ be the projection maps.
 Recall the induced ring-homomorphism\\
   $\tilde{\varphi}^{\sharp}:C^{\infty}(X\times Y)
      \rightarrow C^{\infty}(\End_{\Bbb C}(E))$ over ${\Bbb R}\subset {\Bbb C}$
   and the graph $\tilde{\cal E}_{\varphi}$ of $\varphi$ and
       its support $\Supp(\tilde{\cal E}_{\varphi})$ on $X\times Y$.
 Denote $\pr_Y^{\sharp}(f)\in C^{\infty}(X\times Y)$ still by $f$ when there is no confusion.
 For clarity, we proceed the proof of the Statement in three steps.

 \bigskip
 
 \noindent
 {\it Step $(1)\;$ How $R^f[1]$ is constructed in {\rm [L-Y8: Sec.\ 4] (D(13.1))}}\hspace{2em}
 For  any $p\in U$, let $p\in U^{\prime}\subset U$ be a neighborhood of $p$ in $U$
  over which the Generalized Division Lemma \`{a} la Malgrange is applied to $f$
  on a neighborhood $U^{\prime}\times V^{\prime}$ of
    $(\{p\}\times V \cap \Supp(\tilde{\cal E}_{\varphi}))_{\redscriptsize}
          =:\{q_1,\,\cdots\,,\, q_s\}$ in $(X\times Y)/X$
  with respect to the characteristic polynomials
  $\chi_{\varphi}^{(i)} := \det(y^i\cdot\Id_{r\times r}-\varphi^{\sharp}(y^i))
   \in C^{\infty}(U^{\prime})[y^1,\,\cdots\,,\, y^n]\in C^{\infty}(U^{\prime}\times V)$,
  $i=1,\,\cdots\,,\, n$.
 Passing to a smaller open subset if necessary,
  one may assume that $V^{\prime}$ is a disjoint union
   $V^{\prime}_1\cup\,\cdots\,\cup V^{\prime}_s$
  with $U^{\prime}\times V^{\prime}_k$ a neighborhood of $q_k$
  and the closure $\overline{V_1},\,\cdots\,,\, \overline{V_s}$ are all disjoint from each other.
 Let  $1_{(k)}\in C^{\infty}(Y)$
  be a smooth functions on $Y$ that takes the value $1$ on $V^{\prime}_k$
  and the value zero on $V^{\prime}_{k^{\prime}}$, $k^{\prime}\ne k$, $k=1,\,\cdots\,,\, s$.
 (Cf.$\,$[L-Y8:$\,$Sec.$\,$4.2.3] (D(13.1)).)
 Then
  $$
   f|_{U^{\prime}\times V^{\prime}}\;
     =\;  \sum_{k=1}^s\, 1_{(k)}\cdot	
	        \Big(
	         \sum_{\scriptsizeboldd}c^{f;k}_{\scriptsizeboldd}\boldy^{\scriptsizeboldd}\;
			 +\; \sum_{i,j=1}^n
		               Q^{f;k}_{(i,j)}\chi_{\varphi}^{(i)}\chi_{\varphi}^{(j)}				
			   \Big)
  $$
   for some
   \begin{itemize}
    \item[\LARGE $\cdot$]
	  $c^{f;k}_{\scriptsizeboldd}\in C^{\infty}(U^{\prime})$ for all $k$ and $\boldd$,   and
	  $\sum_{\scriptsizeboldd}c^{f;k}_{\scriptsizeboldd}\boldy^{\scriptsizeboldd}
	     \in C^{\infty}(U^{\prime})[y^1,\,\cdots\,,\, y^n]$ for all $k$,
		
	\item[\LARGE $\cdot$]	
	  $Q^{f;k}_{(i,j)} \in C^{\infty}(U^{\prime}\times V^{\prime}_k)$
	   for all $k$ and $i,j$.
   \end{itemize}
 In terms of this, over $U^{\prime}$,
  $$
    R^f[1]\; =\; \sum_{k=1}^s \varphi^{\sharp}(1_{(k)})
	      \sum_{\scriptsizeboldd}c^{f;k}_{\scriptsizeboldd}\boldy^{\scriptsizeboldd}
  $$
  and
  $$
     \varphi^{\sharp}(f)\big|_{U^{\prime}}\;
	   =\; R^f[1]\big|
			    _{\scriptsizeboldy^{\tinyboldd}
 			                    \rightsquigarrow\,
								  \varphi^{\sharp}(\scriptsizeboldy^{\tinyboldd})\;
								 \mbox{\scriptsize for all multi-degree $\scriptsizeboldd$ in $R^f[1]$}}
  $$
 since
  $\;\tilde{\varphi}^{\sharp}(\chi_{\varphi}^{(i)})=0$ for $i=1,\,\cdots\,,\, n$,
  $\;\varphi^{\sharp}(f)=\tilde{\varphi}^{\sharp}(\pr_Y^{\sharp}(f))$.
  
 \bigskip
 
 \noindent
 {\it Step $(2)\;$
    $\{\varphi^{\sharp}(1_{(k)})\}_{k=1}^s$
	as the maximal complete set of orthogonal $D$-parallel idempotents$/U^{\prime}$}\\
 Since
 $\varphi^{\sharp}(f)$ depends only on the restriction of $f$, regarded  on $X\times Y$,
  to $\Supp(\tilde{\cal E}_{\varphi})$,
 one has
  $$
     \varphi^{\sharp}(1_{(1)}+\,\cdots\,+1_{(s)})\;
	  =\; \tilde{\varphi}^{\sharp}(1_{X\times Y})\;=\;  \Id_{{\cal E}|_{U^{\prime}}}
  $$
  over $U^{\prime}\subset X$.
 Since, in addition,
    $1_{(k)}^{\;2}=1_{(k)}$ for all $k$,
    $1_{(k)}\,1_{(k^{\prime})}=0$ for all $k\ne k^{\prime}$, and
	$s =$ the number of the connected components of $X_{\varphi}|_{U^{\prime}}$
	  for $U^{\prime}$ small enough,
 the collection
   $\{\varphi^{\sharp}(1_{(1)}),\,\cdots\,,\, \varphi^{\sharp}(1_{(s)})   \}$
  gives the maximal complete set of orthogonal idempotents in
   ${\cal A}_{\varphi}|_{U^{\prime}}$.

 Furthermore, since
  $D_{\xi}{\cal A}_{\varphi}\subset \Comm({\cal A}_{\varphi})$
    for all $\xi\in {\cal T}_{\ast}X$,
 $D_{\xi}\varphi^{\sharp}(1_{(k)})$ and $\varphi^{\sharp}(1_{(k^{\prime})})$
  commute for all $k, k^{\prime}=1,\,\cdots\,,\, s$.
 It follows from Lemma 1.6
 that
 $$
   D_{\xi}\varphi^{\sharp}(1_{(1)})\;
   =\; \cdots\;
   =\; D_{\xi}\varphi^{\sharp}(1_{(s)})\;=\;    0
 $$
 for all $\xi$.
 In other words,
   $\varphi^{\sharp}(1_{(1)}),\, \cdots\,,\, \varphi^{\sharp}(1_{(s)})$
  are $D$-parallel over $U^{\prime}$.
  
 \bigskip
 
 \noindent
 {\it Step $(3)\;$
         The evaluation of $(\varphi^{\diamond}(df))(\xi)$ over $U^{\prime}$}\hspace{2em}
 We are now ready to evaluate the ${\cal O}_X^{A\!z}$-valued derivation $D_{\xi}\varphi$ on $f$,
  locally and germwise over $U^{\prime}$.
 Step (1) and Step (2) together imply that, for $\xi\in {\cal T}_{\ast}X$ and over $U^{\prime}$,   
  \begin{eqnarray*}
   (\varphi^{\diamond}(df))(\xi)
	  & :=\: & D_{\xi}(\varphi^{\sharp}(f)) \\[.6ex]
	& =
      &  \sum_{k=1}^s\,\varphi^{\sharp}(1_{(k)})
              \sum_{\scriptsizeboldd}
			      (\xi c^{f;k}_{\scriptsizeboldd})\,\varphi^{\sharp}(\boldy^{\scriptsizeboldd})\;
		  +\; \sum_{k=1}^s\,\varphi^{\sharp}(1_{(k)}) 	
		         \sum_{\scriptsizeboldd}
				   c^{f;k}_{\scriptsizeboldd}\, D_{\xi}(\varphi^{\sharp}(\boldy^{\scriptsizeboldd}))
				   \\[.6ex]
    & =
      & \mbox{Term (I)}	\;+\; \mbox{Term (II)}
  \end{eqnarray*}
 since  $\;D_{\xi}\varphi^{\sharp}(1_{(k)})=0$ for $k=1,\,\cdots\,,\, s$.
 
 Note that
  $$
    \mbox{Term (II)}\;
	   =\; R^f[1]\big|
			    _{\scriptsizeboldy^{\tinyboldd}
 			                    \rightsquigarrow\,
								D_{\xi}
								   (\varphi^{\sharp}(\scriptsizeboldy^{\tinyboldd})),\;
								 \mbox{\scriptsize for all multi-degree $\scriptsizeboldd$ in $R^f[1]$}}\,.
  $$
  It remains to prove that Term (I) vanishes.
  But this is the situation studied in [L-Y8: Proposition 4.2.3.1:Proof] (D(13.1)).
  In essence, since
     $$
       f\;=\; \sum_{k=1}^s 1_{(k)}\cdot	
	               \Big(
				      \sum_{\scriptsizeboldd}
	                     c^{f;k}_{\scriptsizeboldd}\boldy^{\scriptsizeboldd}\;
					  +\; \sum_{i,j=1}^n
					           Q^{f;k}_{(i,j)}\chi_{\varphi}^{(i)}\chi_{\varphi}^{(j)}
					  \Big)					  	
     $$
       on $U^{\prime}\times V^{\prime}$   and
     $f$ on $(X\times Y)/X$ is independent of $X$,
   one has
    $$
      \mbox{Term (I)}\;
	    =\;  \varphi^{\sharp}\Big(\sum_{k}1_{(k)}
		         \sum_{\scriptsizeboldd}
				    \big(\xi  c^{f;k}_{\scriptsizeboldd}\big) \boldy^{\scriptsizeboldd}\Big)
		=\; \varphi^{\sharp}(\xi f)\;=\; 0\,.	
    $$
 Here we denote the canonical lifting of $\xi\in {\cal T}_{\ast}X$ to ${\cal T}_{\ast}(X\times Y)$,
    via the product structure of $X\times Y$,
   by the same notation $\xi$.
   
 This completes the proof.
 
\end{proof}

\bigskip

\begin{lemma}
{\bf [local expression of $\varphi^{\diamond}(df)$
           for $(\ast_1)$-admissible $(\varphi,\nabla)$, II]}$\;$
 Let $(\varphi,\nabla)$ be $(\ast_1)$-admissible.
 Continuing the setting and notations in Lemma~2.1.6.
										
 Then, locally,
  $$
     (\varphi^{\diamond}(df))(\xi)\;
     =\; \sum_{i=1}^n (\varphi^{\diamond}dy^i)(\xi)
	                \otimes \mbox{$\frac{\partial}{\partial y^i}$}f\;
	 =\; \sum_{i=1}^n
	          \Big(
			   D_{\xi}\varphi^{\sharp}(y^i)\,
			   \cdot\,
			   \varphi^{\sharp}\big(\mbox{\Large$\frac{\partial f}{\partial y^i}$}\big)
			  \Big).
  $$
  Here $\,\cdot\,$ is the multiplication in the ring $C^{\infty}(\End_{\Bbb C}(E))$
   (and will be omitted later when there is no sacrifice to clarity).
\end{lemma}

\smallskip

\begin{proof}
 This is a consequence of Lemma~2.1.6.
 Continuing the setup in the Statement and the proof thereof.
 Then,
  since $\tilde{\varphi}^{\sharp}(\chi_{\varphi}^{(i)})=0$
   for $i=1,\,\cdots\,,\, n$,
  one has
  $$
   \varphi^{\sharp}\big(\mbox{\Large$\frac{\partial}{\partial y^i}$}f \big)\;
   =\; \tilde{\varphi}^{\sharp}
          \big(\mbox{$\frac{\Large\partial}{\partial y^i}$} R^f[1] \big)\;
   =\; R^f[1]\big|
			    _{\scriptsizeboldy^{\tinyboldd}
 			           \rightsquigarrow\,
					   \varphi^{\sharp}
					    \big(\frac{\partial}{\partial y^i}\scriptsizeboldy^{\tinyboldd}\big),\;
								 \mbox{\scriptsize for all multi-degree $\scriptsizeboldd$ in $R^f[1]$}}\,.
  $$    		
 Since $(\varphi,\nabla)$	is $(\ast_1)$-admissible,
  $D_{\xi}\varphi^{\sharp}(y^i)$  and $\varphi^{\sharp}(y^j)$ commute
  for $i,j=1,\,\cdots\,,\, n.$
 It follows that
  $$
   \sum_{i=1}^n
     D_{\xi}\varphi^{\sharp}(y^i)
	  \cdot \varphi^{\sharp}(\mbox{$\frac{\partial}{\partial y^i}$}f)\;	
	 =\; R^f[1]\!\!\left.\rule{0ex}{0.8em}\right|
			    _{\scriptsizeboldy^{\tinyboldd}
 			                    \rightsquigarrow\,
								D_{\xi}
								   (\varphi^{\sharp}(\scriptsizeboldy^{\tinyboldd})),\;
								 \mbox{\scriptsize for all multi-degree $\scriptsizeboldd$ in $R^f[1]$}}\,.
  $$
 Which is $\varphi^{\diamond}(df)$ by Lemma~2.1.6.
 This proves the lemma.
 
\end{proof}

\bigskip

\noindent
{\it Proof of Theorem~2.1.5.}
 We now check in two steps that $\varphi^{\diamond}$ is well-defined.
 Note that we only need to do so locally over $X$.
 Thus,
   let $U\subset X$ be the open set in Lemma~2.1.6 such that
    $\varphi(U^{A\!z})$ is contained in a coordinate chart $V$ of $Y$,
    with the coordinate $(y^1,\,\cdots\,,\, y^n)$.
 Lemma~2.1.7 implies then
    that the following assignment is the restriction of $\varphi^{\diamond}$ to over $U$
            and hence is independent of the local coordinate $(y^1,\,\cdots\,,\, y^n)$ on $V\,$:
  $$
   \begin{array}{ccccc}
   \varphi^{\diamond}& : & \Omega_{C^{\infty}(V)}  & \longrightarrow
      & \Omega_{C^{\infty}(U)}
	       \otimes_{C^{\infty}(U)}C^{\infty}(\End_{\Bbb C}(E|_U))\\[.8ex]
	&& f_1df_2  & \longmapsto
	  & 	\varphi^{\sharp}(f_1)\cdot
	        \sum_{i=1}^n
	          \Big(
			   \varphi^{\sharp}\big(\mbox{\Large $\frac{\partial f_2}{\partial y^i}$}\big)\,
			   \cdot   D\varphi^{\sharp}(y^i)
			  \Big).	
   \end{array}	
  $$
 Here, we use again the fact that $(\varphi,\nabla)$ is $(\ast_1)$-admissible so that
  the summand
     $\;D\varphi^{\sharp}(y^i)
          \cdot \varphi^{\sharp}\big(\mbox{$\frac{\partial f}{\partial y^i}$}\big)$
    in Lemma~2.1.7 is equal to
	$\varphi^{\sharp}\big(\mbox{$\frac{\partial f_2}{\partial y^i}$}\big)
	            \cdot   D\varphi^{\sharp}(y^i)$ here, with $f$ replaced by $f_2$.
 It remains to show that
  $\varphi^{\diamond}$ is compatible with (a) the commutative Leibniz rule
  and (b) the chain-rule identities from the $C^{\infty}$-ring structure of $C^{\infty}(V)$.
  
 \bigskip

 \noindent
 {\it $(a)\;$ The commutative Leibniz rule} \hspace{2em}
 For $f_1, f_2\in C^{\infty}\in C^{\infty}(V)$, one has
   $$
     d(f_1f_2)\, -\, f_2\,df_1\,-\, f_1\,df_2\;=\; 0
   $$
   in $\Omega_{C^{\infty}(V)}$.
 Under $\varphi^{\diamond}$, one has
  $$
     \varphi^{\diamond}
	  \Big( d(f_1f_2)\;-\; f_2\,df_1\,-\, f_1\,df_2 \Big)\;\;
      =\;\;	D\,\varphi^{\sharp}(f_1f_2)\,
	              -\, \varphi^{\sharp}(f_2)\,D\varphi^{\sharp}(f_1)\,
                  -\, \varphi^{\sharp}(f_1)\,D\varphi^{\sharp}(f_2)\;\;
      =\;\; 0				
  $$
 since
  $$
    D \big(\varphi^{\sharp}(f_1f_2)\big)\;
	=\;  D \big(\varphi^{\sharp}(f_1)\varphi^{\sharp}(f_2)\big)\;
	=\; (D\varphi^{\sharp}(f_1))\varphi^{\sharp}(f_2)\,
	         +\, \varphi^{\sharp}(f_1)D\varphi^{\sharp}(f_2)\,,
  $$
   which is
    $$
        \varphi^{\sharp}(f_2)	   D\varphi^{\sharp}(f_1)
	          +\, \varphi^{\sharp}(f_1)D\varphi^{\sharp}(f_2)
    $$
	for $(\varphi, \nabla)$ $(\ast_1)$-admissible.
   
 \bigskip
 
 \noindent
 {\it $(b)\;$  The chain-rule identities from the $C^{\infty}$-ring structure}\hspace{2em}
 Let
   $\zeta\in C^{\infty}({\Bbb R}^l)$, $l\in {\Bbb Z}_{\ge 1}$ and
   $f_1,\,\cdots\,,\,f_l\in C^{\infty}(V)$.
 Then, one has
   $$
     d \big(\zeta(f_1,\,\cdots\,,\, f_l)\big)\,
	   -\,  \sum_{k=1}^l(\partial_k\zeta)(f_1,\,\cdots\,,\, f_l)\,df_k\;=\; 0
   $$
  in $\Omega_{C^{\infty}(V)}$.
 Here, $\partial_k\zeta$  is the partial derivative of $\zeta\in C^{\infty}({{\Bbb R}^l})$
   with respect to its $k$-th argument.
 Under $\varphi^{\diamond}$, one has
   \begin{eqnarray*}
    \lefteqn{
	  \varphi^{\diamond} \Big(
	      d \big(\zeta(f_1,\,\cdots\,,\, f_l)\big)\,
	              -\,  \sum_{k=1}^l  (\partial_k\zeta)(f_1,\,\cdots\,,\, f_l)\,df_k
				                                \Big)
   												} \\[-1ex]
     &&	=\;\;
              D\varphi^{\sharp}\big(\zeta(f_1,\,\cdots\,,\, f_l)\big)\,
                   -\, \sum_{k=1}^l	
                           \varphi^{\sharp}\big((\partial_k\zeta)(f_1,\,\cdots\,,\, f_l)\big)\,
						    D\varphi^{\sharp}(f_k)  \;\;=\;\;0
   \end{eqnarray*}
 since, by Lemma~2.1.7 and the $(\ast_1)$-admissibility of $(\varphi,\nabla)$,
  %
  \begin{eqnarray*}
    \lefteqn{
	  D\varphi^{\sharp}\big(\zeta(f_1,\,\cdots\,,\, f_l)\big)\;\;
	    =\;\;  \sum_{i=1}^n
		               D\varphi^{\sharp}(y^i)
		                  \otimes  \mbox{\Large $\frac{\partial}{\partial y^i}$}\zeta(f_1,\,\cdots\,,\,f_l)		
	              }\\
   && =\;\; 				
       \sum_{i=1}^n
		    D\varphi^{\sharp}(y^i)
		        \otimes \sum_{k=1}^l
     				 (\partial_k\zeta  )(f_1,\,\cdots\,,\, f_l)\,
					     \mbox{\Large $\frac{\partial f_k}{\partial y^i}$}   	   \\
    &&=\;\;
      \sum_{k=1}^l D\varphi^{\sharp}(f_k)\otimes (\partial_k\zeta)(f_1,\,\cdots\,,\, f_l)	\\
    && =\;\;
	   \sum_{k=1}^l D\varphi^{\sharp}(f_k)
                    	 \varphi^{\sharp}\big(  (\partial_k\zeta)(f_1,\,\cdots\,,\, f_l)\big)\;\;
	   =\;\;   \sum_{k=1}^l
	                \varphi^{\sharp}\big(  (\partial_k\zeta)(f_1,\,\cdots\,,\, f_l)\big)\,
	                    D\varphi^{\sharp}(f_k)\,.
  \end{eqnarray*}
                
This completes the proof of Theorem~2.1.5

\noindent\hspace{40.8em}$\square$

\bigskip

\begin{flushleft}
{\bf The pull-push $\varphi^{\diamond}$ on
         tensor product
           $\Omega_{C^{\infty}(Y)}\otimes_{C^{\infty}(Y)}\,\cdots\,
               \otimes_{C^{\infty}(Y)}		 \Omega_{C^{\infty}(Y)}$}
\end{flushleft}
Having the well-defined
 $$
   \varphi^{\diamond}\; :\;  \Omega_{C^{\infty}(Y)}\;
   \longrightarrow\;
   \Omega_{C^{\infty}(X)}\otimes_{C^{\infty}(X)}C^{\infty}(\End_{\Bbb C}(E))\,,
 $$
 it is natural to consider the  extension of  $\varphi^{\diamond}$ to a correspondence between tensor products
 $$
  \begin{array}{cccccl}
   \varphi^{\diamond}& :
     & \otimes_{C^{\infty}(Y)}^k \Omega_{C^{\infty}(Y)}
	 & \longrightarrow
     & \big(\otimes_{C^{\infty}(X)}^k \Omega_{C^{\infty}(X)}\big)
	         \otimes_{C^{\infty}(X)}C^{\infty}(\End_{\Bbb C}(E)) \\[1.2ex]
    &&  f_0\,df_1 \otimes\,\cdots\,\otimes df_k	
      & \longmapsto
	    & \varphi^{\sharp}(f_0)
		    D\varphi^{\sharp}(f_1)	 \otimes\,\cdots\, \otimes D\varphi^{\sharp}(f_k)    &.
  \end{array}			
 $$
Here, the tensor
 $D\varphi^{\sharp}(f_1)	 \otimes\,\cdots\, \otimes D\varphi^{\sharp}(f_k)  $
 is defined to the tensors of the underlying $1$-forms in $\Omega_{C^{\infty}(X)}$ and
 multiplication of the coefficients in $C^{\infty}(\End_{\Bbb C}(E))$ from each factor.
Explicitly, in terms of a local coordinate $(x^1,\,\cdots\,,\,x^m)$ on a  chart $U\subset X$,
 \begin{eqnarray*}
   \lefteqn{
    \varphi^{\sharp}(f_0)
		D\varphi^{\sharp}(f_1)	 \otimes\,\cdots\, \otimes D\varphi^{\sharp}(f_k)   }\\
     && =\; \sum_{\mu_1,\,\cdots\,,\, \mu_k=1}^m
                   \varphi^{\sharp}(f_0)
			          D _{\partial/\partial x^{\mu_1}}\varphi^{\sharp}(f_1)\,
                      \cdots\,
				     D_{\partial /\partial x^{\mu_k}}\varphi^{\sharp}(f_k)\,
				     dx^{\mu_1}\otimes\,\cdots\, \otimes dx^{\mu_k}\,.
 \end{eqnarray*}			
    
\bigskip

\begin{lemma} {\bf [pull-push of (covariant) tensor]}$\;$
 For $(\varphi,\nabla)$ $(\ast_1)$-admissible,
 the above extension of $\varphi^{\diamond}$ to covariant tensors is well-defined.
\end{lemma}

\smallskip

\begin{proof}
 In $\otimes_{C^{\infty}(Y)}^k \Omega_{C^{\infty}(Y)}$, one has the identities
  $$
    \begin{array}{l}
     f_0\,df_1 \otimes df_2\otimes \,\cdots\,\otimes df_k\;\;
	   =\;\;   df_1 f_0 \otimes df_2\otimes\,\cdots\,\otimes df_k \\[.8ex]
	\hspace{2em}
      =\;\; 	 df_1 \otimes f_0df_2\otimes\,\cdots\,\otimes df_k\;\;
	  =\;\;   \cdots\cdots      \\[.6ex]
	\hspace{2em}
	  =\;\;  df_1 \otimes\,\cdots\,\otimes df_{k-1}f_0 \otimes df_k\;\;	
      =\;\;  df_1 \otimes\,\cdots\,\otimes f_0 df_k\;\;	
	  =\;\;  df_1 \otimes\,\cdots\,\otimes df_k	f_0\,.
    \end{array}	
  $$
 Since the $(\ast_1)$-Admissible Condition implies that
  $\varphi^{\sharp}(f_0)$ commutes with all of
     $D\varphi^{\sharp}(f_1)$, $\cdots\,$, $D\varphi^{\sharp}(f_k)$,
 the parallel identities	
   $${\small
    \begin{array}{l}
     \varphi^{\sharp}(f_0)D\varphi^{\sharp}(f_1)
	     \otimes D\varphi^{\sharp}(f_2)\otimes \,\cdots\,\otimes D\varphi^{\sharp}(f_k)\;\;
	   =\;\;   D\varphi^{\sharp}(f_1)\varphi^{\sharp}(f_0)
	                \otimes D\varphi^{\sharp}(f_2)
					             \otimes\,\cdots\,\otimes D\varphi^{\sharp}(f_k) \\[1.2ex]
	\hspace{2em}
      =\;\;  D\varphi^{\sharp}(f_1)
	           \otimes \varphi^{\sharp}(f_0)D\varphi^{\sharp}(f_2)
			      \otimes\,\cdots\,\otimes D\varphi^{\sharp}(f_k)\;\;
	  =\;\;   \cdots\cdots      \\[1.2ex]
	\hspace{2em}
	  =\;\;  D\varphi^{\sharp}(f_1) \otimes\,\cdots\,
	                 \otimes D\varphi^{\sharp}(f_{k-1})\varphi^{\sharp}(f_0)
					  \otimes D\varphi^{\sharp}(f_k)\;\;	     \\[1.2ex]
    \hspace{2em}					
      =\;\;  D\varphi^{\sharp}(f_1) \otimes\,\cdots\,
	              \otimes\varphi^{\sharp}(f_0) D\varphi^{\sharp}(f_k)\;\;	
	  =\;\;  D\varphi^{\sharp}(f_1) \otimes\,\cdots\,
	              \otimes D\varphi^{\sharp}(f_k)\varphi^{\sharp}(f_0)
    \end{array}}
  $$
  hold in
  $\big(\otimes_{C^{\infty}(X)}^k \Omega_{C^{\infty}(X)}\big)
   	          \otimes_{C^{\infty}(X)}C^{\infty}(\End_{\Bbb C}(E))$.
 This proves the lemma.		

\end{proof}

\bigskip

Note that for a $(\ast_1)$-admissible map $\varphi:(X^{\!A\!z},{\cal E};\nabla)\rightarrow Y$,
 since $D_{\xi}{\cal A}_{\varphi}\subset Comm({\cal A}_{\varphi})$
     for all $\xi\in{\cal T}_{\ast}X$
 and $\Comm({\cal A}_{\varphi})$ is itself a (possibly noncommutative)
  ${\cal O}_X^{\,\Bbb C}$-subalgebra of ${\cal O}_X^{A\!z}$,
 the pull-push $\varphi^{\diamond}\alpha$ of a (covariant)    tensor $\alpha$ on $Y$ to $X$
  is indeed $\Comm({\cal A}_{\varphi})$-valued.
  
\bigskip

\begin{example} {\bf  [pull-push of $2$-tensor under $(\ast_1)$-admissible $(\varphi,\nabla)$]}$\;$
{\rm
 Let $\varphi:(X^{\!A\!z},{\cal E};\nabla)\rightarrow Y$ be a $(\ast_1)$-admissible map
   and $\alpha=\sum_{i,j}\alpha_{ij}dy^i\otimes y^j$ be a $2$-tensor on $Y$.
 Then,
   with respect to local coordinates $(x^1,\,\cdots\,,\, x^n)$ on $X$ and $(y^1,\,\cdots\,,\, y^n)$ on $Y$,
   $$
     \varphi^{\diamond}\alpha\;
	  =\;   \sum_{\mu,\nu=1}^m \Big(
	                \sum_{i,j=1}^n	
	                 \varphi^{\sharp}(\alpha_{ij})
					    D_{\frac{\partial}{\partial x^{\mu}}}\varphi^{\sharp}(y^i)
						D_{\frac{\partial}{\partial x^{\nu}}}\varphi^{\sharp}(y^j)  \Big)
						dx^{\mu}\otimes dx^{\nu}\,.
   $$
 Since in general
   $D_{\partial/\partial x^{\mu}}\varphi^{\sharp}(y^i)
     D_{\partial/\partial x^{\nu}}\varphi^{\sharp}(y^j)
	  \ne   D_{\partial/\partial x^{\nu}}\varphi^{\sharp}(y^j)
			   D_{\partial/\partial x^{\mu}}\varphi^{\sharp}(y^i)$,
   $\varphi^{\diamond}$ does {\it not} take a symmetric $2$-tensor on $Y$
    to a $\Comm({\cal A}_{\varphi})$-valued symmetric $2$-tensor on $X$,
	nor an antisymmetric $2$-tensor on $Y$
       to a $\Comm({\cal A}_{\varphi})$-valued antisymmetric $2$-tensor on $X$.
 However,  after the post-composition with the trace map
   $\Tr:{\cal O}_X^{\!A\!z}\rightarrow {\cal O}_X^{\,\Bbb C}$,
  $\Tr\varphi^{\diamond}$
   does take a symmetric (resp.\ antisymmetric) $2$-tensor on $X$
    to an ${\cal O}_X^{\,\Bbb C}$-valued symmetric (resp.\ antisymmetric) $2$-tensor on $X$.
}\end{example}

\medskip

\begin{example} {\bf
 [pull-push of higher-rank tensor under $(\ast_1)$-admissible $(\varphi,\nabla)$]}$\;$ {\rm
 Continuing Example~2.1.9.
 For $\alpha$ a (covariant) tensor on $Y$ of rank $\ge 3$,
  the trace map no longer help bring symmetric (resp.\ antisymmetric) tensors
  to symmetric (resp.\ antisymmetric) tensors.
}\end{example}

\bigskip

The situation gets better for a map $\varphi:(X^{\!A\!z},{\cal E};\nabla)\rightarrow Y$
  that satisfies the stronger $(\ast_2)$-Admissible Condition:
  $D_{\xi}\Comm({\cal A}_{\varphi})\subset \Comm({\cal A}_{\varphi})$
   for $\xi\in{\cal T}_{\ast}X$.
   
\bigskip

\begin{lemma} {\bf  [pull-push of tensor under $(\ast_2)$-admissible $(\varphi,\nabla)$]}$\;$
 Let $\varphi:(X^{\!A\!z},{\cal E};\nabla)\rightarrow Y$
  be a $(\ast_2)$-admissible map.
 Then $\varphi^{\diamond}$ takes  a symmetric (resp.\ antisymmetric) tensor on $Y$
  to a $\Comm({\cal A}_{\varphi})$-valued symmetric (resp.\ antisymmetric) tensor on $X$.
\end{lemma}

\smallskip

\begin{proof}
 In terms of local coordinates $(x^1,\,\cdots\,,\, x^n)$ on $X$ and $(y^1,\,\cdots\,,\, y^n)$ on $Y$,
 a (covariant) $k$-tensor
 $$
   \alpha
     = \sum_{i_1,\,\cdots\,i_k}
           \alpha_{i_1\,\cdots\,i_k}dy^{i_1}\otimes\,\cdots\,\otimes dy^{i_k}
 $$
  on $Y$ is pull-pushed to a $\Comm({\cal A}_{\varphi})$-valued $k$-tensor
  {\small
 $$
     \varphi^{\diamond}\alpha\;
	  =\;   \sum_{\mu_1,\,\cdots\,,\, \mu_k=1}^m \Big(
	                \sum_{i_1,\,\cdots\,,\, i_k=1}^n	
	                 \varphi^{\sharp}(\alpha_{i_1\,\cdots\,i_k})
					    D_{\frac{\partial}{\partial x^{\mu_1}}}\varphi^{\sharp}(y^{i_1})
						 \,\cdots\,
						D_{\frac{\partial}{\partial x^{\mu_k}}}\varphi^{\sharp}(y^{i_k})  \Big)
						dx^{\mu_1}\otimes\,\cdots\,\otimes dx^{\mu_k}\,.
 $$}
 on $X$.
 It follows from Lemma~2.1.4
  that, under the $(\ast_2)$-Admissible Condition, all the factors
  $$
    \varphi^{\sharp}(\alpha_{i_1\,\cdots\,i_k})\,,\;\;
		D_{\frac{\partial}{\partial x^{\mu_1}}}\varphi^{\sharp}(y^{i_1})\,,\;\;
						\cdots\,,\;\;
		D_{\frac{\partial}{\partial x^{\mu_k}}}\varphi^{\sharp}(y^{i_k})
   $$
  in a summand commute among themselves.
 This implies, in particular, that
   $\varphi^{\diamond}$ now takes  a symmetric (resp.\ antisymmetric) tensor on $Y$
  to a $\Comm({\cal A}_{\varphi})$-valued symmetric (resp.\ antisymmetric) tensor on $X$.

\end{proof}

\bigskip

Let $\bigwedge^{\mbox{\tiny $\bullet$}}{\cal T}^{\ast}Y$ be the sheaf of differential forms on $Y$.
The same proof of Lemma~2.1.11
  gives also

\bigskip

\begin{lemma} {\bf [$\varphi^{\diamond}$ and $\wedge$]}$\;$
 Let $\varphi:(X^{\!A\!z},{\cal E};\nabla)\rightarrow Y$ be a $(\ast_2)$-admissible map.
 For $\alpha,\beta\in \bigwedge^{\mbox{\tiny $\bullet$}}{\cal T}^{\ast}Y$,
  define the wedge product
   $$
      \varphi^{\diamond}\alpha \wedge \varphi^{\diamond}\beta
   $$
  of $\varphi^{\diamond}\alpha$,
      $\varphi^{\diamond}\beta
	     \in (\bigwedge^{\mbox{\tiny $\bullet$}}{\cal T}^{\ast}X)^{\Bbb C}
		              \otimes_{{\cal O}_X^{\,\Bbb C}}{\cal O}_X^{A\!z} $
   by applying the wedge product to the differential forms on $X$
        and multiplication to the ${\cal O}_X^{A\!z}$-valued coefficients.
 Then,
  $$
      \varphi^{\diamond} (\alpha\wedge \beta)\;
	   =\;  (\varphi^{\diamond}\alpha)\wedge (\varphi^{\diamond}\beta)\,.
  $$
\end{lemma}

%
%
%

\bigskip

\begin{remark} $[$admissible condition and Ramond-Ramond field$\,]\;$ {\rm
 While the current note will take $(\varphi,\nabla)$ to be $(\ast_1)$-admissible most of the time,  
 Example~2.1.9, Example~2.1.10, Lemma~2.1.11 and Lemma~2.1.12
  together suggest that
  when the coupling of D-brane to Ramond-Ramond fields is taken into account,
  the more natural admissible condition on $(\varphi,\nabla)$ is the stronger
   $(\ast_2)$-Admissible Condition.
}\end{remark}

\bigskip

\subsection{Admissible conditions from the aspect of open strings}

We address in this subsection
 the implication of Admissible Condition $(\ast_1)$ on $(\varphi,\nabla)$
  to the mass of the connection $\nabla$
  from the aspect of open strings in the target-space $Y$.
  
Let $\varphi:(X^{\!A\!z},{\cal E};\nabla)\rightarrow Y$  be a $(\ast_1)$-admissible map.
Recall the surrogate $X_{\varphi}:= \boldSpec^{\Bbb R}{\cal A}_{\varphi}$
  of $X^{\!A\!z}$ specified by $\varphi$ and
  the built-in dominant morphism $\pi_{\varphi}:X_{\varphi}\rightarrow X$;
 cf.$\,${\sc Figure} 1-2.
For $x\in X$,
  let $\{e_1,\,\cdots\,,\, e_s\}\subset {\cal A}_{\varphi,\, x}$
  the maximal complete set of orthogonal idempotents in the stalk of ${\cal A}_{\varphi}$ at $x$.
Then, by $(\ast_1)$-admissibility of $(\varphi,\nabla)$ and Lemma~1.6,
   $$
      D_{\xi}e_1\;=\; \cdots\cdots \;=\; D_{\xi}e_s\;=\; 0
   $$
  for all $\xi\in ({\cal T}_{\ast}X)_x$.
It follows that

\bigskip

\begin{lemma} {\bf [covariantly invariant decomposition of stalks of ${\cal E}$]}$\;$
 For any $x\in X$, the decomposition
   $$
     {\cal E}_x\;=\;   e_1{\cal E}_x  + \,\cdots\, + e_s{\cal E}_x
   $$
 is invariant under $\nabla$.
 I.e.\  $\nabla_{\xi}(e_k{\cal E}_x)\subset e_k{\cal E}_x$,
   for $k=1,\,\cdots\,,\, s$ and all $\xi\in ({\cal T}_{\ast}X)_x$.
\end{lemma}
  
\bigskip

As a consequence,
 the connection $\nabla$ on ${\cal E}_x$ induces a connection $\nabla^{(k)}$
   on each direct summand $e_k{\cal E}_x$ of ${\cal E}_x$    and
  one has the direct-sum decomposition
  $$
   ({\cal E}_x, \nabla)\;
      =\; 	(e_1{\cal E}_x, \nabla^{(1)})
	           \oplus \,\cdots\, \oplus (e_s{\cal E}_x, \nabla^{(s)})\,.
  $$

On the other hand,
 the maximal complete set of orthogonal idempotents
 $\{e_1,\,\cdots\,,\, e_s\}\subset {\cal A}_{\varphi,\,x}$
 corresponds canonically and bijectively to the set of connected components of the germ $X_{\varphi,\, x}$
 of $X_{\varphi}$ over $x\in X$:
 $$
   X_{\varphi,\,x}\;=\;
     X_{\varphi,\,x}^{(1)}\sqcup\,\cdots\, \sqcup X_{\varphi,\,x}^{(s)}\,.
 $$
Through the built-in inclusion ${\cal A}_{\varphi,\,x}\subset {\cal O}_{X\,x}^{A\!z}$,
 ${\cal E}_x$ as the fundamental ${\cal O}_{X,\,x}^{A\!z}$-module is canonically
 an ${\cal A}_{\varphi,\,x}$-module as well.
Since $e_ke_l=0$ for $k\ne l$,
 as an ${\cal A}_{\varphi,\,x}$-module
 the direct summand $e_k{\cal E}_x$ is supported exactly on $X_{\varphi,\,x}^{(k)}$,
 for $k=1,\,\cdots\,,\, s$.
The above decomposition of $({\cal E},\nabla)$ says then
  geometrically and in terms of physics terminology
 that the gauge field $\nabla$ on ${\cal E}$ has no components that mixes
     $e_k{\cal E}_x$ on $X_{\varphi,\,x}^{(k)}$  and
  	 $e_l{\cal E}$ on  $X_{\varphi,\,x}^{(l)}$
	for some $k\ne l$;
cf.\ {\sc Figure}~2-2-1.
%

Recall now the string-theory origin of D-branes:
 \begin{itemize}
  \item[\LARGE $\cdot$]
   A D-brane is where the end-points of an open string stick to.
  
  \vspace{-1.6ex}
  \item[\LARGE $\cdot$]
   Excitations of open strings create fields on the D-brane.
   
  \vspace{-1.6ex}
  \item[\LARGE $\cdot$]
  As the tension of open strings are constant, the mass of an open string
    --- and hene fields it creates on  the D-brane--- is proportional to its length.
  Open strings with arbitrarily small length create massless fields on the brane
    while open strings with length bounded away from zero  create massive fields on the brane.
 \end{itemize}
That  the germ $(X_{\varphi,\,x}, {\cal E}_x; \nabla)$ over any $x\in X$
 is decomposable in accordance with the connected-component decomposition of $X_{\varphi,\,x}$
 says that
 $\nabla$ must be created by open-strings of arbitrarily small length,
 rather than by those of length bounded away from zero.
In other words, $\nabla$ is massless.

In summary:
 
\bigskip

\begin{corollary} {\bf [$(\ast_1)$-Admissible Condition implies massless of $\nabla$]}$\;$
 For a $(\ast_1)$-admissible map
  $\varphi:(X^{\!A\!z},{\cal E}; \nabla)\rightarrow Y$,
  the gauge field $\nabla$ on the Chan-Paton sheaf ${\cal E}$ on the D-brane
   (or D-brane world-volume)  $X^{\!A\!z}$
   is massless
   from the aspect of open strings in the taget-space (or target-space-time) $Y$.
\end{corollary}


\begin{figure}[htbp]
 \bigskip
  \centering
  \includegraphics[width=0.80\textwidth]{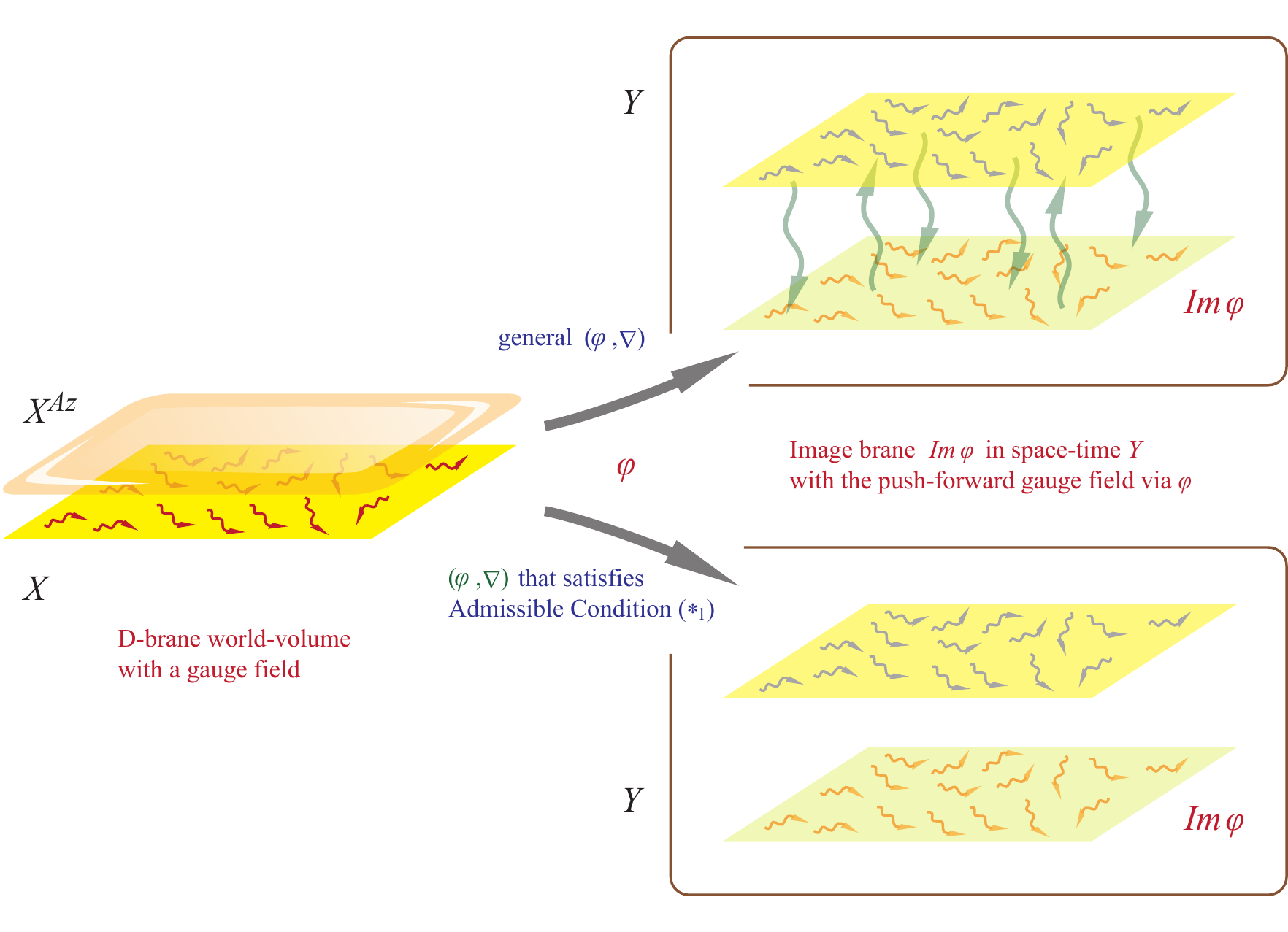}
 
  \bigskip
  \bigskip
 \centerline{\parbox{13cm}{\small\baselineskip 12pt
  {\sc Figure}~2-2-1.
 When $(\varphi,\nabla)$ is $(\ast_1)$-admissible,
  the gauge field $\nabla$ on the Chan-Paton sheaf ${\cal E}$ on any small neighborhood $U$ of $x\in X$
   localizes at each connected branch of $\varphi(U^{\!A\!z})$ from the viewpoint of open strings in $Y$.
 In other words, $\nabla$ is massless from the open-string aspect.
 In the illustration,
  the noncommutative space $X^{\!A\!z}$ is expressed as a noncommutative cloud shadowing
  over its underlying topology $X$,
  the connection $\nabla$ on $E$ over $X$  is indicated by a gauge field on $X$.
 Both the gauge field on $X$   and how open strings ``see" it in $Y$
  are indicated by squiggling arrows $\rightsquigarrow$.
 The situation for a general $(\varphi,\nabla)$  (cf.\ top)
  and a $(\varphi,\nabla)$ satisfying Admissible Condition $(\ast_1)$ (cf.\ bottom)
  are compared.
 From the open-string aspect, in the former situation $\nabla$ can have both massless components
  (which are local fields from the open-string and target-space viewpoint) and massive components
  (which become nonlocal fields from the open-string and target-space viewpoint),
  while in the latter situation $\nabla$ has only massless components.
  }}
\end{figure}		

\bigskip

By Lemma~2.1.2,
   the same holds for $(\ast_2)$-admissible maps and $(\ast_3)$-admissible maps as well.

\bigskip

\section{The differential $d\varphi$ of $\varphi$ and its decomposition,
 the three basic ${\cal O}_X^{\,\Bbb C}$-modules, induced structures, and some covariant calculus}

At the classical level Polyakov string or its generalization, a sigma model,
 is a theory of harmonic maps on the mathematical side.
In this section we construct all the building blocks to generalize the existing theory of harmonic maps
 to a theory of maps $\varphi:(X^{\!A\!z},{\cal E};\nabla)\rightarrow Y$, which describe D-branes.
It will turn out that both the connection $\nabla$ and the Admissible Condition $(\ast_1)$ chosen
 are needed to build up a mathematically sound theory for such maps $\varphi$.

\bigskip

\subsection{The differential $d\varphi$ of $\varphi$ and its decomposition induced by $\nabla$}

Three kinds of differentials, $d\varphi$, $D\varphi$, and $\ad\varphi$,   of a map $\varphi$
    that naturally appear in the setting are defined
and their local expressions are worked out in this subsection.

\bigskip

\begin{flushleft}
{\bf The differential, the covariant differential, and the inner differential of a map $\varphi$}
\end{flushleft}
Let
 $$
    \varphi\;:\;(X, {\cal O}_X^{A\!z},\,{\cal E})\; \longrightarrow\; Y
 $$
 be a map defined contravariantly by an equivalence class of gluing systems of ring-homomorphisms
 $$
   \varphi^{\sharp}\; :\; {\cal O}_Y\; \longrightarrow\; {\cal O}_X^{A\!z}
 $$
 over ${\Bbb R}\subset {\Bbb C}$.
Then, for any derivation $\eta$ on ${\cal O}_X^{A\!z}$,
  the correspondence
  $$
   \begin{array}{cccc}
    {\cal O}_Y      & \longrightarrow    & {\cal O}_X^{A\!z} \\[.6ex]
        f    & \longmapsto    & \eta (\varphi^{\sharp}(f))
   \end{array}
  $$
  defines an ${\cal O}_X^{A\!z}$-valued derivation on ${\cal O}_Y$.
It follows that
 $\varphi$ induces a correspondence
 $$
  \begin{array}{ccc}
  {\cal T}_{\ast}X^{\!A\!z}
     & \longrightarrow
	 & {\cal O}_X^{A\!z}\otimes_{\varphi^{\sharp},{\cal O}_Y}\!\!{\cal T}_{\ast}Y \\[.6ex]
   \eta	 & \longmapsto   & d_{\eta}\varphi
  \end{array}
 $$
 that is ${\cal O}_X^{\,\Bbb C}$-linear.
  
\bigskip

\begin{definition} {\bf [differential $d\varphi$ of $\varphi$]}$\;$ {\rm
 The above ${\cal O}_X^{\,\Bbb C}$-linear correspondence is denoted by $d\varphi$
   and called the {\it differential  of $\varphi$}.
}\end{definition}
   
\bigskip

Recall from Sec.$\,$1 that when ${\cal E}$ is equipped with a connection $\nabla$,
 $\nabla$ induces a connection $D$ on
 ${\cal O}_X^{A\!z}:=\Endsheaf_{{\cal O}_X^{\,\Bbb C}}({\cal E})$,
 which in turn induces a splitting
  $$
    \begin{array}{ccc}
    {\cal T}_{\ast}X^{\Bbb C}  &  \longrightarrow   & {\cal T}_{\ast}X^{\!A\!z}\\[.8ex]
	   \xi            & \longmapsto   & D_{\xi}
	\end{array}
  $$
 of the exact sequence
 $$
   0\; \longrightarrow\;      \Inn({\cal O}_X^{A\!z})\;
         \longrightarrow\;   {\cal T}_{\ast}X^{\!A\!z}\;
		 \longrightarrow\;   {\cal T}_{\ast}X^{\Bbb C}\;          \longrightarrow\; 0 \,.
 $$
 
\bigskip

\begin{definition} {\bf [covariant differential $D\varphi$ of $\varphi$]}$\;$ {\rm	
 Let
   $$
     \varphi^{\ast}{\cal T}_{\ast}Y\;
	   :=\;   {\cal O}_X^{A\!z}\otimes_{\varphi^{\sharp},{\cal O}_Y}{\cal T}_{\ast}Y\,,
   $$
     regarded as a (left) ${\cal O}_X$-module
     via the built-in inclusion ${\cal O}_X\hookrightarrow {\cal O}_X^{A\!z}$,
   be the {\it pull-push} of the tangent sheaf ${\cal T}_{\ast}Y$ of $X$ to $X$.
 The {\it covariant differential}
  $$
     D\varphi \in C^{\infty}(
           {\cal T}^{\ast}X\otimes_{{\cal O}_X}\varphi^{\ast}{\cal T}_{\ast}Y    )
  $$
  of $\varphi$ is the (${\cal O}_X^{A\!z}$-valued-derivation-on-${\cal O}_Y$)-valued 1-form on $X$  
   defined by
   $$
     (D_{\xi}\varphi) f\;   :=\;   D_{\xi}(\varphi^{\sharp}(f))\;
	   \in\; C^{\infty}(\End_{\Bbb C}(E))
   $$
   for $\xi\in C^{\infty}({\cal T}_{\ast}X)=\Der(C^{\infty}(X))$ and $f\in C^{\infty}(Y)$.
 In other words, $D\varphi$ takes a tangent vector field on $X$
  to a $C^{\infty}(\End_{\Bbb C}(E))$-valued derivation on $C^{\infty}(Y)$.
 In the equivalent sheaf format and notations,
   $D_{\xi}\varphi \in \varphi^{\ast}{\cal T}_{\ast}Y$
   for $\xi\in {\cal T}_{\ast}X$.
}\end{definition}

\medskip

\begin{definition} {\bf [inner differential $\ad\varphi$ of $\varphi$]}$\:$ {\rm
 Continuing Definition~3.1.2.
 Represent elements in $\Inn({\cal O}_X^{A\!z})$ by elements $m\in {\cal O}_X^{A\!z}$.
 The {\it inner differential $\ad\varphi$ of $\varphi$}
  is defined by the ${\cal O}_X^{\,\Bbb C}$-linear correspondence
   $$
     \begin{array}{cccccc}
	  ad\varphi  & : & \Inn({\cal O}_X^{A\!z})   & \longrightarrow
	    & \varphi^{\ast}{\cal T}_{\ast}Y 	 \\[.6ex]
	 && m  & \longmapsto
	    &  \ad_m\varphi\;
	        :=\; [m\,,\, \varphi^{\sharp}(\mbox{\Large $\cdot$}\,)] &.
	 \end{array}
   $$
 Since $[{\cal O}_X^{\,\Bbb C},\,\mbox{\Large $\cdot$}\,]=0$,
   $\ad_m\varphi$ depends only on the inner derivation $m$ represents.
}\end{definition}

\bigskip

By construction,
 $$
   d_{\eta}\varphi\; =\; D_{\xi_{\eta}} \,+\, ad_{m_{\eta}}\varphi\,,
 $$
 for $\eta = \xi_{\eta}+m_{\eta}
         \in {\cal T}_{\ast}X^{\!A\!z}
		 \simeq {\cal T}_{\ast}X^{\Bbb C}\oplus \Inn ({\cal O}_X^{A\!z})$
   induced by $D$.

\bigskip

\begin{flushleft}
{\bf Local expressions of the covariant differential $D\varphi$ of $\varphi$
          for $(\ast_1)$-admissible $(\varphi,\nabla)$}
\end{flushleft}
Note that
 if $\varphi:(X^{\!A\!z},{\cal E};\nabla)\rightarrow Y$ is $(\ast_1)$-admissible,
 then $D_{\xi}\varphi$ is a $\Comm({\cal A}_{\varphi})$-valued derivation on ${\cal O}_Y$
  for $\xi\in{\cal T}_{\ast}X$.
In this case, one has the following lemma and corollary that are simply re-writings of
 Lemma~2.1.6 and Lemma~2.1.7 respectively.
						
\bigskip

\begin{lemma}
{\bf [local expression of $D\varphi$ for $(\ast_1)$-admissible $(\varphi, \nabla)$, I]}$\;$
 Let $(\varphi, \nabla)$ be $(\ast_1)$-admissible;
   i.e.\ $D_{\xi}{\cal A}_{\varphi}
              \subset \Comm({\cal A}_{\varphi})$ for all $\xi\in{\cal T}_{\ast}X$.
 Let  $U\subset X$ be a small enough open set
   so that $\varphi(U^{A\!z})$ is contained in a coordinate chart of $Y$,
   with coordinates $\boldy =(y^1, \,\cdots\,,\, y^n)$.
 For $f\in C^{\infty}(Y)$,
  recall the germwise-over-$U$ polynomial $R^f[1]$ in $(y^1,\,\cdots\,,\, y^n)$
   with coefficients in ${\cal O}_U^{A\!z}$ from
   {\rm $\,$[L-Y8: Sec.\ 4 \& Remark/Notation 4.2.3.5] (D.13.1)}.
 Then, for $\xi$ a vector field on $U$ and $f\in C^{\infty}(Y)$, and at the level of germs over $U$,
  $$
     (D_{\xi}\varphi)\,f\;
	   =\; R^f[1]\!\!\left.\rule{0ex}{0.8em}\right|
			    _{\scriptsizeboldy^{\tinyboldd}
 			                    \rightsquigarrow\,
								D_{\xi}
								   (\varphi^{\sharp}(\scriptsizeboldy^{\tinyboldd})),\;
								 \mbox{\scriptsize for all multi-degree $\scriptsizeboldd$ in $R^f[1]$}}\,.
  $$
 Here,
  for a multiple degree $\boldd =(d_1,,\cdots\,,\, d_n)$, $d_i\in {\Bbb Z}_{\ge 0}$,
  $\boldy^{\scriptsizeboldd} := (y^1)^{d_1}\,\cdots\, (y^n)^{d_n}$ and\\
  $\boldy^{\scriptsizeboldd}
 			                    \rightsquigarrow\,
								D_{\xi}
								   (\varphi^{\sharp}(\boldy^{\scriptsizeboldd}))$
    means
     `replacing $\boldy^{\scriptsizeboldd}$
      by	$D_{\xi}(\varphi^{\sharp}(\boldy^{\scriptsizeboldd}))$'.	
\end{lemma}

\medskip

\begin{corollary}
{\bf [local expression of $D\varphi$ for $(\ast_1)$-admissible $(\varphi,\nabla)$, II]}$\;$
 Assume that $(\varphi,\nabla)$ is $(\ast_1)$-admissible.
 Let $d_Y$ be the exterior differential on $Y$.
 Then
  $$
    D\varphi \; =\;  \varphi^{\diamond}d_Y\,.
  $$
 Locally explicitly,
   let $(e^i)_{i=1,\,\cdots\,,\,n}$ be a local frame on $Y$ and
        $(e^i)_{i=1,\,\cdots\,,\,n}$ its dual co-frame.
 In terms of these dual pair of local frames, $d_Y= \sum_{i=1}^n e^i \otimes e_i$
   under the canonical isomorphism
   ${\cal T}^{\ast}Y\otimes_{{\cal O}_Y}{\cal O}_Y \simeq  {\cal T}^{\ast}Y$.
 Then
  $$
    (D_{\xi}\varphi)\, f\;
      =\; \sum_{i=1}^n (\varphi^{\diamond}e^i)(\xi)\otimes e_if\;
	  =\; \sum_{i=1}^n (\varphi^{\diamond}e^i)(\xi)\,\varphi^{\sharp}(e_if)
	  \;\in   {\cal O}_X^{A\!z}
  $$
  under the canonical isomorphism
     ${\cal O}_X^{A\!z}\otimes_{\varphi^{\sharp},{\cal O}_Y}{\cal O}_Y
	     \simeq {\cal O}_X^{A\!z}$.
 In particular,
   let $(y^1,\,\cdots\,, y^n)$ be coordinates of a local chart on $Y$.
 Then, locally,
  $$
   (D_{\xi}\varphi)\, f\;
     =\; \sum_{i=1}^n (\varphi^{\diamond}dy^i)(\xi)
	                \otimes \mbox{$\frac{\partial}{\partial y^i}$}f\;
	 =\; \sum_{i=1}^n
	          \left (
			   D_{\xi}\varphi^{\sharp}(y^i)\,
			   \cdot\,
			   \varphi^{\sharp}(\mbox{$\frac{\partial f}{\partial y^i}$})
			  \right).
  $$
  Here $\,\cdot\,$ is the multiplication in ${\cal O}_X^{A\!z}$
   (and will be omitted later when there is no sacrifice to clarity).
\end{corollary}

\bigskip

\begin{flushleft}
{\bf Local expressions of the inner differential $ad\varphi$ of $\varphi$ for admissible inner derivations}
\end{flushleft}
Though for the purpose of defining an action functional for D-branes,
 $(\varphi,\nabla)$ is the dynamical field of the focus and, hence,
 the induced covariant derivation $D$ on ${\cal O}_X^{A\!z}$ may look to play more roles in our discussion,
mathematically results related to $D$ like Lemma~1.6, Lemma~3.1.4, and Corollary~3.1.5
  involve only the fact that, for $\xi\in{\cal T}_{\xi}X$,
  $D_{\xi}$ satisfies the Leibniz rule on ${\cal O}_X^{A\!z}$
  and some additional commutativity assumption.
This suggest that similar statements hold if one considers inner derivations on ${\cal O}_X^{A\!z}$
  that are compatible to $\varphi$ in an appropriate sense.
Cf.$\:$Lemma~1.6 vs.\ Lemma~1.7.
This motivates the setting of the current theme.

\bigskip

\begin{definition} {\bf [admissible inner derivation]}$\;$ {\rm (Cf.\ Definition~2.1.1.)
  Let $m\in \Inn({\cal O}_X^{A\!z})$ be an inner derivation represented by an element of
   ${\cal O}_X^{A\!z}$. Then
  $m$ is called
   \begin{itemize}
    \item[]
     {\it $(\ast_1)$-admissible to $\varphi$}\hspace{1em}
     if $\;[m,{\cal A}_{\varphi}] \subset \Comm({\cal A}_{\varphi})\,$;

    \vspace{-1.6ex}	
	\item[]	
     {\it $(\ast_2)$-admissible to $\varphi$}\hspace{1em}
     if $\;[m,\Comm({\cal A}_{\varphi})] \subset \Comm({\cal A}_{\varphi})\,$;
	
    \vspace{-1.6ex}	
	\item[]
      {\it $(\ast_3)$-admissible to $\varphi$}\hspace{1em}
      if $\;[m,{\cal A}_{\varphi}]\subset  {\cal A}_{\varphi}\,$.
  \end{itemize}	
 Note that these conditions are independent of the representative chosen in ${\cal O}_X^{A\!z}$
    of the inner derivation.
 The set of  all $(\ast_1)$-admissible-to-$\varphi$ inner derivations on ${\cal O}_X^{A\!z}$
   form an ${\cal O}_X^{\,\Bbb C}$-module,
   which will be denoted by $\Inn^{\varphi}_{(\ast_1)}({\cal O}_X^{A\!z})$.
 Similarly, for $\Inn^{\varphi}_{(\ast_2)}({\cal O}_X^{A\!z})$  and
                       $\Inn^{\varphi}_{(\ast_3)}({\cal O}_X^{A\!z})$.
}\end{definition}
	
\medskip

\begin{lemma} {\bf [hierarchy of admissible conditions on inner derivation]}$\;$
 {\rm (Cf.$\,$Lemma$\,$2.1.2.)}
 $$
   \Inn^{\varphi}_{(\ast_3)}({\cal O}_X^{A\!z})\;
	 \subset\; \Inn^{\varphi}_{(\ast_2)}({\cal O}_X^{A\!z})\;
     \subset\; \Inn^{\varphi}_{(\ast_1)}({\cal O}_X^{A\!z})\,.
 $$				   				
\end{lemma}

\smallskip

\begin{proof}
 Since ${\cal A}_{\varphi}\subset \Comm({\cal A}_{\varphi})$,
 it  is immediate that
   $\Inn^{\varphi}_{(\ast_2)}({\cal O}_X^{A\!z})
       \subset \Inn^{\varphi}_{(\ast_1)}({\cal O}_X^{A\!z})$.		
 For the inclusion
   $\Inn^{\varphi}_{(\ast_3)}({\cal O}_X^{A\!z})
	   \subset \Inn^{\varphi}_{(\ast_2)}({\cal O}_X^{A\!z})$,
  let
   $m\in \Inn^{\varphi}_{(\ast_3)}({\cal O}_X^{A\!z})$
         represented by an element in ${\cal O}_X^{A\!z}$,
   $m^{\prime}\in \Comm({\cal A}_{\varphi})$, and
   $m^{\prime\prime}\in {\cal A}_{\varphi}$.
 Then,
   $$
    [[m,m^{\prime}], m^{\prime\prime}]\;
	=\;  [[m, m^{\prime\prime}], m^{\prime}]\,+\, [m, [m^{\prime}, m^{\prime\prime}]]
   $$
  from either the Jacobi identity of Lie bracket or the Leibniz rule for a derivation.
  The first term vanishes
    since $[m,m^{\prime\prime}]\in {\cal A}_{\varphi}$ and
	  $[{\cal A}_{\varphi},m^{\prime}]=0$.
  The second term also vanishes since $[m^{\prime},m^{\prime\prime}]=0$.	
 Since
   $m^{\prime}\in \Comm({\cal A}_{\varphi})$ and
   $m^{\prime\prime}\in {\cal A}_{\varphi}$
    are arbitrary,
 this shows that $[m,\Comm({\cal A}_{\varphi})]\subset \Comm({\cal A}_{\varphi})$.
 This proves the lemma.
 
\end{proof}

\bigskip

\begin{remark} $[$Lemma~2.1.2 vs.\ Lemma~3.1.7$\,]\;$ {\rm
 Note that in Lemma~2.1.2, the implication $(\ast_3)\Rightarrow (\ast_2)$
   uses $D$-parallel transport properties implied by $(\ast_3)$, which is an analytic technique.
 Indeed, the proof of Lemma~3.1.7 applies there.
 Which says that in both situations, the hierarchy is an algebraic consequence.
}\end{remark}
  
\bigskip

In terms of this setting and with arguments parallel to the proof of Lemma~3.1.4 and Corollary~3.1.5,
 one has the following lemma and corollary:
 
\bigskip

\begin{lemma}
{\bf [local expression of $\ad\varphi$
           for $\Inn^{\varphi}_{(\ast_1)}({\cal O}_X^{A\!z})$, I]}$\;$
 {\rm (Cf.\ Lemma~3.1.4.)}\\
 Let $\varphi: (X^{\!A\!z},{\cal E})\rightarrow Y$ be a map
   and $m\in \Inn^{\varphi}_{(\ast_1)}({\cal O}_X^{A\!z})$
    represented by an element of ${\cal O}_X^{A\!z}$
   i.e.\ $[m,{\cal A}_{\varphi}]\subset \Comm({\cal A}_{\varphi})$.
 Let  $U\subset X$ be a small enough open set
   so that $\varphi(U^{A\!z})$ is contained in a coordinate chart of $Y$,
   with coordinates $\boldy =(y^1, \,\cdots\,,\, y^n)$.
 For $f\in C^{\infty}(Y)$,
  recall the germwise-over-$U$ polynomial $R^f[1]$ in $(y^1,\,\cdots\,,\, y^n)$
   with coefficients in ${\cal O}_U^{A\!z}$ from
   {\rm $\,$[L-Y8: Sec.\ 4 \& Remark/Notation 4.2.3.5] (D.13.1)}.
 Then, at the level of germs over $U$,
  $$
     (\ad_m\varphi)\,f\;
	   =\; R^f[1]\!\!\left.\rule{0ex}{0.8em}\right|
			    _{\scriptsizeboldy^{\tinyboldd}
 			                    \rightsquigarrow\,
								\mbox{\scriptsize\it ad}_m
								   (\varphi^{\sharp}(\scriptsizeboldy^{\tinyboldd})),\;
								 \mbox{\scriptsize for all multi-degree $\scriptsizeboldd$ in $R^f[1]$}}\,.
  $$
 Here,
  for a multiple degree $\boldd =(d_1,,\cdots\,,\, d_n)$, $d_i\in {\Bbb Z}_{\ge 0}$,
  $\boldy^{\scriptsizeboldd} := (y^1)^{d_1}\,\cdots\, (y^n)^{d_n}$ and\\
  $\boldy^{\scriptsizeboldd}
 			                    \rightsquigarrow\,
								\ad_m
								   (\varphi^{\sharp}(\boldy^{\scriptsizeboldd}))$
    means
     `replacing $\boldy^{\scriptsizeboldd}$
      by	$\ad_m(\varphi^{\sharp}(\boldy^{\scriptsizeboldd}))$'.	
\end{lemma}

\smallskip
 
\begin{proof}
 Recall the proof of Lemma~3.1.4 through the proof of Lemma 2.1.6.
 .                          
 With the same setup and notations there,
 note that for $m\in \Inn^{\varphi}_{(\ast_1)}({\cal O}_X^{A\!z})$,
  $$
   [m, \varphi^{\sharp}(1_{(k)})]\;=\; 0\,, \hspace{2em}\mbox{for $k=1,\,\cdots\,,\, s$}
  $$
  by Lemma~1.7.
 Since $[m,{\cal O}_X]=0$ holds automatically,
  the lemma follows.
  
\end{proof}
 
\bigskip

\begin{corollary}
{\bf [local expression of $\ad\varphi$
           for $\Inn^{\varphi}_{(\ast_1)}({\cal O}_X^{A\!z})$, II]}$\;$
 {\rm (Cf.\ Corollary~3.1.5.)}
 Continuing the setting of Lemma~3.1.9.
 Recall that
  $m\in \Inn^{\varphi}_{(\ast_1)}({\cal O}_X^{A\!z})$
     is represented by an element of ${\cal O}_X^{A\!z}$.
 Let $(y^1,\,\cdots\,, y^n)$ be coordinates of a local chart on $Y$.
 Then, locally,
  $$
   (\ad_m\varphi)\, f\;
	 =\; \sum_{i=1}^n
	          \left (
			   \ad_m\varphi^{\sharp}(y^i)\,
			   \cdot\,
			   \varphi^{\sharp}(\mbox{$\frac{\partial f}{\partial y^i}$})
			  \right)\;
	 =\; \sum_{i=1}^n
	          \left (
			   [m, \varphi^{\sharp}(y^i)]\,
			   \cdot\,
			   \varphi^{\sharp}(\mbox{$\frac{\partial f}{\partial y^i}$})
			  \right)\,,
  $$
  where $\,\cdot\,$ is the multiplication in ${\cal O}_X^{A\!z}$
   (and will be omitted later when there is no sacrifice to clarity).
  In other words,
  $$
      \ad_m\varphi\;=\; \sum_{i=1}^n
	     [m, \varphi^{\sharp}(y^i)]\otimes \mbox{\Large $\frac{\partial}{\partial y^i}$}\,.
  $$	
\end{corollary}

\smallskip

\begin{proof}
 The related last part of the proof of Corollary~3.1.5 through the proof of Lemma~2.1.7
   works in verbitum with $D_{\xi}$ replaced by $\ad_m = [m,\,\mbox{\Large $\cdot$}\,]$.						
 This is simply a re-writing of Lemma~3.1.9 above.
 
\end{proof}

\bigskip

\begin{remark} $[$comparison with differential of ordinary map$]\;$ {\rm
 As a comparison,
 let $u:X\rightarrow Y$ be a map between manifolds.
 Then, $du$ defines a bundle map $T_{\ast}X\rightarrow u^{\ast}T_{\ast}Y$
  that satisfies
   $$
      du(\xi)\,f \; =\;  u_{\ast}(\xi)\,f\; =\;  \xi(u\circ f)\; =\; \xi (u^{\sharp}(f))
   $$
	for any $\xi\in T_{\ast}X$ and $f\in C^{\infty}(Y)$.
 In terms of local coordinates
   {\boldmath $x$}$=(x^{\mu})_{\mu=1,\,\cdots\,,\,m}$ on $X$  and
     {\boldmath $y$}$=(y^i)_{i=1,\,\cdots\,,\,n}$ on $Y$,
   $$
     u_{\ast}  \Big(\mbox{\Large $\frac{\partial}{\partial x^{\mu}}$}\Big) \;
	  =\;  \sum_{i=1}^n
	          \mbox{\Large $\frac{dy^i}{dx^{\mu}}\frac{\partial f}{\partial y^i}$}(u(x))\,.
   $$
 The above notion of covariant differential $D\varphi$ of $\varphi$
    is exactly the generalization of the ordinary differential $du$ of $u$,
  taking into account the fact that
    $\varphi$  is now only defined contravariantly through $\varphi^{\sharp}$  and
   that the noncommutative structure sheaf ${\cal O}_X^{A\!z}$ of $X^{A\!z}$
      is no longer naturally trivial as an ${\cal O}_X$-module
    but, rather, is endowed with a natural induced connection $D$ from $\nabla$.
	
 Furthermore,
  when $(\varphi,\nabla)$ is $(\ast_1)$-admissible or
  when considering only $\Inn^{\varphi}_{(\ast_1)}({\cal O}_X^{A\!z})$,
  both the covariant differential $D\varphi$   and the inner differential $\ad\varphi$ takes the same form
   as the chain rule in the commutative case.	
}\end{remark}

\bigskip

\subsection{The
         three basic ${\cal O}_X^{\,\Bbb C}$-modules relevant to $D\varphi$, with induced structures}

Underlying the notion of covariant differential $D\varphi$  of $\varphi$
 are two basic ${\cal O}_X^{\,\Bbb C}$-modules
 \begin{itemize}
  \item[\LARGE $\cdot$]
   the pull-back tangent sheaf
     $\varphi^{\ast}{\cal T}_{\ast}Y
        :=  {\cal O}_X^{A\!z}\otimes_{\varphi^{\sharp},{\cal O}_Y}\!{\cal T}_{\ast}Y$,
     a left ${\cal O}_Y^{A\!z}$-module but now regarded
	 as a (left) ${\cal O}_X^{\,\Bbb C}$-module through the built-in inclusion
     ${\cal O}_X^{\,\Bbb C}\hookrightarrow {\cal O}_X^{A\!z}$,	
    and
  
 \item[\LARGE $\cdot$]
  the ${\cal O}_X^{\,\Bbb C}$-module
	  ${\cal T}^{\ast}X\otimes_{{\cal O}_X}\!\varphi^{\ast}{\cal T}_{\ast}Y$,
	  where $D\varphi$ lives.
 \end{itemize}	
We study them in this subsection after taking a look at another basic but simpler
 ${\cal O}_X^{\,\Bbb C}$-module
 ${\cal T}^{\ast}X\otimes_{{\cal O}_X}\!\!{\cal O}_X^{A\!z}$.
They play fundamental roles in our variational problem.
 
\bigskip

\noindent {\it Remark 3.2.0.1.$\;[$structures
          on the ${\cal O}_X^{\,\Bbb C}$-algebra ${\cal O}_X^{A\!z}]$}$\;$
 Recall the {\it connection} $D$ on the noncommutative structure sheaf
   ${\cal O}_X^{A\!z}:=\Endsheaf_{{\cal O}_X^{\,\Bbb C}}({\cal E})$ of $X^{\!A\!z}$
   induced by the connection $\nabla$ on ${\cal E}$.
 The multiplication $\,\cdot\,$ in the ${\cal O}_X^{\,\Bbb C}$-algebra structure of ${\cal O}_X^{A\!z}$
  defines a {\it nonsymmetric ${\cal O}_X^{A\!z}$-valued inner product} on ${\cal O}_X^{A\!z}$
    that is ${\cal O}_X^{\,\Bbb C}$-bilinear.
 This inner product is $D$-invariant in the sense that
   $$
      D(m^1\cdot m^2)\;=\;  (Dm^1)\cdot m^2\,+\, m^1\cdot Dm^2\,,
   $$
   for $m^1,\, m^2\in {\cal O}_X^{A\!z}$.
 Together with the built-in {\it trace map}
   $$
      \Tr;:\; {\cal O}_X^{A\!z}\;\longrightarrow\; {\cal O}_X^{\,\Bbb C}\,,
   $$
   as an ${\cal O}_X^{\,\Bbb C}$-module-homomorphism,
  one has a {\it symmetric ${\cal O}_X^{\,\Bbb C}$-valued inner product} on ${\cal O}_X^{A\!z}$
   defined by the assignment
   $$
      (m^1,m^2)\;\longmapsto\; \Tr(m^1\cdot m^2) =: \Tr(m^1m^2)\,,
   $$
   for $m^1,\, m^2\in {\cal O}_X^{A\!z}$.
  This inner product is ${\cal O}_X^{\,\Bbb C}$-bilinear;
    and is covariantly constant over $X$ in the sense that
    $$
	  d_X\, (\Tr(m^1m^2))\;=\; \Tr((Dm^1) m^2)\,+\, \Tr(m^1 Dm^2)\,,
	$$
	where $d_X$ is the exterior differential on $X$.

\bigskip

\subsubsection{The ${\cal O}_X^{A\!z}$-valued cotangent sheaf
   ${\cal T}^{\ast}X\otimes_{{\cal O}_X}\!\!{\cal O}_X^{A\!z}$ of $X$, and beyond}
   
Let
  $X$ be endowed with a (Riemannian or Lorentzian) metric $h$ and
  $\nabla^h$ be the Levi-Civita connection on ${\cal T}_{\ast}X$ induced by $h$.
The corresponding inner product on ${\cal T}_{\ast}X$,  its dual ${\cal T}^{\ast}X$,
  and their tensor products
  will be denoted $\langle\,\mbox{\Large $\cdot$}\,,\,\mbox{\Large $\cdot$}\,\rangle_h$.
The induced connection
   on the dual ${\cal T}^{\ast}X$   and
   on the tensor product of copies of ${\cal T}_{\ast}X$ and copies of ${\cal T}^{\ast}X$
 will be denoted also by $\nabla^h$.
The defining features of $\nabla^h$ are
 $$
 \begin{array}{cl}
  \nabla^h\langle \xi_1,\xi_2\rangle_h\;
    =\;    \langle \nabla^h\xi_1,\xi_2\rangle_h\, +\, \langle \xi_1, \nabla^h\xi_2\rangle_h
	& \hspace{2em}\mbox{($h$ be $\nabla^h$-covariantly constant)}\,,\\[1.2ex]
  \Tor_{\nabla^h}(\xi_1,\xi_2)
    := \nabla^h_{\xi_1}\xi_2 - \nabla^h_{\xi_2}\xi_1 -[\xi_1,\xi_2]\;=\; 0
	& \hspace{2em}\mbox{($\nabla^h$ be torsionless)}\,,
 \end{array}
 $$
 for all $\xi_1,\,\xi_2\in {\cal T}_{\ast}X$.

\bigskip

\begin{flushleft}
{\bf The ${\cal O}_X^{A\!z}$-valued cotangent sheaf
         ${\cal T}^{\ast}X\otimes_{{\cal O}_X}\!\!{\cal O}_X^{A\!z}$ of $X$}
\end{flushleft}
The connection $\nabla^h$ on ${\cal T}^{\ast}X$ and the connection $D$ on ${\cal O}_X^{A\!z}$
 together induce a connection
  $$
    \nabla^{(h,D)}\;
	  :=\;  \nabla^h\otimes \Id_{{\cal O}_X^{A\!z}}\,+\,  \Id_{{\cal T}^{\ast}X}\otimes D
  $$
  on ${\cal T}^{\ast}X\otimes_{{\cal O}_X}\!{\cal O}_X^{A\!z}$.
The inner product
   $\langle\,\mbox{\Large $\cdot$}\,,\,\mbox{\Large $\cdot$}\,\rangle_h$ on ${\cal T}_{\ast}X$
    and
  the inner product $\cdot$ on ${\cal O}_X^{A\!z}$
 together induce
  an {\it ${\cal O}_X^{A\!z}$-valued,
             ${\cal O}_X^{\,\Bbb C}$-bilinear (nonsymmetric) inner product} on
	   ${\cal T}^{\ast}X \otimes_{{\cal O}_X}\!{\cal O}_X^{A\!z}$
       by extending ${\cal O}_X^{\,\Bbb C}$-bilinearly
       $$
         \langle  \omega^1\otimes m^1    \,,\,  \omega^2\otimes m^2 \rangle_h\;
    	     :=\;  \langle \omega^1\,,\, \omega^2\rangle_h
		                 \cdot  (m^1_T m^2_T)\,,
       $$
       where $\omega^1,\, \omega^2\in {\cal T}^{\ast}X$
        and $m^1,\, m^2\in {\cal O}_X^{A\!z}$.
    The trace map $\Tr:{\cal O}_X^{A\!z}\rightarrow {\cal O}_X^{\,\Bbb C}$
 	   turns this further to an {\it ${\cal O}_X^{\,\Bbb C}$-valued,
	   ${\cal O}_X^{\,\Bbb C}$-bilinear (symmetric) inner product} on
	   ${\cal T}^{\ast}X \otimes_{{\cal O}_X}\!\!{\cal O}_X^{A\!z}$
       by the post composition with the above inner product
	   $$
	      (\,\mbox{\LARGE $\cdot$}\,,\, \mbox{\LARGE $\cdot$}^{\prime})\;\longmapsto\;
  		 \Tr (\langle\, \mbox{\LARGE $\cdot$}\,,\, \mbox{\LARGE $\cdot$}^{\prime} \rangle_h)\;
		   =:\; 	 \Tr \langle\, \mbox{\LARGE $\cdot$}\,,\, \mbox{\LARGE $\cdot$}^{\prime} \rangle_h
	   $$
	   for {\LARGE $\cdot$}, {\LARGE $\cdot$}$^{\prime} \in
	      {\cal T}^{\ast}X \otimes_{{\cal O}_X}\!\!{\cal O}_X^{A\!z}$.
     By construction, both inner products are covariantly constant with respect to $\nabla^{(h,D)}$   and
	 they satisfy the Leibniz rules
	   \begin{eqnarray*}
	    D  \langle\, \mbox{\LARGE $\cdot$}\,,\, \mbox{\LARGE $\cdot$}^{\prime} \rangle_h
		 & =
		  &  \langle\,   \nabla^{(h,D)}\,\mbox{\LARGE $\cdot$}\,,\,
		                            \mbox{\LARGE $\cdot$}^{\prime} \rangle_h\,
		       +\, \langle\, \mbox{\LARGE $\cdot$}\,,\,
			             \nabla^{(h,D)}\,\mbox{\LARGE $\cdot$}^{\prime} \rangle_g  \,,   \\			   
        d \Tr \langle\, \mbox{\LARGE $\cdot$}\,,\, \mbox{\LARGE $\cdot$}^{\prime} \rangle_h 					 
         & =
		  & \Tr (D  \langle\, \mbox{\LARGE $\cdot$}\,,\,
		                              \mbox{\LARGE $\cdot$}^{\prime} \rangle_h)\;\;
		      =\;\; \Tr \langle\,   \nabla^{(h,D)}\,\mbox{\LARGE $\cdot$}\,,\,
			                          \mbox{\LARGE $\cdot$}^{\prime} \rangle_h\,
		       +\, \Tr \langle\, \mbox{\LARGE $\cdot$}\,,\,
			                    \nabla^{(h,D)}\,\mbox{\LARGE $\cdot$}^{\prime} \rangle_h
       \end{eqnarray*}
	   for {\LARGE $\cdot$}, {\LARGE $\cdot$}$^{\prime} \in
	      {\cal T}^{\ast}X \otimes_{{\cal O}_X}\!\!{\cal O}_X^{A\!z}$.

\bigskip

\begin{flushleft}		
{\bf The sheaf
         $(\bigwedge^{\mbox{\tiny $\bullet$}}{\cal T}^{\ast}X)
		    \otimes_{{\cal O}_X}\!\!{\cal O}_X^{A\!z}$
		 of ${\cal O}_X^{A\!z}$-valued differential forms on $X$}
\end{flushleft}
The setting in the previous theme generalizes to
  the sheaf
    $(\bigwedge^{\mbox{\tiny $\bullet$}}{\cal T}^{\ast}X)
		    \otimes_{{\cal O}_X}\!\!{\cal O}_X^{A\!z}$
		 of ${\cal O}_X^{A\!z}$-valued differential forms on $X$,
 with the $1$-forms $\omega^1$, $\omega^2$ on $X$ there
   replaced by general differential forms $\alpha^1$, $\alpha^2$ on $X$.
We will use the same notations
  $$
    \nabla^h\,,\hspace{2em}
    \nabla^{(h,D)}\,,\hspace{2em}
    \langle\,\mbox{\Large $\cdot$}\,,\,\mbox{\Large $\cdot$}\, \rangle_h\,,\hspace{2em}
    \Tr \langle\,\mbox{\Large $\cdot$}\,,\,\mbox{\Large $\cdot$}\, \rangle_h
  $$	
 to denote
   the {\it connection on $\bigwedge^{\mbox{\tiny $\bullet$}}{\cal T}^{\ast}X$},
   the {\it connection on
                  $(\bigwedge^{\mbox{\tiny $\bullet$}}{\cal T}^{\ast}X)
		                           \otimes_{{\cal O}_X}\!\!{\cal O}_X^{A\!z}$},
   the {\it ${\cal O}_X^{A\!z}$-valued, ${\cal O}_X^{\,\Bbb C}$-bilinear
                (nonsymmetric) inner product on
				 $(\bigwedge^{\mbox{\tiny $\bullet$}}{\cal T}^{\ast}X)
		               \otimes_{{\cal O}_X}\!\!{\cal O}_X^{A\!z}$},    and
   the {\it ${\cal O}_X^{\,\Bbb C }$-valued, ${\cal O}_X^{\,\Bbb C}$-bilinear
                (symmetric) inner product on
				 $(\bigwedge^{\mbox{\tiny $\bullet$}}{\cal T}^{\ast}X)
		               \otimes_{{\cal O}_X}\!\!{\cal O}_X^{A\!z}$}    					   				
 respectively.
They satisfy the same Leibniz rule as in the case of
  ${\cal T}^{\ast}X\otimes_{{\cal O}_X}\!\!{\cal O}_X^{A\!z}$

\bigskip
		
\subsubsection{The pull-back tangent sheaf $\varphi^{\ast}{\cal T}_{\ast}Y$}

This is the main character among the three basic ${\cal O}_X^{\,\Bbb C}$-modules
 and is slightly subtler than $u^{\ast}{\cal T}_{\ast}Y$  in the commutative case
   (cf.$\,$Remark~3.1.11)
 or ${\cal T}^{\ast}X\otimes_{{\cal O}_X}\!\!{\cal O}_X^{A\!z}$ in Sec.~3.2.1.

\bigskip

\begin{flushleft}
{\bf The induced connection and the induced partially-defined inner products}
\end{flushleft}
Let
  $Y$ be endowed with a (Riemannian or Lorentzian) metric $g$ and
  $\nabla^g$ be the Levi-Civita connection on ${\cal T}_{\ast}Y$ induced by $g$.
The corresponding inner product on ${\cal T}_{\ast}Y$ or its dual ${\cal T}^{\ast}Y$
  will be denoted $\langle\,\mbox{\Large $\cdot$}\,,\,\mbox{\Large $\cdot$}\,\rangle_g$.
The induced connection
   on the dual ${\cal T}^{\ast}Y$   and
   on the tensor product of copies of ${\cal T}_{\ast}Y$ and copies of ${\cal T}^{\ast}Y$
 will be denoted also by $\nabla^g$.
The defining features of $\nabla^g$ are
 $$
 \begin{array}{cl}
  \nabla^g\langle v_1,v_2\rangle_g\;
    =\;    \langle \nabla^g v_1, v_2\rangle_g\, +\, \langle v_1, \nabla^g v_2\rangle_g
	& \hspace{2em}\mbox{($g$ be $\nabla^g$-covariantly constant)}\,,\\[1.2ex]
  \Tor_{\nabla^g}(v_1, v_2)\;
    :=\; \nabla^g_{v_1}v_2 - \nabla^g_{v_2}v_1 -[v_1,v_2]\; =\;  0
	& \hspace{2em}\mbox{($\nabla^g$ be torsionless)}\,,
 \end{array}
 $$
 for all $v_1,\, v_2\in {\cal T}_{\ast}Y$.
 
\bigskip

\begin{sslemma}
{\bf [$(D,\nabla^g)$-induced connection on $\varphi^{\ast}{\cal T}_{\ast}Y$
           for $(\ast)_1$-admissible $(\varphi,\nabla)$]}$\;$
 Assume that $(\varphi,\nabla)$ is $(\ast_1)$-admissible.
 Then,
   the connection $D$ on ${\cal O}_X^{A\!z}$   and
   the connection $\nabla^g$ on ${\cal T}_{\ast}Y$ together
	   induce a connection $\nabla^{(\varphi,g)}$ on $\varphi^{\ast}{\cal T}_{\ast}Y$,
	 locally of the form
     $$
       \nabla^{(\varphi,g) }\;
	   =\;  D\otimes \Id_{{\cal T}_{\ast}Y} \,
		       +\,  \Id_{{\cal O}_X^{A\!z}}
		                 \cdot
		                 \sum_{i=1}^n
		                   D\varphi^{\sharp}(y^i)
		                                   \otimes \nabla^g_{\frac{\partial}{\partial y^i}}\,.
     $$
\end{sslemma}	

\smallskip
	
\begin{proof}
 Our construction of an induced connection on $\varphi^{\ast}{\cal T}_{\ast}Y$ is local in nature.
 As long as a construction is independent of coordinates chosen,
   the local construction glues to a global construction.
 Let $U\subset X$ be a small enough open set so that $\varphi(U^{A\!z})$
   is contained in a coordinate chart of $Y$, with coordinate $\boldy:=(y^1,\,\cdots\,,\, y^n)$.
 Then the local expression
  $$
   \nabla^{(\varphi,g);\scriptsizeboldy}\;
	   :=\;  D\otimes \Id_{{\cal T}_{\ast}Y} \,
		       +\,  \Id_{{\cal O}_X^{A\!z}}
		                 \cdot
		                 \sum_{i=1}^n
		                   D\varphi^{\sharp}(y^i)
		                                   \otimes \nabla^g_{\frac{\partial}{\partial y^i}}\,.
  $$
   in the Statement defines a connection
  on $\varphi^{\ast}{\cal T}_{\ast}Y|_U$.
 We only need to show that
  it is independent of the coordinate $(y^1,\,\cdots\,,\, y^n)$ chosen.
  
 Let  $\boldz:=(z^1,\,\cdots\,,\, z^n)$ be another coordinate on the chart.
 Then, for $(\varphi,\nabla)$ $(\ast_1)$-admissible,
  $D\varphi^{\sharp}(y^i)=:(D\varphi)y^i$ has a local expression in terms of $\boldz$
   $$
      (D\varphi)y^i\; := D(\varphi^{\sharp}(y^i))\;
	    =\; \sum_{j=1}^n
		        D(\varphi^{\sharp}(z^j))
		          \otimes \mbox{\Large $\frac{\partial y^i}{\partial z^j}$}
   $$
   by Corollary~3.1.5, for $i=1,\,\cdots\,,\, n$.
 It follows that
   \begin{eqnarray*}
    \nabla^{(\varphi,g);\scriptsizeboldz}
	 & :=\:
	   & D\otimes \Id_{{\cal T}_{\ast}Y} \,
		       +\,  \Id_{{\cal O}_X^{A\!z}}
		                 \cdot
		                 \sum_{j=1}^n
		                   D\varphi^{\sharp}(z^j)
		                                   \otimes \nabla^g_{\frac{\partial}{\partial z^j}}  \\
    & =
      & D\otimes \Id_{{\cal T}_{\ast}Y} \,
		       +\,  \Id_{{\cal O}_X^{A\!z}}
		                 \cdot
		                 \sum_{j=1}^n
		                   D\varphi^{\sharp}(z^j)
		                      \otimes   \sum_{i=1}^n
								                  \mbox{\Large $\frac{\partial y^i}{\partial z^j}$}  
											      \nabla^g_{\frac{\partial}{\partial y^i}}  \\
    & =
     & D\otimes \Id_{{\cal T}_{\ast}Y} \,
		       +\,  \Id_{{\cal O}_X^{A\!z}}
		                 \cdot
		                 \sum_{i=1}^n
		                   D\varphi^{\sharp}(y^i)
		                                   \otimes \nabla^g_{\frac{\partial}{\partial y^i}}  \\ 	
    & \:=:
      & \nabla^{(\varphi,g);\scriptsizeboldy}\,.
   \end{eqnarray*}
  
 This completes the proof.
\end{proof}

\bigskip	

Consider next the induced inner products on $\varphi^{\ast}{\cal T}_{\ast}Y$.
Completely naturally, one may attemp to combine
   the multiplication in ${\cal O}_X^{A\!z}$  and
   the inner product $\langle\,\cdot\,,\,\cdot\,\rangle_g$ on ${\cal T}_{\ast}Y$
  to an ${\cal O}_X^{A\!z}$-valued, ${\cal O}_X^{\,\Bbb C}$-bilinear (nonsymmetric) inner product
   on $\varphi^{\ast}{\cal T}_{\ast}Y$
 by extending ${\cal O}_X^{\,\Bbb C}$-bilinearly
  $$
    \langle  m^1\otimes v_1    \,,\,  m^2\otimes v_2 \rangle_g\;
        :=\;  m^1m^2\otimes \langle v_1\,,\, v_2\rangle_g\;
	     =\;  m^1m^2\varphi^{\sharp}(\langle v_1\,,\, v_2\rangle_g)\,,
  $$
   where
	 $m^1,\, m^2\in \Comm({\cal A}_{\varphi})
		                                    \subset {\cal O}_X^{A\!z}$ and
	 $v_1,\, v_2\in {\cal T}_{\ast}Y$ and
	   the last equality follows from the canonical isomorphism
    $$
      {\cal O}_X^{A\!z}
		          \otimes_{\varphi^{\sharp},\, {\cal O}_Y}\!\!{\cal O}_Y\;
		       \simeq\;  {\cal O}_X^{A\!z}\,.
    $$	
{\it However}, for this to be well-defined, it is required that
  $$
    \langle  m^1\otimes f_1v_1    \,,\,  m^2\otimes f_2v_2 \rangle_g\;
      =\;    \langle  m^1 \varphi^{\sharp}(f_1) \otimes v_1    \,,\,
	                          m^2 \varphi^{\sharp}(f_2) \otimes v_2 \rangle_g\,,
  $$
 i.e.\
  $$
     m^1m^2 \varphi^{\sharp}\big(f_1f_2\langle v_1,v_2\rangle_g\big)\;
	  =\; m^1\varphi^{\sharp}(f_1)m^2\varphi^{\sharp}(f_2)
	        \varphi^{\sharp}(\langle v_1,v_2\rangle_g)\,,
  $$
  for all $m^1,m^2\in {\cal O}_X^{A\!z}$, $v_1,v_2\in{\cal T}_{\ast}Y$,
            and $f_1,f_2\in {\cal O}_Y$.
Which holds if and only if
 $$
   m^2\;  \in\; \Comm({\cal A}_{\varphi})\,.
 $$

What happens if one brings in the trace map
 $\Tr:{\cal O}_X^{A\!z}\rightarrow {\cal O}_X^{\,\Bbb C}\,$?
In this case,
 $$
   \Tr \langle  m^1\otimes f_1v_1    \,,\,  m^2\otimes f_2v_2 \rangle_g\;
   =\;   \Tr\big(m^1m^2 \varphi^{\sharp}(f_1f_2\langle v_1,v_2\rangle_g)\big)\;
   =\;   \Tr\big(\varphi^{\sharp}(f_1)m^1
                   m^2\varphi^{\sharp}(f_2\langle v_1,v_2\rangle_g)\big)
 $$	
 by the cyclic-invariance property of $\Tr$
 while
 $$
   \Tr \langle  m^1 \varphi^{\sharp}(f_1) \otimes v_1    \,,\,
	                          m^2 \varphi^{\sharp}(f_2) \otimes v_2 \rangle_g\;
   =\;  \Tr\big(
             m^1\varphi^{\sharp}(f_1)m^2\varphi^{\sharp}(f_2\langle v_1,v_2\rangle_g)
			   \big)\,.
 $$
The two equal if
 $$
   \mbox{either $\hspace{2em}m_1\in\Comm({\cal A}_{\varphi})\hspace{2em}$
                   or       $\hspace{2em}m_2\in\Comm({\cal A}_{\varphi})\,$.}
 $$
 
\bigskip

\begin{ssdefinition}  {\bf [partially-defined inner product
                $\langle\,\mbox{\Large $\cdot$}\,,\,\mbox{\Large $\cdot$}\,\rangle_g$
                on $\varphi^{\ast}{\cal T}_{\ast}Y$]}$\;$ {\rm
 The multiplication in ${\cal O}_X^{A\!z}$  and
     the inner product $\langle\,\cdot\,,\,\cdot\,\rangle_g$ on ${\cal T}_{\ast}Y$
  together induce
   a {\it partially defined, ${\cal O}_X^{A\!z}$-valued,
              ${\cal O}_X^{\,\Bbb C}$-bilinear (nonsymmetric) inner product}
   on $\varphi^{\ast}{\cal T}_{\ast}Y$
 by extending ${\cal O}_X^{\,\Bbb C}$-bilinearly
   $$
          \langle  m^1\otimes v_1    \,,\,  m^2\otimes v_2 \rangle_g\;
    	   :=\;  m^1m^2\otimes \langle v_1\,,\, v_2\rangle_g\;
		    =\;  m^1m^2\varphi^{\sharp}(\langle v_1\,,\, v_2\rangle_g)\,,
        $$
       where
	      $m^1\in{\cal O}_X^{A\!z}$,
		  $m^2\in \Comm({\cal A}_{\varphi})
		                                    \subset {\cal O}_X^{A\!z}$ and
	      $v_1,\, v_2\in {\cal T}_{\ast}Y$ and
		  the last equality follows from the canonical isomorphism
		   ${\cal O}_X^{A\!z}
		          \otimes_{\varphi^{\sharp},\, {\cal O}_Y}\!\!{\cal O}_Y
		         \simeq  {\cal O}_X^{A\!z}$.	
}\end{ssdefinition}

\medskip

\begin{ssdefinition}  {\bf [partially-defined inner product
                $\Tr\langle\,\mbox{\Large $\cdot$}\,,\,\mbox{\Large $\cdot$}\,\rangle_g$
                on $\varphi^{\ast}{\cal T}_{\ast}Y$]}$\;$ {\rm
 The multiplication in ${\cal O}_X^{A\!z}$,
     the inner product $\langle\,\cdot\,,\,\cdot\,\rangle_g$ on ${\cal T}_{\ast}Y$, and
	 the tarce map $\Tr:{\cal O}_X^{A\!z}\rightarrow {\cal O}_X^{\,\Bbb C}$
  together induce
   a {\it partially defined, ${\cal O}_X^{\,\Bbb C}$-valued,
              ${\cal O}_X^{\,\Bbb C}$-bilinear (symmetric) inner product}
   on $\varphi^{\ast}{\cal T}_{\ast}Y$
 by extending ${\cal O}_X^{\,\Bbb C}$-bilinearly
   $$
     \Tr \langle  m^1\otimes v_1    \,,\,  m^2\otimes v_2 \rangle_g\;
    	   :=\;  \Tr\big(m^1m^2\otimes \langle v_1\,,\, v_2\rangle_g\big)\;
		    =\;  \Tr\big(m^1m^2\varphi^{\sharp}(\langle v_1\,,\, v_2\rangle_g)\big)\,,
   $$
    where
	  either $m^1$ or $m^2$ is in $\Comm({\cal A}_{\varphi})$,
	  $v_1,\, v_2\in {\cal T}_{\ast}Y$, and
      the last equality follows from the canonical isomorphism
		${\cal O}_X^{A\!z}
		          \otimes_{\varphi^{\sharp},\, {\cal O}_Y}\!\!{\cal O}_Y
		     \simeq  {\cal O}_X^{A\!z}$.	
}\end{ssdefinition}

\bigskip

By construction, both inner products, when defined,  are covariantly constant with respect to
  $\nabla^{(\varphi,g)}$  and one has the Leibniz rules
	\begin{eqnarray*}
	    D \langle\, \mbox{--}\,,\, \mbox{--}^{\prime} \rangle_g
		 & =
		  &  \langle\,   \nabla^{(\varphi,g)}\,\mbox{--}\,,\, \mbox{--}^{\prime} \rangle_g\,
		       +\, \langle\, \mbox{--}\,,\,
			             \nabla^{(\varphi,g)}\,\mbox{--}^{\prime} \rangle_g  \,,   \\			   
        d_X \Tr \langle\, \mbox{--}\,,\, \mbox{--}^{\prime} \rangle_g 					
         & =
		  & \Tr (D \langle\, \mbox{--}\,,\, \mbox{--}^{\prime} \rangle_g)\;\;
		      =\;\; \Tr \langle\,   \nabla^{(\varphi,g)}\,\mbox{--}\,,\, \mbox{--}^{\prime} \rangle_g\,
		       +\, \Tr \langle\, \mbox{--}\,,\,
			                    \nabla^{(\varphi,g)}\,\mbox{--}^{\prime} \rangle_g  \,,
    \end{eqnarray*}
   whenever the
    $\langle\, \mbox{--}^{\prime\prime}\,,\, \mbox{--}^{\prime\prime\prime} \rangle_g$
       or
	$\Tr\langle\, \mbox{--}^{\prime\prime}\,,\, \mbox{--}^{\prime\prime\prime} \rangle_g$
	involved are defined.

The followng lemma	 is an immediate consequence of Corollary~3.1.5:
	
\bigskip

\begin{sslemma}
{\bf [sample list of defined inner products for admissible $(\varphi,\nabla)$]}
 %
 $(1)$
 For $(\varphi,\nabla)$ $(\ast_1)$-admissible, the following list of inner products
  $$
     \langle \,\mbox{--}\,, D_{\xi}\varphi \rangle_g\,,\hspace{2em}
	 \Tr\langle\,\mbox{--}\,, D_{\xi}\varphi \rangle_g \,,\hspace{2em}
	 \Tr\langle D_{\xi}\varphi\,,\,\mbox{--}\,\rangle_g
  $$  	
  are defined for $\xi\in{\cal T}_{\ast}X$.
  
 $(2)$
 For $(\varphi,\nabla)$ $(\ast_2)$-admissible, the following additional list of inner products
  $$
     \langle \,\mbox{--}\,,
	   \nabla^{(\varphi,g)}_{\xi_1}\,\cdots\,\nabla^{(\varphi,g)}_{\xi_k}
	     D_{\xi}\varphi \rangle_g\,,\hspace{.8em}
     \Tr\langle \,\mbox{--}\,,
	   \nabla^{(\varphi,g)}_{\xi_1}\,\cdots\,\nabla^{(\varphi,g)}_{\xi_k}
	     D_{\xi}\varphi \rangle_g\,,\hspace{.8em}
     \Tr\langle
	    \nabla^{(\varphi,g)}_{\xi_1}\,\cdots\,\nabla^{(\varphi,g)}_{\xi_k}
	     D_{\xi}\varphi\,,\,
        ,\mbox{--}\, \rangle_g		
  $$  	
  are also defined for $\xi, \xi_1,\,\cdots\,,\, \xi_k \in{\cal T}_{\ast}X$, $k\in{\Bbb Z}_{\ge 0}$.
\end{sslemma}

\bigskip

\begin{flushleft}
{\bf The symmetry properties of the curvature tensor
          of $\nabla^{(\varphi,g)}$ on $\varphi^{\ast}{\cal T}_{\ast}Y$}
\end{flushleft}
Let $\varphi:(X^{\!A\!z},{\cal E};\nabla)\rightarrow Y$ be a $(\ast_2)$-admissible map.
Let  $F_{\nabla^{(\varphi,g)}}$
 be the curvature tensor of the induced connection $\nabla^{(\varphi,g)}$
  on $\varphi^{\ast}{\cal T}_{\ast}Y$
  --- the $\Endsheaf_{{\cal O}_X^{\,\Bbb C}}(\varphi^{\ast}{\cal T}_{\ast}Y)$-valued
       $2$-form on $X$ defined by
   $$
     F_{\nabla^{(\varphi,g)}}(\xi_1,\xi_2)\,s\;
	  :=\;  \big(\nabla^{(\varphi,g)}_{\xi_1}\nabla^{(\varphi,g)}_{\xi_2}\,
	              -\, \nabla^{(\varphi,g)}_{\xi_2}\nabla^{(\varphi,g)}_{\xi_1}\,
                  -\, \nabla^{(\varphi,g)}_{[\xi_1, \xi_2]}					\big)\, s\;\;
				  \in\; \varphi^{\ast}{\cal T}_{\ast}Y
   $$
   for $\xi_1, \xi_2\in {\cal T}_{\ast}X$ and $s\in \varphi^{\ast}{\cal T}_{\ast}Y$.
 (This is ${\cal O}_X$-linear in $\xi_1$, $\xi_2$, and $s$ and, hence, a tensor on $X$).
By construction,
 $$
    F_{\nabla^{(\varphi,g)}}(\xi_1,\xi_2)\;
	 =\;  -\,F_{\nabla^{(\varphi,g)}}(\xi_2,\xi_1)
 $$
 for $\xi_1,\xi_2 \in{\cal T}_{\ast}X$.
From Lemma$\,$3.2.2.4,
 the inner products	
   $$
    \begin{array}{lcl}
     \langle \nabla^{(\varphi,g)}_{\xi_1}D_{\xi_3}\varphi\,,\,
	     D_{\xi_4}\varphi  \rangle_g\,,
      && \langle \nabla^{(\varphi,g)}_{\xi_1}D_{\xi_3}\,,\,
                 \nabla^{(\varphi,g)}_{\xi_2}D_{\xi_4}\varphi  \rangle_g\,,  \\[.8ex]
     \langle
       \nabla^{(\varphi,g)}_{\xi_1}\nabla^{(\varphi,g)}_{\xi_2}D_{\xi_3}\varphi\,,\,
          D_{\xi_4}\varphi  \rangle_g\,,
       && \langle    F_{\nabla^{(\varphi,g)}}(\xi_1,\xi_2)D_{\xi_3}\varphi\,,\,
                               D_{\xi_4}\varphi  \rangle_g
	 \end{array}
   $$
 are all defined.
      
\bigskip

\begin{sslemma} {\bf [symmetry property of the curvature tensor
          of $\nabla^{(\varphi,g)}$ on $\varphi^{\ast}{\cal T}_{\ast}Y$]}$\;$
 For a $(\ast_2)$-admissible pair $(\varphi,\nabla)$,
 \begin{eqnarray*}
  \lefteqn{
    \langle F_{\nabla^{(\varphi,g)}}(\xi_1,\xi_2)D_{\xi_3}\varphi\,,\,
     	  D_{\xi_4}\varphi \rangle_g  }\\
   &&
     =\;\;  -\:   \langle D_{\xi_3}\varphi\,,\,
	                    F_{\nabla^{(\varphi,g)}}(\xi_1,\xi_2)D_{\xi_4}\varphi \rangle_g\:
			  +\: [F_{\nabla}(\xi_1,\xi_2)\,,\,
			         \langle D_{\xi_3}\varphi\,,\, D_{\xi_4}\varphi  \rangle_g]\,.
 \end{eqnarray*}
 And, hence,
  \begin{eqnarray*}
   \lefteqn{
      \Tr \langle F_{\nabla^{(\varphi,g)}}(\xi_1,\xi_2)D_{\xi_3}\varphi\,,\,
     	                  D_{\xi_4}\varphi \rangle_g\;\;
     =\;\;  -\, \Tr \langle D_{\xi_3}\varphi\,,\,
	                    F_{\nabla^{(\varphi,g)}}(\xi_1,\xi_2)D_{\xi_4}\varphi \rangle_g  }\\[.6ex]
    && =\;\;  -\, \Tr \langle
	              F_{\nabla^{(\varphi,g)}}(\xi_1,\xi_2)D_{\xi_4}\varphi\,,\,
				  D_{\xi_3}\varphi \rangle_g\;\;
          =\;\;   \Tr \langle
	              F_{\nabla^{(\varphi,g)}}(\xi_2,\xi_1)D_{\xi_4}\varphi\,,\,
				  D_{\xi_3}\varphi \rangle_g\,.				  	
  \end{eqnarray*}
\end{sslemma}

\smallskip

\begin{proof}
 This is a consequence of the Leibniz rule
  $$
    D \langle\, \mbox{--}\,,\, \mbox{--}^{\prime} \rangle_g\;
      =\;  \langle\,   \nabla^{(\varphi,g)}\,\mbox{--}\,,\, \mbox{--}^{\prime} \rangle_g\,
		       +\, \langle\, \mbox{--}\,,\,
			             \nabla^{(\varphi,g)}\,\mbox{--}^{\prime} \rangle_g  \,,
  $$
  for --, --$^{\prime}\in \varphi^{\ast}{\cal T}_{\ast}Y$,
  when all inner products involved are defined.
 In detail,
   \begin{eqnarray*}
    \lefteqn{
      \langle
	    F_{\nabla^{(\varphi,g)}}(\xi_1,\xi_2)D_{\xi_3}\varphi\,,\,
		    D_{\xi_4}\varphi \rangle_g              }\\[.8ex]
   &&
    =\;\; \langle
	           (\nabla^{(\varphi,g)}_{\xi_1}\nabla^{(\varphi,g)}_{\xi_2}
			       - \nabla^{(\varphi,g)}_{\xi_2}\nabla^{(\varphi,g)}_{\xi_1}
				   - \nabla^{(\varphi,g)}_{[\xi_1,\xi_2]})D_{\xi_3}\varphi\,,\, D_{\xi_4}\varphi
	           \rangle_g      \\[.8ex]
   &&
    =\;\; D_{\xi_1}
	          \langle \nabla^{(\varphi,g)}_{\xi_2}D_{\xi_3}\varphi\,,\,
			                D_{\xi_4}\varphi  \rangle_g\;
		   -\;  \langle \nabla^{(\varphi,g)}_{\xi_2}D_{\xi_3}\varphi\,,\,
			               \nabla^{(\varphi,g)}_{\xi_1} D_{\xi_4}\varphi  \rangle_g  \\[.8ex]			   
   && \hspace{4em}
          -\; D_{\xi_2}
	          \langle \nabla^{(\varphi,g)}_{\xi_1}D_{\xi_3}\varphi\,,\,
			                D_{\xi_4}\varphi  \rangle_g\;
		   +\;  \langle \nabla^{(\varphi,g)}_{\xi_1}D_{\xi_3}\varphi\,,\,
			               \nabla^{(\varphi,g)}_{\xi_2} D_{\xi_4}\varphi  \rangle_g  \\[.8ex]
   && \hspace{4em}
         -\; D_{[\xi_1,\xi_2]}\langle D_{\xi_3}\varphi\,,\, D_{\xi_4}\varphi  \rangle_g\;
		 +\;  \langle D_{\xi_3}\varphi\,,\,
			        \nabla^{(\varphi,g)}_{[\xi_1,\xi_2]} D_{\xi_4}\varphi  \rangle_g  \\[.8ex]	
   &&
    =\;\;  \langle
	   D_{\xi_3}\varphi\,,\,
	           (\nabla^{(\varphi,g)}_{\xi_2}\nabla^{(\varphi,g)}_{\xi_1}
			       - \nabla^{(\varphi,g)}_{\xi_1}\nabla^{(\varphi,g)}_{\xi_2}
				   + \nabla^{(\varphi,g)}_{[\xi_1,\xi_2]})D_{\xi_4}\varphi
		\rangle_g              \\[.6ex]
   && \hspace{12em}		
	  +\; (D_{\xi_1}D_{\xi_2}- D_{\xi_2}D_{\xi_1} -D_{[\xi_1,\xi_2]})
	                     \langle D_{\xi_3}\varphi\,,\, D_{\xi_4}\varphi \rangle_g  \\[.6ex]
   && \hspace{2em}
         \mbox{\small after repeatedly applying the Leibniz rule,} \\[.8ex]
   &&
    =\;\;
	     -\; \langle  D_{\xi_3}\varphi\,,\,
	             F_{\nabla^{(\varphi,g)}}(\xi_1,\xi_2)D_{\xi_4}\varphi  \rangle_g\;
         +\;  F_D(\xi_1,\xi_2) \langle D_{\xi_3}\varphi\,,\, D_{\xi_4}\varphi \rangle_g\,.
  \end{eqnarray*}
 Note that $\; F_D\,=\, [F_{\nabla},\,\mbox{\Large $\cdot$}\,]\;$  on ${\cal O}_X^{A\!z}$.
 The lemma follows.
  
\end{proof}

\bigskip

\begin{flushleft}
{\bf Covariant differentiation and evaluation}
\end{flushleft}
\begin{sslemma} {\bf [$D_{\xi}(m\otimes vf)$
                   vs.\ $\big(\nabla^{(\varphi,g)}_{\xi}(m\otimes v)\big)f\,$]}$\;$
 Let $\xi\in{\cal T}_{\ast}X$, $f\in{\cal O}_Y$,
   and $m\otimes v\in \varphi^{\ast}{\cal T}_{\ast}Y$.
 Then,
  $$
     D_{\xi}(m\otimes vf)\;
	   =\; \big(\nabla^{(\varphi,g)}_{\xi}(m\otimes v)\big)f\,
	          +\, m\,\sum_{i=1}^n
			            D_{\xi}\varphi^{\sharp}(y^i)
						 \otimes \Big(
						    \mbox{\Large $\frac{\partial}{\partial y^i}$}(vf)
							 - \big(\nabla^g_{\frac{\partial}{\partial y^i}} v \big)f
  						              \Big)\,.			
  $$
\end{sslemma}

\smallskip

\begin{proof}
{\small
 $$
  D_{\xi}(m\otimes vf)\;
   =\; D_{\xi}\big( m \varphi^{\sharp}(vf)  \big)\;
   =\; D_{\xi}m \otimes vf \,
         +\, \sum_i D_{\xi}\varphi^{\sharp}(y^i)
		                      \otimes \mbox{\large $\frac{\partial}{\partial y^i}$} (vf)
 $$}
 while
 {\small
  $$
    \big(\nabla^{(\varphi,g)}_{\xi}(m\otimes v) \big)f\;
	 =\;  \Big (D_{\xi}m\otimes v\,
	         +\, \sum_i D_{\xi}\varphi^{\sharp}(y^i)
			                       \otimes \nabla^g_{\frac{\partial}{\partial y^i}}v   \Big)\,f\,.
  $$}
 The lemma follows.
 
\end{proof}

\bigskip
	
\subsubsection{The ${\cal O}_X^{\,\Bbb C}$-module
	  ${\cal T}^{\ast}X\otimes_{{\cal O}_X}\varphi^{\ast}{\cal T}_{\ast}Y$,
	  where $D\varphi$ lives}

 Assume that $(\varphi,\nabla)$ is $(\ast_1)$-admissible and
  recall the metric $h$ on $X$ and the metric $g$ on $Y$.
 Then the  construction in this subsubsection is a combination of the constructions
  in Sec.$\,$3.2.1 and Sec.$\,$3.2.2.
	
 The connection $\nabla^h$ on ${\cal T}^{\ast}X$,
	   the connection $D$ on ${\cal O}_X^{A\!z}$,   and
	   the connection $\nabla^g$ on ${\cal T}_{\ast}Y$
  together induce a connection $\nabla^{(h,\varphi,g)}$ on
	 ${\cal T}^{\ast}X
           \otimes_{{\cal O}_X}\!\varphi^{\ast}{\cal T}_{\ast}Y$,
    locally of the form		   		
	 {\small
     $$
       \nabla^{(h, \varphi,g) }\;=\;
	    \nabla^h\otimes \Id_{{\cal O}_X^{A\!z}}\otimes \Id_{{\cal T}_{\ast}Y}\,
         +\, \Id_{{\cal T}^{\ast}X} \otimes D^T\otimes \Id_{{\cal T}_{\ast}Y}  \,
		 +\,  \Id_{{\cal T}^{\ast}X} \otimes \Id_{{\cal O}_X^{A\!z}}
		              \cdot
		                \sum_{i=1}^n
		                    D\varphi^{\sharp}(y^i)
		                                   \otimes \nabla^g_{\frac{\partial}{\partial y^i}}\,.
     $$}
Lemma~3.2.2.1 implies that
     this is independent of the coordinate $(y^1,\,\cdots\,,\, y^n)$ on coordinate charts chosen
	 and hence well-defined	
    
The inner product $\langle\,\cdot\,,\,\cdot\, \rangle_h$ on ${\cal T}^{\ast}X$,
    the multiplication in ${\cal O}_X^{A\!z}$,  and
    the inner product $\langle\,\cdot\,,\,\cdot\,\rangle_g$ on ${\cal T}_{\ast}Y$
 together induce a {\it partially defined,
	    $\,{\cal O}_X^{A\!z}$-valued, ${\cal O}_X^{\,\Bbb C}$-bilinear
	   (nonsymmetric) inner product}  on
	    ${\cal T}^{\ast}X
             \otimes_{{\cal O}_X}\!\varphi_T^{\ast}{\cal T}_{\ast}Y$
	   by extending ${\cal O}_X^{\,\Bbb C}$-bilinearly
		\begin{eqnarray*}
		 \lefteqn{
          \langle  \omega^1\otimes m^1_T\otimes v_1\,,\,
		                 \omega^2m^2 \otimes v_2 \rangle_{(h,g)} }\\
          &&   :=\;\;   \langle\omega^1\,,\, \omega^2\rangle_h
		               \cdot  m^1_T m^2_T \otimes \langle v_1\,,\, v_2\rangle_g\;\;
		    =\;\;     \langle\omega^1\,,\, \omega^2\rangle_h\,
			            m^1m^2\,\varphi^{\sharp}(\langle v_1\,,\, v_2\rangle_g)\,,
        \end{eqnarray*}
       where
	      $\omega^1,\, \omega^2\in {\cal T}^{\ast}X$,
	      $m^1\in {\cal O}_X^{A\!z}$,
		     $m^2\in \Comm({\cal A}_{\varphi}) \subset {\cal O}_X^{A\!z}$,
	      $v_1,\, v_2\in {\cal T}_{\ast}Y$ and
		  the last equality follows from the canonical isomorphism
		   $ {\cal O}_X^{A\!z}
		          \otimes_{\varphi^{\sharp},\, {\cal O}_Y}\!\!{\cal O}_Y
		       \simeq  {\cal O}_X^{A\!z}$.	
    The trace map $\Tr:{\cal O}_X^{A\!z}\rightarrow {\cal O}_X^{\,\Bbb C}$
	  gives another {\it partially defined, $\,{\cal O}_X^{\,\Bbb C}$-valued,
	   ${\cal O}_X^{\,\Bbb C}$-bilinear (symmetric) inner product} on
	   ${\cal T}^{\ast}X
            \otimes_{{\cal O}_X}\!\varphi^{\ast}{\cal T}_{\ast}Y$
       by extending ${\cal O}_X^{\,\Bbb C}$-bilinearly
		\begin{eqnarray*}
		 \lefteqn{
          \Tr\langle  \omega^1\otimes m^1_T\otimes v_1\,,\,
		                 \omega^2m^2 \otimes v_2 \rangle_{(h,g)} }\\
          &&   :=\;\;   \Tr\big(\langle\omega^1\,,\, \omega^2\rangle_h
		               \cdot  m^1_T m^2_T \otimes \langle v_1\,,\, v_2\rangle_g \big)\;\;
		    =\;\;    \Tr\big( \langle\omega^1\,,\, \omega^2\rangle_h\,
			            m^1m^2\,\varphi^{\sharp}(\langle v_1\,,\, v_2\rangle_g) \big)\,,
        \end{eqnarray*}
       where
	      $\omega^1,\, \omega^2\in {\cal T}^{\ast}X$,
	      either $m^1$ or $m^2$ is in $\Comm({\cal A}_{\varphi})$,
	      $v_1,\, v_2\in {\cal T}_{\ast}Y$.

     By construction, both inner products, when defined,  are covariantly constant with respect to 
        $\nabla^{(h,\varphi,g)}$  and one has the Leibniz rules
	   \begin{eqnarray*}
	    D \langle\, \sim\,,\, \sim^{\prime} \rangle_{(h,g)}
		 & =
		  &  \langle\,   \nabla^{(h,\varphi,g)}\sim\,,\, \sim^{\prime} \rangle_{(h,g)}\,
		       +\, \langle\, \sim\,,\,
			             \nabla^{(h, \varphi, g)}\sim^{\prime} \rangle_{(h,g)}  \,,   \\[.6ex]	   
        d \Tr \langle\, \sim\,,\, \sim^{\prime} \rangle_{(h,g)} 					
         & =
		  & \Tr (D \langle\, \sim\,,\, \sim^{\prime} \rangle_{(h,g)})   \\
		 & =
		  & \Tr \langle\,   \nabla^{(h,\varphi,g)}\sim\,,\,
			                                                                     \sim^{\prime} \rangle_{(h,g)}\,
		       +\, \Tr \langle\, \sim\,,\,
			                    \nabla^{(h,\varphi,g)}\sim^{\prime} \rangle_{(h,g)}  \,,
       \end{eqnarray*}
	  whenever  the $\langle\, \sim^{\prime\prime}\,,\, \sim^{\prime\prime\prime} \rangle_{(h,g)}$ 
	      or $\Tr\langle\, \sim^{\prime\prime}\,,\, \sim^{\prime\prime\prime} \rangle_{(h,g)}$ involved
	   are defined.

\bigskip

With all the preparations in Sec.$\,$1--Sec.$\,$3,
 we are finally ready to construct and study the standard action for D-branes along our line of pursuit.

\bigskip
  
\section{The standard action for D-branes}

We introduce in this section the standard action, which is to D-branes
 as the (Brink-Di Vecchia-Howe/Deser-Zumino/)Polyakov action is to fundamental strings.
Abstractly, it is an enhanced non-Abelian gauged sigma model based on maps
 $\varphi: (X^{\!A\!z},{\cal E};\nabla)\rightarrow Y$.

\bigskip

\begin{flushleft}
{\bf The gauge-symmetry group $C^{\infty}(\Aut_{\Bbb C}(E))$}
\end{flushleft}
Let $\Aut_{\Bbb C}(E)$ be the {\it automorphism bundle}
 of the complex vector bundle $E$ (of rank $r$) over $E$.
$\Aut_{\Bbb  C}(E)\subset \End_{\Bbb C}(E)$ canonically
   as the bundle of invertible endomorphisms;
 it is a principal $\GL_r({\Bbb C})$-bundle over $X$.
The set
 $$
   {\cal G}_{\gaugescriptsize}\; :=\;  C^{\infty}(\Aut_{\Bbb C}(E))
  $$
 of smooth sections of  $\Aut_{\Bbb C}(E)$ forms an infinite-dimensional Lie group and acts
  on the space of  pairs $(\varphi,\nabla)$ as a gauge-symmetry group:
  $$
   \begin{array}{rccccll}
    & g^{\prime}\in {\cal G}_{\gaugescriptsize}\;:
	  && (\varphi, \nabla=d+A) & \longmapsto
	  & (^{g^{\prime}}\!\!\varphi\,,\, ^{g^{\prime}}\!\nabla = d+ \,^{g^{\prime}}\!\!A) \\
    &&&&&
	   \:\::=\; \big(g^{\prime}\varphi {g^{\prime}}^{-1}\,,\,
	             d -(dg^{\prime}){g^{\prime}}^{-1} + g^{\prime} A{g^{\prime}}^{-1}\big)&.	
   \end{array}				
  $$
The induced action of ${\cal G}_{\gaugescriptsize}$ on other basic objects
  are listed in the lemma below:
  
\bigskip

\begin{slemma}
{\bf [induced action of ${\cal G}_{\gaugescriptsize}$ on other basic objects]}$\;$
 (All the ${\cal G}_{\gaugescriptsize}$-actions are denoted by
      a representation $\rho_{\gaugescriptsize}$ of ${\cal G}_{\gaugescriptsize}$, if in need.)
 \begin{itemize}
  \item[\rm (0$_1$)] \parbox[t]{12em}{\it on ${\cal O}_X^{A\!z}\,:$}\hspace{2em}
    $\rho_{\gaugescriptsize}(g^{\prime})(m)\;=\; g^{\prime}m{g^{\prime}}^{-1}\;\;$
	for $m\in{\cal O}_X^{A\!z}$.
	
  \item[\rm (0$_2$)] \parbox[t]{12em}{\it on induced connections$\,:$}\hspace{2em}
    $D=d+[A,\,\mbox{\Large $\cdot$}\,]\; \longmapsto\;
	   \,^{g^{\prime}}\!D\,:=\, d+ [\,^{g^{\prime}}\!\!A\,,\,\mbox{\Large $\cdot$}\,]$.
	        		
  \item[\rm (1)] \parbox[t]{12em}{\it on
           ${\cal T}^{\ast}X^{\,\Bbb C}
		       \otimes_{{\cal O}_X^{\,\Bbb C}}{\cal O}_X^{A\!z}\,:$}\hspace{2em}
    $\rho_{\gaugescriptsize}(g^{\prime})(\omega\otimes m)\;
	   =\;   \omega\otimes( g^{\prime}m{g^{\prime}}^{-1})
	           =: g^{\prime}(\omega\otimes m){g^{\prime}}^{-1}$.
 
  \item[\rm (2)] \parbox[t]{12em}{\it for $\varphi^{\ast}{\cal T}_{\ast}Y\,:$}\hspace{2em}
      $$
	    \begin{array}{ccccc}
	     &\varphi^{\ast}{\cal T}_{\ast}Y
		   & \longrightarrow
		   & ^{g^{\prime}}\!\!\varphi^{\ast}{\cal T}_{\ast}Y \\[.6ex]
         & m\otimes v
		   & \longmapsto
           & (g^{\prime}m {g^{\prime}}^{-1})\otimes v 	\,
		       =:\, g^{\prime}(m\otimes v){g^{\prime}}^{-1} &.	
       \end{array}		
	  $$.	
	
  \item[\rm (3)] \parbox[t]{12em}{\it for
            ${\cal T}^{\ast}X
			    \otimes_{{\cal O}_X}\!\!\varphi^{\ast}{\cal T}_{\ast}Y\,:$}\hspace{2em}
      $$
	    \begin{array}{ccccc}
	     &{\cal T}^{\ast}X
			    \otimes_{{\cal O}_X}\!\!\varphi^{\ast}{\cal T}_{\ast}Y
		   & \longrightarrow
		   & {\cal T}^{\ast}X
			          \otimes_{{\cal O}_X}\!^{g^{\prime}}\!\!\varphi^{\ast}{\cal T}_{\ast}Y \\[.6ex]
         & \omega\otimes m\otimes v
		   & \longmapsto
           & \omega\otimes (g^{\prime}m {g^{\prime}}^{-1})\otimes v 	\,
		       =:\, g^{\prime}(\omega\otimes m\otimes v){g^{\prime}}^{-1} &.	
       \end{array}		
	  $$.
    
  \item[\rm (4)] \parbox[t]{12em}{\it for covariant differential$\,:$}\hspace{2em}
    $D\varphi\; \longmapsto\;
	   \,^{g^{\prime}}\!D\,^{g^{\prime}}\!\!\varphi\,
	     =\,  g^{\prime}D\varphi\,{g^{\prime}}^{-1} $.
		
  \item[\rm (5)] \parbox[t]{12em}{\it for pull-push$\,:$}\hspace{2em}
    $(^{g^{\prime}}\!\!\varphi)^{\diamond}\alpha\;
	   =\;   g^{\prime}\,\varphi^{\diamond}\alpha\,{g^{\prime}}^{-1} $.		 		
 \end{itemize}
\end{slemma}

\smallskip

\begin{proof}
 The proof is elementary. Let us demonstrate Item (2) as an example.
 
 For
  $m\otimes v
    \in \varphi^{\ast}{\cal T}_{\ast}Y
	:= {\cal O}_X^{A\!z}
	        \otimes_{\varphi^{\sharp},{\cal O}_Y}\!\!{\cal T}_{\ast}Y$,
   $$
     \rho_{\gaugescriptsize}(g^{\prime})(m\otimes v)\;
      =\;  \rho_{\gaugescriptsize}(g^{\prime})(m)\otimes v\;
	  =\;(g^{\prime}m {g^{\prime}}^{-1}  )\otimes v
   $$
   since
     ${\cal G}_{\gaugescriptsize}$ acts on ${\cal T}_{\ast}Y$ trivially
    	 (i.e.\ by    by the identity map $\Id_Y$).
 The only issue is: Where does $\,(g^{\prime}m {g^{\prime}}^{-1}  )\otimes v\,$ now  live?		
 To answer this, note that, for $f\in C^{\infty}(Y)$,
  on one hand
   $$
     \rho_{\gaugescriptsize}(g^{\prime})(m\otimes fv)\;
      =\; \rho_{\gaugescriptsize}(g^{\prime})(m\varphi^{\sharp}(f)\otimes v)\;
	  =\;  (g^{\prime}m\varphi^{\sharp}(f){g^{\prime}}^{-1})\otimes v\,,
   $$
 while on the other hand
   $$
      \rho_{\gaugescriptsize}(g^{\prime})(m\otimes fv)\;
	  =\;  (g^{\prime}m {g^{\prime}}^{-1})\otimes fv\,,
   $$
 It follows that
  \begin{eqnarray*}
   \lefteqn{
    (g^{\prime}m {g^{\prime}}^{-1})\otimes fv    }\\
   &&	
	  =\;  \big(g^{\prime}m\varphi^{\sharp}(f){g^{\prime}}^{-1}\big)\otimes v\;
	  =\;  \big(g^{\prime}m {g^{\prime}}^{-1}\cdot
	           g^{\prime}  \varphi^{\sharp}(f){g^{\prime}}^{-1}
			   \big)\otimes v\;
	  =\; \big(g^{\prime}m {g^{\prime}}^{-1}\cdot
	                {\,^{g^{\prime}}\!\!\varphi}^{\sharp}(f)
			\big)\otimes v\,.
  \end{eqnarray*} 			
 Which says that our section
  $\,(g^{\prime}m {g^{\prime}}^{-1}  )\otimes v\,$
   now lives in $\,^{g^{\prime}}\!\!\varphi^{\ast}{\cal T}_{\ast}Y$.

\end{proof}

%
%
%
%
   
\vspace{6em}
\bigskip

\begin{flushleft}
{\bf The standard action for D-branes}
\end{flushleft}
Fix a (dilaton field $\rho$, metric $h$) on the underlying smooth manifold $X$ (of dimension $m$)
 of the Azumaya/matrix manifold with a fundamental module $(X^{\!A\!z},{\cal E})$.
Fix a background (dilaton field $\Phi$, metric $g$ , $B$-field $B$, Ramond-Ramond field $C$)
 on the target space(-time) $Y$ (of dimension $n$).\footnote{For
                              mathematicians,
 							   $\rho$ is a smooth function on $X$,
                               $\Phi$ is a smooth function on $Y$,
                               $B$ is a $2$-form and $C$ is a general differential form on $Y$.
                               Such background fields $(\Phi, g, B,C)$ on $Y$
							     are created by massless excitations of closed superstrings on $Y$.
							   The notations for these particular fields are almost already carved into stone in string-theory literature.
							    Which we adopt here.}
Here, $h$ and $g$ can be either Riemannian or Lorentzian.
  
\bigskip

\begin{sdefinition} {\bf [standard action = enhanced non-Abelian gauged sigma model]}$\;$ {\rm
 With the given background fields $(\rho,h)$ on $X$ and $(\Phi,g,B,C)$ on $Y$,
 the {\it standard action} $S_{\standardscriptsize}^{(\rho,h;\Phi,g,B,C)}(\varphi,\nabla)$
 for $(\ast_1)$-admissible pairs $(\varphi,\nabla)$
 is defined to be the functional
 \begin{eqnarray*}
    S_{\standardscriptsize}^{(\rho, h;\Phi,g,B,C)}(\varphi, \nabla)
    &  :=  &   S_{\nAGSM^+}^{(\rho, h;\Phi,g,B,C)}(\varphi,\nabla)\\[1.2ex]
	&  :=  &   S_{\mapscriptsize:\kineticscriptsize^+}^{(\rho, h;\Phi,g)}(\varphi,\nabla)\;
	                +\; S^{(h;B)}_{\gaugescriptsize/\YMscriptsize}(\varphi, \nabla)\;
		            +\; S^{(C,B)}_{\CSWZ}(\varphi,\nabla)
 \end{eqnarray*}
 with the {\it enhanced kinetic term for maps}
 $$
  S^{(\rho, h;\Phi,g)}_{\mapscriptsize:\kineticscriptsize^+}(\varphi,\nabla)\;
    :=\;
	\mbox{\Large $\frac{1}{2}$}\,  T_{m-1}
        \int_X  \Real\Big( \Tr  \langle  D\varphi\,,\, D\varphi \rangle_{(h,g)}\Big)\, \vol_h\,
     + \, \int_X    \Real\Big(  \Tr  \langle d\rho, \varphi^{\diamond}d\Phi \rangle_h\Big)\,\vol_h\,,
 $$
 the {\it gauge/Yang-Mills term}
 $$
   S^{(h;B)}_{\gaugescriptsize/\YMscriptsize}(\varphi, \nabla)\;
    :=\;   -\,  \mbox{\Large $\frac{1}{2}$}\int_X  \Real \Big( \Tr
	                \| 2\pi\alpha^{\prime}F_{\nabla}+ \varphi^{\diamond}B\|_h^2 \Big)\, \vol_h
 $$
  and the {\it Chern-Simons/Wess-Zumino term}\\
  (if $(\varphi,\nabla)$ is furthermore $(\ast_2)$-admissible, cf.\ Remark 2.1.13)
  {\small
 $$
  S_{\CSWZ}^{(C,B)}(\varphi,\nabla)\;
   \stackrel{\mbox{\tiny{\rm formally}}}{=}\;
     T_{m-1}\,\int_X  \Real \Big( \Tr
    	 \Big(\varphi^{\diamond}C
		            \wedge e^{2\pi\alpha^{\prime}F_{\nabla}+\varphi^{\diamond}B}
	                \wedge  \sqrt{\hat{A}(X^{\!A\!z})/\hat{A}(N_{X^{\!A\!z}/Y})\,}\,
					\Big) \Big)_{(m)}\,.
 $$}
 Here,
 \begin{itemize}
  \item[(0)] {\it On $\Real$}\hspace{2em}
   Note that while eigenvalues of $\varphi^{\sharp}(f)$
     are all real ([L-Y5: Sec.$\,$3.1] (D(11.1))) for $f\in {\cal O}_Y$,
   the eigenvalues of $D_{\xi}\varphi^{\sharp}(f)$, $\xi\in{\cal T}_{\ast}X$, may not be so
    under the $(\ast_1)$-Admissible Condition.
  Thus, 	$\Tr (\cdots\cdots)$ in the integrand of terms in
    $S_{\standardscriptsize}^{(\rho, h;\Phi,g,B,C)}(\varphi, \nabla)$
	are in general ${\Bbb C}$-valued and we take the {\it real part}
	$\Real\Tr(\cdots\cdots)$ of it.
      
  \item[(1)] {\it The enhanced kinetic term for maps}\hspace{2em}
   The first summand of  $S_{\mapscriptsize:\kineticscriptsize^+}^{(\rho, h;\Phi,g)}$
    defines the {\it kinetic energy}
    $$
       E^{\nabla}(\varphi)\;
	   :=\; S_{\mapscriptsize:\kineticscriptsize}^{(h;g)}(\varphi,\nabla)\;
	   :=\;    \mbox{\Large $\frac{1}{2}$}\,  T_{m-1}
                 \int_X \Real \Big( \Tr  \langle  D\varphi\,,\, D\varphi \rangle_{(h,g)}
				                      \Big)\, \vol_h\;
    $$
      of the map $\varphi$ for a given $\nabla$ and, hence, will be called the {\it kinetic term} for maps
	  in the standard action
	  $ S_{\standardscriptsize}^{(\rho, h;\Phi,g,B,C)}(\varphi, \nabla)$.
   When the metric $g$ on $Y$	 is Lorentzian,
    then depending on the convention of its signature $(-,+,\,\cdots\,+)$ vs.\ $(+,-,\,\cdots\,-)$,
	one needs to add an overall  minus $-$ vs.\ plus $+$ sign.
   In this note, for simplicity of presentation, we choose $h$ and $g$ to be both Riemannian
   (i.e.\ for Euclideanized/Wick-rotated D-branes and space-time).

  \item[\LARGE $\cdot$]
   The world-volume $X^{\!A\!z}$ of D-brane is $m$-dimensional;
   $T_{m-1}$ is the tension of $(m-1)$-dimensional D-branes.
   Like the tension of the fundamental string, it is a fixed constant of nature.
	
  \item[\LARGE $\cdot$]	
   The second summand of  $S_{\mapscriptsize:\kineticscriptsize^+}^{(\rho, h;\Phi,g)}$
    $$
	  S_{\dilatonscriptsize}^{(\rho, h;\Phi)}(\varphi)\;
       :=\; 	  \int_X    \Real \Big(
	                   \Tr  \langle d\rho, \varphi^{\diamond}d\Phi \rangle_h
			                              \Big)\,\vol_h\,,
	$$
	will be called the {\it dilaton term} of the standard action
	$ S_{\standardscriptsize}^{(\rho, h;\Phi,g,B,C)}(\varphi,\nabla)$.
	
   \vspace{-.6ex}
   \item[]\hspace{1em}
   Note that if let $U$ be small enough and fix a local trivialization of $E|_U$.
   and assume that $\nabla=d+A$ with respect to this local trivialization.
   Then $D=d+[A,\,\mbox{\Large $\cdot$}\,]$
    and, over $U$ with an orthonormal frame $(e_{\mu})_{\mu}$,
    \begin{eqnarray*}
     \lefteqn{
      \Tr\langle d\rho, \varphi^{\diamond}d\Phi\rangle_h\;\;
       =\;\; \sum_{\mu}
          \Tr \big(d\rho(e_{\mu})D_{e_{\mu}}\varphi^{\sharp}(\Phi) \big)}\\
      && =\;\;
              \sum_{\mu}
              \Tr \Big(d\rho(e_{\mu})
		        \Big(e_{\mu}\varphi^{\sharp}(\Phi)
			            + [A(e_{\mu}), \varphi^{\sharp}(\Phi)] \Big) \Big)\;\;
           =\;  \sum_{\mu}
                  \Tr \Big(d\rho(e_{\mu})
		            \big(e_{\mu}\varphi^{\sharp}(\Phi)\big) \Big)\,.
     \end{eqnarray*}
    Thus,  while $\varphi^{\diamond}d\Phi$ depends on the connection $\nabla$,
	   the integrand $\big( \Tr  \langle d\rho, \varphi^{\diamond}d\Phi \rangle_h\big)\,\vol_h$
	   does not.
	This justifies the dilaton term as a functional of $\varphi$ alone.
	
   \vspace{-.6ex} 	
   \item[]\hspace{1em}
   In contrast, over $U$ with the above setting,
    $\Tr\langle D\varphi,D\varphi\rangle_{(h,g)}$ contains summand
	$$
	  \sum_{\mu}\sum_{i,j}
	    \Tr\Big([A(e_{\mu}), \varphi^{\sharp}(y^i)]\,
		               [A(e_{\mu}), \varphi^{\sharp}(y^j)]\,
					   \varphi^{\sharp}(g_{ij})     \Big)\,,
	$$
   which does not vanish in general.
  Thus,  $\Tr\langle D\varphi,D\varphi\rangle_{(h,g)}$
   does depend on the pair $(\varphi,\nabla)$.
      
  \item[(2)] {\it The gauge/Yang-Mills term
          $S^{(h;B)}_{\gaugescriptsize/\YMscriptsize}(\varphi, \nabla)$}\hspace{2em}
  $\alpha^{\prime}$ is the Regge slope;
  $2\pi\alpha^{\prime}$ is the inverse to the tension of a fundamental string.
  
  \item[\LARGE $\cdot$]
   $F_{\nabla}$ is the curvature tensor of the connection $\nabla$ on $E$;
   $2\pi\alpha^{\prime}F_{\nabla}+\varphi^{\diamond}B$
      is an ${\cal O}_X^{A\!z}$-valued $2$-tensor on $X$; and
   $$
     \| 2\pi\alpha^{\prime}F_{\nabla}+ \varphi^{\diamond}B \|_h^2\;
      :=\;  \langle
	          2\pi\alpha^{\prime}F_{\nabla}+ \varphi^{\diamond}B\,,\,
			  2\pi\alpha^{\prime}F_{\nabla}+ \varphi^{\diamond}B \rangle_h
   $$
   from Sec.~3.2.1.
  Up to the shift by $\varphi^{\diamond}B$,
   this is a norm-squared of the field strength of the gauge field, and hence the name
   {\it Yang-Mills term}.
  Note that in $S^{(h;B)}_{\gaugescriptsize/\YMscriptsize}(\varphi, \nabla)$,
   $\nabla$ couples with $\varphi$ only through the background $B$-field $B$.
  When $B=0$, this is simply a functional
   $S^{(h)}_{\gaugescriptsize/\YMscriptsize}(\nabla)$ of $\nabla$ alone.
	
  \item[\LARGE $\cdot$]
   In the current bosonic case,
   the {\it Yang-Mills functional} for the gauge term
   $S^{(h;B)}_{\gaugescriptsize/\YMscriptsize}(\varphi, \nabla)$
    can be replaced any other standard action functional, e.g.\ Chern-Simons functional,  in gauge theories.
		
  \item[(3)] {\it The Chern-Simons/Wess-Zumino term
          $S^{(C,B)}_{\CSWZ}(\varphi,\nabla)$}\hspace{4em}
   The coupling constant of Ramond-Ramond fields with D-branes is taken to be equal to the D-brane tension
      $T_{m-1}$.
   This choice is adopted from the situation of the Dirac-Born-Infeld action.
   However, in the current bosonic case, one may take a different constant.
   As given here,  $S^{(C,B)}_{\CSWZ}(\varphi,\nabla)$ is only formal;
   the {\it anomaly factor
       $\sqrt{\hat{A}(X^{\!A\!z})/\hat{A}(N_{X^{\!A\!z}/Y})\,}$} in its integrand
	   remains to be understood in the current situation.
  
  \item[\LARGE $\cdot$]
   The wedge product of ${\cal O}_X^{A\!z}$-valued differential forms
    was discussed in [L-Y8: Sec.6.1] (D(13.1)).
   An Ansatz was proposed there in accordance with the notion of `symmetrized determinant'
     for an ${\cal O}_X^{A\!z}$-valued $2$-tensor on $X$
	 in the construction of the non-Abelian Dirac-Born-Infeld action ibidem.
   Here, we no longer have a direct guide from the construction of the kinetic term
     $S_{\mapscriptsize:\kineticscriptsize}^{(h;g)}(\varphi,\nabla)$ for maps
	as to how to define such wedge products.
   However, just like Polyakov string should be thought of as being equivalent to Nambu-Goto string
    (at least at the classical level) but technically more robust,
   here we would think that `standard D-branes'	should be equivalent to `Dirac-Born-Infeld D-branes'
    (at least classically) and, hence, will take the same Ansatz:
  
  \item[]\hspace{1em}\parbox[t]{38em}{
    {\bf Ansatz$\;$[wedge product in the Chern-Simons/Wess-Zumino action]}$\;$
     We interpret the wedge products that appear in the formal expression
	 for the Chern-Simons/Wess-Zumino term
     $S_{\CSWZ}^{(C,B)}(\varphi,\nabla)$
     through the symmetrized determinant that applies to the above defining identities for wedge product;
     namely, we require that
     $$
     (\omega^1\wedge\,\cdots\,\wedge\omega^s)(e_1\wedge\,\cdots\,\wedge e_s)\;=\;
        \SymDet(\omega^i(e_j))
     $$
      for ${\cal O}_X^{A\!z}$-valued $1$-forms $\omega^1,\,\cdots\,,\, \omega^s$ on $X$.
     Denote this generalized wedge product by $\odotwedge$.					
	 }
 
  \item[]
  Then, for lower-dimensional D-branes $m=0,1,2,3$,
    it is reasonable to assume that the anomaly factor is $1$ (i.e.\ no anomaly) and
   $S_{\CSWZ}^{(C,B)}(\varphi,\nabla)$ can be written out precisely.
   
  \item[]\hspace{1.2em}
   Locally in terms of a local frame $(e_{\mu})_{\mu}$ on an open set $U\subset X$
    and a coordinate $(y^1,\,\cdots\,,\, y^n)$ on a local chart of $Y$, one has:
  (Assuming that $B=\sum_{i,j}B_{ij}dy^i\otimes dy^j$, $B_{ji}=-B_{ij}$.)
    \begin{itemize}
     \item[\LARGE $\cdot$]
      For {\it D$(-1)$-brane} world-point $(m=0)\,$:
	   {\small
       $$
	     S_{\tinyCSWZ}^{(C_{(0)})}(\varphi)\;
	     =\; T_{-1}\,\cdot\, \Tr(\varphi^{\diamond}C_{(0)})\;
	     =\; T_{-1}\,\cdot\, \Tr(\varphi^{\sharp}(C_{(0)}))\,.
	   $$}
   
     \medskip
     \item[\LARGE $\cdot$]
      For {\it D-particle} world-line $(m=1)\,$:
	   Assume that  $C_{(1)}=\sum_{i=1}^nC_i\,dy^i$ locally; then
	   {\small
       $$
	     S_{\tinyCSWZ}^{(C_{(1)})}(\varphi)\;
	     =\; T_0 \int_X \Real\big( \Tr(  \varphi^{\diamond}C_{(1)}) \big)\;\;\;	
	     \stackrel{\mbox{\tiny locally}}{=}\;
	     T_0\int_U \Real \Big(\Tr
	        \Big(
		       \sum_{i=1}^n \varphi^{\sharp}(C_i) \cdot D_{e_1}\varphi^{\sharp}(y^i)
			    \Big) \Big) de^1\,.	
	   $$}
     Note that as in the case of the dilaton term
	   $S_{\dilatonscriptsize}^{(\rho,h;\Phi)}(\varphi)$,
	  this is a functional of $\varphi$ alone.
   
     \medskip
     \item[\LARGE $\cdot$]
      For {\it D-string} world-sheet $(m=2)\,$:
	  Assume that $C_{(2)}= \sum_{i,j=1}^nC_{ij}\,dy^i\otimes dy^j$ locally,\\
	   with $C_{ij}=-C_{ji}$;
	  then
	  {\small
      \begin{eqnarray*}
	   \lefteqn{
	    S_{\tinyCSWZ}^{(C_{(0)}, C_{(2)}, B)}(\varphi,\nabla)\;\;
	       = \;\; T_1 \int_X    \Real( \Tr(
	                   \varphi^{\diamond}C_{(2)}\, +\,  \varphi^{\diamond}(C_{(0)}B)
			            +\, 2\pi\alpha^{\prime} \varphi^{\sharp}(C_{(0)})\odot F_{\nabla} ))   }\\
          && \hspace{.7em}=\hspace{1.3em}
	              T_1 \int_X    \Real( \Tr(
	                  \varphi^{\diamond}(C_{(2)}+C_{(0)}B   )\,
			           +\, \pi\alpha^{\prime} \varphi^{\sharp}(C_{(0)})F_{\nabla}\,
			  		   +\, \pi\alpha^{\prime}F_{\nabla}\varphi^{\sharp}(C_{(0)})
				        ))    \\
         && \stackrel{\mbox{\tiny locally}}{=}\;\;\,
	        T_1\int_U
	         \Real    \!\left(\rule{0ex}{1em}\right.\!\!
		        \Tr  \!\left(\rule{0ex}{1em}\right.
		        \sum_{i,j=1}^n\,
		  	        \varphi^{\sharp}(C_{ij}+C_{(0)}B_{ij})\,
				      D_{e_1}\varphi^{\sharp}(y^i)\,D_{e_2}\varphi^{\sharp}(y^j)\,  \\
         && \hspace{6em}					
                 +\, \pi\alpha^{\prime}
				         \varphi^{\sharp}(C_{(0)})\,F_{\nabla}(e_1,e_2)\,
			    +\, \pi\alpha^{\prime}
					     F_{\nabla}(e_1,e_2)\, \varphi^{\sharp}(C_{(0)})		
			   \left.\rule{0ex}{1em}\right)\!\!
			        \left.\rule{0ex}{1em}\right)
		      e^1\wedge e^2   \\
         && \hspace{.7em}=\hspace{1.3em}
	        T_1\int_U
	         \Real    \!\left(\rule{0ex}{1em}\right.\!\!
		        \Tr  \!\left(\rule{0ex}{1em}\right.
		        \sum_{i,j=1}^n\,
		  	        \varphi^{\sharp}(C_{ij}+C_{(0)}B_{ij})\,
				      D_{e_1}\varphi^{\sharp}(y^i)\,D_{e_2}\varphi^{\sharp}(y^j)\,  \\[-2ex]
         && \hspace{20em}					
                 +\, 2\,\pi\alpha^{\prime}
				         \varphi^{\sharp}(C_{(0)})\,F_{\nabla}(e_1,e_2)\,
			   \left.\rule{0ex}{1em}\right)\!\!
			        \left.\rule{0ex}{1em}\right)
		      e^1\wedge e^2\,.
	   \end{eqnarray*}}
	 Here, the last identity comes from the effect of the trace map $\Tr$.
	
	 \medskip
     \item[\LARGE $\cdot$]
      For {\it D-membrane} world-volume $(m=3)\,$:
	  Assume that
	     $C_{(1)}=\sum_{i=1}^nC_i\,dy^i$ and
	     $C_{(3)}=\sum_{i,j,k=1}^nC_{ijk}\,dy^i\otimes dy^j\otimes dy^k$ locally,
	     with $C_{ijk}$ alternating with respect to $ijk$;
	  then
    {\small
  	  \begin{eqnarray*}
	   \lefteqn{
	    S_{\tinyCSWZ}^{(C_{(1)}, C_{(3)}, B)}(\varphi,\nabla)\;\;
	     =\;\;  T_2 \int_X    \Real( \Tr(
	                \varphi^{\diamond}C_{(3)}\,
					+\,  \varphi^{\diamond}(C_{(1)}\wedge B)\,
	                +\, 2\pi\alpha^{\prime}\,
					            \varphi^{\diamond}C_{(1)}\odotwedge F_{\nabla}\,
			                ))     }   \\
       && \stackrel{\mbox{\tiny locally}}{=}\;\;
	     T_2\int_U
	         \Real    \!\left(\rule{0ex}{1em}\right.\!\!
		        \Tr  \!\left(\rule{0ex}{1em}\right.
			     \sum_{i,j,k=1}^n
				    \varphi^{\sharp}(C_{ijk}+C_iB_{jk}+C_jB_{ki}+C_kB_{ij})\,  \\[-2.4ex]
          &&\hspace{22em} \cdot\,	
					   D_{e_1}\varphi^{\sharp}(y^i)\,
					   D_{e_2}\varphi^{\sharp}(y^j)\,
					   D_{e_3}\varphi^{\sharp}(y^k) \\
       && \hspace{4em}					
		      +\, \pi\alpha^{\prime}
		 	        \sum_{(\lambda\mu\nu)\in\scriptsizeSym_3}
			          \sum_{i=1}^n\,
			           (-1)^{(\lambda\mu\nu)}
					     \left(\rule{0ex}{1em}\right.
						     \varphi^{\sharp}(C_i)\,D_{e_{\lambda}}(\varphi^{\sharp}(y^i))\,
						           F_{\nabla}(e_{\mu}, e_{\nu})\,    \\[-2.4ex]
       && \hspace{18em}								
                            +\, F_{\nabla}(e_{\mu}, e_{\nu})\,
							          \varphi^{\sharp}(C_i)\,D_{e_{\lambda}}\varphi^{\sharp}(y^i)
                           \!\left.\rule{0ex}{1em}\right)
			   \!\!\left.\rule{0ex}{1em}\right)\!\!
			        \left.\rule{0ex}{1em}\right)
		      e^1\wedge e^2\wedge e^3                  \\
     && \hspace{.7em}=\hspace{1.3em}
	     T_2\int_U
	         \Real    \!\left(\rule{0ex}{1em}\right.\!\!
		        \Tr  \!\left(\rule{0ex}{1em}\right.
			     \sum_{i,j,k=1}^n
				    \varphi^{\sharp}(C_{ijk}+C_iB_{jk}+C_jB_{ki}+C_kB_{ij})\,  \\[-2.4ex]
          &&\hspace{22em} \cdot\,	
					   D_{e_1}\varphi^{\sharp}(y^i)\,
					   D_{e_2}\varphi^{\sharp}(y^j)\,
					   D_{e_3}\varphi^{\sharp}(y^k) \\
       && \hspace{4em}					
		      +\, 2\pi\alpha^{\prime}
		 	        \sum_{(\lambda\mu\nu)\in\scriptsizeSym_3}
			          \sum_{i=1}^n\,
			           (-1)^{(\lambda\mu\nu)}
					     \left(\rule{0ex}{1em}\right.
						     \varphi^{\sharp}(C_i)\,D_{e_{\lambda}}(\varphi^{\sharp}(y^i))\,
						           F_{\nabla}(e_{\mu}, e_{\nu})
                           \!\left.\rule{0ex}{1em}\right)
			   \!\!\left.\rule{0ex}{1em}\right)\!\!
			        \left.\rule{0ex}{1em}\right)
		      e^1\wedge e^2\wedge e^3\,.		    								
	  \end{eqnarray*}}
	  Here, the last identity comes from the effect of the trace map $\Tr$.
    \end{itemize}
  Their partial study was done in [L-Y8 : Sec.$\,$6.2] (D(13.1)).
    		
  \item[(4)]	{\it The background $B$-field}\hspace{2em}
   The coupling of $(\varphi,\nabla)$ with the background $B$-field $B$ on $Y$ in the part
    $$
	   S^{(h;B)}_{\gaugescriptsize/\YMscriptsize}(\varphi, \nabla)\;
        +\; S_{\CSWZ}^{(C,B)}(\varphi,\nabla)
    $$
    of the standard action		
	means that
    we have to adjust the fundamental module ${\cal E}$ on $X$
	by a compatible ``twisting" governed by $\varphi$ and $B$.
   With this ``twisting",	${\cal E}$ now lives on a gerb over $X$.
   See [L-Y2] (D(5)) for details and further references.
   However,
    since the study of the variational problems in this note is mainly local and focuses on
	 the enhanced kinetic term for maps
	 $S_{\mapscriptsize:\kineticscriptsize^+}^{(\rho, h;\Phi,g)}$,
   we'll ignore	this twisting for the current note to keep the language and expressions simple.  
 \end{itemize}
}\end{sdefinition}

\medskip

\begin{sremark} $[$other effects from $B$-field and Ramond-Ramond field$\,]\;$ {\rm
 There are other effects to D-branes beyond just mentioned above
   from the background $B$-field and Ramond-Ramond field
   that have not yet been taken into account in this project so far; e.g.\ [H-M1], [H-M2], and [H-Y].
 They can influence the action for D-branes as well.
 Such additional effects should be investigated in the future.
}\end{sremark}

\medskip

\begin{stheorem} {\bf [well-defined gauge-symmetry-invariant action]}$\;$
  Except the anomaly factor in the Chen-Simons/Wess-Zumino term, {\rm which is yet to be understood},
  the standard action
   $S_{\standardscriptsize}^{(\rho, h;\Phi,g,B,C)}(\varphi, \nabla)$
   as given in Definition~4.2
   for $(\ast_1)$-admissible pairs $(\varphi,\nabla)$
   (and $S_{\CSWZ}^{(C,B)}(\varphi,\nabla)$ for $(\ast_2)$-admissible $(\varphi,\nabla)$)
   is well-defined.
 Assume that the anomaly factor in the Chen-Simons/ Wess-Zumino term transforms also by conjugation
  as for ${\cal O}_X^{A\!z}$ under a gauge symmetry,
  then $S_{\standardscriptsize}^{(\rho, h;\Phi,g,B,C)}(\varphi, \nabla)$
    is invariant under gauge symmetries:
    $$
	  S_{\standardscriptsize}^{(\rho, h;\Phi,g,B,C)}(\varphi, \nabla)\;
	     =\; S_{\standardscriptsize}^{(\rho, h;\Phi,g,B,C)}
		      (\,^{g^{\prime}}\!\!\varphi, \,^{g^{\prime}}\!\nabla)
    $$
	for $g^{\prime}\in {\cal G}_{\gaugescriptsize}:=C^{\infty}(\Aut_{\Bbb C}(E))$.
\end{stheorem}
 
\smallskip
 
\begin{proof}
 For the kinetic term for maps
  $$
    S_{\mapscriptsize:\kineticscriptsize}^{(h;g)}(\varphi,\nabla)\;
	   :=\;    \mbox{\Large $\frac{1}{2}$}\,  T_{m-1}
                 \int_X \Real \Big( \Tr  \langle  D\varphi\,,\, D\varphi \rangle_{(h,g)}\Big)\, \vol_h\,,
  $$				
 that it is well-defined follows Lemma~3.2.2.4.
 Under a gauge transformation
  $g^{\prime}\in {\cal G}_{\gaugescriptsize}:= C^{\infty}(\Aut_{\Bbb C}(E))$
  and in terms of local coordinates
    $(x^1,\,\cdots\,,\, x^m)$ on $X$ and $(y^1,\,\cdots\,,\, y^n)$ on $Y$,
 {\small	
  $$
    \,^{g^{\prime}}\!\!D\,^{g^{\prime}}\!\!\varphi\;
	=\;
	  \sum_{\mu}dx^{\mu}
	   \otimes
	    \sum_i\,
	         ^{g^{\prime}}\!\!D_{\frac{\partial}{\partial x^{\mu}}}\,
			 ^{g^{\prime}}\!\!\varphi^{\sharp}
			           \big(\mbox{\large $\frac{\partial}{\partial y^i}$}\big)
			 \otimes_{^{g^{\prime}}\!\!\varphi}
			   \mbox{\large $\frac{\partial}{\partial y^i}$}\;
     =\; 	
	   \sum_{\mu} dx^{\mu}
	    \otimes
	     \sum_i\,\Big(
	         g^{\prime}\,
			  \Big(D_{\frac{\partial}{\partial x^{\mu}}}
			     \varphi^{\sharp} \big(\mbox{\large $\frac{\partial}{\partial y^i}$}\big)\Big)\,
			 {g^{\prime}}^{-1}
	           \Big)
			 \otimes_{^{g^{\prime}}\!\!\varphi}
			   \mbox{\large $\frac{\partial}{\partial y^i}$}\,.
  $$}
 Thus,
  {\small
  \begin{eqnarray*}
   \lefteqn{
    \langle\,^{g^{\prime}}\!\!D\,^{g^{\prime}}\!\!\varphi\,,\,
	  ^{g^{\prime}}\!\!D\,^{g^{\prime}}\!\!\varphi\rangle_{(h,g)}
	  }\\
  &&	=\;
    \sum_{\mu, \nu}\sum_{i,j}
	   h^{\mu\nu}
	    \otimes
	     \Big(
	         g^{\prime}\,
			  \Big(D_{\frac{\partial}{\partial x^{\mu}}}
			     \varphi^{\sharp} \big(\mbox{\large $\frac{\partial}{\partial y^i}$}\big)\Big)\,
			 {g^{\prime}}^{-1}\,
			  \cdot\,			
			   g^{\prime}
			    \Big(D_{\frac{\partial}{\partial x^{\nu}}}
			     \varphi^{\sharp} \big(\mbox{\large $\frac{\partial}{\partial y^j}$}\big)\Big)\,
			 {g^{\prime}}^{-1}
	           \Big)
		 \otimes_{^{g^{\prime}}\!\!\varphi} g_{ij}    \\
    && =\;
	  \sum_{\mu, \nu}\sum_{i,j}
	   h^{\mu\nu}\cdot
	     \Big(
	         g^{\prime}\,
			  \Big(D_{\frac{\partial}{\partial x^{\mu}}}
			     \varphi^{\sharp} \big(\mbox{\large $\frac{\partial}{\partial y^i}$}\big)\Big)\,
			 {g^{\prime}}^{-1}\,
			  \cdot\,			
			   g^{\prime}
			    \Big(D_{\frac{\partial}{\partial x^{\nu}}}
			     \varphi^{\sharp} \big(\mbox{\large $\frac{\partial}{\partial y^j}$}\big)\Big)\,
			 {g^{\prime}}^{-1}
	           \Big)
		 \cdot  g^{\prime}\varphi^{\sharp}( g_{ij} ){g^{\prime}}^{-1}\\
    && =\;
	  g^{\prime}
	  \Big(
	  \sum_{\mu, \nu}\sum_{i,j} 	
	     h^{\mu\nu}
		   \cdot
		     D_{\frac{\partial}{\partial x^{\mu}}}
			     \varphi^{\sharp} \big(\mbox{\large $\frac{\partial}{\partial y^i}$}\big)
			    \cdot D_{\frac{\partial}{\partial x^{\nu}}}
			     \varphi^{\sharp} \big(\mbox{\large $\frac{\partial}{\partial y^j}$}\big)
	           \cdot \varphi^{\sharp}( g_{ij} ) \Big)\,{g^{\prime}}^{-1}\\		 			
	&& =\; g^{\prime}\,
	             \langle D\varphi\,,\, D\varphi \rangle_{(h,g)}\,
			  {g^{\prime}}^{-1}\,.
  \end{eqnarray*}}
 It follows that
   $\;\Tr\langle\,^{g^{\prime}}\!\!D\,^{g^{\prime}}\!\!\varphi\,,\,
	  ^{g^{\prime}}\!\!D\,^{g^{\prime}}\!\!\varphi\rangle_{(h,g)}
	=  \Tr\langle D\varphi\,,\, D\varphi \rangle_{(h,g)}\;$
  and, hence,
  $$
     S_{\mapscriptsize:\kineticscriptsize}^{(h;g)}
         (\,^{g^{\prime}}\!\!\varphi, \,^{g^{\prime}}\!\nabla)\;
      =\; S_{\mapscriptsize:\kineticscriptsize}^{(h;g)}(\varphi,\nabla)\,.
  $$	
   	
 The other terms in
  $S_{\standardscriptsize}^{(\rho, h;\Phi,g,B,C)}(\varphi, \nabla)$
  do not involve a partially-defined inner product and hence are all defined.
 That the integrand inside $\Tr$
  all transform by conjugation under a gauge symmetry as for ${\cal O}_X^{A\!z}$
  follows Lemma~4.1.
 
 This proves the theorem.
 
\end{proof}

\bigskip

\begin{sremark} $[$gauge-fixing condition$]\;$ {\rm
 As in any gauge field theory (e.g.\ [P-S]),
 understanding how to fix the gauge is an important part of understanding
 $S_{\standardscriptsize}^{(\rho,h;\Phi,g,B,C)}(\varphi,\nabla)$.
}\end{sremark}

\bigskip

\begin{flushleft}
{\bf The standard action as an enhanced non-Abelian gauged sigma model}
\end{flushleft}
Recall that, in an updated language and in a form for easy comparison,
a {\it sigma model} ($\sigma$-model, SM) on a (Riemannian or Lorentzian) manifold $(Y,g)$
  (of dimension $n$)
 is a field theory on a (Riemannian or Lorentzian) manifold $(X,h)$ (of some dimension $m$) with
   \begin{itemize}
    \item[\LARGE $\cdot$] {\it Field$\,:$} \hspace{2em}
	 differentiable maps $\; f:X\rightarrow Y$,
	
    \item[\LARGE $\cdot$]	{\it Action functional$\,:$}\hspace{2em}
      \begin{eqnarray*}
        S^{(h,g)}_{\mbox{\scriptsize\it sigma model }}(f)
		 &:=
		   & \pm\, \mbox{\Large $\frac{1}{2}$}\, \int_X
		             \langle  df\,,\, df \rangle_{(g,h)}\,\vol_h\;\;
             =\;\;  \pm \mbox{\Large $\frac{1}{2}$}\, \int_X  \| f^{\ast}g\|^2_h\,\vol_h   \\
		& :=
		  & \pm\, \mbox{\Large $\frac{1}{2}$}\, \int_X
		         \sum_{\mu,\nu=1}^m\sum_{i,j=1}^n
				   h^{\mu\nu}(\boldx)g_{ij}(f(\boldx))
				     \mbox{\Large $\frac{\partial f^i}{\partial x^{\mu}}$}(\boldx)\,
					 \mbox{\Large $\frac{\partial f^j}{\partial x^{\nu} }$}(\boldx)\,
					\sqrt{|\determinant h (\boldx)|}\: d^{\,m}\boldx\,,
      \end{eqnarray*}
   \end{itemize}
   in terms of local coordinates
    $\boldx=(x^1,\,\cdots\,,\, x^m)$ on $X$ and $\boldy=(y^1,\,\cdots\,,\, y^n)$ on $Y$;
  cf.\ [GM-L] and see e.g.\ [C-T] for modern update and further references.
 (The $\pm$ sign depends on the signature of the metric.)
 At the classical level, this is a theory of harmonic maps;
  cf.\ [E-L], [E-S], [Ma], [Sm].

Back to our situation.
To begin with, the kinetic term
  $$
    S_{\mapscriptsize:\kineticscriptsize}^{(h;g)}(\varphi,\nabla)\;
	   :=\;    \mbox{\Large $\frac{1}{2}$}\,  T_{m-1}
                 \int_X \Real \big( \Tr  \langle  D\varphi\,,\, D\varphi \rangle_{(h,g)}
				                      \big)\, \vol_h
  $$
 qualifies the standard action
  $S_{\standardscriptsize}^{(\rho, h;\Phi,g,B,C)}(\varphi, \nabla)$
  to be regarded as a sigma model,
  now based on
  \begin{itemize}
   \item[\LARGE $\cdot$] {\it Field$\,:$}\hspace{2em}
    $(\ast_1)$-admissible differentiable maps
	$\;\varphi:(X^{\!A\!z},{\cal E};{\nabla})\rightarrow Y$.
  \end{itemize}
The fact that $S_{\standardscriptsize}^{(\rho, h;\Phi,g,B,C)}(\varphi, \nabla)$
  is invariant under the gauge symmetry group\\
    ${\cal G}_{\gaugescriptsize}:= C^{\infty}(\Aut_{\Bbb C}(E))$
  and that the latter is non-Abelian justify that
   this sigma model is indeed a {\it non-Abelian gauged sigma model} (nAGSM).
However,
 compared with, for example, the well-studied $d=2$, $N=(2,2)$ (Abelian) gauged linear sigma model,
     e.g.\ [H-V] and [Wi1],
 the gauge symmetry of $S_{\standardscriptsize}^{(\rho, h;\Phi,g,B,C)}(\varphi, \nabla)$
  does not arise from gauging a global group-action on the target space $Y$.
(For this reason, one may call
    $S_{\standardscriptsize}^{(\rho, h;\Phi,g,B,C)}(\varphi, \nabla)$
	 a {\it sigma model with non-Abelian gauge symmetry} as well.)
For D-branes, its additional coupling to the background Ramond-Ramond field $C$ on $Y$ is essential ([Po1])
   and, hence, the Chern-Simons/Wess-Zumino term $S^{(C,B)}_{\CSWZ}(\varphi,\nabla)$.
 Also, we like our dynamical field $(\varphi,\nabla)$
     coupled to the background dilaton field $\Phi$ on $Y$ as well
   so that the essence of the other important action --- the Dirac-Born-Infeld action ---  for D-branes
   can be retained as much as we can.
 This motivates the dilaton term $S_{\dilatonscriptsize}^{(\rho, h;\Phi)}(\varphi)$.
 In summary,
  \begin{eqnarray*}
   S_{\standardscriptsize}^{(\rho, h;\Phi,g,B,C)}(\varphi, \nabla)
      & :=\:
	  &   S_{\nAGSM}^{(\rho, h;\Phi,g,B)}(\varphi,\nabla)\;
	            +\;     S^{(C,B)}_{\CSWZ}(\varphi,\nabla)\;
				+\;     S_{\dilatonscriptsize}^{(\rho, h;\Phi)}(\varphi)   \\[.8ex]
    & \:=:
	  &   S_{\nAGSM^+}^{(\rho, h;\Phi,g,B,C)}(\varphi,\nabla)\,,
  \end{eqnarray*}
  which explains the name {\it enhanced non-Abelian gauged sigma model} (nAGSM$^+$).

\bigskip

\section{Admissible family of admissible pairs $(\varphi_T,\nabla^T)$}

In this section we introduce the notion of one-parameter admissible families of admissible pairs
 and rephrase the basic settings and results in Sec.$\,$3.2 in a relative format for such a family.
Some curvature tensor computations are given for later use.
The natural generalization (without work) to two-parameter admissible families of admissible pairs is remarked
 in the last theme of the section.
This prepares us for the study of the variational problem of the enhanced kinetic term for maps
 $S_{\mapscriptsize:\kineticscriptsize^+}^{(\rho, h;\Phi,g)}(\varphi,\nabla)$
 in the standard action $S_{\standardscriptsize}^{(\rho, h;\Phi,g,B,C)}(\varphi, \nabla)$
  for D-branes.

\bigskip

\begin{flushleft}
{\bf Basic setup and the notion of admissible families of admissible pairs $(\varphi_T, \nabla^T)$}
\end{flushleft}
Let
 $T=(-\varepsilon, \varepsilon)\subset {\Bbb R}^1$, with coordinate $t$
   and $\varepsilon>0$ small, be the one-parameter space  and
$\partial_t := \partial/\partial t$ and $dt$ be respectively the tangent vector field and the $1$-form
 determined by the coordinate $t$ on $T$.
Let
 $(X,E)$ be a manifold $X$ of dimension $m$ with a complex vector bundle $E$ of rank $r$.
Recall the structure sheaf ${\cal O}_X$ of $X$  and
 the ${\cal O}_X$-module ${\cal E}$ from $E$.

Consider the following families of objects over $T\,$:
 \begin{itemize}
  \item[\LARGE $\cdot$]
   $X_T:= X\times T $, with the structure sheaf ${\cal O}_{X_T}$  and
      regarded as the constant family of manifolds over $T$ determined by $X$.
   $X_T$ is equipped with the built-in projection maps
      $\pr_X: X_T\rightarrow X$ and $\pr_T:X_T\times T\rightarrow T$.
   For $U\subset X$ an open set, we will denote by $U_T$
    the corresponding open set $U\times T\subset X\times T$ over $T$.
	
  \item[\LARGE $\cdot$]	
   $T_{\ast}X_T :=$ the {\it tangent bundle} of $X_T\;$ and
     $\;{\cal T}_{\ast}X_T :=$ the {\it tangent sheaf} of $X_T$;\\
   $T^{\ast}X_T :=$ the {\it cotangent bundle} of $X_T\;$  and
     $\;{\cal T}^{\ast}X_T :=$ the {\it cotangent sheaf} of $X_T$;\\
   $T_{\ast}(X_T/T) :=$ the {\it relative tangent bundle} of $X_T$ over $T\;$   and\\
   ${\cal T}_{\ast}(X_T/T) :=$ the {\it relative tangent sheaf} of $X_T$ over $T$;\\
   $T^{\ast}(X_T/T) :=$ the {\it relative cotangent bundle} of $X_T$ over $T\;$ and\\
   ${\cal T}^{\ast}(X_T/T) :=$ the {\it relative cotangent sheaf} of $X_T$ over $T$.\\
   When $X$ is endowed with a (Riemannain or Lorentzian) metric $h$,
    $h$ induces canonically an inner-product structure on fibers of $T_{\ast}(X_T/T)$
    and its dual, $T^{\ast}(X_T/T)$, over $T$.
   These induced inner-product structure will be denoted by $\langle\,\cdot\,\,,\,\cdot\rangle_h$.
   
  \item[\LARGE $\cdot$]
   $E_T := \pr_X^{\ast}E$ the pull-back vector bundle of $E$ to $X_T$,
       regarded as the constant $T$-family of vector bundles over $X$ determined by $E$;  and
   ${\cal E}_T := \pr_X^{\ast}{\cal E}$ the corresponding ${\cal O}_{X_T}$-module,
       regarded as the constant $T$-family of ${\cal O}_X$-modules determined by ${\cal E}$.\\
   The projection map $\pr_X:X_T\rightarrow X$ induces a projection map
    $\pr_E: E_T\rightarrow E$ between the total space of bundles in question.
  $T_{\ast}E_T$ (resp.\ ${\cal T}_{\ast}E_T$) denotes the tangent space
    (resp.\ the tangent sheaf) of the total space of $E_T$. 	

  \item[\LARGE $\cdot$]
   $(X_T^{\!A\!z}, {\cal E}_T)
       \,:=\,  (X_T,
	            {\cal O}_{X_T}^{A\!z}
		           \!:= \Endsheaf_{{\cal O}_{X_T}^{\,\Bbb C}}\!\!({\cal E}_T),
		      {\cal E}_T)$,
    regarded as the constant $T$-family of Azumaya/matrix manifolds with a fundamental module
	determined by $(X^{\!A\!z},{\cal E})$.
 There is a {\it trace map}
   $$
      \Tr\; :\; {\cal O}_{X_T}^{A\!z}\;\longrightarrow\; {\cal O}_{X_T}^{\,\Bbb C}
   $$
   as ${\cal O}_{X_T}$-modules, which takes $\Id_{{\cal E}_T}$ to $r$.
 \end{itemize}
and take the following notational conventions:
 \begin{itemize}
  \item[\LARGE $\cdot$]	
   Through the product structure $X_T=X\times T$,
    a vector field $\xi$ (resp.\ $1$-form $\omega$) on $X$ and the vector field $\partial_t$ on $T$
	lift canonically to a vector field (resp.\ $1$-form) on $X_T$,
	which will still be denoted by $\xi$ (resp.\ $\omega$) and $\partial_t$ respectively.
	
  \item[\LARGE $\cdot$]	
  For referral, the restriction of $X_T, X^{\!A\!z}_T$, $E_T$, $\cdots_T$ to over $t\in T$ will be denoted
   $X_t$, $X^{\!A\!z}_t$, $E_t$, $\cdots_t$ respectively.
 \end{itemize}	

\bigskip

\begin{sdefinition} {\bf [connection/covariant derivation trivially flat over $T$]}$\;$ {\rm
 A connection $\nabla^T$ on $E_T$
   (equivalently, connection/covariant derivative $\nabla^T$ on ${\cal E}_T$)
  is said to be {\it trivially flat over $T$}
  if the horizontal lifting of $\partial_t$ to $T_{\ast}E_T$ lies in the kernel of the map
  ${\pr_E}_{\ast}:T_{\ast}E_T \rightarrow T_{\ast}E$.
 For such a $\nabla^T$,
   we will denote the covariant derivative $\nabla^T_{\partial_t}$ simply by $\partial_t$.
 The {\it curvature tensor} of $\nabla^T$  will be denoted by $F_{\nabla^T}$.
}\end{sdefinition}

\bigskip

Note that any connection on $E_T$ is flat over $T$ and hence, due to the topology of $T$,
 can be made trivially flat over $T$ after a bundle-isomorphism.
Thus  the notion of `trivially flat' is only a notational convenience for our variational problem, not a true constraint.
However, caution that
 while $\nabla^T$ is always flat over $T$,
   its restriction $\nabla^t$ to $X_t$ varies as $t$ varies in $T$.
Thus, in general, $F_{\nabla^T}(\partial_t, \,\cdot\,)\ne 0$.
 
\bigskip

\begin{sdefinition} {\bf [admissible family of admissible pairs $(\varphi_T,\nabla^T)$]}$\;$ {\rm
 A {\it T-family of maps with varying connections} from $(X^{\!A\!z},{\cal E})$ to $Y$
   is a pair $(\varphi_T,\nabla^T)$, where
  $$
   \varphi_T\; :\; (X^{\!A\!z},{\cal E}_T)\;\longrightarrow\; Y
  $$
  is a map from $(X_T^{\!A\!z},{\cal E}_T)$ to $Y$
  defined contravariantly by a ring-homomorphism
  $$
   \varphi_T^{\sharp}\; :\;
    C^{\infty}(Y)\;  \longrightarrow\; C^{\infty}(\End_{\Bbb C}(E_T))
  $$
   over ${\Bbb R}\subset {\Bbb C}$
   and
  $\nabla^T$ is a connection on ${\cal E}_T$ that is trivially flat over $T$.
 $\varphi_T^{\sharp}$ induces a homomorphism
  $$
    {\cal O}_Y\; \longrightarrow\; {\cal O}_{X_T}^{A\!z}
  $$
   between equivalence classes of gluing systems of rings,
 which will still be denoted by $\varphi_T^{\sharp}$.
 
 Let
  ${\cal A}_{\varphi_T}\subset {\cal O}_{X_T}^{A\!z}
   ={\cal O}_{X_T}\langle\Image {\varphi_T^{\sharp}}\rangle$.
 Then $(\varphi_T,\nabla^T)$ is said to be a
  {\it $(\ast_i)$-admissible $T$-family of $(\ast_j)$-admissible pairs}
  if $(\varphi_T,\nabla^T)$ satisfies Admissible Condition $(\ast_i)$ along $T$
    and Admissible Condition $(\ast_j)$ along $X$, for $i,j=1,2,3$. 	
}\end{sdefinition}

\medskip

\begin{sexample}
 {\bf [$(\ast_2)$-admissible $T$-family of $(\ast_1)$-admissible pairs]}$\;${\rm
 A $(\ast_2)$-admissible $T$-family of $(\ast_1)$-admissible pairs $(\varphi_T,\nabla^T)$
  is a $T$-family of maps $\varphi_T$ with a varying connection $\nabla^T$ trivially flat over $T$ such that
  $$
    (\ast_2)\;:\;\;
	    \partial_t \Comm({\cal A}_{\varphi_T})\,\subset\,\Comm({\cal A}_{\varphi_T})
     \hspace{2em}\mbox{and}\hspace{2em}
	(\ast_1)\;:\;\;
	   \nabla^T_{\xi}{\cal A_{\varphi_T}}\,\subset\,\Comm({\cal A}_{\varphi_T})
  $$
  for all $\xi\in {\cal T}_{\ast}(X_T/T)$.
 Here,  $\Comm({\cal A}_{\varphi_T})$ is the commutant of ${\cal A}_{\varphi_T}$
  in ${\cal O}_{X_T}^{A\!z}$.
}\end{sexample}

\bigskip

\begin{flushleft}
{\bf Three basic ${\cal O}_{X_T}$-modules with induced structures}
\end{flushleft}
Let
 $X$ be endowed with a (Riemannian or Lorentzian) metric $h$  and
 $Y$ be endowed with a (Riemannian or Lorentzian) metric $g$.
Denote the canonically induced inner-product structure from $h$ and $g$ on whatever bundle applicable
 by $\langle\,\cdot\,,\,\cdot\,\rangle_h$ and $\langle\,\cdot\,,\,\cdot\,\rangle_g$ respectively.
Denote the induced connection on ${\cal T_{\ast}}(X_T/T)$ and ${\cal T}^{\ast}(X_T/T)$
 by $\nabla^h$ and the Levi-Civita connection on ${\cal T}_{\ast}Y$ by $\nabla^g$.
The associated Riemann curvature tensor is denoted by $R^h$ and $R^g$ respectively.

Let $(\varphi_T,\nabla^T)$ be a $(\ast_1)$-admissible $T$-family of $(\ast_1)$-admissible pairs.
The basic ${\cal O}_{X_T}^{\,\Bbb C}$-modules with induced structures from the setting,
  as in Sec.$\,$3.2,
 are listed below to fix notations.
  \begin{itemize}
   \item[(0)]
 {\it ${\cal O}_{X_T}^{A\!z}\,:$ $\;\;$the noncommutative structure sheaf on $X_T$}
	\vspace{-.6ex}
	\begin{itemize}
	 \item[\LARGE $\cdot$]
      The induced  connection $D^T$ from $\nabla^T$, which is also trivially flat over $T$,
	
	 \item[\LARGE $\cdot$]
	  An ${\cal O}_{X_T}^{A\!z}$-valued, ${\cal O}_X^{\,\Bbb C}$-bilinear
	  (nonsymmetric) inner product from the multiplication in ${\cal O}_{X_T}^{A\!z}$; \\
	   an ${\cal O}_X^{\,\Bbb C}$-valued,
	   ${\cal O}_X^{\,\Bbb C}$-bilinear (symmetric) inner product after the post-composition with $\Tr$.
   
      \item[\LARGE $\cdot$]
      Both inner products are covariantly constant with respect to $D^T$ and one has the Leizniz rules
	    \begin{eqnarray*}
	     D^T (m_T^1m_T^2)
	        & =  &  (D^Tm_T^1)  m_T^2\,+\,   m_T^1\, D^T m_T^2 \,; \\[.6ex]
         d \Tr(m_T^1m_T^2)
	        & =  &  \Tr D^T(m_T^1m_T^2) \\
	       & = &  \Tr \big((D^Tm_T^1)  m_T^2\big)\,+\, \Tr\big(m_T^1\, D^T m_T^2\big)\,.
        \end{eqnarray*}
    \end{itemize}
 
   \item[(1)]
   {\it ${\cal T}^{\ast}(X_T/T)
                    \otimes_{{\cal O}_{X_T}}\!\!{\cal O}_{X_T}^{A\!z}\,:$
            $\;\;{\cal O}_{X_T}^{A\!z}$-valued relative $1$-forms on $X_T/T$}
	\vspace{-.6ex}
	\begin{itemize}
	 \item[\LARGE $\cdot$]
     The induced connection $\nabla^{T, (h,D^T) }\;:=\; \nabla^h\otimes \Id \,+  \Id\otimes D^T$,
	  trivially flat over $T$.
    
	 \item[\LARGE $\cdot$]	
     An ${\cal O}_{X_T}^{A\!z}$-valued, ${\cal O}_{X_T}^{\,\Bbb C}$-bilinear
	   (nonsymmetric) inner product
	   $\langle\,\mbox{\Large $\cdot$}\,,\,\mbox{\Large $\cdot$}\,\rangle_h$;\\
	 an ${\cal O}_X^{\,\Bbb C}$-valued,
	   ${\cal O}_X^{\,\Bbb C}$-bilinear (symmetric) inner product
       $\Tr\langle\,\mbox{\Large $\cdot$}\,,\,\mbox{\Large $\cdot$}\,\rangle_h$.
		
     \item[\LARGE $\cdot$]	
     Both inner products are covariantly constant with respect to
        $\nabla^{T,(h,D^T)}$  and one has the Leibniz rules
	   \begin{eqnarray*}
	    D^T   \langle\, \mbox{\LARGE $\cdot$}\,,\, \mbox{\LARGE $\cdot$}^{\prime} \rangle_h
		 & =
		  &  \langle\,   \nabla^{T,(h,D^T)}\,\mbox{\LARGE $\cdot$}\,,\,
		                            \mbox{\LARGE $\cdot$}^{\prime} \rangle_h\,
		       +\, \langle\, \mbox{\LARGE $\cdot$}\,,\,
			             \nabla^{T,(h,D^T)}\,\mbox{\LARGE $\cdot$}^{\prime} \rangle_g  \,,   \\			   
        d \Tr \langle\, \mbox{\LARGE $\cdot$}\,,\, \mbox{\LARGE $\cdot$}^{\prime} \rangle_h 					 
         & =
		  & \Tr (D^T   \langle\, \mbox{\LARGE $\cdot$}\,,\,
		                              \mbox{\LARGE $\cdot$}^{\prime} \rangle_h)\;\;
		      =\;\; \Tr \langle\,   \nabla^{T,(h,D^T)}\,\mbox{\LARGE $\cdot$}\,,\,
			                          \mbox{\LARGE $\cdot$}^{\prime} \rangle_h\,
		       +\, \Tr \langle\, \mbox{\LARGE $\cdot$}\,,\,
			                    \nabla^{T,(h,D^T)}\,\mbox{\LARGE $\cdot$}^{\prime} \rangle_h
       \end{eqnarray*}
	   for {\LARGE $\cdot$}, {\LARGE $\cdot$}$^{\prime} \in
	      {\cal T}^{\ast}(X_T/T) \otimes_{{\cal O}_{X_T}}\!\!{\cal O}_{X_T}^{A\!z}$.
     \end{itemize}	
  	
    \item[(2)]
     {\it $\varphi_T^{\ast}{\cal T}_{\ast}Y
                 := {\cal O}_{X_T}^{A\!z}
	               \otimes_{\varphi_T^{\sharp},\,{\cal O}_Y}\!\!{\cal T}_{\ast}Y\,:$
             $\;\;{\cal O}_{X_T}^{A\!z}$-valued derivations on ${\cal O}_Y$}
	 \vspace{-.6ex}
	 \begin{itemize}
	  \item[\LARGE $\cdot$]
	   The induced connection
        $\nabla^{T, (\varphi_T,g) } :=
             D^T\otimes \Id
		        +  \Id\cdot
		                 \sum_{i=1}^n
		                     D^T \varphi_T^{\sharp}(y^i)
		                                    \otimes \nabla^g_{\frac{\partial}{\partial y^i}}$
			(in local expression), trivially flat over $T$.
	
	  \item[\LARGE $\cdot$]
	   A {\it partially defined}
	     $\,{\cal O}_{X_T}^{A\!z}$-valued, ${\cal O}_X^{\,\Bbb C}$-bilinear
	     (nonsymmetric) inner product
	     $\langle\,\mbox{\Large $\cdot$}\,,\,\mbox{\Large $\cdot$}\,\rangle_g$;\\
	   a {\it partially defined} $\,{\cal O}_X^{\,\Bbb C}$-valued,
	     ${\cal O}_X^{\,\Bbb C}$-bilinear (symmetric) inner product
	     $\Tr\langle\,\mbox{\Large $\cdot$}\,,\,\mbox{\Large $\cdot$}\,\rangle_g$.
	
	  \item[\LARGE $\cdot$]   	
       Both inner products, when defined,  are covariantly constant with respect to
        $\nabla^{T,(\varphi_T,g)}$  and one has the Leibniz rules
	     \begin{eqnarray*}
	      D^T   \langle\, \mbox{--}\,,\, \mbox{--}^{\prime} \rangle_g
		   & =
		    &  \langle\,   \nabla^{T,(\varphi_T,g)}\,\mbox{--}\,,\, \mbox{--}^{\prime} \rangle_g\,
		       +\, \langle\, \mbox{--}\,,\,
			             \nabla^{T,(\varphi_T,g)}\,\mbox{--}^{\prime} \rangle_g  \,,   \\			   
          d \Tr \langle\, \mbox{--}\,,\, \mbox{--}^{\prime} \rangle_g 					
           & =
		    & \Tr (D^T   \langle\, \mbox{--}\,,\, \mbox{--}^{\prime} \rangle_g)\;\;
		      =\;\; \Tr \langle\,   \nabla^{T,(\varphi_T,g)}\,\mbox{--}\,,\,
			                                    \mbox{--}^{\prime} \rangle_g\,
		       +\, \Tr \langle\, \mbox{--}\,,\,
			                    \nabla^{T,(\varphi_T,g)}\,\mbox{--}^{\prime} \rangle_g  \,,
        \end{eqnarray*}
	   whenever all $\langle\, \mbox{--}^{\prime\prime}\,,\, \mbox{--}^{\prime\prime\prime} \rangle_g$
	     and $\Tr\langle\, \mbox{--}^{\prime\prime}\,,\, \mbox{--}^{\prime\prime\prime} \rangle_g$
		 involved are defined.
	 \end{itemize}
	
    \item[(3)]
    {\it ${\cal T}^{\ast}(X_T/T)
        \otimes_{{\cal O}_{X_T}}\!\varphi_T^{\ast}{\cal T}_{\ast}Y\,:$
         $\;\;$(${\cal O}_{X_T}^{A\!z}$-valued relative $1$-form)-valued
		     derivations on ${\cal O}_Y$}\\[1.2ex]
    This is a combination of the construction in Item (1) and in Item (2).
	 \begin{itemize}
	  \item[\LARGE $\cdot$]
	   The induced connection 	
       $$
         \nabla^{T, (h, \varphi_T,g) }\;=\;
	      \nabla^h\otimes \Id\otimes \Id\,
          +\, \Id \otimes D^T\otimes \Id  \,
		  +\,  \Id \otimes \Id
		              \cdot
		                \sum_{i=1}^n
		                    D^T \varphi_T^{\sharp}(y^i)
		                                   \otimes \nabla^g_{\frac{\partial}{\partial y^i}}
       $$
	  (in local expression), trivially flat over $T$.
	
	  \item[\LARGE $\cdot$]
	   A {\it partially defined}
	     $\,{\cal O}_{X_T}^{A\!z}$-valued, ${\cal O}_X^{\,\Bbb C}$-bilinear
	     (nonsymmetric) inner product
		  $\langle\,\mbox{\Large $\cdot$}\,,\,\mbox{\Large $\cdot$}\,\rangle_{(h,g)}$;\\
	   a {\it partially defined} $\,{\cal O}_X^{\,\Bbb C}$-valued,
	     ${\cal O}_X^{\,\Bbb C}$-bilinear (symmetric) inner product
		 $\Tr\langle\,\mbox{\Large $\cdot$}\,,\,\mbox{\Large $\cdot$}\,\rangle_{(h,g)}$.

      \item[\LARGE $\cdot$]
      Both inner products, when defined,  are covariantly constant with respect to
        $\nabla^{T,(h,\varphi_T,g)}$  and one has the Leibniz rules
	   \begin{eqnarray*}
	    D^T   \langle\, \sim\,,\, \sim^{\prime} \rangle_{(h,g)}
		 & =
		  &  \langle\,   \nabla^{T,(h,\varphi_T,g)}\sim\,,\, \sim^{\prime} \rangle_{(h,g)}\,
		       +\, \langle\, \sim\,,\,
			             \nabla^{T,(h,\varphi_T,g)}\sim^{\prime} \rangle_{(h,g)}  \,,   \\[.6ex]	
        d \Tr \langle\, \sim\,,\, \sim^{\prime} \rangle_{(h,g)} 					
         & =
		  & \Tr (D^T   \langle\, \sim\,,\, \sim^{\prime} \rangle_{(h,g)})   \\
		 & =
		  & \Tr \langle\,   \nabla^{T,(h,\varphi_T,g)}\sim\,,\,
			                                                                     \sim^{\prime} \rangle_{(h,g)}\,
		       +\, \Tr \langle\, \sim\,,\,
			                    \nabla^{T,(h,\varphi_T,g)}\sim^{\prime} \rangle_{(h,g)}  \,,
       \end{eqnarray*}
	  whenever the
	  $\langle\, \sim^{\prime\prime}\,,\, \sim^{\prime\prime\prime} \rangle_{(h,g)}$ and
      $\Tr\langle\, \sim^{\prime\prime}\,,\, \sim^{\prime\prime\prime} \rangle_{(h,g)}$
	  involved are defined.
   \end{itemize}	
\end{itemize}

\vspace{4em}
\bigskip

\begin{flushleft}
{\bf Curvature tensors with $\partial_t$ and other order-switching formulae}
\end{flushleft}
Let $(\varphi_T,\nabla^T)$ be a $(\ast_1)$-admissible $T$-family of $(\ast_1)$-admissible pairs.
A very basic step in (particularly the second) variational problem involves passing $\partial_t$
 over a differential operator on $X_t$'s.
In general, a curvature term appears whenever such passing occurs.
In this theme, we collect and prove such formulae we need.
 
\bigskip

First, passing $\partial_t$ over a differential operator usually means the appearance  of a curvature term
 by the very definition of a curvature tensor:

\bigskip

\begin{slemma} {\bf [curvature tensor with $\partial_t$]}$\;$
 Let $(\varphi_T,\nabla^T)$ be a $(\ast_1)$-admissible $T$-family of $(\ast_1)$-admissible pairs.
 Let  $\xi$ be a vector field on an open set $U\subset X$ small enough so that
   $\varphi_T(U_T^{A\!z})$ is contained in a coordinate chart on $Y$,
   with coordinates $(y^1,\,\cdots\,, y^n)$.
 The standard lifting of $\xi$ to $U_T$ is denoted also by $\xi$.
 Note that, by construction, $[\partial_t, \xi]=0$ and all our connection $\nabla^{\mbox{\Large $\cdot$}}$
  in Theme `{\sl Three basic ${\cal O}_X^{A\!z}$-modules with induced structures}' are trivially flat;
 hence,
   $F_{\nabla^{\mbox{\Large $\cdot$}}}(\partial_t,\xi)
      = \partial_t\nabla^{\mbox{\Large $\cdot$}}_{\xi}
	        - \nabla^{\mbox{\Large $\cdot$}}_{\xi}\partial_t$.
 One has the following curvature expressions with $\partial_t$ on the basic ${\cal O}_{X_T}$-modules:
 (Below we adopt the convention that
        the Riemann curvature tensor from a metric is denoted  by $R$
        while the curvature tensor of a connection in all other bundle situations is denoted by $F$.)
 \begin{itemize}
   \item[$(0_1)$]
    For sections $\omega_T$ of ${\cal T}^{\ast}(X_T/T)\,$:	
    $\;R_{\nabla^h}(\partial_t,\xi)\,\omega_T\;=\; 	
	   \partial_t\nabla^h_{\xi}\omega_T\, -\, \nabla^h_{\xi}\partial_t\omega_T\; =\; 0$.

   \item[$(0_2)$]
    For sections $m_T$ of ${\cal O}_{X_T}^{A\!z}\,$:
	 $\;F_{D^T}(\partial_t,\xi)\, m_T\;
	   =\;  \partial_t D^T_{\xi}m_T\, - \, D_{\xi}^T \partial_t m_T\;
	   =\; [(\partial_t\nabla^T)(\xi), m_T]\,$.\\
    $\mbox{\hspace{1em}}$
	As a consequence of this,
      if $(\varphi_T,\nabla^T)$ is furthermore a $(\ast_2)$-admissible $T$-family
	    of $(\ast_2)$-admissible pairs, then
		$$
		   (\partial_t\nabla^T)(\xi)\;\in\;
		       \Inn^{\varphi}_{(\ast_1)}({\cal O}_X^{A\!z})\,\;
			 \hspace{2em}\mbox{i.e.}\hspace{1em}
           [(\partial_t\nabla^T)(\xi), {\cal A}_{\varphi_T}]\;
		      \subset\; \Comm({\cal A}_{\varphi_T})\,.
		$$
     		  	 	
   \item[$(1)$]	
    For sections $\omega_T\otimes m_T$ of
	 ${\cal T}^{\ast}(X_T/T)\otimes_{{\cal O}_{X_T}}{\cal O}_{X_T}^{A\!z}\,$:
	 \begin{eqnarray*}
	  \lefteqn{
	    F_{\nabla^{T,(h,D^T)}}(\partial_t,\xi)\,(\omega_T\otimes m_T)   }\\
	   && =\;  \partial_t\nabla^{T,(h,D^T)}_{\xi}
	                (\omega_T\otimes m_T)\,
		             -\,  \nabla^{T,(h,D^T)}_{\xi}\partial_t (\omega_T\otimes m_T)\;
             =\;  \omega_T\otimes [(\partial_t\nabla^T)(\xi), m_T]\,.
	 \end{eqnarray*}
 
  \item[$(2)$]
   For sections $m_T\otimes v$ of
     $\varphi_T^{\ast}{\cal T}_{\ast}Y
	     := {\cal O}_{X_T}^{A\!z}
		         \otimes_{\varphi_T^{\sharp},{\cal O}_{Y}}{\cal T}_{\ast}Y\,$: \\
    ($v$ on the coordinate chart of $Y$ above, with coordinates $(y^1,\,\cdots\,,\, y^n)$)				 
	{\small
	 \begin{eqnarray*}
	  \lefteqn{
	   F_{\nabla^{T,(\varphi_T,g)}}(\partial_t,\xi)(m_T\otimes v)\;\;
	   =\;\;
       \partial_t \nabla^{T,(\varphi_T,g)}_{\xi} (m_T\otimes v)\;
	    -\;  \nabla^{T,(\varphi_T, g)}_{\xi}\partial_t (m_T\otimes v)   }\\[.6ex]
       && 					
         =\, [(\partial_t\nabla^T)(\xi), m_T]\otimes v\;\;
             +\;\; m_T \sum_{i=1}^n
			         [(\partial_t\nabla^T)(\xi), \varphi_T^{\sharp}(y^i)]
					                                            \otimes \nabla^g_{\frac{\partial}{\partial y^i}}v  \\[-1.2ex]
		&& \hspace{1.6em}
		     +\;\; m_T\,\sum_{i,j=1}^n   \left(
			                 D^T_{\xi}\varphi_T^{\sharp}(y^i)\,\partial_t\varphi_T^{\sharp}(y^j)
							    \otimes \nabla^g_{\frac{\partial}{\partial y^j}}
								               \nabla^g_{\frac{\partial}{\partial y^i}}\, v\:
						 -\: \partial_t\varphi_T^{\sharp}(y^j)\,D^T_{\xi}\varphi_T^{\sharp}(y^i)
							    \otimes \nabla^g_{\frac{\partial}{\partial y^i}}
								               \nabla^g_{\frac{\partial}{\partial y^j}}\, v
			                                                      \right)\!.
    \end{eqnarray*}}
   $\mbox{\hspace{1em}}$
   If $(\varphi_T,\nabla^T)$ is furthermore
     a $(\ast_2)$-admissible $T$-family of $(\ast_1)$-admissible pairs, 		
   then the last term has a $Y$-coordinate-free form
   $$
      \mbox{\it The last term}\;
	    =\;  m_T\,\sum_{i,j}
		            \partial_t \varphi_T^{\sharp}(y^j)\,D_{\xi}\varphi_T^{\sharp}(y^i)
					  \otimes R^g\big(\mbox{$\frac{\partial}{\partial y^j}$},
					                            \mbox{$\frac{\partial}{\partial y^i}$}\big)\,v\;
        =\; m_T\,\big((\varphi_T^{\diamond} R^g)(\partial_t,\xi)\big)v\,.
   $$
   	
  \item[$(3)$]	
   For sections $\omega_T\otimes m_T\otimes v$ of
     ${\cal T}^{\ast}(X_T/T)\otimes \varphi_T^{\ast}{\cal T}_{\ast}Y
	        :=   {\cal T}^{\ast}(X_T/T) \otimes_{{\cal O}_{X_T}}{\cal O}_{X_T}^{A\!z}
                    \otimes_{\varphi_T^{\sharp},{\cal O}_Y}{\cal T}_{\ast}Y\,$:\\
  ($v$ on the coordinate chart of $Y$ above, with coordinates $(y^1,\,\cdots\,,\, y^n)$ )			
    \begin{eqnarray*}
	  \lefteqn{
	    F_{\nabla^{T,(h,\varphi_T,g)}}(\partial_t,\xi)(\omega_T\otimes m_T\otimes v) }\\[.6ex]
	   && =\;\;
	            \partial_t \nabla^{T,(h,\varphi_T,g)}_{\xi} (\omega_T\otimes m_T\otimes v)\;
		       -\;  \nabla^{T,(h,\varphi_T,g)}\partial_t(\omega_T\otimes m_T\otimes v)     \\
       && =\;\;
	      \omega_T\otimes \left(\rule{0ex}{1em}\right.\!
				           F_{\nabla^{T,(\varphi_T,g)}}(\partial_t,\xi)(m_T\otimes v)
    				                                       \!\left.\rule{0ex}{1em}   \right).
    \end{eqnarray*}
 \end{itemize}
\end{slemma}

\smallskip

\begin{proof}
 Statement$\,(0_1)$ follows from the fact that $X_T$ is a constant family over $T$.
 Statement$\,(0_2)$, First Part, follows from a computation with respect to an induced local trivialization
  of ${\cal E}_T$ from a local trivialization of ${\cal E}$
   {\small
   \begin{eqnarray*}
    \partial_tD^T_{\xi}m_T
	 & =
	   & \partial_t \big(\xi m_T+[A_{\nabla^T}(\xi), m_T ]\big)  \\
   & = &  \xi\partial_t m_T
	                 + [\partial_tA_{\nabla^T}(\xi), m_T]\,+\,[A_{\nabla^T}(\xi),\partial_tm_T]\;\; 
             =\;\;  D^T_{\xi}\partial_t m_T + [(\partial_t\nabla^T)(\xi), m_T]\,.	
   \end{eqnarray*}}
 For Second Part,
   if $(\varphi_T,\nabla^T)$ is furthermore a $(\ast_2)$-admissible T-family
    of $(\ast_2)$-admissible pairs,
  then for $f_1,f_2\in {\cal O}_Y$,
    by First Part and the $(\ast_2)$-Admissible Condition,
  {\small
   $$
     \big[  [(\partial_t \nabla^T)(\xi),\varphi_T^{\sharp}(f_1)],
	            \varphi_T^{\sharp}(f_2)  \big]\;
	 =\;   [\partial_tD^T_{\xi}\varphi_T^{\sharp}(f_1), \varphi_T^{\sharp}(f_2)]\,
	         -\, [D^T_{\xi}\partial_t\varphi_T^{\sharp}(f_1), \varphi_T^{\sharp}(f_2)]\;
     =\; 0\,.			
   $$}
  Which says that
    $(\partial_t\nabla^T)(\xi)
	    \in \Inn^{\varphi_T}_{(\ast_1)}({\cal O}_{X_T}^{A\!z})$.
   
 Statement$\,(1)$ is a consequence of Statement$\,(0_1)$ and Statement$\,(0_2)$.
 Statement$\,(3)$ is a consequence of Statement$\,(0_1)$ and a property of the induced connection
    on a tensor product of ${\cal O}_{X_T}^{\,\Bbb C}$-modules with a connection.
 Let us carry out  Statement$\,(2)$ as a demonstration of the covariant differential calculus involved.
 
 Let $m_T\otimes v\in \varphi_T^{\ast}{\cal T}_{\ast}Y$.
 Then, by Statement$\,(0_2)$,
  {\small
  \begin{eqnarray*}
   \lefteqn{
      \partial_t \nabla^{T,(\varphi_T,g)}_{\xi}(m_T\otimes v)\;\;
	     =\;\; \partial_t
		            \Big(
					   D^T_{\xi}m_T\otimes v \;
					   +\;  m_T\sum_i
					            D^T_{\xi}\varphi_T^{\sharp}(y^i)
					              \otimes \nabla^g_{\frac{\partial}{\partial y^i}}v
					   \Big )                   }\\
    && =\;\;
	  \big(D^T_{\xi}\partial_t m_T+ [(\partial_t\nabla^T)(\xi), m_T]\big) \otimes v\;
	  +\; (D^T_{\xi}m_T)
	         \sum_i \partial_t\varphi_T^{\sharp}(y^i)
			                          \otimes \nabla^g_{\frac{\partial}{\partial y^i}}v   \\
    && \hspace{2em}
	  +\; (\partial_t m_T)
	           \sum_i	D^T_{\xi}\varphi_T^{\sharp}(y^i)
			                       \otimes \nabla^g_{\frac{\partial}{\partial y^i}}v \;
      +\; m_T \sum_i
	         \big(D^T_{\xi}\partial_t\varphi_T^{\sharp}(y^i)
			          + [(\partial_t\nabla^T)(\xi), \varphi_T^{\sharp}(y^i)]  \big)
                            \otimes \nabla^g_{\frac{\partial}{\partial y^i}} v  \\
    && \hspace{18em}
      +\; m_T \sum_{i,j} 							
	          D^T_{\xi}\varphi_T^{\sharp}(y^i)\partial_t\varphi_T^{\sharp}(y^j)
			    \otimes \nabla^g_{\frac{\partial}{\partial y^j}}
				                 \nabla^g_{\frac{\partial}{\partial y^i}} v
  \end{eqnarray*}}
 while
 {\small
  \begin{eqnarray*}
   \lefteqn{
     \nabla^{T,(\varphi_T,g)}_{\xi}\partial_t (m_T\otimes v)\;\;
     =\;\; 	\nabla^{T,(\varphi_T,g)}_{\xi}
	                \Big( \partial_tm_T\otimes v  \;
					       +\;  m_T \sum_i \partial_t\varphi_T^{\sharp}(y^i)
					                   \otimes \nabla^g_{\frac{\partial}{\partial y^i}}v   \Big)  }\\
   && =\;\;
     D^T_{\xi}\partial_t m_T\otimes v\;
     +\; (\partial_t m_T)
	          \sum_i  D^T_{\xi}\varphi_T^{\sharp}(y^i)
			          \otimes \nabla^g_{\frac{\partial}{\partial y^i}}v\;
	 +\;  (D^T_{\xi}m_T)
	           \sum_i \partial_t\varphi_T^{\sharp}(y^i)
			          \otimes \nabla_{\frac{\partial}{\partial y^i}}v    \\
  && \hspace{2em}					
     +\; m_T \sum_i	 D^T_{\xi}\partial_t\varphi_T^{\sharp}(y^i)
                                    \otimes 	\nabla^g_{\frac{\partial}{\partial y^i}}v\;
	 +\; m_T\sum_{i,j}\partial_t\varphi_T^{\sharp}(y^i)
	                    D^T_{\xi}\varphi_T^{\sharp}(y^j)
						  \otimes \nabla^g_{\frac{\partial}{\partial y^j}}
                                          \nabla^g_{\frac{\partial}{\partial y^i}} v\,.					
  \end{eqnarray*}}
 Thus,
  {\small
 \begin{eqnarray*}
  \lefteqn{
   F_{\nabla^{T,(\varphi_T,g)}}(\partial_t,\xi)(m_T\otimes v)\;\;
     =\;\;  (\partial_t\nabla^{T,(\varphi_T,g)}- \nabla^{T,(\varphi_T,g)}\partial_t)
	               (m_T\otimes v)    }\\[.6ex]
   && =\;\;
	 [(\partial_t\nabla^T)(\xi), m_T] \otimes v\;
      +\; m_T \sum_i
			             [(\partial_t\nabla^T)(\xi), \varphi_T^{\sharp}(y^i)]
                              \otimes \nabla^g_{\frac{\partial}{\partial y^i}} v  \\
    && \hspace{2em}
      +\; m_T \sum_{i,j} 	\Big(						
	          D^T_{\xi}\varphi_T^{\sharp}(y^i)\partial_t\varphi_T^{\sharp}(y^j)
			    \otimes \nabla^g_{\frac{\partial}{\partial y^j}}
				                 \nabla^g_{\frac{\partial}{\partial y^i}} v\;
           -\;  \partial_t\varphi_T^{\sharp}(y^j)
	                    D^T_{\xi}\varphi_T^{\sharp}(y^i)
						  \otimes \nabla^g_{\frac{\partial}{\partial y^i}}
                                          \nabla^g_{\frac{\partial}{\partial y^j}} v		\Big)	
 \end{eqnarray*}}
 as claimed, after a relabeling of $i,j$.
 
 If $(\varphi_T,\nabla^T)$ is furthermore a $(\ast_2)$-admissible $T$-family of
  $(\ast_1)$-admissible pairs,
 then
  $D^T_{\xi}\varphi_T^{\sharp}(y^i)$ and $\partial_t\varphi_T^{\sharp}(y^j)$  commute
  since
   $[D^T_{\xi}\varphi_T^{\sharp}(y^i), \varphi_T^{\sharp}(y^j)]=0$
      by the $(\ast_1)$-Admissible Condition along $X$ and, hence,
{\small
   \begin{eqnarray*}
      0 & =  & \partial_t  [D^T_{\xi}\varphi_T^{\sharp}(y^i), \varphi_T^{\sharp}(y^j)] \\
	   & = & [ \partial_tD^T_{\xi}\varphi_T^{\sharp}(y^i), \varphi_T^{\sharp}(y^j)]\,
		           +\,  [D^T_{\xi}\varphi_T^{\sharp}(y^i), \partial_t\varphi_T^{\sharp}(y^j)]\;\;
          =\;\;  [D^T_{\xi}\varphi_T^{\sharp}(y^i), \partial_t\varphi_T^{\sharp}(y^j)]	
   \end{eqnarray*}}
  by the $(\ast_1)$-Admissible Condition along $X$ and the $(\ast_2)$-Admissible Condition along $T$.
 The last summand of $F_{\nabla^{T,(\varphi_T,g)}}(\partial_t,\xi)(m_T\otimes v)$
  is then equal to
  {\small
  $$
    m_T \sum_{i,j} 	
	          \partial_t\varphi_T^{\sharp}(y^j)
	          D^T_{\xi}\varphi_T^{\sharp}
			    \otimes \Big(
				   \nabla^g_{\frac{\partial}{\partial y^j}}\nabla^g_{\frac{\partial}{\partial y^i}}\,
                    -\,  \nabla^g_{\frac{\partial}{\partial y^i}}
                                          \nabla^g_{\frac{\partial}{\partial y^j}}
							   \Big) v \;
    =\; m_T\, \big((\varphi_T^{\diamond}R^g)(\partial_t,\xi)\big)\,v\,.							   
  $$}
  
 This proves the lemma.
 
\end{proof}

\bigskip

The following lemma addresses the issue of passing $\partial_t$  over the covariant differential
  $D\varphi_T$ of $\varphi_T$.
Though such passing is not a curvature issue in the conventional sense,
 it does carry a taste of curvature calculations.

\bigskip

\begin{slemma}
 {\bf [$\partial_t D^T\varphi_T$ versus $\nabla^{T,(\varphi_T,\,g)}\partial_t\varphi_T$]}$\;$
 Let $(\varphi_T,\nabla^T)$ be a $(\ast_1)$-admissible $T$-family of $(\ast_1)$-admissible pairs.
 With the above notation and convention,
  let $\xi$ be a vector field on $X$.
 Then, for a chart of $Y$ with coordinates $(y^1,\,\cdots\,,\, y^n)$,
 one has
 $$
  \partial_t D^T_{\xi}\varphi_T\;\;
    =\;\; \nabla^{T,(\varphi_T,\,g)}_{\xi}\partial_t\varphi_T\;
           -\; (ad\otimes\nabla^g)_{\partial_t\varphi_T} D^T_{\xi}\varphi_T\;
           +\, \sum_{i=1}^n
		            [(\partial_t\nabla^T)(\xi), \varphi_T^{\sharp}(y^i)]
					    \otimes \mbox{$\frac{\partial}{\partial y^i}$}\,.		
 $$
 Here, only as a compact notation,
  \begin{eqnarray*}
   \lefteqn{
    (ad\otimes\nabla^g)_{\partial_t\varphi_T} D^T_{\xi}\varphi_T\;\;
	  :=\;\; \sum_{i,j=1}^n
	            [\partial_t\varphi^{\sharp}(y^i), D^T_{\xi}\varphi^{\sharp}(y^j)]
		         \otimes \nabla^g_{\frac{\partial}{\partial y^i}}
			                            \mbox{\Large $\frac{\partial}{\partial y^j}$}     }\\
    && =\;\;  -\, \sum_{i,j=1}^n
                  [D^T_{\xi}\varphi^{\sharp}(y^j), \partial_t\varphi^{\sharp}(y^i)]
                  \otimes \nabla^g_{\frac{\partial}{\partial y^j}}
                                 \mbox{\Large $\frac{\partial}{\partial y^i}$}\;\;
				=:\;\; -\,   (\ad\otimes\nabla^g)_{D^T_{\xi}\varphi_T}\partial_t\varphi_T\,.
  \end{eqnarray*}
 
 If $(\varphi_T,\nabla^T)$ is furthermore
   a $(\ast_2)$-admissible $T$-family of $(\ast_2)$-admissible pairs,
 then the last term has a $Y$-coordinate-free expression
   $$
      \ad_{\,(\partial_t\nabla^T)(\xi)}\varphi_T\,.
   $$
\end{slemma}

\smallskip

\begin{proof}
 Under the given setting and by Lemma~5.4 $(0_2)$,
  %
  {\small
   \begin{eqnarray*}
    \lefteqn{
	 \partial_tD^T_{\xi}\varphi_T\;\;
	  =\;\; \partial_t\Big( \sum_i D^T_{\xi}\varphi_T^{\sharp}(y^i)
	                 \otimes \mbox{\large $\frac{\partial}{\partial y^i}$}    \Big)   }\\
    &&=\;\; \sum_i 	\big(D^T_{\xi}\partial_t\varphi_T^{\sharp}(y^i)
	                                        + [(\partial_t\nabla^T)(\xi),\varphi_T^{\sharp}(y^i)]       \big)
                                      \otimes \mbox{\large $\frac{\partial}{\partial y^i}$}\;
                 +\; \sum_{i,j}D^T_{\xi}\varphi_T^{\sharp}(y^i)
			                               \partial_t\varphi_T^{\sharp}(y^j)
                                           \otimes \nabla^g_{\frac{\partial}{\partial y^j}}
										                      \mbox{\large $\frac{\partial}{\partial y^i}$}
   \end{eqnarray*}}
 while
 {\small
   \begin{eqnarray*}
    \lefteqn{
	   \nabla^{T,(\varphi_T,g)}_{\xi}\partial_t\varphi_T\;\;
	     =\;\;  \nabla^{T,(\varphi_T,g)}_{\xi}
		          \Big( \sum_i \partial_t\varphi_T^{\sharp}(y^i)
				                               \otimes \mbox{\large $\frac{\partial}{\partial y^i}$}\Big) }\\
    && =\;\; \sum_i D^T_{\xi}\partial_t\varphi_T^{\sharp}(y^i)
	                               \otimes \mbox{\large $\frac{\partial}{\partial y^i}$} \;
			+\; \sum_{i,j}\partial_t\varphi_T^{\sharp}(y^i)
			                            D^T_{\xi}\varphi_T^{\sharp}(y^j)
										\otimes \nabla^g_{\frac{\partial}{\partial y^j}}
										  \mbox{\large $\frac{\partial}{\partial y^i}$}    \,.
   \end{eqnarray*}}
 Thus,
 {\small
   \begin{eqnarray*}
    \lefteqn{
	  \partial_tD^T_{\xi}\varphi\,-\, \nabla^{T,(\varphi_T,g)}_{\xi}\partial_t\varphi_T  }\\
	   && = \;\; \sum_i 	[(\partial_t\nabla^T)(\xi),\varphi_T^{\sharp}(y^i)]
                                     \otimes \mbox{\large $\frac{\partial}{\partial y^i}$}            \\
       && \hspace{2em}
   	          +\; \sum_{i,j}D^T_{\xi}\varphi_T^{\sharp}(y^i)
			                               \partial_t\varphi_T^{\sharp}(y^j)
                                           \otimes \nabla^g_{\frac{\partial}{\partial y^j}}
										                      \mbox{\large $\frac{\partial}{\partial y^i}$}   \;
             -\; \sum_{i,j}\partial_t\varphi_T^{\sharp}(y^i)
			                            D^T_{\xi}\varphi_T^{\sharp}(y^j)
										\otimes \nabla^g_{\frac{\partial}{\partial y^j}}
										  \mbox{\large $\frac{\partial}{\partial y^i}$}
   \end{eqnarray*}}
 Either apply the identity
   $\nabla^g_{\frac{\partial}{\partial y^j}}\frac{\partial}{\partial y^i}
      = \nabla^g_{\frac{\partial}{\partial y^i}}\frac{\partial}{\partial y^j}$
	to the second term and relabeling $i,j$ of the third,
  or apply the identity
   $\nabla^g_{\frac{\partial}{\partial y^j}}\frac{\partial}{\partial y^i}
      = \nabla^g_{\frac{\partial}{\partial y^i}}\frac{\partial}{\partial y^j}$
	to the third term and relabeling $i,j$ of the second,
 {\small
   \begin{eqnarray*}
    \lefteqn{
   	    \sum_{i,j}D^T_{\xi}\varphi_T^{\sharp}(y^i)
			                               \partial_t\varphi_T^{\sharp}(y^j)
                                           \otimes \nabla^g_{\frac{\partial}{\partial y^j}}
										                      \mbox{\large $\frac{\partial}{\partial y^i}$}   \;
             -\; \sum_{i,j}\partial_t\varphi_T^{\sharp}(y^i)
			                            D^T_{\xi}\varphi_T^{\sharp}(y^j)
										\otimes \nabla^g_{\frac{\partial}{\partial y^j}}
										  \mbox{\large $\frac{\partial}{\partial y^i}$}     }\\
    && =\;\;  \sum_{i,j}[D^T_{\xi}\varphi_T^{\sharp}(y^i)\,,\,
			                                 \partial_t\varphi_T^{\sharp}(y^j)]
                                             \otimes \nabla^g_{\frac{\partial}{\partial y^i}}
										                      \mbox{\large $\frac{\partial}{\partial y^j}$}
                      \hspace{4.1em}\Big(\;
					    :=\;\; (\ad\otimes\nabla^g)_{D^T_{\xi}\varphi}\partial_t\varphi_T\;
                                            \hspace{1.7em}\Big)\\
    && =\;\;  -\, \sum_{i,j}[\partial_t\varphi_T^{\sharp}(y^i)\,,\,
	                                         D^T_{\xi}\varphi_T^{\sharp}(y^j)]
                                             \otimes \nabla^g_{\frac{\partial}{\partial y^i}}
										                      \mbox{\large $\frac{\partial}{\partial y^j}$}
					  \hspace{3em}\Big(\;
                       :=\;\;  -\, (\ad\otimes\nabla^g)_{\partial_t\varphi_T}D^T_{\xi}\varphi_T\;  \Big)\,.
   \end{eqnarray*}}
 This proves the First Statement in Lemma.
   
 The Second Statement in Lemma is a consequence of Corollary~3.1.10 and Lemma~5.4 $(0_2)$.
							
 This proves the lemma.
							
\end{proof}

\bigskip

Before continuing the discussion, we introduce a notion that is needed in the next lemma.

\bigskip

\begin{sdefinition}
  {\bf [half-torsion tensor $\Tor_{\nabla^g}^{\frac{1}{2}, \,\bullet\,}$]}$\;$ {\rm
 Recall the torsion tensor $\Tor_{\nabla^{\prime}}$ of a connection $\nabla^{\prime}$ on $Y$
   $$
     \Tor_{\nabla^{\prime}}(v_1,v_2)\;
	  :=\;  \nabla^{\prime}_{v_1}\,v_2\,-\, \nabla^{\prime}_{v_2}\,v_1\,-\, [v_1,v_2]
   $$
   for $v_1,v_2\in{\cal T}_{\ast}Y$.
  For the Levi-Civita connection $\nabla^g$ associated to a metric $g$ on $Y$,
    $\Tor_{\nabla^g}\equiv 0$ by construction.
 Thus, in this case, for a $\Phi\in C^{\infty}(Y)$,
   $$
      (\nabla^g_{v_1}\,v_2\, -\,  v_1v_2)\Phi
	   =  (\nabla^g_{v_2}\,v_1\, -\, v_2v_1)\Phi
   $$
   for $v_1,v_2\in{\cal T}_{\ast}Y$.
 This defines a symmetric $2$-tensor on $Y$
  $$
   \begin{array}{ccccc}
    \Tor_{\nabla^g}^{\frac{1}{2},\,\Phi}
	  & : & {\cal T}_{\ast}Y \times_Y {\cal T}_{\ast}Y & \longrightarrow  & {\cal O}_Y \\[.6ex]
	&& (v_1, v_2)   & \longmapsto
      &   	  (\nabla^g_{v_1}\,v_2\, -\,  v_1v_2)\,\Phi
   \end{array}\,,
  $$
  called the {\it half-torsion tensor of} (the torsion-free connection)
      {\it $\nabla^g$ associated to $\Phi\in C^{\infty}(Y)$}.
}\end{sdefinition}

\bigskip

The following lemma addresses the issue of passing $\partial_t$ over
  `evaluation of an ${\cal O}_{X_T}^{A\!z}$-valued derivation on $C^{\infty}(Y)$',
  and another similar situation:

\bigskip

\begin{slemma}
 {\bf [$\partial_t ((D^T_{\xi}\varphi_T)\Phi)$
              versus $(\partial_t D^T_{\xi}\varphi_T)\Phi\,$;
			$\; D^T_{\xi}((\partial_t\varphi_T)\Phi)$
			 versus $(\nabla^{T,(\varphi_T,\,g)}_{\xi}\partial_t\varphi_T)\Phi$]}$\;$\\
 Let $(\varphi_T,\nabla^T)$ be a $(\ast_1)$-admissible $T$-family of $(\ast_1)$-admissible pairs.
 Continue the notation and convention in Lemma~5.4.
 Under the canonical isomorphism	
  ${\cal O}_{X_T}^{A\!z}\otimes_{\varphi_T^{\sharp},{\cal O}_Y} {\cal O}_Y
     \simeq   {\cal O}_{X_T}^{A\!z}$,	
 \begin{eqnarray*}
   \lefteqn{\partial_t \big((D^T_{\xi}\varphi_T)\Phi\big)  }\\[-.6ex]
    && =\;\;   \big(\partial_t D^T_{\xi}\varphi_T \big)\Phi\;
                +\; \sum_{i, j=1}^n
				        D^T_{\xi}\varphi_T^{\sharp}(y^i)\,
						\partial_t\varphi_T^{\sharp}(y^j)\,
						\otimes \Big(
						   \mbox{\large $\frac{\partial}{\partial y^i}
						                             \frac{\partial}{\partial y^j}$}\Phi\,
						     -\,  \Big(
							     \nabla^g_{\frac{\partial}{\partial y^i}}
							             \mbox{\large $\frac{\partial}{\partial y^j}$}
										\Big)\Phi
						               \Big)  \\
    &&	=\;\;  \big(\partial_t D^T_{\xi}\varphi_T \big)\Phi\;
	                  -\; \big(\varphi_T^{\diamond}\Tor_{\nabla^g}^{\frac{1}{2},\Phi}
					       \big) (\xi,\partial_t)\,;
 \end{eqnarray*}  		
 and
  \begin{eqnarray*}
   \lefteqn{D^T_{\xi} \big((\partial_t\varphi_T)\Phi\big)  }\\[-.6ex]
    && =\;\;   \big(\nabla^{T,(\varphi_T,\,g)}_{\xi}\partial_t \varphi_T\big)\Phi\;
                +\; \sum_{i, j=1}^n
					    \partial_t\varphi_T^{\sharp}(y^i)\,
				        D^T_{\xi}\varphi_T^{\sharp}(y^j)\,
						\otimes  \Big(
						   \mbox{\large $\frac{\partial}{\partial y^i} \frac{\partial}{\partial y^j}$}
						   \Phi\,
						     -\,  \Big(
							     \nabla^g_{\frac{\partial}{\partial y^i}}
							             \mbox{\large $\frac{\partial}{\partial y^j}$}
										 \Big)\Phi
						               \Big)  \\
    && =\;\; \big(\nabla^{T,(\varphi_T,\,g)}_{\xi}\partial_t \varphi_T\big)\Phi\;	
		        -\; \big(\varphi_T^{\diamond}\Tor_{\nabla^g}^{\frac{1}{2},\Phi}
					   \big) (\partial_t, \xi)\,.
 \end{eqnarray*}  	
\end{slemma}

\smallskip

\begin{proof}
 For the first identity,
{\small
 \begin{eqnarray*}
  \lefteqn{
   \partial_t \big((D^T_{\xi}\varphi_T)\Phi\big)\;\;
	 =\;\; \partial_t\Big(
	              \sum_i D^T_{\xi}\varphi_T^{\sharp}(y^i)
				                    \otimes \mbox{\large $\frac{\partial}{\partial y^i}$}\Phi    \Big)  }\\
  &&=\;\; \sum_i
                   \partial_tD^T_{\xi}\varphi_T^{\sharp}
                     \otimes \mbox{\large $\frac{\partial}{\partial y^i}$}\Phi \;
            +\; \sum_{i,j} D^T_{\xi}\varphi_T^{\sharp}(y^i)
                                          \partial_t\varphi_T^{\sharp}(y^j)
                                          \otimes \mbox{\large $\frac{\partial}{\partial y^j}$}
										                 \mbox{\large $\frac{\partial}{\partial y^i}$}\Phi
 \end{eqnarray*}}
 while
{\small
 \begin{eqnarray*}
  \lefteqn{
    \big(\partial_t D^T_{\xi}\varphi_T \big)\Phi\;\;
	 =\;\; \Big( \partial_t \sum_i D^T_{\xi}\varphi_T^{\sharp}(y^i)
                                                               \otimes \mbox{\large $\frac{\partial}{\partial y^i}$}	
		      \Big)\Phi     }\\
  && =\;\; \Big (
                   \sum_i \partial_t D^T_{\xi}\varphi_T^{\sharp}(y^i)
				                   \otimes \mbox{\large $\frac{\partial}{\partial y^i}$}\;
			     +\; \sum_{i,j}D^T_{\xi}\varphi_T^{\sharp}(y^i)
				                              \partial_t\varphi_T^{\sharp}(y^j)
											  \otimes \nabla^g_{\frac{\partial}{\partial y^j}}
											                     \mbox{\large $\frac{\partial}{\partial y^i}$}				
                 \Big) \Phi\,.
 \end{eqnarray*}}
 Thus,
 {\small
  \begin{eqnarray*}
   \lefteqn{
      \partial_t \big((D^T_{\xi}\varphi_T)\Phi\big)\,
	   -\,   \big(\partial_t D^T_{\xi}\varphi_T \big)\Phi   }\\
   && =\;\;  \sum_{i,j} D^T_{\xi}\varphi_T^{\sharp}(y^i)
                                       \partial_t\varphi_T^{\sharp}(y^j)
                                       \otimes \Big(
									       \mbox{\large $\frac{\partial}{\partial y^j}$}
										                 \mbox{\large $\frac{\partial}{\partial y^i}$}\,
										 -\, \nabla^g_{\frac{\partial}{\partial y^j}}
											                     \mbox{\large $\frac{\partial}{\partial y^i}$}
												    \Big)\Phi   \\
   && =\;\;  \sum_{i,j} D^T_{\xi}\varphi_T^{\sharp}(y^i)
                                       \partial_t\varphi_T^{\sharp}(y^j)
                                       \otimes \Big(
									       \mbox{\large $\frac{\partial}{\partial y^i}$}
										                 \mbox{\large $\frac{\partial}{\partial y^j}$}\,
										 -\, \nabla^g_{\frac{\partial}{\partial y^i}}
											                     \mbox{\large $\frac{\partial}{\partial y^j}$}
												    \Big)\Phi \;\;
        =\;\;   -\, \big(\varphi_T^{\diamond}\Tor_{\nabla^g}^{\frac{1}{2},\Phi}
					   \big) (\xi,\partial_t)													
  \end{eqnarray*}}
 and the first identity follows.
 
 For the second identity,
 {\small
 \begin{eqnarray*}
  \lefteqn{
   D^T_{\xi} \big((\partial_t\varphi_T)\Phi\big)\;\;
	 =\;\;  D^T_{\xi}\Big(
	                   \sum_i \partial_t\varphi_T^{\sharp}(y^i)
					                  \otimes \mbox{\large $\frac{\partial}{\partial y^i}$}\Phi
	                                  \Big)      }\\
  && =\;\; \sum_i D^T_{\xi}\partial_t\varphi_T^{\sharp}(y^i)
                             \otimes \mbox{\large $\frac{\partial}{\partial y^i}$}\Phi\;
           +\; \sum_{i,j} \partial_t\varphi_T^{\sharp}(y^i)
		                             D^T_{\xi}\varphi_T^{\sharp}(y^j)
                                     \otimes \mbox{\large $\frac{\partial}{\partial y^j}$}	
									              \mbox{\large $\frac{\partial}{\partial y^i}$} \Phi
 \end{eqnarray*}}
 while
 {\small
 \begin{eqnarray*}
  \lefteqn{
    \big(\nabla^{T,(\varphi_T,\,g)}_{\xi}\partial_t \varphi_T\big)\Phi\;\;
	 =\;\; \Big(
	           \nabla^{T,(\varphi_T,g)}_{\xi}
			    \sum_i \partial_t\varphi_T^{\sharp}(y^i)
				              \otimes \mbox{\large $\frac{\partial}{\partial y^i}$}\Big) \Phi    }\\
  && =\:\:	\Big (\sum_i D^T_{\xi} \partial_t\varphi_T^{\sharp}(y^i)
                              \otimes \mbox{\large $\frac{\partial}{\partial y^i}$}\;
                   +\; \sum_{i,j} \partial_t\varphi_T^{\sharp}(y^i)
                                              D^T_{\xi}\varphi_T^{\sharp}(y^j)
                                              \otimes \nabla^g_{\frac{\partial}{\partial y^j}}
                                                           \mbox{\large $\frac{\partial}{\partial y^i}$}
				\Big) \Phi  \,.
 \end{eqnarray*}}
 Thus,
 {\small
 \begin{eqnarray*}
  \lefteqn{
    D^T_{\xi} \big((\partial_t\varphi_T)\Phi\big)\,
	 -\,  \big(\nabla^{T,(\varphi_T,\,g)}_{\xi}\partial_t \varphi_T\big)\Phi } \\
   && =\;\; \sum_{i,j} \partial_t\varphi_T^{\sharp}(y^i)
		                             D^T_{\xi}\varphi_T^{\sharp}(y^j)
                                     \otimes \Big( \mbox{\large $\frac{\partial}{\partial y^j}$}	
									              \mbox{\large $\frac{\partial}{\partial y^i}$}\,
                                             -\, 	 \nabla^g_{\frac{\partial}{\partial y^j}}
                                                        \mbox{\large $\frac{\partial}{\partial y^i}$}											  
 												  \Big)\Phi  \\	
   && =\;\; \sum_{i,j} \partial_t\varphi_T^{\sharp}(y^i)
		                             D^T_{\xi}\varphi_T^{\sharp}(y^j)
                                     \otimes \Big( \mbox{\large $\frac{\partial}{\partial y^i}$}	
									              \mbox{\large $\frac{\partial}{\partial y^j}$}\,
                                             -\, 	 \nabla^g_{\frac{\partial}{\partial y^i}}
                                                        \mbox{\large $\frac{\partial}{\partial y^j}$}											  
 												  \Big)\Phi\;\;
        =\;\;   -\, \big(\varphi_T^{\diamond}\Tor_{\nabla^g}^{\frac{1}{2},\Phi}
					   \big) (\partial_t, \xi)
 \end{eqnarray*}}
 and the second identity follows.
 
 This proves the lemma.
     
\end{proof}
 
\bigskip

\begin{sremark} $[$for $(\ast_2)$-admissible family of $(\ast_1)$-admissible pairs$\,]\;$
{\rm
 If $(\varphi_T,\nabla^T)$ is furthermore
   a $(\ast_2)$-admissible $T$-family of $(\ast_1)$-admissible pairs,
 then, as in the proof of  Lemma~5.4 (2),
  $D^T_{\xi}\varphi_T^{\sharp}(y^i)$ and $\partial_t\varphi_T^{\sharp}(y^j)$	
   commute for all $i,j$.
 In this case,
  $$
    \big( \varphi^{\diamond}\,
                     \Tor_{\nabla^g}^{\frac{1}{2},\Phi}\big)(\xi, \partial_t)\;
    =\;   \big( \varphi^{\diamond}\,
                     \Tor_{\nabla^g}^{\frac{1}{2},\Phi}\big)(\partial_t,\xi)\,.
  $$
}\end{sremark}

\bigskip

\begin{flushleft}
{\bf Two-parameter admissible families of admissible pairs}
\end{flushleft}
Let $T = (-\varepsilon, \varepsilon)^2 \subset {\Bbb R}^2$, $\varepsilon>0$ small,
  be a two-parameter space with coordinates $(s,t)$.
The setting and results above for one-parameter admissible families of admissible pairs
 generalizes without work to two-parameter admissible of admissible pairs.
In particular,
  
\bigskip

\begin{sdefinition} {\bf [two-parameter admissible family of admissible pairs]}$\;$ {\rm
 A {\it $(\ast_2)$-admissible $T$-family of $(\ast_1)$-admissible maps}
  is a $(\ast_1)$-admissible map
   $\varphi_T:(X_T^{\!A\!z},{\cal E}_T;\nabla^T)\rightarrow  Y$,
     where ${\cal E}_T$ is trivially flat over $T$,
   such that
   $\partial_s \Comm{\cal A}_{\varphi_T}\subset \Comm({\cal A }_{\varphi_T})$  and
   $\partial_t  \Comm{\cal A}_{\varphi_T}\subset \Comm({\cal A }_{\varphi_T})$.
}\end{sdefinition}
  
\bigskip

The following is a consequence of the proof of Lemma~3.2.2.5:
      
\bigskip

\begin{slemma} {\bf [symmetry property of
    $\Tr\langle F_{\nabla^{T,(\varphi_T,g)}}(\partial_s,\xi_2)\partial_t\varphi_T,
	                      D^T_{\xi_4}\varphi_T \rangle$]}$\;$
 Let\\ $\varphi_T:(X_T^{\!A\!z},{\cal E}_T;\nabla^T)\rightarrow Y$
   be a $(\ast_2)$-admissible $T$-family of $(\ast_1)$-admissible maps.
 Let $\xi_2,\xi_4\in {\cal T}_{\ast}X$ and denote the same for their respective lifting
   to  ${\cal T}_{\ast}(X_T/T)$.
 Then,
  %
  \begin{eqnarray*}
   \lefteqn{
      \Tr \langle F_{\nabla^{T,(\varphi_T,g)}}(\partial_s,\xi_2) \partial_t\varphi_T\,,\,
     	                  D^T_{\xi_4}\varphi_T \rangle_g\;\;
     =\;\;  -\, \Tr \langle \partial_t\varphi_T\,,\,
	                   F_{\nabla^{T,(\varphi_T,g)}}
					                  (\partial_s,\xi_2)D_{\xi_4}\varphi_T \rangle_g  }\\[.6ex]
    && =\;\;  -\, \Tr \langle
	              F_{\nabla^{T,(\varphi_T,g)}}(\partial_s,\xi_2)D^T_{\xi_4}\varphi_T\,,\,
				  \partial_t\varphi_T \rangle_g\;\;
          =\;\;   \Tr \langle
	              F_{\nabla^{T,(\varphi_T,g)}}(\xi_2,\partial_s)D^T_{\xi_4}\varphi_T\,,\,
				  \partial_t\varphi_T \rangle_g\,.				  	
  \end{eqnarray*}
\end{slemma}

\smallskip

\begin{proof}
 Let $\xi$ be $\xi_2$ or $\xi_4$.
 Since
     $\partial_s\Comm({\cal A}_{\varphi_T})\subset \Comm({\cal A}_{\varphi_T})$
	 and and $\partial_{\xi}{\cal A}_{\varphi_T}\subset \Comm({\cal A}_{\varphi_T})$,
  both $\partial_sD^T_{\xi}\varphi_T$  and $\partial_s\partial_t\varphi_T$
	 lie in $\Comm({\cal A}_{\varphi_T})
	              \otimes_{\varphi^{\sharp},{\cal O}_Y}\!\!{\cal T}_{\ast}Y$.
 Locally explicitly,
   \begin{eqnarray*}
      \partial_sD^T_{\xi}\varphi_T
	   & = &\sum_i
		         \partial_s D^T_{\xi}\varphi_T^{\sharp}(y^i)
		           \otimes  \mbox{\Large $\frac{\partial}{\partial y^i}$}\,
			   +\, \sum_{i,j}
		               D^T_{\xi}\varphi_T^{\sharp}(y^i)\,\partial_s\varphi_T^{\sharp}(y^j)
		               \otimes   \nabla^g_{\frac{\partial}{\partial y^j}}
					                          \mbox{\Large $\frac{\partial}{\partial y^i}$}\,; \\
      \partial_s\partial_t\varphi_T
	   & = &\sum_i
		         \partial_s \partial_t\varphi_T^{\sharp}(y^i)
		           \otimes  \mbox{\Large $\frac{\partial}{\partial y^i}$}\,
			   +\, \sum_{i,j}
		               \partial_t \varphi_T^{\sharp}(y^i)\,\partial_s\varphi_T^{\sharp}(y^j)
		               \otimes   \nabla^g_{\frac{\partial}{\partial y^j}}
					                          \mbox{\Large $\frac{\partial}{\partial y^i}$}\,.
   \end{eqnarray*}
 
 Now follow the proof of Lemma~3.2.2.5,
  but under only the $(\ast_1)$-Admissible Condition on $(\varphi_T,\nabla^T)$,
    to convert
    $\;\Tr \langle F_{\nabla^{T,(\varphi_T,g)}}(\partial_s,\xi_2) \partial_t\varphi_T\,,\,
     	                  D^T_{\xi_4}\varphi_T \rangle_g\;$
    to
	$\;\Tr \langle \partial_t\varphi_T\,,\,
	                   F_{\nabla^{T,(\varphi_T,g)}}
					                  (\partial_s,\xi_2)D_{\xi_4}\varphi_T \rangle_g\,$.
 Since $\Tr\langle\,\mbox{--}   \,,\,\mbox{--}^{\prime}\,\rangle_g$ is defined as long as one of
  --, --$^{\prime}$ is in
    $\Comm({\cal A}_{\varphi_T})
         \otimes_{\varphi_T^{\sharp},{\cal O}_Y}{\cal T}_{\ast}Y$,
  one realizes that all the terms that appear in the process via the Leibniz rule are defined {\it except}		
   $$
     -\,\Tr \langle \nabla^{T,(\varphi_T,g)}_{\xi_2}\partial_t\varphi_T\,,\,
                 \partial_s D^T_{\xi_4}\varphi_T\rangle_g\;
	 +\; \Tr \langle \partial_s\partial_t\varphi_T\,,\,
                   \nabla^{T,(\varphi_T,g)}_{\xi_2}D^T_{\xi_4}\varphi_T\rangle_g\,.
   $$
 Under the additional $(\ast_2)$-Admissible Condition on $(\varphi_T,\nabla^T)$ along $T$,
   both $\partial_sD^T_{\xi_4}\varphi_T$  and $\partial_s\partial_t\varphi_T$
	 now lie in $\Comm({\cal A}_{\varphi_T})
	              \otimes_{\varphi^{\sharp},{\cal O}_Y}\!\!{\cal T}_{\ast}Y$;
	and the above two exceptional terms become defined. 			
	
 The lemma follows.									
	
\end{proof}

\bigskip

\section{The first variation of the enhanced kinetic term for maps and ......   }

Let $(\varphi,\nabla)$ be a $(\ast_1)$-admissible pair.
Recall the setup in Sec.$\,$5.
Let
 $T=(-\varepsilon,\varepsilon)\subset {\Bbb R}^1$, for some $\varepsilon>0$ small,  and
 $(\varphi_T,\nabla^T)$ be a $(\ast_1)$-admissible $T$-family of $(\ast_1)$-admissible pairs
 that deforms $(\varphi,\nabla)=(\varphi_T,\nabla^T)|_{t=0}$.
We derive in Sec.$\,$6.1 and Sec.$\,$6.2
 the first variation formula of
 the newly introduced enhanced kinetic term for maps
 $$
   S^{(\rho, h;\Phi,g)}_{\mapscriptsize:\kineticscriptsize^+}(\varphi,\nabla)\;
    :=\;   \frac{1}{2}\,  T_{m-1}
                 \int_X \Real \Tr  \langle  D\varphi\,,\, D\varphi \rangle_{(h,g)}\, \vol_h\;
             + \; \int_X \Real \Tr  \langle d\rho, \varphi^{\diamond}d\Phi \rangle_h\,\vol_h
 $$
 in the standard action for D-branes.
As the `{\it taking the real part}' operation $\Real(\cdots\cdots)$
  is a ${\cal O}_X$-linear operation and can always be added back in the end,
 we will consider
 $$
   S^{(\rho, h;\Phi,g)}_{\mapscriptsize:\kineticscriptsize^+}
    (\varphi,\nabla)^{\Bbb C}\;
    :=\;   \frac{1}{2}\,  T_{m-1}
                 \int_X  \Tr  \langle  D\varphi\,,\, D\varphi \rangle_{(h,g)}\, \vol_h\;
             + \; \int_X      \Tr  \langle d\rho, \varphi^{\diamond}d\Phi \rangle_h\,\vol_h
 $$
 so that we don't have to carry $\Real$ around.
 
The first variation of the gauge/Yang-Mills term is analogous to that in the ordinary Yang-Mills theory   and
the first variation of the Chern-Simons/Wess-Zumino term is an update from [L-Y8: Sec.$\,$6] (D(13.1)).
Both are given in Sec.$\,$6.3 under the stronger $(\ast_2)$-Admissible Condition.

\bigskip

\subsection{The first variation of the kinetic term for maps}

Recall the (complexified) kinetic energy
 $E^{\nabla^t}(\varphi_t)^{\Bbb C}$ of $\varphi_t$ for a given $\nabla^t$,
   $t\in T:=(-\varepsilon, \varepsilon)$,
 $$
  E^{\nabla^t}(\varphi_t)^{\Bbb C}\;
    :=\; S_{\mapscriptsize:\kineticscriptsize}^{(h;g)}(\varphi_t,\nabla^t)^{\Bbb C}\;
    :=\; \frac{1}{2}\, T_{m-1} \int_X \Tr
	        \langle D^t\varphi_t\,,\, D^t\varphi^t \rangle_{(h,g)}\, \vol_h\,.
 $$
As $t$ varies, with a slight abuse of notation, denote the resulting function of $t$ by
 $$
  E^{\nabla^T}(\varphi_T)^{\Bbb C}\;
    :=\; S_{\mapscriptsize:\kineticscriptsize}^{(h;g)}(\varphi_T,\nabla^T)^{\Bbb C}\;
    :=\; \frac{1}{2}\, T_{m-1} \int_X \Tr
	        \langle D^T\varphi_T\,,\, D^T\varphi_T \rangle_{(h,g)}\, \vol_h\,,
 $$
 with the understanding that all expressions are taken on $X_t$ with $t$ varying in $T$.
  
Let $U\subset X$ be an open set
  with an orthonormal frame $(e_{\mu})_{\mu=1,\,\cdots\,,\, m}$.
Let $(e^{\mu})_{\mu=1,\,\cdots\,,\,m}$ be the dual co-frame.
Assume that $U$ is small enough
 so that $\varphi_T(U_T^{A\!z})$ is contained in a coordinate chart of $Y$,
   with coordinates $(y^1,\,\cdots\,,\, y^n)$.
Then, over $U$,
 {\small
 \begin{eqnarray*}
   \frac{d}{dt} E^{\nabla^T}(\varphi_T)^{\Bbb C}
   & =
	 & \frac{1}{2}\,T_{m-1} \int_U  \partial_t \Tr
	               \langle  D^T\varphi_T\,,\,D^T\varphi_T \rangle_{(h,g)} \,
				   \vol_h          \\[.6ex]
   & =
     & \frac{1}{2}\, T_{m-1} \int_U  \Tr  \partial_t
	               \langle D^T\varphi_T\,,\, D^T\varphi_T \rangle_{(h,g)}	
                   \vol_h           \\[.6ex]
   & =
     &  \frac{1}{2}\, T_{m-1} \int_U  \Tr  \partial_t
	               \sum_{\mu=1}^m
				      \langle D^T_{e_{\mu}}\varphi_T\,,\, D^T_{e_{\mu}}\varphi_T \rangle_g
                   \vol_h \\[.6ex]
   & =
     &  T_{m-1} \int_U  \Tr
	               \sum_{\mu=1}^m
				      \langle  \partial_t D^T_{e_{\mu}}\varphi_T\,,\,
					        D^T_{e_{\mu}}\varphi_T \rangle_g\,
                    \vol_h                                                                            \\[.6ex]
   & =
     & T_{m-1}\int_U \Tr  \sum_{\mu}
	                \langle  \nabla^{T,(\varphi_T,g)}_{e_{\mu}} \partial_t \varphi_T\,,\,
         		          D^T_{e_{\mu}}\varphi_T\rangle_g\,\vol_h   \\
   && \hspace{4em}
         +\;   T_{m-1} \int_U \Tr \sum_{\mu}
		          \langle
			     (\ad\otimes\!\nabla^g)_{D^T_{e_{\mu}}\varphi_T}\partial_t\varphi_T\,,\,
				         D^T_{e_{\mu}}\varphi_T  \rangle_g\,   \vol_h \\[.6ex]
   && \hspace{8em}
          +\;  T_{m-1} \int_U \sum_{\mu}
		           \langle
				       \sum_{i=1}^n
					       [(\partial_t\nabla^T)(e_{\mu}), \varphi_T^{\sharp}(y^i)]
					         \otimes \mbox{$\frac{\partial}{\partial y^i}$}\,,\,
				       D^T_{e_{\mu}} \varphi_T \rangle_g\, \vol_h    \\[.6ex]
   &
     =  & (\mbox{I}.1) \;+\; (\mbox{I}.2)\;+\; (\mbox{I}.3)\,.									
 \end{eqnarray*}}
  
 {\small
 \begin{eqnarray*}
  \lefteqn{
   (\mbox{I.1})\;\;
     =\;\;  T_{m-1} \int_U  \sum_{\mu}\Tr \left(
	   D^T_{e_{\mu}}\langle
	         \partial_t\varphi_T\,,\,  D^T_{e_{\mu}}\varphi_T \rangle_g\;
		-\; \langle  \partial_t \varphi_T\,,\,
               \nabla^{T,(\varphi_T,g)}_{e_{\mu}}D^T_{e_{\mu}}\varphi_T		\rangle_g	  	          
	                     \right)\,\vol_h
	                             }\\[.6ex]
   & & =\;\;
    T_{m-1} \int_U  \sum_{\mu}  e_{\mu} \Tr
             \langle  \partial_t\varphi_T \,,\, D^T_{e_{\mu}}\varphi_T\rangle_g\,\vol_h\;
     +\; T_{m-1}\int_U \Tr
             \langle \partial_t\varphi_T\,,\,
			     -\, \mbox{$\sum$}_{\mu}
				       \nabla^{T,(\varphi_T,g)}_{e_{\mu}}D^T_{e_{\mu}}\varphi_T
      					   \rangle_g\, \vol_h                          \\[.6ex]	
   & &
   =\;\;  (\mbox{I}.1.1)\;+\; (\mbox{I}.1.2)\,.
 \end{eqnarray*}}
  
Summand (I.1.1) suggests a boundary term.
To really extract the boundary term from it,
  consider the $T$-family of ${\Bbb C}$-valued $1$-forms on $U$
  $$
     \alpha^T_{(\mbox{\scriptsize I},\, \partial_t\varphi_T)}\;
	 :=\;  \Tr \langle \partial_t\varphi_T  \,,\, D^T\varphi_T \rangle_g\,,
  $$
  which depends $C^{\infty}(U)^{\Bbb C}$-linearly on $\partial_t\varphi_T$.
 Let
  $$
    \xi^T_{(\mbox{\scriptsize I},\,\partial_t\varphi_T)}\;
	  :=\;    \sum_{\mu=1}^m
	              \left(\rule{0ex}{1em}\right.\!\!\!
				    \Tr \langle \partial_t\varphi_T  \,,\, D^T_{e_{\mu}}\varphi_T \rangle_g
				   \!\!\left.\rule{0ex}{1em}\right) e_{\mu}
  $$
   be the $T$-family of dual ${\Bbb C}$-valued vector fields of
   $\alpha^T_{(\mbox{\scriptsize I},\,\partial_t\varphi_T)}$ on $U$
   with respect to the metric $h$.
 Note that
  $\xi^T_{(\mbox{\scriptsize I},\,\partial_t\varphi_T)}$
  depends $C^{\infty}(U)^{\Bbb C}$-linearly on $\partial_t\varphi_T$ as well.
Then
 \begin{eqnarray*}
  \lefteqn{ (\mbox{I}.1.1)\;\;
   =\;\;   T_{m-1}\int_U \sum_{\mu}
                 e_{\mu}\langle \xi^T_{(\mbox{\scriptsize I},\,\partial_t\varphi_T)}\,,\,
				                e_{\mu}\rangle_h\,\vol_h                   } \\[.6ex]
   & &
    =\;\;  T_{m-1} \int_U  \sum_{\mu}
                \langle \nabla^h_{e_{\mu}}\xi^T_{(\mbox{\scriptsize I},\,\partial_t\varphi_T)}\,,\, 
				              e_{\mu}\rangle_h\,\vol_h\;
              +\; T_{m-1} \int_U
			          \langle  \xi^T_{(\mbox{\scriptsize I},\,\partial_t\varphi_T)}   \,,\,
					      \mbox{$\sum_{\mu}$}\,\nabla^h_{e_{\mu}} e_{\mu}\rangle_h \,\vol_h\,.
 \end{eqnarray*}
 The first term is equal to
  $$
    T_{m-1} \int_U
	  (- \,\divv\, \xi^T_{(\mbox{\scriptsize I},\,\partial_t\varphi_T)})\,\vol_h\;\;
	=\;\; T_{m-1} \int_U d\, i_{\xi^T_{(\mbox{\tiny I},\,\partial_t\varphi_T)}} \vol_h\;\;
	=\;\; T_{m-1} \int_{\partial U} i_{\xi^T_{(\mbox{\tiny I},\,\partial_t\varphi_T)}}\vol_h\,,
  $$
   which is the sought-for boundary term, whose integrand satisfies the requirement that
    it be $C^{\infty}(U)^{\Bbb C}$-linear on $\partial_t\varphi_T$.
 The second term is equal to
  $$
    T_{m-1} \int_U  \Tr
	  \langle \partial_t\varphi_T\,,\,
	                D^T_{\sum_{\mu}\nabla^h_{e_{\mu}}e_{\mu}}\varphi_T \rangle_g\,\vol_h
  $$
  by construction, which is $C^{\infty}(U)^{\Bbb C}$-linear in $\partial_t\varphi_T$
  and hence in a final form.
  
The integrand of Summand (I.1.2)
 is already $C^{\infty}(U)^{\Bbb C}$-linear in $\partial_t\varphi_T$
 and hence in a final form.
 
Summand (I.2) can be re-written as
$$
 (\mbox{I.2})\;\;=\;\;
    -\, T_{m-1} \int_U \Tr \sum_{\mu}
	       \langle
		      (\ad\otimes\!\nabla^g)_{\partial_t\varphi_T} D^T_{e_{\mu}}\varphi_T\,,\,
			  D^T_{e_{\mu}}\varphi_T \rangle_g\,\vol_h\,.
$$
Thus, its integrand is already $C^{\infty}(U)^{\Bbb C}$-linear in $\partial_t\varphi_T$
 and hence in a final form.

Finally,
 since the built-in inclusion ${\cal O}_U^{\,\Bbb C} \subset {\cal O}_U^{A\!z}$
 identifies ${\cal O}_U^{\,\Bbb C}$ with the center of ${\cal O}_U^{A\!z}$,
 Summand (I.3) is $C^{\infty}(U)^{\Bbb C}$-linear and hence in its final fom.
 
\bigskip
 
Altogether, we almost complete the calculation except the issue of whether all the inner products
 $\Tr\langle\,\mbox{\LARGE $\cdot$}\,,\,\mbox{\LARGE $\cdot$}\,\rangle_g$
 that appear in the procedure are truly defined.
For this, one notices that wherever such an inner product appears above,
 at least one of its arguments is either $\partial_t\varphi_T$ or $D^T_{e_{\mu}}\varphi_T$, for some $\mu$.
It follows from Lemma 3.2.2.4 that they are indeed defined.
 
In summary,

\bigskip

\begin{proposition} {\bf [first variation of kinetic term for maps]}$\;$
 Let $(\varphi_T,\nabla^T)$ be a $(\ast_1)$-admissible $T$-family
  of $(\ast_1)$-admissible pairs.
 Then,
 \begin{eqnarray*}
  \frac{d}{dt}E^{\nabla^T}(\varphi_T)^{\Bbb C}
   & =
    & \frac{d}{dt}
          \left(
		    \frac{1}{2}\, T_{m-1} \int_U \Tr
	        \langle D^T\varphi_T\,,\, D^T\varphi_T \rangle_{(h,g)}\, \vol_h
		   \right)                     \\[.6ex]
   & =
     & T_{m-1} \int_{\partial U}
            	 i_{\xi^T_{(\mathrm{I},\,\partial_t\varphi_T)}}\vol_h \\[.6ex]
   &&
     +\;\;   T_{m-1} \int_U  \Tr
	             \big\langle \partial_t\varphi_T\,,\,
	                \big(D^T_{\sum_{\mu=1}^m \nabla^h_{e_{\mu}}e_{\mu}}\,
                         -\, \mbox{$\sum$}_{\mu=1}^m
				                  \nabla^{T,(\varphi_T,g)}_{e_{\mu}}D^T_{e_{\mu}}		
						\big )\, \varphi_T \big\rangle_g\,\vol_h  \\[.6ex]
   &&	\hspace{2em}
	  -\;\; T_{m-1} \int_U \Tr \sum_{\mu=1}^m
	                   \langle (
					     \ad\otimes\!\nabla^g)_{\partial_t\varphi_T} D^T_{e_{\mu}}\varphi_T\,,\,
			             D^T_{e_{\mu}}\varphi_T \rangle_g\,\vol_h   \\[.6ex]
    &&
	  +\;\;  T_{m-1}\int_U \sum_{\mu=1}^m\,
		           \langle
				    \sum_{i=1}^n
					   [(\partial_t\nabla^T)(e_{\mu}), \varphi_T^{\sharp}(y^i)]
					       \otimes \mbox{$\frac{\partial}{\partial y^i}$} \,,\,
					 D^T_{e_{\mu}}\varphi_T     \rangle_g\, \vol_h\,.
 \end{eqnarray*}
 Here,
  the first summand is the boundary term with
  $\,\xi^T_{(\mathrm{I},\,\partial_t\varphi_T)}
	   := \sum_{\mu=1}^m
	        (\Tr \langle \partial_t\varphi_T, D^T_{e_{\mu}}\varphi_T \rangle_g) e_{\mu}\,$
  $C^{\infty}(U)^{\Bbb C}$-linear in $\partial_t\varphi_T$;
 the integrand of the second and the third terms are $C^{\infty}(U)^{\Bbb C}$-linear
     in $\partial_t\varphi_T$   and
 their real part contribute first-order and second-order terms
     to the equations of motion for $(\varphi, \nabla)$;
 the integrand of the last term is $C^{\infty}(U)^{\Bbb C}$-linear in $\partial_t\nabla^T$   and
     its real part contributes terms,
	 first order in $\varphi$ but zeroth order in the connection $1$-from of $\nabla$, 	
	 to the equations of motion for $(\varphi, \nabla)$
	 in addition to those from the first variation of
	 the rest part of $S_{\standardscriptsize}^{(\rho,h;\Phi,g,B,C)}(\varphi,\nabla)$.
 These lower-order terms contribute to the equations of motion for $(\varphi,\nabla)$
   but do not change the signature of the system.
\end{proposition}
 
\bigskip
 
\begin{remark} $[$for  $(\ast_2)$-admissible $T$-family of $(\ast_2)$-admissible pairs$]\;$
{\rm
 If furtheremore $(\varphi,\nabla)$ is $(\ast_2)$-admissible and
    $(\varphi_T,\nabla^T)$ is a $(\ast_2)$-admissible $T$-family of $(\ast_2)$-admissible pairs
	that deforms $(\varphi,\nabla)$,	
  then
   the third summand of the first variation formula in Proposition~6.1.1 vanishes  and
   the fourth/last summand has a $Y$-coordinate-free form
  $$
      T_{m-1}\int_U \sum_{\mu=1}^m\,
		           \langle
				    \ad_{(\partial_t\nabla^T)(e_{\mu})}\varphi_T\,,\,
					 D^T_{e_{\mu}}\varphi_T     \rangle_g\, \vol_h\,.
   $$
 In this case, the first variation with respect to $\varphi$ alone (i.e.\ setting $\partial_t\nabla^T=0$),
 cf.\ the first two summands,
  takes the form of a direct formal generalization of the first variation formula in the study of harmonic maps;
  e.g.\ [E-L], [E-S], [Ma], [Sm].
}\end{remark}

\bigskip

\subsection{The first variation of the dilaton term}

We now turn to the (complexified) dilaton term
 in $S_{\standardscriptsize}^{(\rho,h;\Phi,g,B,C)}(\varphi,\nabla)^{\Bbb C}$.
 
Let $\varphi_T:(X^{\!A\!z},{\cal E}_T;\nabla^T)\rightarrow Y$
 be a $(\ast_1)$-admissible $T$-family of $(\ast_1)$-admissible pairs.
Then, over an open set $U\subset X$,
\begin{eqnarray*}
  S^{(\rho, h;\Phi)}_{\dilatonscriptsize} (\varphi_T)^{\Bbb C}
    &  =   &  \int_U \Tr \langle d\rho \,,\, \varphi_T^{\diamond}d\Phi\rangle_h\,\vol_h  \\[.6ex]
  & =
    &  \int_U\Tr \sum_{\mu=1}^m
	        \left(
			  d\rho(e_{\mu})\,
			  \sum_{i=1}^n
			    D^T_{e_{\mu}}\varphi_T^{\sharp}(y^i)\,
				  \varphi_T^{\sharp}
				    \left(\rule{0ex}{1em} \right.\!
					   \frac{\partial\Phi}{\partial y^i}
					\!\left.\rule{0ex}{1em}\right)
            \right) \vol_h     \\[.6ex]
   & =   & \int_U\Tr \left(\sum_{\mu}
	              d\rho(e_{\mu})\, ((D^T_{e_{\mu}}\varphi_T)\Phi)\right) \vol_h\,.
\end{eqnarray*}
 
{\small
\begin{eqnarray*}
 \lefteqn{
   \frac{d}{dt}  S^{(\rho, h;\Phi)}_{\dilatonscriptsize}(\varphi_T)^{\Bbb C}\;\;
    =\;\;   \int_U \Tr  \sum_{\mu=1}^m
	            d\rho(e_{\mu})\,
			     \partial_t \!\left((D^T_{e_{\mu}}\varphi_T)\Phi  \right) \vol_h  }\\
   && =\;\;
	  \int_U \Tr \sum_{\mu}
            d\rho(e_{\mu})
			  \left((\partial_t D^T_{e_{\mu}}\varphi_T)\Phi  \right)
            \vol_h    \\
    && \hspace{4em}		
	  +\; \int_U \Tr \sum_{\mu}
	          d\rho(e_{\mu})\,
			   \sum_{i,j=1}^n
			    D^T_{e_{\mu}}\varphi_T^{\sharp}(y^i) \partial_t\varphi_T^{\sharp}(y^j)
			      \otimes \left(
						   \mbox{$\frac{\partial}{\partial y^j} \frac{\partial}{\partial y^i}$}\Phi
						     - \left(\rule{0ex}{1em}\right.\!
							     \nabla^g_{\frac{\partial}{\partial y^j}}
							             \mbox{$\frac{\partial}{\partial y^i}$}
										\!\left.\rule{0ex}{1em}\right)\Phi
						               \right)
	         \vol_h	          \\
   && =\;\;  (\mbox{II}.1)\;+\; (\mbox{II}.2)\,. 			
\end{eqnarray*}}

\noindent
The integrand of Summand (II.2) is $C^{\infty}(U)^{\Bbb C}$-linear in $\partial_t\varphi_T$
  and hence in a final form.
 
{\small
\begin{eqnarray*}
 (\mbox{II}.1) & =
    & \int_U \Tr \sum_{\mu}
	      d\rho(e_{\mu})
	         \left( \big(\nabla^{T,(\varphi_T,\,g)}_{e_{\mu}}\partial_t\varphi_T\big)
			  \Phi \right)
		 \vol_h	  \\
    && \hspace{4em}		
		-\;\;
           \int_U \Tr \sum_{\mu}
	         d\rho(e_{\mu})
	          \left(
			   \big((\ad\otimes \nabla^g)_{\partial_t\varphi_T}
			                 D^T_{e_{\mu}}\varphi_T \big)
			       \Phi \right)
		     \vol_h	  \\
    && \hspace{8em}			
        +\;\;
		 \int_U \Tr \sum_{\mu}
	       d\rho(e_{\mu})
	         \left(\rule{0ex}{1em}\right.\!\!
			  \left(\rule{0ex}{1em}\right.\!
			     \sum_i
			       \big[(\partial_t\nabla^T)(e_{\mu})\,,\, \varphi_T^{\sharp}(y^i)\big]
			                  \otimes \mbox{$\frac{\partial}{\partial y^i}$}
				 \!\left.\rule{0ex}{1em}\right)\Phi
			  \!\left.\rule{0ex}{1em}\right)
		  \vol_h	\;\;		 \\[.6ex]
    & = & (\mbox{II}.1.1)\;+\;    (\mbox{II}.1.2)\; +\; 	(\mbox{II}.1.3)\,.	
\end{eqnarray*}}

\noindent
Both Summand (II.1.2)  and Summand (II.1.3) vanish since
 $$
   \Tr([a,b]c)\;=\; 0   \hspace{2em}\mbox{if $\;[b,c]\;=\;0$}
 $$
 for $r\times r$ matrices $a, b, c$.

{\small
\begin{eqnarray*}
 (\mbox{II}.1.1) & =
    & \int_U \Tr \sum_{\mu}
	     d\rho(e_{\mu})\,
		    D^T_{e_{\mu}}((\partial_t\varphi_T)\Phi))\,
			\vol_h \\
   &&  \hspace{2em}
     -\;\;
      \int_U \Tr \sum_{\mu}
	     d\rho(e_{\mu})\,
		   \sum_{i,j=1}^n
				 \partial_t\varphi_T^{\sharp}(y^i)\,
				 D^T_{e_{\mu}}\varphi_T^{\sharp}(y^j)\,
						\otimes \left(
						   \mbox{$\frac{\partial}{\partial y^i} \frac{\partial}{\partial y^j}$}\Phi\,
						     -\,  \left(\rule{0ex}{1em}\right.\!
							     \nabla^g_{\frac{\partial}{\partial y^i}}
							             \mbox{$\frac{\partial}{\partial y^j}$}
										\!\left.\rule{0ex}{1em}\right)\Phi
						               \right)				
			\vol_h \\[.6ex]
   &=  & (\mbox{II}.1.1.1)\;  + \; (\mbox{II}.1.1.2)\,.			
\end{eqnarray*}}

\noindent
The integrand of Summand (II.1.1.2) is
 $C^{\infty}(U)^{\Bbb C}$-linear in $\partial_t\varphi_T$
and hence in a final form. It can be combined with Summand (II.2) to give
{\small
 \begin{eqnarray*}
  \lefteqn{
  (\mbox{II}.1.1.2)\,+\, (\mbox{II}.2) }\\
    && =\;
	 -\,  \int_U \Tr \sum_{\mu}
	        d\rho(e_{\mu})\,
		      \sum_{i,j=1}^n
				 [\partial_t\varphi_T^{\sharp}(y^i)\,,\,
				   D^T_{e_{\mu}}\varphi_T^{\sharp}(y^j)]
						\otimes \left(
						   \mbox{$\frac{\partial}{\partial y^i} \frac{\partial}{\partial y^j}$}\Phi\,
						     -\,  \left(\rule{0ex}{1em}\right.\!
							     \nabla^g_{\frac{\partial}{\partial y^i}}
							             \mbox{$\frac{\partial}{\partial y^j}$}
										\!\left.\rule{0ex}{1em}\right)\Phi
						               \right)				
			\vol_h\,,
 \end{eqnarray*}}
which again vanishes due to $\Tr$.
 
{\small
\begin{eqnarray*}
 (\mbox{II}.1.1.1)& =
  & \int_U \sum_{\mu} d\rho(e_{\mu})
        \Tr D^T_{e_{\mu}}((\partial_t\varphi_T)\Phi)\,\vol_h  \\
 &=
  & \int_U \sum_{\mu} d\rho(e_{\mu})\,
          e_{\mu}\Tr((\partial_t\varphi_T)\Phi)\, \vol_h  \\
 & =
  & \int_U \sum_{\mu} e_{\mu}
       \left(\rule{0ex}{1em}\right.\!
	      d\rho(e_{\mu})\,\Tr ((\partial_t\varphi_T)\Phi)
	   \!\left.\rule{0ex}{1em}\right) \vol_h\;
     -\;  \int_U
	          \left(\rule{0ex}{1em}\right.\!
  			   \sum_{\mu}e_{\mu}d\rho(e_{\mu})
			  \!\left.\rule{0ex}{1em}\right)
			   \Tr((\partial_t\varphi_T)\Phi)\,\vol_h \\[.6ex]
  & = &  (\mbox{II}.1.1.1.1)\;+\; (\mbox{II}.1.1.1.2)
\end{eqnarray*}}

\noindent
The integrand of Summand (II.1.1.1.2)
  is $C^{\infty}(U)^{\Bbb C}$-linear in $\partial_t\varphi_T$
  and hence in a final form.
To extract the boundary term from Summand (II.1.1.1.1),
 consider the $T$-family of ${\Bbb C}$-valued $1$-forms on $U$
 $$
    \alpha^T_{(\mbox{\scriptsize II}, \partial_t\varphi_T)}\;
	   :=\;  d\rho\,\Tr ((\partial_t\varphi_T)\Phi)\,,
 $$
 which depends $C^{\infty}(U)^{\Bbb C}$-linearly on $\partial_t\varphi_T$.
Let
 $$
   \xi^T_{(\mbox{\scriptsize II},\partial_t\varphi_T)}\;
    =\;   \sum_{\mu=1}^m
	         \left(\rule{0ex}{1em}\right.\!
			    d\rho(e_{\mu})\,\Tr ((\partial_t\varphi_T)\Phi)
		     \!\left.\rule{0ex}{1em}\right) e_{\mu}
 $$
 be the T-family of dual ${\Bbb C}$-valued vector fields of
 $\alpha^T_{(\mbox{\scriptsize II},\partial_t\varphi_T)}$ on $U$ with respect to the metric $h$.
Note that $\xi^T_{(\mbox{\scriptsize II},\partial_t\varphi_T)}$
 depends $C^{\infty}(U)^{\Bbb C}$-linearly on $\partial_t\varphi_T$ as well.
Then

 {\small
\begin{eqnarray*}
 (\mbox{II}.1.1.1.1)
  & =
    & \int_U \sum_{\mu}
	     e_{\mu}
		   \langle \xi^T_{(\mbox{\scriptsize II},\partial_t\varphi_T)}\,,\,  e_{\mu}  \rangle_h\,
		    \vol_h  \\
  & =
    & \int_U \sum_{\mu}
		   \langle  \nabla^h_{e_{\mu}} \xi^T_{(\mbox{\scriptsize II},\partial_t\varphi_T)}\,,\,
  		                   e_{\mu}  \rangle_h\,  \vol_h\;
		 +\;    \int_U
		              \langle \xi^T_{(\mbox{\scriptsize II},\partial_t\varphi_T)}\,,\,
					      \mbox{$\sum_{\mu}$}\,
						     \nabla^h_{e_{\mu}}e_{\mu}  \rangle_h\,   \vol_h 		\\
\end{eqnarray*}}
 
\noindent
The first term is equal to
 $$
     \int_U (- \,\divv\, \xi^T_{(\mbox{\scriptsize II},\,\partial_t\varphi_T)})\,\vol_h\;\;
	=\;\;  \int_U d\, i_{\xi^T_{(\mbox{\tiny II},\,\partial_t\varphi_T)}} \vol_h\;\;
	=\;\;  \int_{\partial U} i_{\xi^T_{(\mbox{\tiny II},\,\partial_t\varphi_T)}}\vol_h\,,
 $$
which is the sought-for boundary term, whose integrand satisfies the requirement that it be
 $C^{\infty}(U)^{\Bbb C}$-linear in $\partial_t\varphi_T$ .
The second term is equal to
 $$
   \int_U
   d\rho\big(\mbox{$\sum_{\mu}\!\nabla^h_{e_{\mu}}e_{\mu}$}\big)\,
    \Tr ((\partial_t\varphi_T)\Phi)\,\vol_h
 $$
by construction, which is $C^{\infty}(U)^{\Bbb C}$-linear in $\partial_t\varphi_T$
 and hence in a final form.
  
In summary,
 
\bigskip

\begin{proposition} {\bf [first variation of dilaton term]}$\;$
 Let $(\varphi_T,\nabla^T)$ be a $(\ast_1)$-admissible $T$-family of $(\ast_1)$-admissible pairs.
 Then,
 {\small
  \begin{eqnarray*}
   \lefteqn{
     \frac{d}{dt}S^{(\rho, h;\Phi)}_{\dilatonscriptsize}(\varphi_T)^{\Bbb C}\;\;
      =\;\;   \frac{d}{dt}
	                \int_U \Tr \langle d\rho \,,\, \varphi_T^{\diamond}d\Phi\rangle_h\,\vol_h   }\\[.6ex]	
   &&=\;\;
    	 \int_{\partial U} i_{\xi^T_{(\mathrm{II},\,\partial_t\varphi_T)}}\vol_h  \\[.6ex]
   && \hspace{2em}
      +\;\;  \int_U
		        \Big(
                    d\rho
					  \big(\mbox{$\sum_{\mu=1}^m\!\nabla^h_{e_{\mu}}e_{\mu}$}\big)\,
					     -\, \mbox{$\sum_{\mu=1}^m$}e_{\mu}d\rho(e_{\mu})
				 \Big)
                      \Tr ((\partial_t\varphi_T)\Phi)\,\vol_h\,.
  \end{eqnarray*}}
 
 \noindent
 Here,
  the first summand is the boundary term with
  $\,\xi^T_{(\mathrm{II},\partial_t\varphi_T)}
      :=   \sum_{\mu=1}^m
	         (d\rho(e_{\mu})\,\Tr ((\partial_t\varphi_T)\Phi) ) e_{\mu}\,$
  $C^{\infty}(U)^{\Bbb C}$-linear in $\partial_t\varphi_T$;
 the integrand of the second summand $C^{\infty}(U)^{\Bbb C}$-linear
     in $\partial_t\varphi_T$   and
 they contribute additional zeroth-order terms to the equations of motion for $(\varphi, \nabla)$.
 In particular,
   while the dilaton term of the standard action modifies the equations of motion for $(\varphi,\nabla)$,
   it does not change the signature of the system.
\end{proposition}

\bigskip

\subsection{The first variation of the gauge/Yang-Mills term and the Chern-Simons/ Wess-Zumino term}
   
To make sure that differential forms on $Y$ of rank $\ge 2$  are pull-pushed to
 (${\cal O}_X^{A\!z}$-valued-)differential forms on $X$
 (cf.\ Lemma 2.1.11),
 we assume in this subsection that
 $\varphi_T:(X_T^{\!A\!z},{\cal E}_T; \nabla^T)\rightarrow Y$
  is a $(\ast_2)$-family of $(\ast_2)$-admissible maps.
(Note that as the gauge/Yang-Mills term is defined through a norm-squared,
    $(\ast_1)$-admissible family of $(\ast_1)$-admissible $(\varphi_T,\nabla^T)$
	is enough for the derivation of the first variation formula of the gauge/Yang-Mills term but the result will be slightly messier.)

\bigskip
   
\subsubsection{The first variation of the gauge/Yang-Mills term}

Let $(e_1,\,\cdots\,,\, e_m)$ be an orthonormal frame on $U$.
Then, over $U$,
 \begin{eqnarray*}
  \lefteqn{
   S^{(h;B)}_{\gaugescriptsize/\YMscriptsize}(\varphi_T, \nabla^T)^{\Bbb C}\;\;
       :=\;\;   -\,  \mbox{\Large $\frac{1}{2}$}\int_U   \Tr
	                \| 2\pi\alpha^{\prime}F_{\nabla^T}+ \varphi_T^{\diamond}B\|_h^2 \, \vol_h   }\\
   && =\;\;   -\,  \mbox{\Large $\frac{1}{2}$}\int_U   \Tr
                     \sum_{\mu, \nu}
					  \Big((2\pi\alpha^{\prime}F_{\nabla^T}
					                  + \varphi_T^{\diamond}B)(e_{\mu}, e_{\nu}) \Big)^2
					   \vol_h   \,.
 \end{eqnarray*}
Applying the following basic identities:
 {\small
 \begin{eqnarray*}
   \partial_t F_{\nabla^T}(e_{\mu},e_{\nu})
    & =  & D^T_{e_{\mu}}\big((\partial_t \nabla^T)(e_{\nu})\big)\;
                -\; D^T_{e_{\nu}} \big( (\partial_t\nabla^T )(e_{\mu}) \big)\;
		        -\; (\partial_t\nabla^T)([e_{\mu}, e_{\nu}])\,,           \\[1.2ex]
   \partial_t\big((\varphi_T^{\diamond}B)(e_{\mu},e_{\nu})\big)
     & = & \sum_{i,j}\partial_t\big(\varphi_T^{\sharp}(B_{ij}) \big)
                     D^T_{e_{\mu}}\varphi_T^{\sharp}(y^i)
					 D^T_{e_{\nu}} \varphi_T^{\sharp}(y^j)   \\[-1ex]
     && \hspace{2em}					
			  +\; \sum_{i,j}\varphi_T^{\sharp}(B_{ij})
			          \Big(D^T_{e_{\mu}}\partial_t\varphi_T^{\sharp}(y^i)
					             + \big[(\partial_t\nabla^T)(e_{\mu}), \varphi_T^{\sharp}(y^i)\big] \Big)
                                                   D^T_{e_{\nu}} \varphi_T^{\sharp}(y^j)	   \\[-1ex]
     && \hspace{4em}				
			  +\; \sum_{i,j}\varphi_T^{\sharp}(B_{ij})
			          D^T_{e_{\mu}}\varphi_T^{\sharp}(y^i)
                       \Big( D^T_{e_{\nu}} \partial_t\varphi_T^{\sharp}(y^j)
                                       + \big[ (\partial_t\nabla^T)(e_{\nu}), \varphi_T^{\sharp}(y^j)\big]	\Big)\,.
 \end{eqnarray*}}
 and proceeding similarly to Sec.~6.1, one has the following results.

{\small
 \begin{eqnarray*}
  \lefteqn{
   \frac{d}{dt}\,
     S^{(h;B)}_{\gaugetiny/\YMtiny}(\varphi_T, \nabla^T)^{\Bbb C}\;\;
      =\;\;   -\,  \mbox{\Large $\frac{1}{2}$}\int_U   \Tr
                     \partial_t \sum_{\mu, \nu}
					  \Big((2\pi\alpha^{\prime}F_{\nabla^T}
					                  + \varphi_T^{\diamond}B)(e_{\mu}, e_{\nu}) \Big)^2
					   \vol_h         }\\
  &&  =\;\;   -\, \int_U   \Tr
                     \sum_{\mu, \nu}
					  \partial_t
					   \Big((2\pi\alpha^{\prime}F_{\nabla^T}
					                  + \varphi_T^{\diamond}B)(e_{\mu}, e_{\nu}) \Big)
                       \cdot									
					   \Big((2\pi\alpha^{\prime}F_{\nabla^T}
					                  + \varphi_T^{\diamond}B)(e_{\mu}, e_{\nu}) \Big)
					   \vol_h           \\					
  &&  =\;\;   -\, \int_U   \Tr
                     \sum_{\mu, \nu}
					   2\pi\alpha^{\prime}
					         \partial_t \big(F_{\nabla^T}(e_{\mu}, e_{\nu})\big)
					  \cdot
					   \Big((2\pi\alpha^{\prime}F_{\nabla^T}
					                  + \varphi_T^{\diamond}B)(e_{\mu}, e_{\nu}) \Big)
					   \vol_h           \\[-.6ex]					
 &&  \hspace{6em}
         -\, \int_U   \Tr
                     \sum_{\mu, \nu}
					   \partial_t\big(   \varphi_T^{\diamond}B(e_{\mu}, e_{\nu})\big)
                       \cdot									
					   \Big((2\pi\alpha^{\prime}F_{\nabla^T}
					                  + \varphi_T^{\diamond}B)(e_{\mu}, e_{\nu}) \Big)
					   \vol_h           \\					    					
  && =\;\; (\mathrm{III}.1)\; +\; (\mathrm{III}.2)\,.
 \end{eqnarray*}}

 {\small
 \begin{eqnarray*}
 (\mathrm{III}.1)
   & :=\: &
     -\, \int_U   \Tr
                     \sum_{\mu, \nu}
					   2\pi\alpha^{\prime}
					         \partial_t \big(F_{\nabla^T}(e_{\mu}, e_{\nu})\big)
					  \cdot
					   \Big((2\pi\alpha^{\prime}F_{\nabla^T}
					                  + \varphi_T^{\diamond}B)(e_{\mu}, e_{\nu}) \Big)
					   \vol_h           \\
   & = &
    -\, 2\pi\alpha^{\prime}
    \int_U   \Tr
                     \sum_{\mu, \nu}
					  \Big(D^T_{e_{\mu}}\big((\partial_t \nabla^T)(e_{\nu})\big)\;
                           -\; D^T_{e_{\nu}} \big( (\partial_t\nabla^T )(e_{\mu}) \big)\;
		                  -\; (\partial_t\nabla^T)([e_{\mu}, e_{\nu}])\Big)		\\[-1.2ex]
   && \hspace{20em}	\cdot
					   \Big((2\pi\alpha^{\prime}F_{\nabla^T}
					                  + \varphi_T^{\diamond}B)(e_{\mu}, e_{\nu}) \Big)
					   \vol_h           \\			
   & = &  -\,4\pi\alpha^{\prime}
               \int_{\partial U}	i_{
			                  \xi^T_{(\mathrm{III}, \partial_t\nabla^T)} }	\vol_h \\
   &&  - \; 4\pi\alpha^{\prime}  \int_U \Tr \sum_{\nu}
			            (\partial_t\nabla^T)(e_{\nu})
						   \cdot
						   \Big(\;
						     (2\pi\alpha^{\prime}F_{\nabla^T}+ \varphi_T^{\diamond}B)
							 (\mbox{$\sum$}_{\mu}\nabla^h_{e_{\mu}}e_{\mu}, e_{\nu}) \\
   && \hspace{14em}												
						    -\, \sum_{\mu}
						           D^T_{e_{\mu}}
							        \big((2\pi\alpha^{\prime}F_{\nabla^T}+ \varphi_T^{\diamond}B)
									             (e_{\mu}, e_{\nu})\big)    \\
   && \hspace{14em}
                 -\, \mbox{\large $\frac{1}{2}$} \sum_{\mu,\lambda}
				       e^{\nu}([e_{\mu}, e_{\lambda}  ])
					 (2\pi\alpha^{\prime}F_{\nabla^T}+ \varphi_T^{\diamond}B)
					                                                                                        (e_{\mu}, e_{\lambda})
						  \;\Big) \vol_h  \,.
 \end{eqnarray*}}
Here,
 $$
    \xi^T_{(\mathrm{III},\partial_t\nabla^T)}\;
	  :=\; \sum_{\mu,\nu}
	          \Tr\Big(
			       (\partial_t\nabla^T)(e_{\nu})
				    \cdot
					(2\pi\alpha^{\prime}F_{\nabla^T}
					                  + \varphi_T^{\diamond}B)(e_{\mu}, e_{\nu})
			        \Big)\, e_{\mu}
    \hspace{2em}\in\; {\cal T}_{\ast}(U_T/T)^{\Bbb C}					
 $$
 is ${\cal O}_U^{\,\Bbb C}$-linear in $\partial_t\nabla^T$;
 and the second summand contributes to the equations of motion for $(\varphi,\nabla)$.
The latter are standard terms from non-Abelian Yang-Mills theory with additional terms from $\varphi^{\diamond}B$.

{\small
\begin{eqnarray*}
 (\mathrm{III}.2)
  & :=\: &   -\, \int_U   \Tr
                     \sum_{\mu, \nu}
					   \partial_t\big(   \varphi_T^{\diamond}B(e_{\mu}, e_{\nu})\big)
                       \cdot									
					   \Big((2\pi\alpha^{\prime}F_{\nabla^T}
					                  + \varphi_T^{\diamond}B)(e_{\mu}, e_{\nu}) \Big)
					   \vol_h           \\					
  & = &  -\, \int_U   \Tr
                     \sum_{\mu, \nu}
			   \bigg(
			      \sum_{i,j}\partial_t\big(\varphi_T^{\sharp}(B_{ij}) \big)
                     D^T_{e_{\mu}}\varphi_T^{\sharp}(y^i)
					 D^T_{e_{\nu}} \varphi_T^{\sharp}(y^j)   \\[-1ex]
     && \hspace{8em}					
			  +\; \sum_{i,j}\varphi_T^{\sharp}(B_{ij})
			          \Big(D^T_{e_{\mu}}\partial_t\varphi_T^{\sharp}(y^i)
					             + \big[(\partial_t\nabla^T)(e_{\mu}), \varphi_T^{\sharp}(y^i)\big] \Big)
                                                   D^T_{e_{\nu}} \varphi_T^{\sharp}(y^j)	   \\[-1ex]
     && \hspace{8em}				
			  +\; \sum_{i,j}\varphi_T^{\sharp}(B_{ij})
			          D^T_{e_{\mu}}\varphi_T^{\sharp}(y^i)
                       \Big( D^T_{e_{\nu}} \partial_t\varphi_T^{\sharp}(y^j)
                                       + \big[ (\partial_t\nabla^T)(e_{\nu}), \varphi_T^{\sharp}(y^j)\big]	\Big)	
                         \bigg)									   \\
  && \hspace{6em}					   					   					
                       \cdot									
					   \Big((2\pi\alpha^{\prime}F_{\nabla^T}
					                  + \varphi_T^{\diamond}B)(e_{\mu}, e_{\nu}) \Big)
					   \vol_h           \\							
  & = &  (\mathrm{III}.2.1)\; +
               +\; \big(  (\mathrm{III}.2.2.1)\,+\, (\mathrm{III}.2.2.2)\big)\;
			   +\; (\mathrm{III}.2.3.1)\,+\, (\mathrm{III}.2.3.2)
\end{eqnarray*}}
in the order of the appearance of the five summands after the expansion.

{\small
\begin{eqnarray*}
  (\mathrm{III}.2.1)
   & :=\: &    -\, \int_U   \Tr
                     \sum_{\mu, \nu}
			      \sum_{i,j}\partial_t\big(\varphi_T^{\sharp}(B_{ij}) \big)
                     D^T_{e_{\mu}}\varphi_T^{\sharp}(y^i)
					 D^T_{e_{\nu}} \varphi_T^{\sharp}(y^j)   \\[-1ex]
   && \hspace{16em}					   					   					
                       \cdot									
					   \Big((2\pi\alpha^{\prime}F_{\nabla^T}
					                  + \varphi_T^{\diamond}B)(e_{\mu}, e_{\nu}) \Big)
					   \vol_h           \\	
   & :=\: &    -\, \int_U   \Tr
                     \sum_{\mu, \nu}
			      \sum_{i,j} \big((\partial_t\varphi_T)B_{ij} \big)
                     D^T_{e_{\mu}}\varphi_T^{\sharp}(y^i)
					 D^T_{e_{\nu}} \varphi_T^{\sharp}(y^j)   \\[-1ex]
   && \hspace{16em}					   					   					
                       \cdot									
					   \Big((2\pi\alpha^{\prime}F_{\nabla^T}
					                  + \varphi_T^{\diamond}B)(e_{\mu}, e_{\nu}) \Big)
					   \vol_h
\end{eqnarray*}}
has an integrand ${\cal O}_U^{\,\Bbb C}$-linear in $\partial_t\varphi_T$ and hence in a final form.

{\small
\begin{eqnarray*}
 \lefteqn{(\mathrm{III}.2.2.1)\,+\, (\mathrm{III}.2.3.1) }\\
  && :=\,\;
    -\, \int_U   \Tr
                     \sum_{\mu, \nu}					 					
			   \Big(
			     \sum_{i,j}\varphi_T^{\sharp}(B_{ij})
			                             D^T_{e_{\mu}}\partial_t\varphi_T^{\sharp}(y^i)
					                     D^T_{e_{\nu}} \varphi_T^{\sharp}(y^j)	   \\[-1ex]
     && \hspace{12em}				
			  +\; \sum_{i,j}\varphi_T^{\sharp}(B_{ij})
			          D^T_{e_{\mu}}\varphi_T^{\sharp}(y^i)
                      D^T_{e_{\nu}} \partial_t\varphi_T^{\sharp}(y^j)									   
                         \Big)									   \\
   && \hspace{8 em}					   					   					
                       \cdot									
					   \Big((2\pi\alpha^{\prime}F_{\nabla^T}
					                  + \varphi_T^{\diamond}B)(e_{\mu}, e_{\nu}) \Big)
					   \vol_h           \\			
   &&=\;\;
    -\, 2\,\int_U   \Tr
             \sum_{\mu, \nu}					 					
			     \sum_{i,j} D^T_{e_{\mu}}\partial_t\varphi_T^{\sharp}(y^i)\,
										 \varphi_T^{\sharp}(B_{ij})\,	
					                     D^T_{e_{\nu}} \varphi_T^{\sharp}(y^j)	   \\[-1ex]
  && \hspace{16 em}					   					   					
                       \cdot									
					   \big((2\pi\alpha^{\prime}F_{\nabla^T}
					                  + \varphi_T^{\diamond}B)(e_{\mu}, e_{\nu}) \big)
					   \vol_h           \\											
  && =\;\;
     -\, 2\, \int_{\partial U} i_{\xi^T_{(\mathrm{III},\partial_t\varphi_T)}} \vol_h \\
  && \hspace{1.6em}	
     -\, 2\, \int_U   \Tr  \sum_{\nu}\sum_{i,j}
	                  \partial_t\varphi_T^{\sharp}(y^i)  \\[-1ex]
  &&  \hspace{6em}
          \bigg(\, \varphi_T^{\sharp}(B_{ij})\,	
					                     D^T_{e_{\nu}} \varphi_T^{\sharp}(y^j)	
                               \cdot									
					           \big((2\pi\alpha^{\prime}F_{\nabla^T}
					                        + \varphi_T^{\diamond}B)
											  (\mbox{$\sum$}_{\mu}
											     \nabla^h_{e_{\mu}}e_{\mu}, e_{\nu}) \big)	  \\[-1ex]
  && \hspace{8em}		
		    -\, \sum_{\mu} D^T_{e_{\mu}}
					    \Big(
                            \varphi_T^{\sharp}(B_{ij})\,	
					                     D^T_{e_{\nu}} \varphi_T^{\sharp}(y^j)	
                       \cdot									
					   \big((2\pi\alpha^{\prime}F_{\nabla^T}
					                  + \varphi_T^{\diamond}B)(e_{\mu}, e_{\nu}) \big)						
						\Big)  \bigg)  \vol_h                \,.
\end{eqnarray*}}
Here,
 $$
  \xi^T_{(\mathrm{III}, \partial_t\varphi_T)}\;
    :=\;   \sum_{\mu}\Big(
	            \sum_{\nu}					 					
			     \sum_{i,j} \partial_t\varphi_T^{\sharp}(y^i)\,
										 \varphi_T^{\sharp}(B_{ij})\,	
					                     D^T_{e_{\nu}} \varphi_T^{\sharp}(y^j)	
                       \cdot									
					   \big((2\pi\alpha^{\prime}F_{\nabla^T}
					                  + \varphi_T^{\diamond}B)(e_{\mu}, e_{\nu}) \big) 	\Big) e_{\mu}
 $$									
  in ${\cal T}_{\ast}(U_T/T)^{\Bbb C}$
    is ${\cal O}_U^{\,\Bbb C}$-linear in $\partial_t\varphi_T$;
 and the second summand contributes to\\
 $\delta S_{\standardscriptsize}^{(\rho,h;\Phi,g,B,C)}(\varphi,\nabla)/\delta \varphi$-part
 of the equations of motion for $(\varphi,\nabla)$.

{\small
\begin{eqnarray*}
 \lefteqn{(\mathrm{III}.2.2.2)\,+\, (\mathrm{III}.2.3.2)  }\\
  && :=\;
   -\, \int_U   \Tr
                     \sum_{\mu, \nu}
			   \Big(	
			      \sum_{i,j}\varphi_T^{\sharp}(B_{ij})
			                             \big[(\partial_t\nabla^T)(e_{\mu}), \varphi_T^{\sharp}(y^i)\big]
                                                   D^T_{e_{\nu}} \varphi_T^{\sharp}(y^j)	   \\[-1ex]
     && \hspace{12em}				
			  +\; \sum_{i,j}\varphi_T^{\sharp}(B_{ij})
			          D^T_{e_{\mu}}\varphi_T^{\sharp}(y^i)
                       \big[ (\partial_t\nabla^T)(e_{\nu}), \varphi_T^{\sharp}(y^j)\big]
                         \Big)								   \\
  && \hspace{8em}					   					   					
                       \cdot									
					   \Big((2\pi\alpha^{\prime}F_{\nabla^T}
					                  + \varphi_T^{\diamond}B)(e_{\mu}, e_{\nu}) \Big)
					   \vol_h           \,.		
\end{eqnarray*}}
has an integrand ${\cal O}_U^{\,\Bbb C}$-linear in $\partial_t\nabla^T$ and hence in a final form.

In summary,

\begin{ssproposition} {\bf [first variation of gauge/Yang-Mills term]}$\;$
 Let $(\varphi_T,\nabla^T)$ be a $(\ast_2)$-admissible family of $(\ast_2)$-admissible pairs.
 Then
  {\small
 \begin{eqnarray*}
  \lefteqn{
    \mbox{\Large $\frac{d}{dt}$}\,
     S^{(h;B)}_{\gaugetiny/\YMtiny}(\varphi_T, \nabla^T)^{\Bbb C}\;\;
	 =\;\;  -\,  \mbox{\Large $\frac{1}{2}$}\,
	                 \mbox{\Large $\frac{d}{dt}$} \int_U   \Tr
	                \| 2\pi\alpha^{\prime}F_{\nabla^T}+ \varphi_T^{\diamond}B\|_h^2 \, \vol_h    }\\
  & = &  -\,4\pi\alpha^{\prime}
               \int_{\partial U}	i_{
			                  \xi^T_{(\mathrm{III}, \partial_t\nabla^T)} }	\vol_h\,
             -\, 2\, \int_{\partial U} i_{\xi^T_{(\mathrm{III},\partial_t\varphi_T)}} \vol_h    \\
   &&  - \; 4\pi\alpha^{\prime}  \int_U \Tr \sum_{\nu}
			            (\partial_t\nabla^T)(e_{\nu})
						   \cdot
						   \Big(\;
						     (2\pi\alpha^{\prime}F_{\nabla^T}+ \varphi_T^{\diamond}B)
							 (\mbox{$\sum$}_{\mu}\nabla^h_{e_{\mu}}e_{\mu}, e_{\nu}) \\
   && \hspace{14em}												
						    -\, \sum_{\mu}
						           D^T_{e_{\mu}}
							        \big((2\pi\alpha^{\prime}F_{\nabla^T}+ \varphi_T^{\diamond}B)
									             (e_{\mu}, e_{\nu})\big)    \\
   && \hspace{14em}
                 -\, \mbox{\large $\frac{1}{2}$} \sum_{\mu,\lambda}
				       e^{\nu}([e_{\mu}, e_{\lambda}  ])
					 (2\pi\alpha^{\prime}F_{\nabla^T}+ \varphi_T^{\diamond}B)
					                                                                                        (e_{\mu}, e_{\lambda})
						  \;\Big) \vol_h  \,.					\\					
   &&
     -\, \int_U   \Tr
                     \sum_{\mu, \nu}
			   \Big(	
			      \sum_{i,j}\varphi_T^{\sharp}(B_{ij})
			                             \big[(\partial_t\nabla^T)(e_{\mu}), \varphi_T^{\sharp}(y^i)\big]
                                                   D^T_{e_{\nu}} \varphi_T^{\sharp}(y^j)	   \\[-1ex]
     && \hspace{12em}				
			  +\; \sum_{i,j}\varphi_T^{\sharp}(B_{ij})
			          D^T_{e_{\mu}}\varphi_T^{\sharp}(y^i)
                       \big[ (\partial_t\nabla^T)(e_{\nu}), \varphi_T^{\sharp}(y^j)\big]
                         \Big)								   \\
   && \hspace{8em}					   					   					
                       \cdot									
					   \Big((2\pi\alpha^{\prime}F_{\nabla^T}
					                  + \varphi_T^{\diamond}B)(e_{\mu}, e_{\nu}) \Big)
					   \vol_h             \\
   &&    -\, \int_U   \Tr
                     \sum_{\mu, \nu}
			      \sum_{i,j} \big((\partial_t\varphi_T)B_{ij} \big)
                     D^T_{e_{\mu}}\varphi_T^{\sharp}(y^i)
					 D^T_{e_{\nu}} \varphi_T^{\sharp}(y^j)   \\[-1ex]
   && \hspace{16em}					   					   					
                       \cdot									
					   \Big((2\pi\alpha^{\prime}F_{\nabla^T}
					                  + \varphi_T^{\diamond}B)(e_{\mu}, e_{\nu}) \Big)
					   \vol_h          						  \\	
   &&
     -\, 2\, \int_U   \Tr  \sum_{\nu}\sum_{i,j}
	                  \partial_t\varphi_T^{\sharp}(y^i)  \\[-1ex]
   &&  \hspace{6em}
          \bigg(\, \varphi_T^{\sharp}(B_{ij})\,	
					                     D^T_{e_{\nu}} \varphi_T^{\sharp}(y^j)	
                               \cdot									
					           \big((2\pi\alpha^{\prime}F_{\nabla^T}
					                        + \varphi_T^{\diamond}B)
											  (\mbox{$\sum$}_{\mu}
											     \nabla^h_{e_{\mu}}e_{\mu}, e_{\nu}) \big)	  \\[-1ex]
   && \hspace{8em}		
		    -\, \sum_{\mu} D^T_{e_{\mu}}
					    \Big(
                            \varphi_T^{\sharp}(B_{ij})\,	
					                     D^T_{e_{\nu}} \varphi_T^{\sharp}(y^j)	
                       \cdot									
					   \big((2\pi\alpha^{\prime}F_{\nabla^T}
					                  + \varphi_T^{\diamond}B)(e_{\mu}, e_{\nu}) \big)						
						\Big)  \bigg)  \vol_h                \,.					
 \end{eqnarray*}}
 Here,
  \begin{eqnarray*}
    \xi^T_{(\mathrm{III},\partial_t\nabla^T)}
	 & := & \sum_{\mu,\nu}
	               \Tr\Big(
   	 		       (\partial_t\nabla^T)(e_{\nu})
				    \cdot
					(2\pi\alpha^{\prime}F_{\nabla^T}
					                  + \varphi_T^{\diamond}B)(e_{\mu}, e_{\nu})
			        \Big)\, e_{\mu}\,, \\
    \xi^T_{(\mathrm{III}, \partial_t\varphi_T)}
      & := & \sum_{\mu}\Big(
	               \sum_{\nu}					 					
			         \sum_{i,j} \partial_t\varphi_T^{\sharp}(y^i)\,
										 \varphi_T^{\sharp}(B_{ij})\,	
					                     D^T_{e_{\nu}} \varphi_T^{\sharp}(y^j)	
                       \cdot									
					   \big((2\pi\alpha^{\prime}F_{\nabla^T}
					                  + \varphi_T^{\diamond}B)(e_{\mu}, e_{\nu}) \big) 	\Big) e_{\mu}
  \end{eqnarray*}
   in ${\cal T}_{\ast}(U_T/T)^{\Bbb C}$,
   with the first ${\cal O}_U^{\,\Bbb C}$-linear in $\partial_t\nabla^T$  and
          the second ${\cal O}_U^{\,\Bbb C}$-linear in $\partial_t\varphi_T$.
\end{ssproposition}

\bigskip

\subsubsection{The first variation of the Chern-Simons/Wess-Zumino term for lower dimensional D-branes}

This is an update of [L-Y8: Sec.6.2] (D(13.1)) in the current setting.
Let $\varphi_T:(X^{\!A\!z},{\cal E}_T;\nabla^T)\rightarrow Y$
 be an $(\ast_2)$-family of $(\ast_2)$-admissible maps.
We work out the first variation of the Chern-Simons/Wess-Zumino term
 $S_{\CSWZ}^{(C,B)}(\varphi,\nabla)$ for the cases where $m:= \dimm X=0,1,2,3$.
As the details involve no identities or techniques that have not yet been used in
 Sec.$\,$6.1, Sec.$\,$6.2, and/or Sec.$\,$6.3.1,
 we only summarize the final results below.

\vspace{4em}
\bigskip
\bigskip

\begin{flushleft}
{\bf 6.3.2.1\hspace{1.1em} D$(-1)$-brane world-point $(m=0)$}
\end{flushleft}
For a D$(-1)$-brane world-point,
 $\dimm X = 0$, $\nabla=0$, and
 $S_{\CSWZ}^{(C_{(0)})}(\varphi_T)
    = T_{-1}\,\cdot\, \Tr(\varphi_T^{\sharp}(C_{(0)}))$.
It follows that
{\small
 $$
   \mbox{\Large $\frac{d}{dt}$}
	 S_{\tinyCSWZ}^{(C_{(0)})}(\varphi_T)\;\;
     =\;\; 	
	 T_{-1}\,
	    \Tr \partial_t(\varphi_T^{\sharp}(C_{(0)}))    \;\;
	 =\;\;
	  T_{-1}\,	
	    \Tr\big((\partial_t\varphi_T)C_{(0)}\big)\,.
  $$}

\bigskip
\bigskip

\begin{flushleft}
{\bf 6.3.2.2\hspace{1.1em} D-particle world-line $(m=1)$}
\end{flushleft}
For a D-particle world-line,  $\dimm X=1$.
Let $e_1$ be the orthonormal frame on an open set  $U\subset X$; $\;e^1$ its dual co-frame.
Then, over $U$,
 {\small
  $$
    S_{\tinyCSWZ}^{(C_{(1)})}(\varphi_T)^{\Bbb C}\;\;
	 =\;\;	
  	 T_0 \int_U \Tr  \varphi_T^{\diamond}C_{(1)}\;\;	
	 =\;\;
  	 T_0\int_U\Tr
	      \Big(
		     \sum_{i=1}^n \varphi_T^{\sharp}(C_i) \cdot D^T_{e_1}\varphi_T^{\sharp}(y^i)
			 \Big)\, e^1\,.
  $$}
It follows that
{\small
 \begin{eqnarray*}
  \mbox{\Large $\frac{d}{dt}$}S_{\tinyCSWZ}^{(C_{(1)})}(\varphi)
   & = &    T_0 \Big( \Tr \sum_i
	                         \partial_t\varphi_T^{\sharp}(y^i) \varphi_T^{\sharp}(C_i)\Big)
							 \big|_{\partial U}  \\
   &&  -\; T_0 \int_U \Tr \Big(\sum_i
                               \partial_t\varphi_T^{\sharp}(y^i)
                               D^T_{e_1}\varphi_T^{\sharp}(C_i)\Big) e^1\;
          +\;  T_0 \int_U \Tr\Big(\sum_i
		                       D_{e_1}\varphi_T^{\sharp}(y^i)
		                       \cdot (\partial_t\varphi_T)C_i \Big) e^1\,.
\end{eqnarray*}}

\bigskip
\bigskip

\begin{flushleft}
{\bf 6.3.2.3\hspace{1.1em} D-string world-sheet $(m=2)$}
\end{flushleft}
Denote
 $$
  \breve{C}_{(2)}\;
   :=\;  C_{(2)}+C_{(0)}B\;
   =\; \sum_{ij}(C_{ij}+C_{(0)}B_{ij})\, dy^i\otimes dy^j\;
   =\; \sum_{i,j}\breve{C}_{ij}dy^i\otimes dy^j
 $$
  in a local coordinate $(y^1,\,\cdots\,,\, y^n)$ of $Y$.
For a D-string world-sheet, $\dimm X=2$.
Let $(e_1, e_2)$ be an orthonormal frame on an open set $U\subset X$; $\; (e^1,e^2)$ its dual co-frame.
Then, over $U$,
{\small
  \begin{eqnarray*}
    \lefteqn{S_{\tinyCSWZ}^{(C_{(0)},C_{(2)},B)}(\varphi_T,\nabla^T)^{\Bbb C} }\\
	 && =\;\;
     	 T_1\int_U
		      \Tr   \Big(
		      \sum_{i,j=1}^n\,
			      \varphi_T^{\sharp}(\breve{C}_{ij})\,
				    D^T_{e_1}\varphi_T^{\sharp}(y^i)\,D^T_{e_2}\varphi_T^{\sharp}(y^j)\,  \\[-1.6ex]
     && \hspace{10em}					
                +\, \pi\alpha^{\prime}
				         \varphi_T^{\sharp}(C_{(0)})\,F_{\nabla^T}(e_1,e_2)\,
			    +\, \pi\alpha^{\prime}
					     F_{\nabla^T}(e_1,e_2)\, \varphi_T^{\sharp}(C_{(0)})		
			          \Big)\,
		      e^1\wedge e^2  \\
     && =\;\;
   	   T_1\int_U
		      \Tr   \Big(
		      \sum_{i,j=1}^n\,
			      \varphi_T^{\sharp}(\breve{C}_{ij})\,
				    D^T_{e_1}\varphi_T^{\sharp}(y^i)\,D^T_{e_2}\varphi_T^{\sharp}(y^j)\,
                +\, 2\, \pi\alpha^{\prime}
				         \varphi_T^{\sharp}(C_{(0)})\,F_{\nabla^T}(e_1,e_2)
			          \Big)\,
		      e^1\wedge e^2\,.
  \end{eqnarray*}}
It follows that
{\footnotesize
\begin{eqnarray*}
 \lefteqn{
  \mbox{\large $\frac{d}{dt}$}
   S_{\tinyCSWZ}^{(C_{(0)}, C_{(2)}, B)}(\varphi_T,\nabla^T)^{\Bbb C}   }\\
  && =\;\;
   	 T_1\int_U
		      \Tr\,   \partial_t \Big(
		      \sum_{i,j=1}^n\,
			      \varphi_T^{\sharp}(\breve{C}_{ij})\,
				    D^T_{e_1}\varphi_T^{\sharp}(y^i)\,D^T_{e_2}\varphi_T^{\sharp}(y^j)\,
                +\, 2\, \pi\alpha^{\prime}
				         \varphi_T^{\sharp}(C_{(0)})\,F_{\nabla^T}(e_1,e_2)
			          \Big)\,
		      e^1\wedge e^2           \\
  && =\;\;   T_1 \int_{\partial U}
                            i_{\xi^T_{(\mathrm{IV},\partial_t\varphi_T)}} (e^1\wedge e^2)\;
                   +\; 2\pi\alpha^{\prime} T_1 \int_{\partial U}
				            i_{\xi^T_{(\mathrm{IV},\partial_t\nabla^T)}} (e^1\wedge e^2)   \\
   && \hspace{1.6em}
     +\;  T_1 \int_U  \Tr \bigg(
	          \sum_{i,j=1}^n
			     \partial_t\varphi^{\sharp}(y^i)
				  \Big(
				     D^T_{e_2}\varphi_T^{\sharp}(y^j)
					 \varphi_T^{\sharp}(\breve{C}_{ij}) e^1\,
					 -\, D^T_{e_1}\varphi_T^{\sharp}(y^j)
					      \varphi_T^{\sharp}(\breve{C}_{ij})e^2
				  \Big)(\nabla^h_{e_1}e_1+\nabla^h_{e_2}e_2)   \bigg)\,e^1\wedge e^2  \\
   && \hspace{4em}
     -\; T_1 \int \int_U \Tr \bigg(  \sum_{i,j}   \partial_t\varphi_T^{\sharp}(y^i)
	                  \Big(  D^T_{e_1}
					                  \big(D^T_{e_2}\varphi_T^{\sharp}(y^j)
						                          \cdot \varphi_T^{\sharp}(\breve{C}_{ij})\big)\,
					               -\, D^T_{e_2}
                                         \big(D^T_{e_1}\varphi_T^{\sharp}(y^j)
										            \cdot \varphi_T^{\sharp}(\breve{C}_{ij})
										 \big)								
					  \Big)  \bigg)\, e^1\wedge e^2	 \\
   && \hspace{4em}				
      +\;   T_1 \int_U \Tr \Big(   \sum_{i,j}
	                           \partial_t\varphi_T^{\sharp}(\breve{C}_{ij})
							   D^T_{e_1}\varphi_T^{\sharp}(y^i)
							   D^T_{e_2}\varphi_T^{\sharp}(y^j) 	
	                                      \Big)\, e^1\wedge e^2    \\	
   && \hspace{4em}
      +\; 2\pi\alpha^{\prime}\, T_1 \int_U \Tr \Big(
	             \partial_t\varphi_T^{\sharp}(C_{(0)})\cdot F_{\nabla^T}(e_1,e_2)
	                  \Big)\,e^1\wedge e^2    \\
   && \hspace{1.6em}	
      +\; 2\pi\alpha^{\prime}\, T_1 \int_U \Tr   \bigg(
               \varphi_T^{\sharp}(C_{(0)})\,
			      \Big( \big((\partial_t\nabla^T)(e_2)\,  e^1 \,-\, (\partial_t\nabla^T )(e_1)\,e^2  \big)
				            (\nabla^h_{e_1}e_1 + \nabla^h_{e_2} e_2) \,
					        - \, (\partial_t\nabla^T)([e_1,e_2])
					\Big)  \bigg)\, e^1\wedge e^2    \\	  	
  && \hspace{4em}
      -\; 2\pi\alpha^{\prime}\, T_1 \int_U \Tr \Big(
	              D^T_{e_1}\varphi_T^{\sharp}(C_{(0)})
				  \cdot (\partial_t\nabla^T)(e_2)\,
                 -\, D^T_{e_2}\varphi_T^{\sharp}(C_{(0)})
                      \cdot (\partial_t\nabla^T)(e_1)				   \Big)\, e^1\wedge e^2\,.
\end{eqnarray*}}
Here,
 {\small
 \begin{eqnarray*}
    \xi^T_{(\mathrm{IV},\partial_t\varphi_T)}
	  & := &  \Tr \big(\mbox{$\sum$}_{i,j}
	                        \partial_t\varphi_T^{\sharp}(y^i)
							D^T_{e_2}\varphi_T^{\sharp}(y^j)
							\varphi_T^{\sharp}(\breve{C}_{ij})
                        	  \big)\, e_1\;
                    - \; \Tr \big( \mbox{$\sum$}_{i,j}
					       \partial_t\varphi_T^{\sharp}(y^i)
						   D^T_{e_1}\varphi_T^{\sharp}(y^j)
						   \varphi_T^{\sharp}(\breve{C}_{ij})
					           \big)\, e_2\,,	           \\
    \xi^T_{(\mathrm{IV},\partial_t\nabla^T)}
	  & := &  \Tr \big( \varphi_T^{\sharp}(C_{(0)})
	                                      \cdot (\partial_t\nabla^T)(e_2)   \big)\, e_1\;
	              -\; \Tr \big(  \varphi_T^{\sharp}(C_{(0)})
				                          \cdot (\partial_t\nabla^T)(e_1)   \big)\, e_2
 \end{eqnarray*}}
 in ${\cal T}_{\ast}(U_T/T)^{\Bbb C}$,
  with the first ${\cal O}_U^{\,\Bbb C}$-linear in $\partial_t\varphi_T$  and
         the second ${\cal O}_U^{\,\Bbb C}$-linear in $\partial_t\nabla^T$.

\bigskip
\bigskip

\begin{flushleft}
{\bf 6.3.2.4\hspace{1.1em} D-membrane world-volume $(m=3)$}
\end{flushleft}
Denote
 \begin{eqnarray*}
  \breve{C}_{(3)}
   & :=\:   &  C_{(3)}+C_{(1)}\wedge B  \\
   & =
      & \sum_{i,j,k}
	       (C_{ijk}+C_iB_{jk}+C_jB_{ki}+C_kB_{ij})\, dy^i\otimes dy^j\otimes dy^k\;\;
       =\;\; \sum_{i,j,k}\breve{C}_{ijk}dy^i\otimes dy^j\otimes dy^k
 \end{eqnarray*}
 in a local coordinate $(y^1,\,\cdots\,,\, y^n)$ of $Y$.
For D-membrane world-volume, $\dimm X=3$.
Let $(e_1,e_2,e_3)$ be an orthonormal frame on an open set $U\subset X$;
 $\;(e^1,e^2,e^3)$ its dual co-frame.
Then, over $U$,
{\small
 \begin{eqnarray*}
  \lefteqn{
  S_{\tinyCSWZ}^{(C_{(1)}, C_{(3)},B)}(\varphi_T,\nabla^T)^{\Bbb C}    } \\
   &&  =\;\;
	   T_2\int_U
		      \Tr  \Big(
			     \sum_{i,j,k=1}^n
				    \varphi^{\sharp}(\breve{C}_{ijk})\,
					   D^T_{e_1}\varphi_T^{\sharp}(y^i)\,
					   D^T_{e_2}\varphi_T^{\sharp}(y^j)\,
					   D^T_{e_3}\varphi_T^{\sharp}(y^k) \\[-2ex]
     && \hspace{7em}					
		      +\, 2\pi\alpha^{\prime}
			        \sum_{(\lambda\mu\nu)\in\scriptsizeSym_3}
			          \sum_{i=1}^n\,
			           (-1)^{(\lambda\mu\nu)}
					      \big(
						     \varphi_T^{\sharp}(C_i)\,
							 D^T_{\lambda}\varphi_T^{\sharp}(y^i)\,
							 F_{\nabla^T}(e_{\mu}, e_{\nu})\,							   							
                             \big)					
			   \Big)\,
		      e^1\wedge e^2\wedge e^3
 \end{eqnarray*}}
It follows that

{\footnotesize
\begin{eqnarray*}
 \lefteqn{
   \mbox{\large $\frac{d}{dt}$}\,
    S_{\tinyCSWZ}^{(C_{(1)}, C_{(3)}, B)}(\varphi_T,\nabla^T)^{\Bbb C}  }\\
  &&  =\;\;
	   T_2\int_U
		      \Tr\,  \partial_t \Big(
			     \sum_{i,j,k=1}^n
				    \varphi^{\sharp}(\breve{C}_{ijk})\,
					   D^T_{e_1}\varphi_T^{\sharp}(y^i)\,
					   D^T_{e_2}\varphi_T^{\sharp}(y^j)\,
					   D^T_{e_3}\varphi_T^{\sharp}(y^k) \\[-2ex]
   && \hspace{7em}					
		      +\, 2\pi\alpha^{\prime}
			        \sum_{(\lambda\mu\nu)\in\scriptsizeSym_3}
			          \sum_{i=1}^n\,
			           (-1)^{(\lambda\mu\nu)}
					      \big(
						     \varphi_T^{\sharp}(C_i)\,
							 D^T_{e_{\lambda}}\varphi_T^{\sharp}(y^i)\,
							 F_{\nabla^T}(e_{\mu}, e_{\nu})\,							   							
                             \big)					
			   \Big)\,
		      e^1\wedge e^2\wedge e^3   \\
  && =\;\; T_2 \int_{\partial U}
           i_{\xi^T_{(\mathrm{IV}, \partial_t\varphi_T; \breve{C}_{(3)})}}
		        (e^1\wedge e^2\wedge e^3)\;
		+\;   4\pi\alpha^{\prime}\, T_2 \int_{\partial U}
		           i_{\xi^T_{(\mathrm{IV},\partial_t\varphi_T;C_{(1)})}}
                   (e^1\wedge e^2\wedge e^3)\;    \\
  &&	 \hspace{8em}			
         +\;  2\pi\alpha^{\prime}\,T_2 \int_{\partial U}
                    i_{\xi^T_{(\mathrm{IV,\partial_t\nabla^T})}} (e^1\wedge e^2 \wedge e^3)\\
  && \hspace{1.6em}					
     +\; T_2 \int_U \bigg( 	
	         \Big(\Tr \sum_{i,j,k}
	                  \varphi_T^{\sharp}(\breve{C}_{ijk})
					  \partial_t\varphi_T^{\sharp}(y^i)
					  D^T_{e_2}\varphi_T^{\sharp}(y^j)
					  D^T_{e_3}\varphi_T^{\sharp}(y^k)  \Big)\, e^1  \\
     && \hspace{7em}					
	     -\; \Big(\Tr \sum_{i,j,k}
		               \varphi_T^{\sharp}(\breve{C}_{ijk})
		               \partial_t\varphi_T^{\sharp}(y^i)
					   D^T_{e_1}\varphi_T^{\sharp}(y^j)
					   D^T_{e_3}\varphi_T^{\sharp}(y^k)   \Big)\, e^2  \\
     && \hspace{7em}					
         -\; \Big( \Tr \sum_{i,j,k}
		               \varphi_T^{\sharp}(\breve{C}_{ijk})
					   \partial_t\varphi_T^{\sharp}(y^i)
					   D^T_{e_2}\varphi_T^{\sharp}(y^j)
					   D^T_{e_1}\varphi_T^{\sharp}(y^k)
		               \Big)\, e^3		
	       \bigg)\,(\mbox{$\sum$}_{\mu=1}^3\nabla^h_{e_{\mu}}e_{\mu} )\,
            e^1\wedge e^2\wedge e^3		   \\	
   && \hspace{2.6em}
         -\; T_2 \int_U \Tr \sum_{i,j,k}\partial_t\varphi_T^{\sharp}(y^i)
		       \Big(
			      D^T_{e_1}\big( \varphi_T^{\sharp}(\breve{C}_{ijk})
                                                       D^T_{e_2} \varphi_T^{\sharp}(y^j)
													   D^T_{e_3} \varphi_T^{\sharp}(y^k) \big)\,
				  -\, D^T_{e_2} \big(\varphi_T^{\sharp}(\breve{C}_{ijk})
                                                           D^T_{e_1}\varphi_T^{\sharp}(y^j)
														   D^T_{e_3}\varphi_T^{\sharp}(y^k) \big)  \\
     && \hspace{24em}		
				  -\, D^T_{e_3} \big(\varphi_T^{\sharp}(\breve{C}_{ijk})
                                                           D^T_{e_2}\varphi_T^{\sharp}(y^j)
                                                           D^T_{e_1}\varphi_T^{\sharp}(y^k) \big)
			   \Big)\, e^1\wedge e^2\wedge e^3     \\  			
  && \hspace{2.6em}			
        +\; T_2 \int_U \Tr \sum_{i,j,k}
		        \partial_t\varphi_T^{\sharp}(\breve{C}_{ijk})
             	D^T_{e_1}\varphi_T^{\sharp}(y^i)
				D^T_{e_2}\varphi_T^{\sharp}(y^j)
				D^T_{e_3}\varphi_T^{\sharp}(y^k)  \,   e^1\wedge e^2 \wedge e^3\\
  && \hspace{2.6em}
     +\; 4\pi\alpha^{\prime}\,T_2 \int_U
	        \bigg(
			  \Big( \Tr \sum_i
	                       \varphi_T^{\sharp}(C_i)  \partial_t\varphi_T^{\sharp}(y^i)
						   F_{\nabla^T}(e_2,e_3)  \Big)\, e^1     \\
      && \hspace{11em}	
           -\; \Big(\Tr \sum_i
		                   \varphi_T^{\sharp}(C_i)  \partial_t\varphi_T^{\sharp}(y^i)
						   F_{\nabla^T}(e_1,e_3)  \Big)\,e^2      \\
      && \hspace{11em}						
		  +\; \Big(\Tr \sum_i
		                   \varphi_T^{\sharp}(C_i) \partial_t\varphi_T^{\sharp}(y^i)
						   F_{\nabla^T}(e_1,e_2)   \Big)\,e^3					
	        \bigg)\,(\mbox{$\sum$}_{\mu=1}^3\nabla^h_{e_{\mu}}e_{\mu} )\,
            e^1\wedge e^2\wedge e^3		   \\	
  && \hspace{2.6em}
        -\; 4\pi\alpha^{\prime}\, T_2 \int_U  \Tr
		          \sum_i  \partial_t\varphi_T^{\sharp}(y^i)
				    \Big(
					 D^T_{e_1}\big( \varphi_T^{\sharp}(C_i) F_{\nabla^T}(e_2,e_3) \big)\,
					  -\, D^T_{e_2} \big( \varphi_T^{\sharp}(C_i) F_{\nabla^T}(e_1,e_3) \big) \\
     && \hspace{27em}					
					  +\, D^T_{e_3}\big( \varphi_T^{\sharp}(C_i) F_{\nabla^T}(e_1,e_2) \big)
					\Big)\, e^1\wedge e^2 \wedge e^3   \\
  && \hspace{2.6em}
     +\; 2\pi\alpha^{\prime}\, T_2 \int_U \Tr
	            \sum_{(\lambda\mu\nu)\in\scriptsizeSym_3}
			          \sum_i\,
			           (-1)^{(\lambda\mu\nu)}
					         \partial_t\varphi_T^{\sharp}(C_i)\,
							 D^T_{e_{\lambda}}\varphi_T^{\sharp}(y^i)\,
							 F_{\nabla^T}(e_{\mu}, e_{\nu})\,							   							
		      e^1\wedge e^2\wedge e^3   \\
  && \hspace{1.6em}
          +\; 2\pi\alpha^{\prime}\, T_2 \int_U
		        \bigg(
				   \Big(\Tr \sum_i
	                 \varphi_T^{\sharp}(C_i)
			         \Big(D^T_{e_3}\varphi_T^{\sharp}(y^i)   (\partial_t\nabla^T)(e_2)\,
					        	-\, D^T_{e_2}\varphi_T^{\sharp}(y^i) (\partial_t\nabla^T) (e_3)
	                   \Big)\Big)\, e^1   \\
      && \hspace{9em}			
	      +\; \Big(\Tr \sum_i \varphi_T^{\sharp}(C_i)
		            \Big(D^T_{e_3}\varphi_T^{\sharp}(y^i) (\partial_t\nabla^T)(e_1)\,
				    +\, D^T_{e_1}\varphi_T^{\sharp}(y^i) (\partial_t\nabla^T)(e_3)
		             \Big) \Big)\, e^2   \\
      && \hspace{9em}					
		  +\; \Big( \Tr \sum_i \varphi_T^{\sharp}(C_i)
		           \Big(D^T_{e_1}\varphi_T^{\sharp}(y^i) (\partial_t\nabla^T)(e_2)\,
				      -\, D^T_{e_2}\varphi_T^{\sharp}(y^i) (\partial_t\nabla^T)(e_1)
		             \Big) \Big)\,e^3						
                \bigg)\,(\mbox{$\sum$}_{\mu=1}^3\nabla^h_{e_{\mu}}e_{\mu} )\,  \\
      && \hspace{36em}                e^1\wedge e^2\wedge e^3		  \\
  && \hspace{2.6em}
   +\;  2\pi\alpha^{\prime}\,T_2 \int_U \Tr
           \sum_{(\lambda\mu\nu)\in\scriptsizeSym_3}
			          \sum_i\,
			           (-1)^{(\lambda\mu\nu)}
					      \Big(
						     \varphi_T^{\sharp}(C_i)\,
							  \big[(\partial_t\nabla^T)(e_{\lambda}), \varphi_T^{\sharp}(y^i)\big]\,
							  F_{\nabla^T}(e_{\mu}, e_{\nu})\,   \\
     && \hspace{24em}							
						   -\,   \varphi_T^{\sharp}(C_i)\,
							     D^T_{e_{\lambda}}\varphi_T^{\sharp}(y^i)\,
							     (\partial_t\varphi_T)([e_{\mu}, e_{\nu}])
                             \Big)\, 		e^1\wedge e^2\wedge e^3\,.
\end{eqnarray*}}
Here,
{\footnotesize
 \begin{eqnarray*}
  \xi^T_{(\mathrm{IV},\partial_t\varphi_T;\breve{C}_{(3)})}
   & :=
     & \Big(\Tr \sum_{i,j,k}
	                  \varphi_T^{\sharp}(\breve{C}_{ijk})
					  \partial_t\varphi_T^{\sharp}(y^i)
					  D^T_{e_2}\varphi_T^{\sharp}(y^j)
					  D^T_{e_3}\varphi_T^{\sharp}(y^k)  \Big)\, e_1\; \\
    && \hspace{4em}					
	     -\; \Big(\Tr \sum_{i,j,k}
		               \varphi_T^{\sharp}(\breve{C}_{ijk})
		               \partial_t\varphi_T^{\sharp}(y^i)
					   D^T_{e_1}\varphi_T^{\sharp}(y^j)
					   D^T_{e_3}\varphi_T^{\sharp}(y^k)   \Big)\, e_2\;  \\
     && \hspace{4em}					
         -\; \Big( \Tr \sum_{i,j,k}
		               \varphi_T^{\sharp}(\breve{C}_{ijk})
					   \partial_t\varphi_T^{\sharp}(y^i)
					   D^T_{e_2}\varphi_T^{\sharp}(y^j)
					   D^T_{e_1}\varphi_T^{\sharp}(y^k)
		               \Big)\, e_3\;, \\
   \xi^T_{(\mathrm{IV},\partial_t\varphi_T;C_{(1)})}					
     & :=
	   & \Big( \Tr \sum_i
	                       \varphi_T^{\sharp}(C_i)  \partial_t\varphi_T^{\sharp}(y^i)
						   F_{\nabla^T}(e_2,e_3)  \Big)\, e_1\;   \\
    && \hspace{1em}	
           -\; \Big(\Tr \sum_i
		                   \varphi_T^{\sharp}(C_i)  \partial_t\varphi_T^{\sharp}(y^i)
						   F_{\nabla^T}(e_1,e_3)  \Big)\,e_2\;
		  +\; \Big(\Tr \sum_i
		                   \varphi_T^{\sharp}(C_i) \partial_t\varphi_T^{\sharp}(y^i)
						   F_{\nabla^T}(e_1,e_2)   \Big)\,e_3\;,    \\
    \xi^T_{(\mathrm{IV},\partial_t\nabla^T)}
	 & :=
	  &  \Big(\Tr \sum_i
	         \varphi_T^{\sharp}(C_i)
			 \Big(D^T_{e_3}\varphi_T^{\sharp}(y^i)   (\partial_t\nabla^T)(e_2)\,
						-\, D^T_{e_2}\varphi_T^{\sharp}(y^i) (\partial_t\nabla^T) (e_3)
	           \Big)\Big)\, e_1\;    \\
      && \hspace{4em}			
	      +\; \Big(\Tr \sum_i \varphi_T^{\sharp}(C_i)
		            \Big(D^T_{e_3}\varphi_T^{\sharp}(y^i) (\partial_t\nabla^T)(e_1)\,
				    +\, D^T_{e_1}\varphi_T^{\sharp}(y^i) (\partial_t\nabla^T)(e_3)
		             \Big) \Big)\, e_2\;   \\
      && \hspace{4em}					
		  +\; \Big( \Tr \sum_i \varphi_T^{\sharp}(C_i)
		           \Big(D^T_{e_1}\varphi_T^{\sharp}(y^i) (\partial_t\nabla^T)(e_2)\,
				      -\, D^T_{e_2}\varphi_T^{\sharp}(y^i) (\partial_t\nabla^T)(e_1)
		             \Big) \Big)\,e_3
 \end{eqnarray*}}
in ${\cal T}_{\ast}(U_T/T)^{\Bbb C}$,
 with the first two ${\cal O}_U^{\,\Bbb C}$-linear in $\partial_t\varphi_T$ and
        the third ${\cal O}_U^{\,\Bbb C}$-linear in $\partial_t\nabla^T$.

\bigskip

\section{The second variation of the enhanced kinetic term for maps}

Let
 $T=(-\varepsilon,\varepsilon)^2 \subset {\Bbb R}^2$, with coordinate $(s,t)$, and
 $(\varphi_T,\nabla^T)$ be an $(\ast_2)$-admissible family of $(\ast_1)$-admissible pairs,
   with $(\varphi_{(0,0)},\nabla_{(0,0)})=(\varphi,\nabla)$.
Assume further that
 $$
  D_{\xi} \partial_s{\cal A}_{\varphi_T}\; \subset\; \Comm({\cal A}_{\varphi_T})
    \hspace{2em}\mbox{for all $\xi\in {\cal T}_{\ast}(X_T/T)$}\,.
 $$
We work out in this section the second variation formula of
 the enhanced kinetic term\\
 $S_{\mapscriptsize:\kineticscriptsize^+}^{(\rho, h;\Phi,g)}(\varphi,\nabla)$
 in the standard action $S_{\standardscriptsize}^{(\rho, h;\Phi,g,B,C)}(\varphi, \nabla)$.

\bigskip

\subsection{The second variation of the kinetic term for maps}
	
Recall
 $$
  E^{\nabla^T}\!(\varphi_T)\;
    :=\; S_{\mapscriptsize:\kineticscriptsize}^{(h;g)}(\varphi_T,\nabla^T)\;
    :=\; \frac{1}{2}\, T_{m-1} \int_X \Tr
	        \big\langle D^T\varphi_T\,,\, D^T\varphi_T  \big\rangle_{(h,g)}\, \vol_h\,,
 $$
 with the understanding that all expressions are taken on $X_{(s,t)}$ with $(s,t)$ varying in $T$.

Let $U\subset X$ be an open set
  with an orthonormal frame $(e_{\mu})_{\mu=1,\,\cdots\,,\, m}$.
Let $(e^{\mu})_{\mu=1,\,\cdots\,,\,m}$ be the dual co-frame.
Assume that $U$ is small enough
 so that $\varphi_T(U_T^{A\!z})$ is contained in a coordinate chart of $Y$,
   with coordinates $(y^1,\,\cdots\,,\, y^n)$.
Then, as in Sec.$\,$6.1, over $U$,
 {\small
 \begin{eqnarray*}
   \frac{\partial}{\partial t} E^{\nabla^T}\!(\varphi_T)
   & =
     & T_{m-1}\int_U \Tr  \sum_{\mu=1}^m
	               \Big\langle  \nabla^{T,(\varphi_T,g)}_{e_{\mu}} \partial_t \varphi_T\,,\,
         		          D^T_{e_{\mu}}\varphi_T  \Big\rangle_g\,\vol_h   \\
   && \hspace{4em}
         +\;   T_{m-1} \int_U \Tr \sum_{\mu=1}^m
		          \Big\langle
			     (\ad\otimes\!\nabla^g)_{D^T_{e_{\mu}}\varphi_T}\partial_t\varphi_T\,,\,
				         D^T_{e_{\mu}}\varphi_T  \Big\rangle_g\,   \vol_h \\[.6ex]
   && \hspace{8em}
          +\;  T_{m-1} \int_U \sum_{\mu=1}^m
		           \Big\langle
				       \sum_{i=1}^n
					       \big[(\partial_t\nabla^T)(e_{\mu}), \varphi_T^{\sharp}(y^i)\big]
					         \otimes \mbox{$\frac{\partial}{\partial y^i}$}\,,\,
				       D^T_{e_{\mu}} \varphi_T   \Big\rangle_g\, \vol_h    \\[.6ex]
   & \:=:  & (\mathrm{I}^{\,2}.1) \;+\; (\mathrm{I}^{\,2}.2)\;+\; (\mathrm{I}^{\,2}.3)\,;
 \end{eqnarray*}}
and
$$
  \frac{\partial^2}{\partial s\,\partial t}\, E^{\nabla^T}\!(\varphi_T)  \;\;
   =\;\;  \frac{\partial}{\partial s}\,(\mathrm{I}^{\,2}.1) \;
             +\; \frac{\partial}{\partial s}\,(\mathrm{I}^{\,2}.2)\;
			 +\; \frac{\partial}{\partial s}\,(\mathrm{I}^{\,2}.3)\,.
$$
Which we now compute term by term.

\bigskip

\begin{flushleft}
{{\bf The term} $\,\frac{\partial}{\partial s}\,(\mathrm{I}^{\,2}.1) $}
\end{flushleft}
\begin{eqnarray*}
  \frac{\partial}{\partial s}\,(\mathrm{I}^{\,2}.1)
  & =  &  T_{m-1}\,\frac{\partial}{\partial s}
                  \int_U \Tr  \sum_{\mu=1}^m
	                \Big\langle  \nabla^{T,(\varphi_T,g)}_{e_{\mu}} \partial_t \varphi_T\,,\,
         		          D^T_{e_{\mu}}\varphi_T  \Big\rangle_g\,\vol_h         \\
  & = &
   T_{m-1}\int_U \Tr \sum_{\mu}
        \partial_s \Big\langle
		   \nabla^{T,(\varphi_T,g)}_{e_{\mu}}\partial_t\varphi_T\,,\,
		            D^T_{e_{\mu}}\varphi_T    \Big\rangle_g\, \vol_h         \\
  & = &
    T_{m-1}\int_U\Tr\sum_{\mu}
      \Big\langle
         \partial_s \nabla^{T,(\varphi_T,g)}_{e_{\mu}}\partial_t \varphi_T\,,\,
	      D^T_{e_{\mu}}\varphi_T
      \Big\rangle_g\, \vol_h	   \\
 && \hspace{6em}	
	  +\;  T_{m-1}\int_U\Tr\sum_{\mu}
             \Big\langle
			   \nabla^{T,(\varphi_T,g)}_{e_{\mu}}\partial_t\varphi_T\,,\,
			       \partial_s D^T_{e_{\mu}}\varphi_T
			 \Big\rangle_g\, \vol_h    \\	  					
 &  = &
    (\mathrm{I}^{\,2}.1.1)\;+\; (\mathrm{I}^{\,2}.1.2)\,.
\end{eqnarray*}

\bigskip

\noindent
{\it $(a)\,$ Term $(\mathrm{\rm I}^{\,2}.1.1)$}
\begin{eqnarray*}
 (\mathrm{I}^{\,2}.1.1)
 & :=\:  &
   T_{m-1}\int_U\Tr\sum_{\mu}
      \Big\langle
         \partial_s \nabla^{T,(\varphi_T,g)}_{e_{\mu}}\partial_t \varphi_T\,,\,
	      D^T_{e_{\mu}}\varphi_T
      \Big\rangle_g\, \vol_h	   \\
 & = &
   T_{m-1}\int_U\Tr\sum_{\mu}
      \Big\langle
        \nabla^{T,(\varphi_T,g)}_{e_{\mu}}\partial_s \partial_t \varphi_T\,,\,
	      D^T_{e_{\mu}}\varphi_T
      \Big\rangle_g\, \vol_h	   \\
 &&	 \hspace{1em}
     +\;  T_{m-1}\int_U\Tr\sum_{\mu}
        \Big\langle
          F_{\nabla^{T,(\varphi_T,g)}}(\partial_s, e_{\mu})\, \partial_t \varphi_T\,,\,
	      D^T_{e_{\mu}}\varphi_T
        \Big\rangle_g\, \vol_h	   \\  	 		
 &  = &
    (\mathrm{I}^{\,2}.1.1.1)\;+\; (\mathrm{I}^{\,2}.1.1.2)\,.
\end{eqnarray*}

For Term $(\mathrm{I}^{\,2}.1.1.1)$, as in Sec.$\,$6.1 for Summand $(\mathrm{I}.1.1)$,
consider the $1$-form on $U_T/T$
 $$
   \alpha^T_{(\mathrm{I}^2, \partial_s\partial_t\varphi_T)}\;
    :=\; \Tr \big\langle \partial_s\partial_t\varphi_T\,,\, D^T\varphi_T\big\rangle_g
 $$
 and let
 $$
   \xi^T_{(\mathrm{I}^2,\partial_s\partial_t\varphi_T)}\;
    :=\; \sum_{\mu=1}^n \Tr
	         \big\langle \partial_s\partial_t\varphi_T\,,\,
			     D^T_{e_{\mu}}\varphi_T  \big\rangle_g\,   e_{\mu}
 $$
 be its dual on $U_T/T$ with respect to $h$.
Then,
 \begin{eqnarray*}
  (\mathrm{I}^{\,2}.1.1.1)
     & =  &   T_{m-1} \int_{\partial U}
	                   i_{\xi^T_{(\mbox{\tiny I}^2, \partial_s\partial_t\varphi_T)}} \vol_h \\
   && \hspace{2em}
     +\;\; T_{m-1} \int_U \Tr
	           \Big\langle
			     \partial_s\partial_t\varphi_T\,,\,
				   \big(
				      D^T_{\sum_{\mu}\nabla^h_{e_{\mu}}e_{\mu}}\,
					   -\, \mbox{$\sum_{\mu}$}
						   \nabla^{T,(\varphi_T,g)}_{e_{\mu}}D^T_{e_{\mu}}
				     \big)\varphi_T
			    \Big\rangle_g\, \vol_h\,.
 \end{eqnarray*}

For Term $(\mathrm{I}^{\,2}.1.1.2)$, recall Lemma~3.2.2.5.
Then,
 \begin{eqnarray*}
   (\mathrm{I}^{\,2}.1.1.2)
    & = &
	 -\; T_{m-1}\int_U\Tr
        \Big\langle
		  \partial_t\varphi_T\,,\,
		   \sum_{\mu}
           F_{\nabla^{T,(\varphi_T,g)}}(\partial_s, e_{\mu}) D^T_{e_{\mu}}\varphi_T
        \Big\rangle_g\, \vol_h	   \\  	 	
    && \hspace{4em}
	  +\; T_{m-1}\int_U\Tr\sum_{\mu}
          \big[
		   F_{\nabla}(\partial_s, e_{\mu})\,,\,
		    \langle  \partial_t\varphi_T , D^T_{e_{\mu}}\varphi_T    \rangle_g
		   \big]\, \vol_h	   \\  	
    & = &
	 -\; T_{m-1}\int_U\Tr
        \Big\langle
		  \partial_t\varphi_T\,,\,
		   \sum_{\mu}
           F_{\nabla^{T,(\varphi_T,g)}}(\partial_s, e_{\mu}) D^T_{e_{\mu}}\varphi_T
        \Big\rangle_g\, \vol_h	  \,.
 \end{eqnarray*}
Here,
{\small
	 \begin{eqnarray*}
	  \lefteqn{
	   F_{\nabla^{T,(\varphi_T,g)}}(\partial_s, e_{\mu})\,
	      D^T_{e_{\mu}}\varphi_T \;\;
	   =\;\;
        \big(\partial_s \nabla^{T,(\varphi_T,g)}_{e_{\mu}} \;
	      -\;  \nabla^{T,(\varphi_T, g)}_{e_{\mu}}\partial_s \big)\,
		    \mbox{\large $\sum$}_{i=1}^n
			   D_{e_{\mu}}\varphi_T^{\sharp}(y^i)
			      \otimes \mbox{\large $\frac{\partial}{\partial y^i}$}           }\\[.6ex]
       && 					
         =\,
		  \sum_{i=1}^n
		    [(\partial_s\nabla^T)(e_{\mu}), D_{e_{\mu}}\varphi_T^{\sharp}(y^i)]
			   \otimes \mbox{\large $\frac{\partial}{\partial y^i}$}\;\;
             +\;\; \sum_{i=1}^n
			          D_{e_{\mu}}\varphi_T^{\sharp}(y^i) \sum_{j=1}^n
			          [(\partial_s\nabla^T)(e_{\mu}), \varphi_T^{\sharp}(y^j)]
					    \otimes \nabla^g_{\frac{\partial}{\partial y^j}}
						                                                 \mbox{\large $\frac{\partial}{\partial y^i}$}  \\[-1.2ex]
		&& \hspace{1.6em}
		     +\;\; \sum_{i=1}^n D_{e_{\mu}}\varphi_T^{\sharp}(y^i)\,
			          \sum_{j,k=1}^n   \Big(
			                 D^T_{e_{\mu}}\varphi_T^{\sharp}(y^j)\,\partial_s\varphi_T^{\sharp}(y^k)
							    \otimes \nabla^g_{\frac{\partial}{\partial y^k}}
								               \nabla^g_{\frac{\partial}{\partial y^j}}\,
											      \mbox{\large $\frac{\partial}{\partial y^i}$}\:     \\[-1.2ex]
        && \hspace{20em}												
						 -\: \partial_s\varphi_T^{\sharp}(y^k)\,D^T_{e_{\mu}}\varphi_T^{\sharp}(y^j)
							    \otimes \nabla^g_{\frac{\partial}{\partial y^j}}
								               \nabla^g_{\frac{\partial}{\partial y^k}}\,
											   \mbox{\large $\frac{\partial}{\partial y^i}$}
			                                                      \Big)
    \end{eqnarray*}}
 explicitly.

\bigskip

\noindent
{\it $(b)\,$ Term $(\mathrm{I}^{\,2}.1.2)$}
\begin{eqnarray*}
(\mathrm{I}^{\,2}.1.2)
  & :=\:  &
    T_{m-1}\int_U\Tr\sum_{\mu}
             \Big\langle
			   \nabla^{T,(\varphi_T,g)}_{e_{\mu}}\partial_t\varphi_T\,,\,
			       \partial_s D^T_{e_{\mu}}\varphi_T
			 \Big\rangle_g\, \vol_h    \\	  	
  & = &
    T_{m-1}\int_U \Tr\sum_{\mu}
        \Big\langle
		  \nabla^{T,(\varphi_T,g)}_{e_{\mu}} \partial_t\varphi_T\,,\,
		      \nabla^{T,(\varphi_T,g)}_{e_{\mu}} \partial_s\varphi_T
        \Big\rangle_g\,  \vol_h \\		
  && + \;\;
         T_{m-1}\int_U\Tr\sum_{\mu}
          \Big\langle
		   \nabla^{T,(\varphi_T,g)}_{e_{\mu}} \partial_t\varphi_T\,,\,
		      (\ad\otimes\nabla^g)_{D^T_{e_{\mu}}\varphi_T}  \partial_s\varphi_T
        \Big\rangle_g\,  \vol_h \\		
  && +\;\;
        T_{m-1}\int_U\Tr\sum_{\mu}
          \Big\langle
		   \nabla^{T,(\varphi_T,g)}_{e_{\mu}} \partial_t\varphi_T\,,\,
		     \mbox{\large $\sum_{i=1}^n$}
			  \big[ (\partial_s\nabla^T)(e_{\mu}), \varphi_T^{\sharp}(y^i)
			   \big]
			  \otimes\mbox{\large $\frac{\partial}{\partial y^i}$}
        \Big\rangle_g\,  \vol_h\,.
\end{eqnarray*}
As in Sec.$\,$6.1, consider the $1$-forms on $U_T/T$,
 \begin{eqnarray*}
  \alpha^T_{(\mathrm{I}^2,\partial_t\varphi_T, \nabla^{T,{(\varphi_T,g)}})}
   &= & \Tr
              \big\langle
		        \partial_t\varphi_T\,,\,
		            \nabla^{T,(\varphi_T,g)} \partial_s\varphi_T
              \big\rangle_g\,,            \\		
  \alpha^T_{(\mathrm{I}^2,\partial_t\varphi_T, D^T\varphi_T)}			
   & = & \Tr
                \big\langle
		          \partial_t\varphi_T\,,\,
		            (\ad\otimes\nabla^g)_{D^T\varphi_T}  \partial_s\varphi_T
                \big\rangle_g\,,   \\		
  \alpha^T_{(\mathrm{I}^2,\partial_t\varphi_T, \partial_s\nabla^T)}				
   &= & \Tr
               \big\langle
		          \partial_t\varphi_T\,,\,
		           \mbox{\large $\sum_{i=1}^n$}
         			  [ \partial_s\nabla^T,  \varphi_T^{\sharp}(y^i)]
		        	  \otimes\mbox{\large $\frac{\partial}{\partial y^i}$}
               \big\rangle_g
 \end{eqnarray*}
 and let
 \begin{eqnarray*}
  \xi^T_{(\mathrm{I}^2,\partial_t\varphi_T, \nabla^{T,{(\varphi_T,g)}})}
   &= & \sum_{\mu}\Tr
              \big\langle
		        \partial_t\varphi_T\,,\,
		            \nabla^{T,(\varphi_T,g)}_{e_{\mu}} \partial_s\varphi_T
              \big\rangle_g\, e_{\mu}\,,           \\		
  \xi^T_{(\mathrm{I}^2,\partial_t\varphi_T, D^T\varphi_T)}			
   & = & \sum_{\mu}\Tr
                \big\langle
		          \partial_t\varphi_T\,,\,
		            (\ad\otimes\nabla^g)_{D^T_{e_{\mu}}\varphi_T}  \partial_s\varphi_T
                \big\rangle_g\, e_{\mu} \,, \\		
  \xi^T_{(\mathrm{I}^2,\partial_t\varphi_T, \partial_s\nabla^T)}				
   &= & \sum_{\mu}\Tr
               \big\langle
		          \partial_t\varphi_T\,,\,
		           \mbox{\large $\sum_{i=1}^n$}
         			  [ (\partial_s\nabla^T)(e_{\mu}),  \varphi_T^{\sharp}(y^i)]
		        	  \otimes\mbox{\large $\frac{\partial}{\partial y^i}$}
               \big\rangle_g\, e_{\mu}
 \end{eqnarray*}
 be their respective dual on $U_T/T$ with respect to $h$.
Then,
{\small
\begin{eqnarray*}
 \mbox{$(\mbox{I}^{\,2}.1.2)$}
  & = &
   T_{m-1}\int_{\partial U}
    i_{\,\xi^T_{(\mathrm{I}^2,\partial_t\varphi_T, \nabla^{T,{(\varphi_T,g)}})}\,
	         +\, \xi^T_{(\mathrm{I}^2,\partial_t\varphi_T, D^T\varphi_T)}\,				
             +\, \xi^T_{(\mathrm{I}^2,\partial_t\varphi_T, \partial_s\nabla^T)}
           }\, \vol_h			 \\[.8ex]
  &&
    +\;\;
    T_{m-1}\int_U \Tr
        \Big\langle
		  \partial_t\varphi_T\,,\,
		    \Big(
			  \nabla^{T,(\varphi_T,g)}_{\sum_{\mu}\nabla^h_{e_{\mu}}e_{\mu}}\,
		      -\,  \mbox{\large $\sum$}_{\mu}
			         \nabla^{T,(\varphi_T,g)}_{e_{\mu}}
			         \nabla^{T,(\varphi_T,g)}_{e_{\mu}}
			   \Big) \partial_s\varphi_T
        \Big\rangle_g\,  \vol_h \\		
  && + \;\;
         T_{m-1}\int_U\Tr
          \Big\langle
		   \partial_t\varphi_T\,,\,
		     \Big(
			 (\ad\otimes\nabla^g)
			      _{D^T_{\sum_{\mu}\nabla^h_{e_{\mu}}e_{\mu}}\varphi_T} \,   \\
  && \hspace{14em}				
		     -\, \mbox{\large $\sum$}_{\mu}\nabla^{T,(\varphi_T,g)}_{e_{\mu}}
		      (\ad\otimes\nabla^g)_{D^T_{e_{\mu}}\varphi_T}
             \Big)\partial_s\varphi_T
        \Big\rangle_g\,  \vol_h \\		
  && +\;\;
        T_{m-1}\int_U\Tr
          \Big\langle
		    \partial_t\varphi_T\,,\,
			 \mbox{\large $\sum_{i=1}^n$}
              \Big(	
			   \big[ (\partial_s\nabla^T)
			                (\mbox{$\sum$}_{\mu} \nabla^h_{e_{\mu}} e_{\mu}),
							\varphi_T^{\sharp}(y^i) \big] \,		   \\[-.6ex]
   && \hspace{14em}							
		     -\, \mbox{\large $\sum$}_{\mu}
			      \nabla^{T,(\varphi_T,g)}_{e_{\mu}}		
			      \big[ (\partial_s\nabla^T)(e_{\mu}), \varphi_T^{\sharp}(y^i) \big]
               \Big)	\otimes\mbox{\large $\frac{\partial}{\partial y^i}$}
        \Big\rangle_g\,  \vol_h\,.     \\
\end{eqnarray*}}

\bigskip

\begin{flushleft}
{{\bf The term} $\,\frac{\partial}{\partial s}\,(\mathrm{I}^{\,2}.2) $}
\end{flushleft}
\begin{eqnarray*}
  \frac{\partial}{\partial s}\,(\mathrm{I}^{\,2}.2)
   & =  &   T_{m-1}\, \frac{\partial}{\partial s} \int_U \Tr \sum_{\mu=1}^m
		          \Big\langle
			     (\ad\otimes\!\nabla^g)_{D^T_{e_{\mu}}\varphi_T}\partial_t\varphi_T\,,\,
				         D^T_{e_{\mu}}\varphi_T  \Big\rangle_g\,   \vol_h             \\
  & = &
    T_{m-1} \int_U \Tr \sum_{\mu}
	    \Big\langle
		  \partial_s \big(
		     (\ad\otimes \nabla^g)_{D_{e_{\mu}}\varphi_T} \partial_t\varphi_T \big)\,,
			 D^T_{e_{\mu}}\varphi_T
		\Big\rangle_g\,\vol_h \\
  && \hspace{4em}
    +\;\;  T_{m-1} \int_U \Tr \sum_{\mu}
	            \Big\langle
		         (\ad\otimes \nabla^g)_{D_{e_{\mu}}\varphi_T} \partial_t\varphi_T\,,
			       \partial_s D^T_{e_{\mu}}\varphi_T
		        \Big\rangle_g\,\vol_h \\
   & = &  (\mathrm{I}^{\,2}.2.1)\;+\; (\mathrm{I}^{\,2}.2.2)\,.		
\end{eqnarray*}

\bigskip

\noindent
{\it $(a)\,$ Term $(\mathrm{I}^{\,2}.2.1)$}
{\small
\begin{eqnarray*}
 (\mathrm{I}^{\,2}.2.1)
  & = &
    T_{m-1} \int_U \Tr\sum_{\mu}
       \Big\langle
	     (\ad\otimes \nabla^g)_{D^T_{e_{\mu}}}(\partial_s\partial_t\varphi_T)\,,\,
		      D^T_{e_{\mu}}\varphi_T    	
       \Big\rangle_g\, \vol_h	   \\
  && +\;\;
        T_{m-1}\int_U\Tr\sum_{\mu}
         \Big\langle
      \sum_{i,j}
	    \big[ D^T_{e_{\mu}}\partial_s\varphi_T^{\sharp}(y^j),
		           \partial_t\varphi_T^{\sharp}(y^i)   \big]
          \otimes\nabla^g_{\frac{\partial}{\partial y^j}}
		                                  \mbox{\Large $\frac{\partial}{\partial y^i}$}\,,\,			     										  
           D^T_{e_{\mu}}\varphi_T
         \Big\rangle_g\, \vol_h		  \\
  && +\;\;
      T_{m-1} \int_U  \Tr \sum_{\mu}
     \Big\langle
	   \sum_{i,j,k} \Big(
	       \big[ D^T_{e_{\mu}}\varphi_T^{\sharp}(y^j),
		         \partial_t\varphi_T^{\sharp}(y^i)  \big]  \partial_s\varphi_T^{\sharp}(y^k)
				\otimes R^g\big(\mbox{\large $\frac{\partial}{\partial y^k}$},
				                          \mbox{\large $\frac{\partial}{\partial y^j}$}\big)
									\mbox{\large $\frac{\partial}{\partial y^i}$}\;  \\[-1.2ex]
   && \hspace{12em}									
          -\; \partial_t\varphi_T^{\sharp}(y^i)
		        \big[ D^T_{e_{\mu}}\varphi_T^{\sharp}(y^j),
				         \partial_s\varphi_T^{\sharp}(y^k)\big]
                  \otimes \nabla^g_{\frac{\partial}{\partial y^j}}
				                 \nabla^g_{\frac{\partial}{\partial y^k}}
				                      \mbox{\large $\frac{\partial}{\partial y^i}$}				   
	       \Big)\,,\,   \\
   && \hspace{30em}		
		    D^T_{e_{\mu}}\varphi_T
     \Big\rangle_g\, \vol_h	  \\
  && +\;\;  T_{m-1} \int_U \Tr \sum_{\mu}
                 \Big\langle
				  \sum_{i,j}
				   \big[ \ad_{(\partial_s\nabla^T)(e_{\mu})}\varphi_T^{\sharp}(y^j),
					         \partial_t\varphi_T^{\sharp}(y^i) \big]
                     \otimes \nabla^g_{\frac{\partial}{\partial y^j}}
					                      \mbox{\large $\frac{\partial}{\partial y^i}$}\,,\,
                  D^T_{e_{\mu}}\varphi_T		
				 \Big\rangle_g\,  \vol_h\,.
\end{eqnarray*}}

\noindent
The integrand of the first summand captures a related part in the system of equations of motion
 for $(\varphi,\nabla)$.
The integrand of the second summand is tensorial in $\partial_t\varphi_T$  and
        first-order differential operatorial in $\partial_s\varphi_T$.
The integrand of the third summand is tensorial in both $\partial_t\varphi_T$ and $\partial_s\varphi_T$.
The integrand of the fourth summand is tensorial in $\partial_t\varphi_T$ and $\partial_s\nabla^T$.

\bigskip

\noindent
{\it $(b)\,$ Term $(\mathrm{I}^{\,2}.2.2)$}
{\small
\begin{eqnarray*}
 (\mathrm{I}^{\,2}.2.2)
     & = &
	   T_{m-1} \int_U \Tr \sum_{\mu}
	            \Big\langle
		         (\ad\otimes \nabla^g)_{D_{e_{\mu}}\varphi_T} \partial_t\varphi_T\,,\,
			       \partial_s D^T_{e_{\mu}}\varphi_T
		        \Big\rangle_g\,\vol_h \\
    &= &
	   T_{m-1} \int_U \Tr \sum_{\mu}
	            \Big\langle
		         (\ad\otimes \nabla^g)_{D_{e_{\mu}}\varphi_T} \partial_t\varphi_T\,,\,
			       \nabla^{T,(\varphi_T,g)}_{e_{\mu}}\partial_s\varphi_T
		        \Big\rangle_g\,\vol_h \\
    && \hspace{1.2em}-\;\;
	   T_{m-1} \int_U \Tr \sum_{\mu}
	            \Big\langle
		         (\ad\otimes \nabla^g)_{D_{e_{\mu}}\varphi_T} \partial_t\varphi_T\,,\,
			       (\ad\otimes\nabla^g)_{\partial_s\varphi_T} D^T_{e_{\mu}}\varphi_T
		        \Big\rangle_g\,\vol_h \\
    && \hspace{1.2em}+\;\;
	   T_{m-1} \int_U \Tr \sum_{\mu}
	            \Big\langle
		         (\ad\otimes \nabla^g)_{D_{e_{\mu}}\varphi_T} \partial_t\varphi_T\,,\,
			      \sum_i
				    \big[ (\partial_s\nabla^T)(e_{\mu}), \varphi_T^{\sharp}(y^i) \big]
					  \otimes \mbox{\large $\frac{\partial}{\partial y^i}$}
		        \Big\rangle_g\,\vol_h\,. 		
\end{eqnarray*}}

\noindent
The integrand of the first summand is tensorial in $\partial_t\varphi_T$  and
        first-order differential operatorial in $\partial_s\varphi_T$.
The integrand of the second summand is tensorial in both $\partial_t\varphi_T$ and $\partial_s\varphi_T$.
The integrand of the third summand is tensorial in $\partial_t\varphi_T$ and $\partial_s\nabla^T$.

\bigskip

\begin{flushleft}
{{\bf The term} $\,\frac{\partial}{\partial s}\,(\mathrm{I}^{\,2}.3) $}
\end{flushleft}
\begin{eqnarray*}
  \frac{\partial}{\partial s}\,(\mathrm{I}^{\,2}.3)
   & = &  T_{m-1} \, \frac{\partial}{\partial s} \int_U \sum_{\mu=1}^m
		           \Big\langle
				       \sum_{i=1}^n
					     \big[(\partial_t\nabla^T)(e_{\mu}), \varphi_T^{\sharp}(y^i)\big]
					         \otimes \mbox{$\frac{\partial}{\partial y^i}$}\,,\,
				       D^T_{e_{\mu}} \varphi_T  \Big\rangle_g\, \vol_h                          \\ 
 & = &  T_{m-1} \, \int_U \sum_{\mu=1}^m
		           \Big\langle
				       \sum_{i=1}^n
					     \partial_s\Big(
						   \big[(\partial_t\nabla^T)(e_{\mu}), \varphi_T^{\sharp}(y^i)\big]
					         \otimes \mbox{$\frac{\partial}{\partial y^i}$}  \Big)\,,\,
				       D^T_{e_{\mu}} \varphi_T  \Big\rangle_g\, \vol_h                          \\ 
 && \hspace{1.2em}+\;\;  T_{m-1} \,  \int_U \sum_{\mu=1}^m
		           \Big\langle
				       \sum_{i=1}^n
					     \big[(\partial_t\nabla^T)(e_{\mu}), \varphi_T^{\sharp}(y^i)\big]
					         \otimes \mbox{$\frac{\partial}{\partial y^i}$}\,,\,
				      \partial_s D^T_{e_{\mu}} \varphi_T  \Big\rangle_g\, \vol_h                          \\ 					
  & = &  (\mathrm{I}^{\,2}.3.1)\; +\; (\mathrm{I}^{\,2}.3.2)\,.
\end{eqnarray*}

\bigskip

\noindent
{\it $(a)\,$ Term $(\mathrm{I}^{\,2}.3.1)$}
{\small
\begin{eqnarray*}
 (\mathrm{I}^{\,2}.3.1)
   & = &   T_{m-1} \, \int_U \sum_{\mu=1}^m
		           \Big\langle
				       \sum_{i=1}^n
					     \partial_s\Big(
						   \big[(\partial_t\nabla^T)(e_{\mu}), \varphi_T^{\sharp}(y^i)\big]
					         \otimes \mbox{\large $\frac{\partial}{\partial y^i}$}  \Big)\,,\,
				       D^T_{e_{\mu}} \varphi_T  \Big\rangle_g\, \vol_h                          \\ 
  & = &   T_{m-1} \, \int_U \sum_{\mu=1}^m
		           \Big\langle
				       \sum_{i=1}^n
						   \big[(\partial_s\partial_t\nabla^T)(e_{\mu}), \varphi_T^{\sharp}(y^i)\big]
					         \otimes \mbox{\large $\frac{\partial}{\partial y^i}$}  \,,\,
				       D^T_{e_{\mu}} \varphi_T  \Big\rangle_g\, \vol_h                          \\ 
  && +\;\;
          T_{m-1} \, \int_U \sum_{\mu=1}^m
		           \Big\langle
				       \sum_{i=1}^n
					     \Big(
						   \big[(\partial_t\nabla^T)(e_{\mu}), \partial_s\varphi_T^{\sharp}(y^i)\big]
					         \otimes \mbox{\large $\frac{\partial}{\partial y^i}$}  \,   \\
  && \hspace{10em}
						+\, \big[ (\partial_t\nabla^T)(e_{\mu}), \varphi_T^{\sharp}(y^i)\big]
						         \mbox{\large$\sum$}_j\partial_s\varphi_T^{\sharp}(y^j)
								     \otimes \nabla^g_{\frac{\partial}{\partial y^j}}
									                   \mbox{\large $\frac{\partial}{\partial y^i}$}\,\Big)\,,\,
	   	     		             D^T_{e_{\mu}} \varphi_T  \Big\rangle_g\, \vol_h\,.                          
\end{eqnarray*}}

\noindent
The integrand of the first summand captures a related part in the system of equations of motion
 for $(\varphi,\nabla)$.
The integrand of the second summand is tensorial in $\partial_s\varphi_T$ and $\partial_t\nabla^T$.

\bigskip

\noindent
{\it $(b)\,$ Term $(\mathrm{I}^{\,2}.3.2)$}
{\small
\begin{eqnarray*}
 (\mathrm{I}^{\,2}.3.2) & = &
    T_{m-1} \, \int_U \sum_{\mu=1}^m
		           \Big\langle
				       \sum_{i=1}^n
					     \big[(\partial_t\nabla^T)(e_{\mu}), \varphi_T^{\sharp}(y^i)\big]
					         \otimes \mbox{\large $\frac{\partial}{\partial y^i}$}\,,\,
				      \partial_s D^T_{e_{\mu}} \varphi_T  \Big\rangle_g\, \vol_h                       \\ 			
   & = & 					
   T_{m-1} \, \int_U \sum_{\mu=1}^m
		           \Big\langle
				       \sum_{i=1}^n
					     \big[(\partial_t\nabla^T)(e_{\mu}), \varphi_T^{\sharp}(y^i)\big]
					         \otimes \mbox{\large $\frac{\partial}{\partial y^i}$}\,,\,
				      \nabla^{T,(\varphi_T,g)}_{e_{\mu}}
					                                   \partial_s\varphi_T  \Big\rangle_g\, \vol_h                       \\ 		
   &&-\;\;
     T_{m-1} \, \int_U \sum_{\mu=1}^m
		           \Big\langle
				      \mbox{\large $\sum$}_{i=1}^n
					     \big[(\partial_t\nabla^T)(e_{\mu}), \varphi_T^{\sharp}(y^i)\big]
					         \otimes \mbox{\large $\frac{\partial}{\partial y^i}$}\,,\,
 			  	    (\ad\otimes\nabla^g)_{\partial_s\varphi_T}D^T_{e_{\mu}}\varphi_T
					  \Big\rangle_g\, \vol_h                 \\
   &&+\;\;
     T_{m-1} \, \int_U \sum_{\mu=1}^m\sum_{i,j=1}^n
		           \Big\langle
					     \big[(\partial_t\nabla^T)(e_{\mu}), \varphi_T^{\sharp}(y^i)\big]
					         \otimes \mbox{\large $\frac{\partial}{\partial y^i}$}\,,\,
				        \big[ (\partial_s\nabla^T)(e_{\mu}), \varphi_T^{\sharp}(y^j)\big]
				            \otimes \mbox{\large $\frac{\partial}{\partial y^j}$}
					  \Big\rangle_g\, \vol_h\,.                       	   					
\end{eqnarray*}}

\noindent
The integrand of the first summand is tensorial in $\partial_t\nabla^T$  and
        first-order differential operatorial in $\partial_s\varphi_T$.
The integrand of the second summand is tensorial in $\partial_s\varphi_T$ and $\partial_t\nabla^T$.
The integrand of the third summand is tensorial in both $\partial_t\nabla^T$ and $\partial_s\nabla^T$. 		

\bigskip

Finally, recall Lemma 3.2.2.4
 and note that with the additional assumption at the beginning of this section,
 all the inner products $\Tr\langle\,\mbox{\LARGE $\cdot$}\,,\,\mbox{\LARGE $\cdot$}\,\rangle_g$
  that appear in the calculation above are defined.
 
\bigskip

In summary,

\begin{proposition} {\bf [second variation of kinetic term for maps]}$\;$
 Let $(\varphi_T,\nabla^T)$ be a $(\ast_2)$-admissible $T$-family
  of $(\ast_1)$-admissible pairs with the additional assumption that\\
   $\;D_{\xi} \partial_s{\cal A}_{\varphi_T} \subset \Comm({\cal A}_{\varphi_T})$
    for all $\xi\in {\cal T}_{\ast}(X_T/T)$.
 Then,
 {\footnotesize
 \begin{eqnarray*}
  \lefteqn{
    \mbox{\large $\frac{\partial}{\partial s}
	                             \frac{\partial}{\partial t}$}E^{\nabla^T}(\varphi_T)^{\Bbb C}\;\;
     = \;\; \mbox{\large $\frac{\partial}{\partial s}\frac{\partial}{\partial t}$}
          \Big(
		    \mbox{\large $\frac{1}{2}$}\, T_{m-1} \int_U \Tr
	        \langle D^T\varphi_T\,,\, D^T\varphi_T \rangle_{(h,g)}\, \vol_h
		   \Big)                     }\\
   && =\;\;  
        T_{m-1} \int_{\partial U}
	                   i_{\xi^T_{(\mbox{\tiny I}^2, \partial_s\partial_t\varphi_T)}} \vol_h \\
   && \hspace{2.6em}
       +\; T_{m-1}\int_{\partial U}
              i_{\,\xi^T_{(\mathrm{I}^2,\partial_t\varphi_T, \nabla^{T,{(\varphi_T,g)}})}\,
	                   +\, \xi^T_{(\mathrm{I}^2,\partial_t\varphi_T, D^T\varphi_T)}\,				 
                       +\, \xi^T_{(\mathrm{I}^2,\partial_t\varphi_T, \partial_s\nabla^T)}
                     }\, \vol_h			 \\					
   && \hspace{1.6em} 
       +\;\; T_{m-1} \int_U \Tr
	           \Big\langle
			     \partial_s\partial_t\varphi_T\,,\,
				   \big(
				      D^T_{\sum_{\mu}\nabla^h_{e_{\mu}}e_{\mu}}\,
					   -\, \mbox{$\sum_{\mu}$}
						   \nabla^{T,(\varphi_T,g)}_{e_{\mu}}D^T_{e_{\mu}}
				     \big)\varphi_T
			    \Big\rangle_g\, \vol_h  		   \\
  &&  \hspace{2.6em}
    +\;\; T_{m-1} \int_U \Tr\sum_{\mu}
       \Big\langle
	     (\ad\otimes \nabla^g)_{D^T_{e_{\mu}}}(\partial_s\partial_t\varphi_T)\,,\,
		      D^T_{e_{\mu}}\varphi_T    	
       \Big\rangle_g\, \vol_h	  \\	
   && \hspace{2.6em}
      +\;\;   T_{m-1} \, \int_U \sum_{\mu=1}^m
		           \Big\langle
				       \sum_{i=1}^n
						   \big[(\partial_s\partial_t\nabla^T)(e_{\mu}), \varphi_T^{\sharp}(y^i)\big]
					         \otimes \mbox{\large $\frac{\partial}{\partial y^i}$}  \,,\,
				       D^T_{e_{\mu}} \varphi_T  \Big\rangle_g\, \vol_h                          \\ 	   
  && \hspace{1.6em}
    +\;\;
    T_{m-1}\int_U \Tr
        \Big\langle
		  \partial_t\varphi_T\,,\,
		    \Big(
			  \nabla^{T,(\varphi_T,g)}_{\sum_{\mu}\nabla^h_{e_{\mu}}e_{\mu}}\,
		      -\,  \mbox{\large $\sum$}_{\mu}
			         \nabla^{T,(\varphi_T,g)}_{e_{\mu}}
			         \nabla^{T,(\varphi_T,g)}_{e_{\mu}}
			   \Big) \partial_s\varphi_T
        \Big\rangle_g\,  \vol_h \\		
  && \hspace{2.6em}
       + \;\;
         T_{m-1}\int_U\Tr
          \Big\langle
		   \partial_t\varphi_T\,,\,
		     \Big(
			 (\ad\otimes\nabla^g)
			      _{D^T_{\sum_{\mu}\nabla^h_{e_{\mu}}e_{\mu}}\varphi_T} \,   \\
  && \hspace{16.6em}				
		     -\, \mbox{\large $\sum$}_{\mu}\nabla^{T,(\varphi_T,g)}_{e_{\mu}}
		      (\ad\otimes\nabla^g)_{D^T_{e_{\mu}}\varphi_T}
             \Big)\partial_s\varphi_T
        \Big\rangle_g\,  \vol_h \\		
 && \hspace{2.6em}
    +\;\;
        T_{m-1}\int_U\Tr\sum_{\mu}
         \Big\langle
      \sum_{i,j}
	    \big[ D^T_{e_{\mu}}\partial_s\varphi_T^{\sharp}(y^j),
		           \partial_t\varphi_T^{\sharp}(y^i)   \big]
          \otimes\nabla^g_{\frac{\partial}{\partial y^j}}
		                                  \mbox{\Large $\frac{\partial}{\partial y^i}$}\,,\,			     										  
           D^T_{e_{\mu}}\varphi_T
         \Big\rangle_g\, \vol_h		 \\
  && \hspace{2.6em}
    +\;\;
      T_{m-1} \int_U  \Tr \sum_{\mu}
     \Big\langle
	   \sum_{i,j,k} \Big(
	       \big[ D^T_{e_{\mu}}\varphi_T^{\sharp}(y^j),
		         \partial_t\varphi_T^{\sharp}(y^i)  \big]  \partial_s\varphi_T^{\sharp}(y^k)
				\otimes R^g\big(\mbox{\large $\frac{\partial}{\partial y^k}$},
				                          \mbox{\large $\frac{\partial}{\partial y^j}$}\big)
									\mbox{\large $\frac{\partial}{\partial y^i}$}     \\[-1.2ex]
   && \hspace{14.6em}									
          -\; \partial_t\varphi_T^{\sharp}(y^i)
		        \big[ D^T_{e_{\mu}}\varphi_T^{\sharp}(y^j),
				         \partial_s\varphi_T^{\sharp}(y^k)\big]
                  \otimes \nabla^g_{\frac{\partial}{\partial y^j}}
				                 \nabla^g_{\frac{\partial}{\partial y^k}}
				                      \mbox{\large $\frac{\partial}{\partial y^i}$}				   
	       \Big)\,,\,    \\
   && \hspace{32.6em}		
		    D^T_{e_{\mu}}\varphi_T
     \Big\rangle_g\, \vol_h	   \\		
  && \hspace{2.6em}
	 +\;\;  T_{m-1} \int_U \Tr \sum_{\mu}
	            \Big\langle
		         (\ad\otimes \nabla^g)_{D_{e_{\mu}}\varphi_T} \partial_t\varphi_T\,,\,
			       \nabla^{T,(\varphi_T,g)}_{e_{\mu}}\partial_s\varphi_T
		        \Big\rangle_g\,\vol_h    \\
    && \hspace{2.6em}
	  -\;\;
	   T_{m-1} \int_U \Tr \sum_{\mu}
	            \Big\langle
		         (\ad\otimes \nabla^g)_{D_{e_{\mu}}\varphi_T} \partial_t\varphi_T\,,\,
			       (\ad\otimes\nabla^g)_{\partial_s\varphi_T} D^T_{e_{\mu}}\varphi_T
		        \Big\rangle_g\,\vol_h        \\	
  && \hspace{2.6em}
	 -\;\; T_{m-1}\int_U\Tr
        \Big\langle
		  \partial_t\varphi_T\,,\,
		   \sum_{\mu}
           F_{\nabla^{T,(\varphi_T,g)}}(\partial_s, e_{\mu}) D^T_{e_{\mu}}\varphi_T
        \Big\rangle_g\, \vol_h	     \\
 && \hspace{1.6em} 
     +\;\;
        T_{m-1}\int_U\Tr
          \Big\langle
		    \partial_t\varphi_T\,,\,
			 \mbox{\large $\sum_{i=1}^n$}
              \Big(	
			   \big[ (\partial_s\nabla^T)
			                (\mbox{$\sum$}_{\mu} \nabla^h_{e_{\mu}} e_{\mu}),
							\varphi_T^{\sharp}(y^i) \big] \,		   \\[-.6ex]
   && \hspace{15.6em}							
		     -\, \mbox{\large $\sum$}_{\mu}
			      \nabla^{T,(\varphi_T,g)}_{e_{\mu}}		
			      \big[ (\partial_s\nabla^T)(e_{\mu}), \varphi_T^{\sharp}(y^i) \big]
               \Big)	\otimes\mbox{\large $\frac{\partial}{\partial y^i}$}
        \Big\rangle_g\,  \vol_h\,.     \\
  && \hspace{2.6em}
     +\;\;  T_{m-1} \int_U \Tr \sum_{\mu}
                 \Big\langle
				  \sum_{i,j}
				   \big[ \ad_{(\partial_s\nabla^T)(e_{\mu})}\varphi_T^{\sharp}(y^j),
					         \partial_t\varphi_T^{\sharp}(y^i) \big]
                     \otimes \nabla^g_{\frac{\partial}{\partial y^j}}
					                      \mbox{\large $\frac{\partial}{\partial y^i}$}\,,\,
                  D^T_{e_{\mu}}\varphi_T		
				 \Big\rangle_g\,  \vol_h       \\
  && \hspace{2.6em}
      +\;\;
	   T_{m-1} \int_U \Tr \sum_{\mu}
	            \Big\langle
		         (\ad\otimes \nabla^g)_{D_{e_{\mu}}\varphi_T} \partial_t\varphi_T\,,\,
			      \sum_i
				    \big[ (\partial_s\nabla^T)(e_{\mu}), \varphi_T^{\sharp}(y^i) \big]
					  \otimes \mbox{\large $\frac{\partial}{\partial y^i}$}
		        \Big\rangle_g\,\vol_h      \\
  && \hspace{1.6em}
      +\;\;
          T_{m-1} \, \int_U \sum_{\mu=1}^m
		           \Big\langle
				       \sum_{i=1}^n
					     \Big(
						   \big[(\partial_t\nabla^T)(e_{\mu}), \partial_s\varphi_T^{\sharp}(y^i)\big]
					         \otimes \mbox{\large $\frac{\partial}{\partial y^i}$}  \,   \\
    && \hspace{11.6em}
						+\, \big[ (\partial_t\nabla^T)(e_{\mu}), \varphi_T^{\sharp}(y^i)\big]
						         \mbox{\large$\sum$}_j\partial_s\varphi_T^{\sharp}(y^j)
								     \otimes \nabla^g_{\frac{\partial}{\partial y^j}}
									                   \mbox{\large $\frac{\partial}{\partial y^i}$}\,\Big)\,,\,
	   	     		             D^T_{e_{\mu}} \varphi_T  \Big\rangle_g\, \vol_h   \\
  && \hspace{2.6em} 					
   +\;\; T_{m-1} \, \int_U \sum_{\mu=1}^m
		           \Big\langle
				       \sum_{i=1}^n
					     \big[(\partial_t\nabla^T)(e_{\mu}), \varphi_T^{\sharp}(y^i)\big]
					         \otimes \mbox{\large $\frac{\partial}{\partial y^i}$}\,,\,
				      \nabla^{T,(\varphi_T,g)}_{e_{\mu}}
					                                   \partial_s\varphi_T  \Big\rangle_g\, \vol_h                       \\ 		
   && \hspace{2.6em}
     -\;\;  T_{m-1} \, \int_U \sum_{\mu=1}^m
		           \Big\langle
				      \mbox{\large $\sum$}_{i=1}^n
					     \big[(\partial_t\nabla^T)(e_{\mu}), \varphi_T^{\sharp}(y^i)\big]
					         \otimes \mbox{\large $\frac{\partial}{\partial y^i}$}\,,\,
 			  	    (\ad\otimes\nabla^g)_{\partial_s\varphi_T}D^T_{e_{\mu}}\varphi_T
					  \Big\rangle_g\, \vol_h                 \\								
  && \hspace{1.6em}
    +\;\; T_{m-1} \, \int_U \sum_{\mu=1}^m\sum_{i,j=1}^n
		           \Big\langle
					     \big[(\partial_t\nabla^T)(e_{\mu}), \varphi_T^{\sharp}(y^i)\big]
					         \otimes \mbox{\large $\frac{\partial}{\partial y^i}$}\,,\,
				        \big[ (\partial_s\nabla^T)(e_{\mu}), \varphi_T^{\sharp}(y^j)\big]
				            \otimes \mbox{\large $\frac{\partial}{\partial y^j}$}
					  \Big\rangle_g\, \vol_h\,.
 \end{eqnarray*}}
 Here,
  {\small
  \begin{eqnarray*}
     \xi^T_{(\mathrm{I}^2,\partial_s\partial_t\varphi_T)}
     & :=
	   &  \sum_{\mu=1}^n \Tr
	          \big\langle \partial_s\partial_t\varphi_T\,,\,
			     D^T_{e_{\mu}}\varphi_T  \big\rangle_g\,   e_{\mu}\;,   \\
     \xi^T_{(\mathrm{I}^2,\partial_t\varphi_T, \nabla^{T,{(\varphi_T,g)}})}
      &= & \sum_{\mu}\Tr
              \big\langle
		        \partial_t\varphi_T\,,\,
		            \nabla^{T,(\varphi_T,g)}_{e_{\mu}} \partial_s\varphi_T
              \big\rangle_g\, e_{\mu}          \;, \\		
     \xi^T_{(\mathrm{I}^2,\partial_t\varphi_T, D^T\varphi_T)}			
      & = & \sum_{\mu}\Tr
                \big\langle
		          \partial_t\varphi_T\,,\,
		            (\ad\otimes\nabla^g)_{D^T_{e_{\mu}}\varphi_T}  \partial_s\varphi_T
                \big\rangle_g\, e_{\mu} \;, \\		
     \xi^T_{(\mathrm{I}^2,\partial_t\varphi_T, \partial_s\nabla^T)}				
      &= & \sum_{\mu}\Tr
               \big\langle
		          \partial_t\varphi_T\,,\,
		           \mbox{\large $\sum_{i=1}^n$}
         			  [ (\partial_s\nabla^T)(e_{\mu}),  \varphi_T^{\sharp}(y^i)]
		        	  \otimes\mbox{\large $\frac{\partial}{\partial y^i}$}
               \big\rangle_g\, e_{\mu}	
  \end{eqnarray*}} 
  and
{\small
	 \begin{eqnarray*}
	  \lefteqn{
	   F_{\nabla^{T,(\varphi_T,g)}}(\partial_s, e_{\mu})\,
	      D^T_{e_{\mu}}\varphi_T \;\;
	   =\;\;
        \big(\partial_s \nabla^{T,(\varphi_T,g)}_{e_{\mu}} \;
	      -\;  \nabla^{T,(\varphi_T, g)}_{e_{\mu}}\partial_s \big)\,
		    \mbox{\large $\sum$}_{i=1}^n
			   D_{e_{\mu}}\varphi_T^{\sharp}(y^i)
			      \otimes \mbox{\large $\frac{\partial}{\partial y^i}$}           }\\[.6ex]
       && 					
         =\,
		  \sum_{i=1}^n
		    [(\partial_s\nabla^T)(e_{\mu}), D_{e_{\mu}}\varphi_T^{\sharp}(y^i)]
			   \otimes \mbox{\large $\frac{\partial}{\partial y^i}$}\;\;
             +\;\; \sum_{i=1}^n
			          D_{e_{\mu}}\varphi_T^{\sharp}(y^i) \sum_{j=1}^n
			          [(\partial_s\nabla^T)(e_{\mu}), \varphi_T^{\sharp}(y^j)]
					    \otimes \nabla^g_{\frac{\partial}{\partial y^j}}
						                                                 \mbox{\large $\frac{\partial}{\partial y^i}$}  \\[-1.2ex]
		&& \hspace{1.6em}
		     +\;\; \sum_{i=1}^n D_{e_{\mu}}\varphi_T^{\sharp}(y^i)\,
			          \sum_{j,k=1}^n   \Big(
			                 D^T_{e_{\mu}}\varphi_T^{\sharp}(y^j)\,\partial_s\varphi_T^{\sharp}(y^k)
							    \otimes \nabla^g_{\frac{\partial}{\partial y^k}}
								               \nabla^g_{\frac{\partial}{\partial y^j}}\,
											      \mbox{\large $\frac{\partial}{\partial y^i}$}\:     \\[-1.2ex]
        && \hspace{20em}												
						 -\: \partial_s\varphi_T^{\sharp}(y^k)\,D^T_{e_{\mu}}\varphi_T^{\sharp}(y^j)
							    \otimes \nabla^g_{\frac{\partial}{\partial y^j}}
								               \nabla^g_{\frac{\partial}{\partial y^k}}\,
											   \mbox{\large $\frac{\partial}{\partial y^i}$}
			                                                      \Big)
    \end{eqnarray*}}
 The summands
  \begin{eqnarray*}
	&& \hspace{1.6em} 
       +\;\; T_{m-1} \int_U \Tr
	           \Big\langle
			     \partial_s\partial_t\varphi_T\,,\,
				   \big(
				      D^T_{\sum_{\mu}\nabla^h_{e_{\mu}}e_{\mu}}\,
					   -\, \mbox{$\sum_{\mu}$}
						   \nabla^{T,(\varphi_T,g)}_{e_{\mu}}D^T_{e_{\mu}}
				     \big)\varphi_T
			    \Big\rangle_g\, \vol_h  		   \\
  &&  \hspace{2.6em}
    +\;\; T_{m-1} \int_U \Tr\sum_{\mu}
       \Big\langle
	     (\ad\otimes \nabla^g)_{D^T_{e_{\mu}}}(\partial_s\partial_t\varphi_T)\,,\,
		      D^T_{e_{\mu}}\varphi_T    	
       \Big\rangle_g\, \vol_h	 \\	
   && \hspace{2.6em}
      +\;\;   T_{m-1} \, \int_U \sum_{\mu=1}^m
		           \Big\langle
				       \sum_{i=1}^n
						   \big[(\partial_s\partial_t\nabla^T)(e_{\mu}), \varphi_T^{\sharp}(y^i)\big]
					         \otimes \mbox{\large $\frac{\partial}{\partial y^i}$}  \,,\,
				       D^T_{e_{\mu}} \varphi_T  \Big\rangle_g\, \vol_h
   \end{eqnarray*}					   	
  will vanish when imposing the equations of motion for $(\varphi,\nabla)$.

 \vspace{30em}
 If $(\varphi_T,\nabla^T)$ is furthermore a $(\ast_2)$-admissible $T$-family
  of $(\ast_2)$-admissible pairs,
 Then, the above expression reduces to
 {\small
 \begin{eqnarray*}
  \lefteqn{
    \mbox{\large $\frac{\partial}{\partial s}
	                             \frac{\partial}{\partial t}$}E^{\nabla^T}(\varphi_T)^{\Bbb C}\;\;
     = \;\; \mbox{\large $\frac{\partial}{\partial s}\frac{\partial}{\partial t}$}
          \Big(
		    \mbox{\large $\frac{1}{2}$}\, T_{m-1} \int_U \Tr
	        \langle D^T\varphi_T\,,\, D^T\varphi_T \rangle_{(h,g)}\, \vol_h
		   \Big)                     }\\
   && =\;\;  
        T_{m-1} \int_{\partial U}
	                   i_{\xi^T_{(\mbox{\tiny I}^2, \partial_s\partial_t\varphi_T)}} \vol_h \\
   && \hspace{2.6em}
       +\; T_{m-1}\int_{\partial U}
              i_{\,\xi^T_{(\mathrm{I}^2,\partial_t\varphi_T, \nabla^{T,{(\varphi_T,g)}})}\,
	                   +\, \xi^T_{(\mathrm{I}^2,\partial_t\varphi_T, D^T\varphi_T)}\,				 
                       +\, \xi^T_{(\mathrm{I}^2,\partial_t\varphi_T, \partial_s\nabla^T)}
                     }\, \vol_h			 \\					
   && \hspace{1.6em} 
       +\;\; T_{m-1} \int_U \Tr
	           \Big\langle
			     \partial_s\partial_t\varphi_T\,,\,
				   \big(
				      D^T_{\sum_{\mu}\nabla^h_{e_{\mu}}e_{\mu}}\,
					   -\, \mbox{$\sum_{\mu}$}
						   \nabla^{T,(\varphi_T,g)}_{e_{\mu}}D^T_{e_{\mu}}
				     \big)\varphi_T
			    \Big\rangle_g\, \vol_h  		   \\	
   && \hspace{2.6em}
      +\;\;   T_{m-1} \, \int_U \sum_{\mu=1}^m
		           \Big\langle
				       \sum_{i=1}^n
						   \big[(\partial_s\partial_t\nabla^T)(e_{\mu}), \varphi_T^{\sharp}(y^i)\big]
					         \otimes \mbox{\large $\frac{\partial}{\partial y^i}$}  \,,\,
				       D^T_{e_{\mu}} \varphi_T  \Big\rangle_g\, \vol_h                          \\ 	   
  && \hspace{1.6em}
    +\;\;
    T_{m-1}\int_U \Tr
        \Big\langle
		  \partial_t\varphi_T\,,\,
		    \Big(
			  \nabla^{T,(\varphi_T,g)}_{\sum_{\mu}\nabla^h_{e_{\mu}}e_{\mu}}\,
		      -\,  \mbox{\large $\sum$}_{\mu}
			         \nabla^{T,(\varphi_T,g)}_{e_{\mu}}
			         \nabla^{T,(\varphi_T,g)}_{e_{\mu}}
			   \Big) \partial_s\varphi_T
        \Big\rangle_g\,  \vol_h \\		
  && \hspace{2.6em}
       + \;\;
         T_{m-1}\int_U\Tr
          \Big\langle
		   \partial_t\varphi_T\,,\,
		     \Big(
			 (\ad\otimes\nabla^g)
			      _{D^T_{\sum_{\mu}\nabla^h_{e_{\mu}}e_{\mu}}\varphi_T} \,   \\
  && \hspace{16.6em}				
		     -\, \mbox{\large $\sum$}_{\mu}\nabla^{T,(\varphi_T,g)}_{e_{\mu}}
		      (\ad\otimes\nabla^g)_{D^T_{e_{\mu}}\varphi_T}
             \Big)\partial_s\varphi_T
        \Big\rangle_g\,  \vol_h \\					
  && \hspace{2.6em}
	 -\;\; T_{m-1}\int_U\Tr
        \Big\langle
		  \partial_t\varphi_T\,,\,
		   \sum_{\mu}
           F_{\nabla^{T,(\varphi_T,g)}}(\partial_s, e_{\mu}) D^T_{e_{\mu}}\varphi_T
        \Big\rangle_g\, \vol_h	     \\
 && \hspace{1.6em} 
     +\;\;
        T_{m-1}\int_U\Tr
          \Big\langle
		    \partial_t\varphi_T\,,\,
			 \mbox{\large $\sum_{i=1}^n$}
              \Big(	
			   \big[ (\partial_s\nabla^T)
			                (\mbox{$\sum$}_{\mu} \nabla^h_{e_{\mu}} e_{\mu}),
							\varphi_T^{\sharp}(y^i) \big] \,		   \\[-.6ex]
   && \hspace{15.6em}							
		     -\, \mbox{\large $\sum$}_{\mu}
			      \nabla^{T,(\varphi_T,g)}_{e_{\mu}}		
			      \big[ (\partial_s\nabla^T)(e_{\mu}), \varphi_T^{\sharp}(y^i) \big]
               \Big)	\otimes\mbox{\large $\frac{\partial}{\partial y^i}$}
        \Big\rangle_g\,  \vol_h\,.     \\    				
  && \hspace{1.6em}
      +\;\;
          T_{m-1} \, \int_U \sum_{\mu=1}^m
		           \Big\langle
				       \sum_{i=1}^n
					     \Big(
						   \big[(\partial_t\nabla^T)(e_{\mu}), \partial_s\varphi_T^{\sharp}(y^i)\big]
					         \otimes \mbox{\large $\frac{\partial}{\partial y^i}$}  \,   \\
    && \hspace{11.6em}
						+\, \big[ (\partial_t\nabla^T)(e_{\mu}), \varphi_T^{\sharp}(y^i)\big]
						         \mbox{\large$\sum$}_j\partial_s\varphi_T^{\sharp}(y^j)
								     \otimes \nabla^g_{\frac{\partial}{\partial y^j}}
									                   \mbox{\large $\frac{\partial}{\partial y^i}$}\,\Big)\,,\,
	   	     		             D^T_{e_{\mu}} \varphi_T  \Big\rangle_g\, \vol_h   \\
  && \hspace{2.6em} 					
   +\;\; T_{m-1} \, \int_U \sum_{\mu=1}^m
		           \Big\langle
				       \sum_{i=1}^n
					     \big[(\partial_t\nabla^T)(e_{\mu}), \varphi_T^{\sharp}(y^i)\big]
					         \otimes \mbox{\large $\frac{\partial}{\partial y^i}$}\,,\,
				      \nabla^{T,(\varphi_T,g)}_{e_{\mu}}
					                                   \partial_s\varphi_T  \Big\rangle_g\, \vol_h                       \\ 		
   && \hspace{2.6em}
     -\;\;  T_{m-1} \, \int_U \sum_{\mu=1}^m
		           \Big\langle
				      \mbox{\large $\sum$}_{i=1}^n
					     \big[(\partial_t\nabla^T)(e_{\mu}), \varphi_T^{\sharp}(y^i)\big]
					         \otimes \mbox{\large $\frac{\partial}{\partial y^i}$}\,,\,
 			  	    (\ad\otimes\nabla^g)_{\partial_s\varphi_T}D^T_{e_{\mu}}\varphi_T
					  \Big\rangle_g\, \vol_h                 \\								
  && \hspace{1.6em}
    +\;\; T_{m-1} \, \int_U \sum_{\mu=1}^m\sum_{i,j=1}^n
		           \Big\langle
					     \big[(\partial_t\nabla^T)(e_{\mu}), \varphi_T^{\sharp}(y^i)\big]
					         \otimes \mbox{\large $\frac{\partial}{\partial y^i}$}\,,\,
				        \big[ (\partial_s\nabla^T)(e_{\mu}), \varphi_T^{\sharp}(y^j)\big]
				            \otimes \mbox{\large $\frac{\partial}{\partial y^j}$}
					  \Big\rangle_g\, \vol_h\,.
 \end{eqnarray*}}
 
 \vspace{30em}
 If one further imposes the equations of motion on  $(\varphi,\nabla)$,
  then the expression reduces further to
 {\small
 \begin{eqnarray*}
  \lefteqn{
    \mbox{\large $\frac{\partial}{\partial s}
	                             \frac{\partial}{\partial t}$}E^{\nabla^T}(\varphi_T)^{\Bbb C}\;\;
     = \;\; \mbox{\large $\frac{\partial}{\partial s}\frac{\partial}{\partial t}$}
          \Big(
		    \mbox{\large $\frac{1}{2}$}\, T_{m-1} \int_U \Tr
	        \langle D^T\varphi_T\,,\, D^T\varphi_T \rangle_{(h,g)}\, \vol_h
		   \Big)                     }\\
   && =\;\;  
        T_{m-1} \int_{\partial U}
	                   i_{\xi^T_{(\mbox{\tiny I}^2, \partial_s\partial_t\varphi_T)}} \vol_h \\
   && \hspace{2.6em}
       +\; T_{m-1}\int_{\partial U}
              i_{\,\xi^T_{(\mathrm{I}^2,\partial_t\varphi_T, \nabla^{T,{(\varphi_T,g)}})}\,
	                   +\, \xi^T_{(\mathrm{I}^2,\partial_t\varphi_T, D^T\varphi_T)}\,				 
                       +\, \xi^T_{(\mathrm{I}^2,\partial_t\varphi_T, \partial_s\nabla^T)}
                     }\, \vol_h			 \\							
  && \hspace{1.6em}
    +\;\;
    T_{m-1}\int_U \Tr
        \Big\langle
		  \partial_t\varphi_T\,,\,
		    \Big(
			  \nabla^{T,(\varphi_T,g)}_{\sum_{\mu}\nabla^h_{e_{\mu}}e_{\mu}}\,
		      -\,  \mbox{\large $\sum$}_{\mu}
			         \nabla^{T,(\varphi_T,g)}_{e_{\mu}}
			         \nabla^{T,(\varphi_T,g)}_{e_{\mu}}
			   \Big) \partial_s\varphi_T
        \Big\rangle_g\,  \vol_h \\		
  && \hspace{2.6em}
       + \;\;
         T_{m-1}\int_U\Tr
          \Big\langle
		   \partial_t\varphi_T\,,\,
		     \Big(
			 (\ad\otimes\nabla^g)
			      _{D^T_{\sum_{\mu}\nabla^h_{e_{\mu}}e_{\mu}}\varphi_T} \,   \\
  && \hspace{16.6em}				
		     -\, \mbox{\large $\sum$}_{\mu}\nabla^{T,(\varphi_T,g)}_{e_{\mu}}
		      (\ad\otimes\nabla^g)_{D^T_{e_{\mu}}\varphi_T}
             \Big)\partial_s\varphi_T
        \Big\rangle_g\,  \vol_h \\					
  && \hspace{2.6em}
	 -\;\; T_{m-1}\int_U\Tr
        \Big\langle
		  \partial_t\varphi_T\,,\,
		   \sum_{\mu}
           F_{\nabla^{T,(\varphi_T,g)}}(\partial_s, e_{\mu}) D^T_{e_{\mu}}\varphi_T
        \Big\rangle_g\, \vol_h	     \\
 && \hspace{1.6em} 
     +\;\;
        T_{m-1}\int_U\Tr
          \Big\langle
		    \partial_t\varphi_T\,,\,
			 \mbox{\large $\sum_{i=1}^n$}
              \Big(	
			   \big[ (\partial_s\nabla^T)
			                (\mbox{$\sum$}_{\mu} \nabla^h_{e_{\mu}} e_{\mu}),
							\varphi_T^{\sharp}(y^i) \big] \,		   \\[-.6ex]
   && \hspace{15.6em}							
		     -\, \mbox{\large $\sum$}_{\mu}
			      \nabla^{T,(\varphi_T,g)}_{e_{\mu}}		
			      \big[ (\partial_s\nabla^T)(e_{\mu}), \varphi_T^{\sharp}(y^i) \big]
               \Big)	\otimes\mbox{\large $\frac{\partial}{\partial y^i}$}
        \Big\rangle_g\,  \vol_h\,.     \\    				
  && \hspace{1.6em}
      +\;\;
          T_{m-1} \, \int_U \sum_{\mu=1}^m
		           \Big\langle
				       \sum_{i=1}^n
					     \Big(
						   \big[(\partial_t\nabla^T)(e_{\mu}), \partial_s\varphi_T^{\sharp}(y^i)\big]
					         \otimes \mbox{\large $\frac{\partial}{\partial y^i}$}  \,   \\
    && \hspace{11.6em}
						+\, \big[ (\partial_t\nabla^T)(e_{\mu}), \varphi_T^{\sharp}(y^i)\big]
						         \mbox{\large$\sum$}_j\partial_s\varphi_T^{\sharp}(y^j)
								     \otimes \nabla^g_{\frac{\partial}{\partial y^j}}
									                   \mbox{\large $\frac{\partial}{\partial y^i}$}\,\Big)\,,\,
	   	     		             D^T_{e_{\mu}} \varphi_T  \Big\rangle_g\, \vol_h   \\
  && \hspace{2.6em} 					
   +\;\; T_{m-1} \, \int_U \sum_{\mu=1}^m
		           \Big\langle
				       \sum_{i=1}^n
					     \big[(\partial_t\nabla^T)(e_{\mu}), \varphi_T^{\sharp}(y^i)\big]
					         \otimes \mbox{\large $\frac{\partial}{\partial y^i}$}\,,\,
				      \nabla^{T,(\varphi_T,g)}_{e_{\mu}}
					                                   \partial_s\varphi_T  \Big\rangle_g\, \vol_h                       \\ 		
   && \hspace{2.6em}
     -\;\;  T_{m-1} \, \int_U \sum_{\mu=1}^m
		           \Big\langle
				      \mbox{\large $\sum$}_{i=1}^n
					     \big[(\partial_t\nabla^T)(e_{\mu}), \varphi_T^{\sharp}(y^i)\big]
					         \otimes \mbox{\large $\frac{\partial}{\partial y^i}$}\,,\,
 			  	    (\ad\otimes\nabla^g)_{\partial_s\varphi_T}D^T_{e_{\mu}}\varphi_T
					  \Big\rangle_g\, \vol_h                 \\								
  && \hspace{1.6em}
    +\;\; T_{m-1} \, \int_U \sum_{\mu=1}^m\sum_{i,j=1}^n
		           \Big\langle
					     \big[(\partial_t\nabla^T)(e_{\mu}), \varphi_T^{\sharp}(y^i)\big]
					         \otimes \mbox{\large $\frac{\partial}{\partial y^i}$}\,,\,
				        \big[ (\partial_s\nabla^T)(e_{\mu}), \varphi_T^{\sharp}(y^j)\big]
				            \otimes \mbox{\large $\frac{\partial}{\partial y^j}$}
					  \Big\rangle_g\, \vol_h\,.
 \end{eqnarray*}}
\end{proposition}

\bigskip

\subsection{The second variation of the dilaton term}

We now work out the second variation of the (complexified) dilaton term
\begin{eqnarray*}
  S^{(\rho, h;\Phi)}_{\dilatonscriptsize} (\varphi_T)^{\Bbb C}
     &  =   &  \int_U \Tr \langle d\rho \,,\, \varphi_T^{\diamond}d\Phi\rangle_h\,\vol_h  \\[.6ex]
     & =   & \int_U\Tr \Big(\sum_{\mu}
	                d\rho
				    \big(e_{\mu})\, ((D^T_{e_{\mu}}\varphi_T)\Phi\big)\Big) \vol_h\,.
\end{eqnarray*}
for an $(\ast_1)$-admissible family of $(\ast_1)$-admissible pairs $(\varphi_T,\nabla^T)$,
$T:=(-\varepsilon,\varepsilon)^2\subset {\Bbb R}^2$ with coordinate $(s,t)$.

It follows from Sec.~6.2 that, due to the effect of the trace map $\Tr$,
\begin{eqnarray*}
 \lefteqn{
   \frac{\partial}{\partial t}  S^{(\rho, h;\Phi)}_{\dilatontiny}
        (\varphi_T)^{\Bbb C}\;\;
    =\;\;   \int_U \Tr  \sum_{\mu=1}^m
	            d\rho(e_{\mu})\,
			     \partial_t  \big((D^T_{e_{\mu}}\varphi_T)\Phi  \big) \vol_h  }\\						 
  && =\;\; \int_U \Tr\sum_{\mu}d\rho(e_{\mu})
       D^T_{e_{\mu}} \big((\partial_t\varphi_T)\Phi \big)\vol_h\,.
\end{eqnarray*}
Thus, due to the effect of the trace map $\Tr$ again,
{\small
\begin{eqnarray*}
 \lefteqn{
   \frac{\partial}{\partial s}
   \frac{\partial}{\partial t}  S^{(\rho, h;\Phi)}_{\dilatontiny}
    (\varphi_T)^{\Bbb C}\;\;
    =\;\;
	 \int_U \Tr\sum_{\mu}d\rho(e_{\mu})
	   \partial_s
       D^T_{e_{\mu}} \big((\partial_t\varphi_T)\Phi \big)\vol_h     }\\
  && =\;\;
     \int_U \Tr\sum_{\mu}d\rho(e_{\mu})
       D^T_{e_{\mu}}\partial_s \big((\partial_t\varphi_T)\Phi \big)\vol_h      \\
  && =\;\;
     \int_U \Tr\sum_{\mu}d\rho(e_{\mu})
       D^T_{e_{\mu}}\big((\partial_s\partial_t\varphi_T)\Phi \big)\vol_h \\
  && \hspace{6em}	
        +\; \int_U\Tr \sum_{\mu} d\rho(e_{\mu})
                D^T_{e_{\mu}}
				 \Big(\sum_{i,j}
				   \partial_t\varphi_T^{\sharp}(y^i)
				   \partial_s\varphi_T^{\sharp}(y^j)
                   \otimes \Big(
				      \mbox{\large $\frac{\partial}{\partial y^i}$}
	                  \mbox{\large $\frac{\partial}{\partial y^j}$}				
					  - \nabla^g_{\frac{\partial}{\partial y^i}}
					      \mbox{\large $\frac{\partial}{\partial y^j}$}				
				                \Big)\Phi				   \Big) \vol_h
    				  \\   	
  && =\;\;  (\mathrm{II}^{\,2}.1)\;+\; (\mathrm{II}^{\,2}.2)\,.	
\end{eqnarray*}}

For Summand $(\mathrm{II}^{\,2}.1)$,
 repeating the same argument in Sec.$\,$6.1 for Summand $(\mathrm{I}.1.1)$,
 one concludes that
 \begin{eqnarray*}
   (\mathrm{II}^{\,2}.1)
     & = & \int_{\partial U}
	     i_{\xi^T _{(\mathrm{II}^2, \partial_s\partial_t\varphi_T)}}\vol_h  \\
     && \hspace{2em}
	    +\; \int_U \Big(\sum_{\mu} \nabla^h_{e_{\mu}}e_{\mu}
                                 - \sum_{\mu}e_{\mu}d\rho(e_{\mu})		\Big)
						\Tr \big((\partial_s\partial_t\varphi_T)\Phi \big) \vol_h\,,
 \end{eqnarray*}
 where
  $$
   \xi^T _{(\mathrm{II}^2, \partial_s\partial_t\varphi_T)}\;
    :=\;  \sum_{\mu}\Big(
	            d\rho(e_{\mu}) \Tr\big( (\partial_s\partial_t\varphi_T)\Phi \big)
	                                \Big) e_{\mu}\hspace{2em}
     \in\; {\cal T}_{\ast}(U_T/T)\,.									
  $$
 The second summand of Summand $(\mathrm{II}^{\,2}.1)$ above is the term
   that captures the $S_{\dilatonscriptsize}^{(\rho,h;\Phi)}(\varphi)$-contribution
   to the system of equations of motion for $(\varphi, \nabla)$.

With $\partial_s\partial_t\varphi_T$ replaced by
 $\mbox{\large $\sum$}_{i,j}
	 \partial_t\varphi_T^{\sharp}(y^i) \partial_s\varphi_T^{\sharp}(y^j)
      \otimes \big(
				      \mbox{\large $\frac{\partial}{\partial y^i}$}
	                  \mbox{\large $\frac{\partial}{\partial y^j}$}				
					  - \nabla^g_{\frac{\partial}{\partial y^i}}
					      \mbox{\large $\frac{\partial}{\partial y^j}$}				
				                \big)\Phi$,
 one has similarly
 \begin{eqnarray*}
   (\mathrm{II}^{\,2}.2)
     & = & \int_{\partial U}
	     i_{\xi^T _{(\mathrm{II}^2, \partial_t\varphi_T, \partial_s\varphi_T)}}\vol_h  \\
     && \hspace{2em}
	    +\; \int_U \Big(\sum_{\mu} \nabla^h_{e_{\mu}}e_{\mu}
                                 - \sum_{\mu}e_{\mu}d\rho(e_{\mu})		\Big)\cdot \\
     && \hspace{6em}								
						\Tr  \Big(\sum_{i,j}
				            \partial_t\varphi_T^{\sharp}(y^i)
				            \partial_s\varphi_T^{\sharp}(y^j)
                            \otimes \Big(
				            \mbox{\large $\frac{\partial}{\partial y^i}$}
	                        \mbox{\large $\frac{\partial}{\partial y^j}$}				
					       - \nabla^g_{\frac{\partial}{\partial y^i}}
					         \mbox{\large $\frac{\partial}{\partial y^j}$}				
				                \Big)\Phi				   \Big) \vol_h\,,
 \end{eqnarray*}
 where
  $$
   \xi^T _{(\mathrm{II}^2, \partial_t\varphi_T, \partial_s\varphi_T)}\;
    :=\;  \sum_{\mu}\Big(
	        d\rho(e_{\mu}) \Tr				
			 \Big(\sum_{i,j}
				   \partial_t\varphi_T^{\sharp}(y^i)
				   \partial_s\varphi_T^{\sharp}(y^j)
                   \otimes \Big(
				      \mbox{\large $\frac{\partial}{\partial y^i}$}
	                  \mbox{\large $\frac{\partial}{\partial y^j}$}				
					  - \nabla^g_{\frac{\partial}{\partial y^i}}
					      \mbox{\large $\frac{\partial}{\partial y^j}$}				
				                \Big)\Phi				   \Big)				
	                                \Big) e_{\mu}
 $$									
 in ${\cal T}_{\ast}(U_T/T)$.
 The second summand of Summand $(\mathrm{II}^{\,2}.2)$ above
  contributes to the zeroth order terms of the differential operator on
  $(\partial_s\varphi_T, \partial_t\varphi_T )$
  from the second variation of
  $S^{(\rho,h;\Phi,g,B,C)}_{\standardscriptsize}(\varphi,\nabla)$.
 
In summary,

\bigskip

\begin{proposition}{\bf [second variation of
             $S_{\dilatonscriptsize}^{(\rho,h;\Phi)}(\varphi)^{\Bbb C}$]}$\;$
 For the (complexified) dilaton term
  $$
   S^{(\rho, h;\Phi)}_{\dilatonscriptsize} (\varphi)^{\Bbb C}\;
    :=\;   \int_U \Tr \langle d\rho \,,\, \varphi^{\diamond}d\Phi\rangle_h\,\vol_h\,,
  $$	
 its second variation
 for a $(\ast_1)$-admissible family of $(\ast_1)$-admissible pairs $(\varphi_T,\nabla^T)$,
 $T:=(-\varepsilon,\varepsilon)^2\subset {\Bbb R}^2$ with coordinate $(s,t)$,
 is given by
 {\small
 \begin{eqnarray*}
  \lefteqn{
    \frac{\partial}{\partial s}
    \frac{\partial}{\partial t}  S^{(\rho, h;\Phi)}_{\dilatontiny}
     (\varphi_T)^{\Bbb C} }\\
   && =\;\; \int_{\partial U}
	        i_{\xi^T _{(\mathrm{II}^2, \partial_s\partial_t\varphi_T)}
			      + \xi^T _{(\mathrm{II}^2, \partial_t\varphi_T\partial_s\varphi_T)}}
			\vol_h   \\
   && \hspace{2em}
	    +\; \int_U \Big(\sum_{\mu} \nabla^h_{e_{\mu}}e_{\mu}
                                 - \sum_{\mu}e_{\mu}d\rho(e_{\mu})		\Big)
						\Tr \big((\partial_s\partial_t\varphi_T)\Phi \big) \vol_h \\
   && \hspace{2em}
	    +\; \int_U \Big(\sum_{\mu} \nabla^h_{e_{\mu}}e_{\mu}
                                 - \sum_{\mu}e_{\mu}d\rho(e_{\mu})		\Big)\cdot \\
   && \hspace{6em}								
						\Tr  \Big(\sum_{i,j}
				            \partial_t\varphi_T^{\sharp}(y^i)
				            \partial_s\varphi_T^{\sharp}(y^j)
                            \otimes \Big(
				            \mbox{\large $\frac{\partial}{\partial y^i}$}
	                        \mbox{\large $\frac{\partial}{\partial y^j}$}				
					       - \nabla^g_{\frac{\partial}{\partial y^i}}
					         \mbox{\large $\frac{\partial}{\partial y^j}$}				
				                \Big)\Phi				   \Big) \vol_h\,,
 \end{eqnarray*}}
 where
 {\small
 \begin{eqnarray*}
   \xi^T _{(\mathrm{II}^2, \partial_s\partial_t\varphi_T)}
    & := &  \sum_{\mu}\Big(
	            d\rho(e_{\mu}) \Tr\big( (\partial_s\partial_t\varphi_T)\Phi \big)
	                                \Big) e_{\mu}  \\
   \xi^T _{(\mathrm{II}^2, \partial_t\varphi_T, \partial_s\varphi_T)}
    & := & \sum_{\mu}\Big(
	        d\rho(e_{\mu}) \Tr				
			 \Big(\sum_{i,j}
				   \partial_t\varphi_T^{\sharp}(y^i)
				   \partial_s\varphi_T^{\sharp}(y^j)
                   \otimes \Big(
				      \mbox{\large $\frac{\partial}{\partial y^i}$}
	                  \mbox{\large $\frac{\partial}{\partial y^j}$}				
					  - \nabla^g_{\frac{\partial}{\partial y^i}}
					      \mbox{\large $\frac{\partial}{\partial y^j}$}				
				                \Big)\Phi				   \Big)				
	                                \Big) e_{\mu}
 \end{eqnarray*}}
 in ${\cal T}_{\ast}(U_T/T)$.
 The integral
  $$
    \int_U \Big(\sum_{\mu} \nabla^h_{e_{\mu}}e_{\mu}
                                 - \sum_{\mu}e_{\mu}d\rho(e_{\mu})		\Big)
						\Tr \big((\partial_s\partial_t\varphi_T)\Phi \big) \vol_h
  $$
  would vanish when imposing the equations of motion of $(\varphi,\nabla)$
  after the combination with other Equations-of-Motion capturing parts from the second variation
  of other terms in
  $S_{\standardscriptsize}^{(\rho,h;\Phi,g,B,C)}(\varphi,\nabla)^{\Bbb C}$.
\end{proposition}

\newpage
{\small
\begin{flushleft}
{\bf\Large Where we are}
\end{flushleft}
The following table summarizes where we are, following the similar steps of fundamental strings
%
		
 \bigskip

 \centerline{
{\footnotesize\it
 \begin{tabular}{|l||l|}\hline
   \hspace{7em}{\bf string theory}\rule{0ex}{1.2em}
       & \hspace{8em}{\bf D-brane theory}\\[.6ex] \hline\hline
     $\begin{array}{l}
	    \mbox{fundamental objects}: \\
 		 \hspace{1em}
		 \mbox{open or closed string}
	   \end{array}$\hspace{.6em}
	   \raisebox{-1.6ex}{\includegraphics[width=0.18\textwidth]{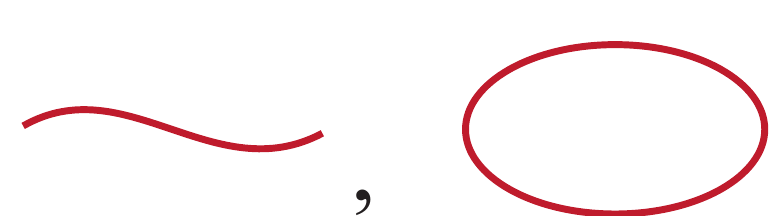}}
       &  $\begin{array}{l}
	           \mbox{fundamental objects}:\rule{0ex}{1.2em}\\
			      \hspace{1em}\mbox{\scriptsize Azumaya/matrix}\\
				  \hspace{3em}\mbox{\scriptsize $(m-1)$-manifold}\\
 				  \hspace{1em}\mbox{\scriptsize with a fundamental module}\\
				  \hspace{3em}\mbox{\scriptsize with a connection}
			   \end{array}$
	           \raisebox{-6ex}{\includegraphics[width=0.25\textwidth]{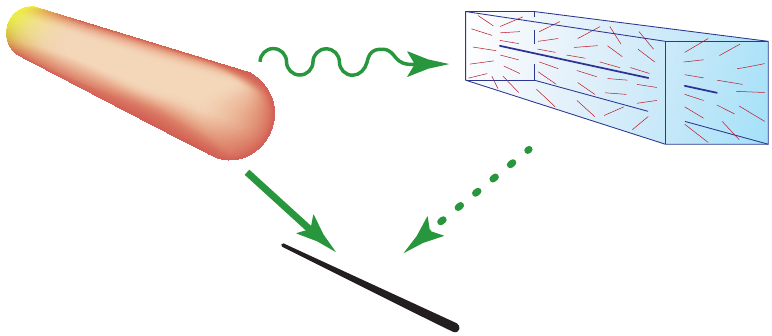}}			   
			   \\[3.8ex] \hline	   	
     $\begin{array}{l}
	    \mbox{string world-sheet}: \\
 		 \hspace{3em}
		 \mbox{$2$-manifold $\,\Sigma$}
	   \end{array}$
       &  $\begin{array}{l}
	           \mbox{D-brane world-volume}:\rule{0ex}{1.2em}\\
			      \hspace{3em}\mbox{Azumaya/matrix $m$-manifold}\\
 				  \hspace{3em}\mbox{with a fundamental module with a connection}\\
				  \hspace{4em}\mbox{$(X^{\!A\!z},{\cal E};\nabla)$}
			   \end{array}$	  				                                                                      \\[3.8ex] \hline
	 $\begin{array}{l}
	    \mbox{string moving in space-time $Y$}:\rule{0ex}{1.2em} \\
 		 \hspace{3em}
		 \mbox{differentiable map $f:\Sigma \rightarrow Y$}
	   \end{array}$
       &  $\begin{array}{l}\rule{0ex}{1.2em}
	           \mbox{D-brane moving in space-time $Y$}:\\
			      \hspace{1em}
	              \mbox{$($admissible$)$ differentiable map
				       $\varphi:(X^{\!A\!z},{\cal E};\nabla)\rightarrow Y$}
			   \end{array}$	  				                                                                             \\[2.4ex] \hline		
       $\begin{array}{c}
	        \mbox{Nambu-Goto action $\,S_{\tinyNG}\,$ for $f$'s}\end{array}$\rule{0ex}{1.2em}
           &  $\begin{array}{c}\rule{0ex}{1.2em}
	                \mbox{Dirac-Born-Infeld action $\,S_{\tinyDBI}\,$ for $(\varphi,\nabla)$'s}\end{array}$
                                                                     	                                                           \\[1ex] \hline
       $\begin{array}{l}\rule{0ex}{1.2em}
	        \mbox{Polyakov action $\,S_{\tinyPolyakov}\,$ for $f$'s}
	       \end{array}$
	        &   
		        $\begin{array}{l}\rule{0ex}{1.2em}
				 \mbox{standard action $\,S_{\mbox{\tiny standard}}\,$
                 				 for $(\varphi,\nabla)$'s}\end{array}$\\[1ex] \hline
       $\begin{array}{l}\rule{0ex}{1.2em}
	        \mbox{action for Ramond-Neveu-Schwarz superstrings}
	       \end{array}$
	       & \hspace{1.2em}
		      $\begin{array}{l}\rule{0ex}{1.2em}
			       \mbox{\rm ???,$\;$ cf.\ [L-Y6: Sec.\ 5.1] (D(11.2))}\end{array}$
			                                                                                                           \\[1ex] \hline		
       $\begin{array}{l}\rule{0ex}{1.2em}
	        \mbox{action for Green-Schwarz superstrings}
	       \end{array}$
	       & \hspace{1.2em}
		       $\begin{array}{l}\rule{0ex}{1.2em}
			     \mbox{\rm ???,$\;$ cf.\ [L-Y6: Sec.\ 5.1] (D(11.2))}\end{array}$\\[1ex] \hline		   
       $\begin{array}{l}\rule{0ex}{1.2em}
	        \mbox{quantization}
	       \end{array}$
	       & \hspace{1.2em}
		      $\begin{array}{l}\rule{0ex}{1.2em}???\end{array}$\\[1ex] \hline		
 \end{tabular}
  }
 }
 
\bigskip

\noindent
(Cf.\ [L-Y8: Remark 3.2.4: second table]) (D(13.1)).
It's by now a history that as the built-in structure of a string is far richer than that for a point,
 a physical theory that takes strings as fundamental objects has brought us to where
 a physical theory that takes only point-particles as fundamental objects cannot reach.
Now that {\it  a D-brane carries even more built-in structures,
 are these even-richer-than-string structures all just in vain?
Or is a physical theory that takes D-branes as fundamental objects going to lead us to somewhere
 beyond that from string theories?}

Besides a theory in its own right, a theory that takes D-branes as fundamental objects
 has deep connection with other themes outside.
In particular, at low dimensions, that there should be the following connections are ``obvious"
 \begin{itemize}
  \item[(0)] $\;\;(m=0)\,\Longrightarrow$
   a new class of matrix models; cf.\ [L-Y8: {\sc Figure} 2-1-2] (D(13.1))
  
  \vspace{-1.2ex}
  \item[(1)] $\;\;(m=1)\,\Longrightarrow$
   nature of non-Abelian Ramond-Ramond fields; cf.\ $e^-$ vs.\ EM field, [Ja]
  
  \vspace{-1.2ex}
  \item[(2)] $\;\;(m=2)\,\Longrightarrow$
   a new Gromov-Witten type theory; cf.\ [L-Y3] (D(10.1)), [L-Y4] (D(10.2))
 \end{itemize}
 but most details to realize these connections remain far from reach at the moment.

%
%
%
%
%
%
%
%
%

%
%
%
%
%
%
%
%
%
%
%
%

\bigskip
\bigskip
\bigskip
\noindent
{\tiny $\bullet$}
{\sl A reflection at the end of the first decade of the D-project}$\,$ since spring 2007:

\smallskip

\centerline{\includegraphics[width=0.60\textwidth]{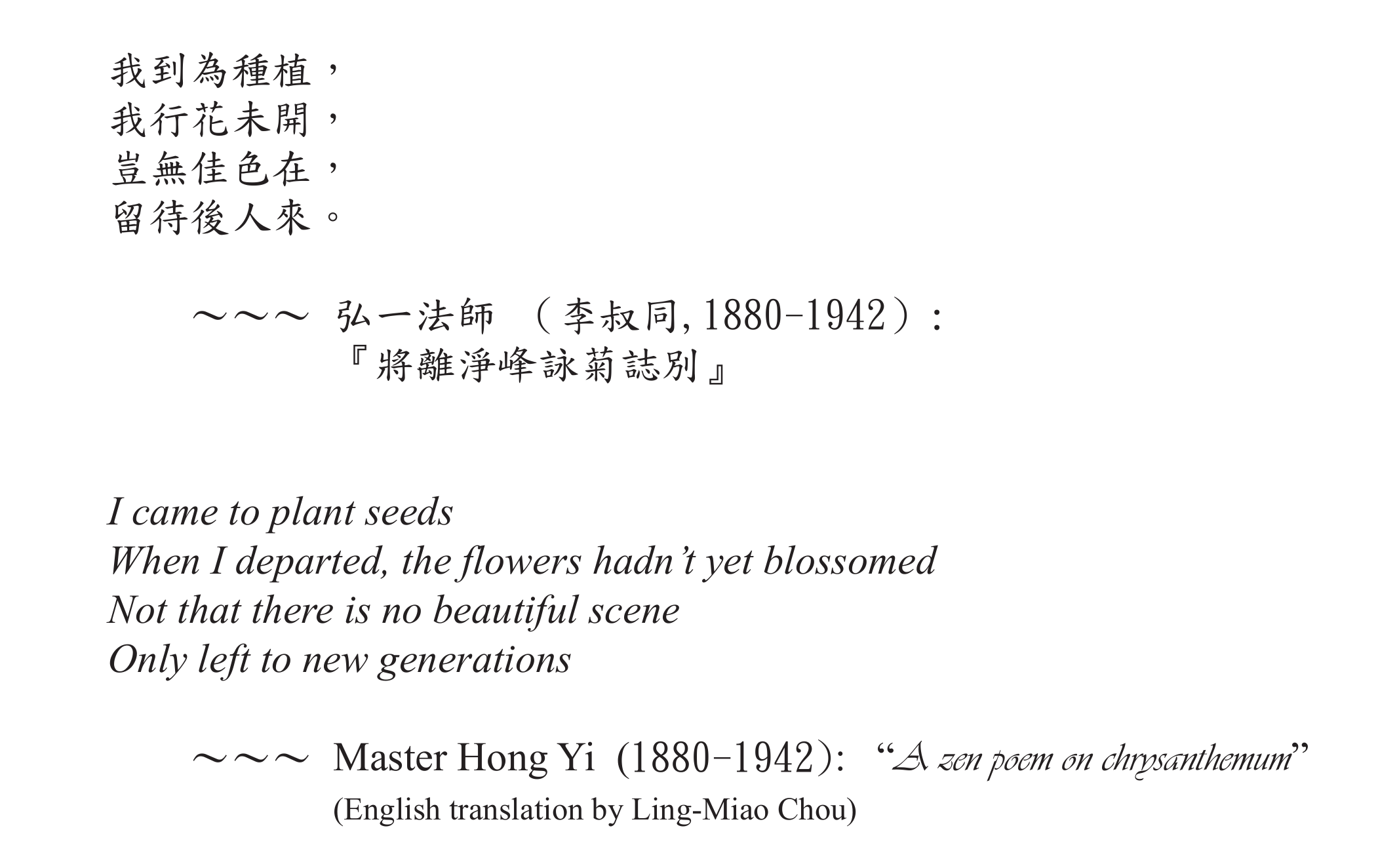}}

} 

\newpage
\baselineskip 13pt
{\footnotesize

\vspace{1em}

\noindent
chienhao.liu@gmail.com, 
chienliu@cmsa.fas.harvard.edu; \\
yau@math.harvard.edu

}

\end{document}